\pdfoutput=1 
\providecommand{\main}{.}
\documentclass[twocolumn]{aastex631}

\newif\ifjournal
\journalfalse

\graphicspath{{\main/figures/}{\main/figures/multiplot/}{\main/figures/imaging/}}
\makeatletter
\def\input@path{{\main/tables/}{\main/figures/}{\main/figures/multiplot/}{\main/figures/imaging/}}
\makeatother

\usepackage{xspace}
\usepackage{amsmath}
\usepackage{mathrsfs}
\usepackage{catchfile}
\usepackage{pgffor}
\usepackage{readarray}
\usepackage{graphicx}
\usepackage{graphbox}
\usepackage{ifthen}
\usepackage[countmax]{subfloat}

\usepackage[caption=false]{subfig}

\DeclareSubrefFormat{parentfig}{#1}
\usepackage{booktabs}
\usepackage{afterpage}
\newenvironment{rotatepage}%
    {\clearpage\pagebreak[4]\global\pdfpageattr\expandafter{\the\pdfpageattr/Rotate 90}}%
    {\clearpage\pagebreak[4]\global\pdfpageattr\expandafter{\the\pdfpageattr/Rotate 0}}%

\makeatletter
\newcommand{\pauselinenumbers}{%
  \@ifclasswith{aastex631}{linenumbers}{\nolinenumbers}{}%
}
\newcommand{\resumelinenumbers}{%
  \@ifclasswith{aastex631}{linenumbers}{\linenumbers}{}%
}
\makeatother

\makeatletter
\renewcommand{\paragraph}{\@startsection{paragraph}{4}{\z@}%
  {0ex}{0pt}{\small\itshape}}
\makeatother

\def\fnum@table{{\bf\tablename~\thetable}}
\def\tablecommentsmod#1{\vskip1pt
\parbox{\textwidth}{
{\small\vskip1sp\indent\vrule height 11pt depth 2pt
width 0pt\currtabletypesize{\sc Note}---{#1}\vskip1pt}}}

\def\tablecommentsmodline#1{\vskip1pt
\parbox{\linewidth}{
{\small\vskip1sp\indent\vrule height 11pt depth 2pt
width 0pt\currtabletypesize{\sc Note}---{#1}\vskip1pt}}}
\def\tablerefsmod#1{\vskip1pt
\parbox{\textwidth}{
{\small\vskip1sp\indent\vrule height 11pt depth 2pt
width 0pt\currtabletypesize{\sc Note}---{#1}\vskip1pt}}}
\def\tablenotetextmod#1#2{\vskip1pt\parbox{\linewidth}{\currtabletypesize\vskip1pt\indent\vrule
height 11pt depth
2pt width0pt\relax$^{\hbox to 5pt{$#1$}}$#2\vskip1pt}}

\newcommand\SPOCdetectedTOIs{2382,4487,5386}
\newcommand\LCOTOIs{2031,2169,2346,2382,2886,3160,3486,3523,3593,3682,3877,4734,5386,5592}
\newcommand\PESTTOIs{2346,2876,3464,3474,4214}

\newcommand\MuscatTOIs{3523,3593,3856,3980,5181,5386}
\newcommand\TRAPPISTNorthTOIs{2886,3523,3856,5592}
\newcommand\TRAPPISTSouthTOIs{3464,4961}
\newcommand\GeminiNorthTOIs{2031,3980}
\newcommand\GeminiSouthTOIs{3160,4734}
\newcommand\SAISpeckleTOIs{2169,3523,3682,3856,3877,3980,4487,4734,5181,5210,5322,5340,5592}
\newcommand\ShaneTOIs{2169,2986,3877,4214}
\newcommand\SOARTOIs{2169,2346,2382,2876,2886,2986,2992,3135,3160,3464,3474,3486,3682,4214,4794,4961,5181,5210}
\newcommand\NESSITOIs{3682,3856,3877,4487,4734,5181,5210,5322,5340,5386}
\newcommand\PalomarTOIs{2169,3523,3593,3682,3856,5181,5386}
\newcommand\TRESTOIs{2031,2169,2346,2382,2876,2886,2986,3523,3593,3682,3856,3877,3980,4214,4734,4794,5181,5210,5322,5340,5386,5592}
\newcommand\TRESorbitTOIs{2031,2886,5340}
\newcommand\HIRESTOIs{2031,2346,3523,3593,3856,3877,4487,4734,5181,5210,5322,5386,5592}
\newcommand\NEIDTOIs{2169,2986,3682,3980,4734,5340,5386}
\newcommand\NEIDoffsetTOIs{2169,2986}
\newcommand\CHIRONTOIs{2382,2876,2992,3135,3464,3474,3486,4214,4794,4961}
\newcommand\CHIRONreconTOIs{2876,3486,4961}
\newcommand\PFSTOIs{2876,3160,3486,4214,4961}
\newcommand\eccentricTOIs{3593,4961}
\newcommand\grazingTOIs{3980,5592}
\newcommand\fitslopeTOIs{2346,2886,3464,3474,3593,3856,3877,4214,4734}

\newcommand{\Kepler}{\textit{Kepler}\xspace}
\newcommand{\TESS}{TESS\xspace}
\newcommand{\Gaia}{\textit{Gaia}\xspace}

\newcommand{\Exofast}{\texttt{EXOFASTv2}\space}

\newcommand{\Rstar}{\ensuremath{R_{\star}}\xspace} 
\newcommand{\Mstar}{\ensuremath{M_{\star}}\xspace}
\newcommand{\Rjup}{\ensuremath{R_\mathrm{J}}\xspace} 
\newcommand{\Mjup}{\ensuremath{M_\mathrm{J}}\xspace}
\newcommand{\Rearth}{\ensuremath{R_\oplus}\xspace} 

\newcommand{\Rsun}{\ensuremath{R_\odot}\xspace} 
\newcommand{\Msun}{\ensuremath{M_\odot}\xspace}
\newcommand{\Rp}{\ensuremath{R_p}\xspace}
\newcommand{\Mp}{\ensuremath{M_p}\xspace}
\newcommand{\rhop}{\ensuremath{\rho_p}\xspace}
\newcommand{\Teff}{\ensuremath{T_\mathrm{eff}}\xspace}
\newcommand{\logg}{\ensuremath{\log{g}}\xspace}
\newcommand{\feh}{\ensuremath{\mathrm{[Fe/H]}}\xspace}
\newcommand{\vsini}{\ensuremath{v\sin{i}}\xspace}
\newcommand{\vmac}{\ensuremath{v_\mathrm{mac}}\xspace}

\newcommand{\tcirc}{\ensuremath{\tau_{\mathrm{circ}}\xspace}}

\newcommand{\ms}{\ensuremath{\mathrm{m}\,\mathrm{s}^{-1}}\xspace}
\newcommand{\msday}{\ensuremath{\mathrm{m}\,\mathrm{s}^{-1}\,\mathrm{day}^{-1}}\xspace}
\newcommand{\kms}{\ensuremath{\mathrm{km}\,\mathrm{s}^{-1}}\xspace}
\newcommand{\gcc}{\ensuremath{\mathrm{g}\,\mathrm{cm}^{-3}}\xspace}

\defcitealias{Yee2022}{I}
\defcitealias{Yee2023}{II}

\newcommand*{\ntois}{}
\foreach \toi [count=\nt] in 2031
2169
2346
2382
2876
2886
2986
2992
3135
3160
3464
3474
3486
3523
3593
3682
3856
3877
3980
4214
4487
4734
4794
4961
5181
5210
5322
5340
5386
5592%
 {\xdef\ntois{\nt}}

\makeatletter
\newcommand{\countTOIs}[1]{%
  \count@=0
  \@for\@element:=#1\do{\advance\count@ by 1}
  \the\count@
}%
\makeatother

\newcounter{arraycount}

\newcommand{\expandTOIs}[1][]{%
\def\toiSuffix{{#1}}%
\expandTOIsRelay%
}

\newcommand{\expandTOIsRelay}[2][TOI-]{%
 \setcounter{arraycount}{0}%
 \foreach \x in #2{%
   \stepcounter{arraycount}%
 }%
  \foreach \num [count=\ntoi] in #2{%
    \if\relax\detokenize{\toiSuffix}\relax%
      #1\num%
    \else%
      #1\num\toiSuffix%
    \fi%
    \ifnum\ntoi<\numexpr\value{arraycount}-1 , %
    \else
      \ifnum\ntoi=\numexpr\value{arraycount}-1\xspace and \fi%
    \fi%
  }%
}

\newcount\numtois\numtois=\number\ntois
\newcount\widetablenumcols\widetablenumcols=5
\divide\numtois by \widetablenumcols
\advance\numtois-1 
\newcommand{\numTables}{\number\the\numtois}
\newcommand{\lastTableNum}{\numexpr\numTables+1}

\usepackage{subfiles}
\usepackage{xr}
\begin{document}
\title{The TESS Grand Unified Hot Jupiter Survey.\ III. Thirty More Giant Planets.
\protect{\footnote{This paper includes data gathered with the 6.5 meter Magellan Telescopes located at Las Campanas Observatory, Chile}}}
\shorttitle{Thirty TESS Giant Planets}

\author[0000-0001-7961-3907]{Samuel W.\ Yee}
\altaffiliation{51 Pegasi b Fellow}
\affiliation{Center for Astrophysics \textbar \ Harvard \& Smithsonian, 60 Garden St, Cambridge, MA 02138, USA}
\affiliation{Department of Astrophysical Sciences, Princeton University, 4 Ivy Lane, Princeton, NJ 08544, USA}
\author[0000-0002-4265-047X]{Joshua N.\ Winn}
\affiliation{Department of Astrophysical Sciences, Princeton University, 4 Ivy Lane, Princeton, NJ 08544, USA}
\author[0000-0001-8732-6166]{Joel D.\ Hartman}
\affiliation{Department of Astrophysical Sciences, Princeton University, 4 Ivy Lane, Princeton, NJ 08544, USA}
\author[0000-0001-8812-0565]{Joseph~E.~Rodriguez}
\affiliation{Center for Data Intensive and Time Domain Astronomy, Department of Physics and Astronomy, Michigan State University, East Lansing, MI 48824, USA}
\author[0000-0002-4891-3517]{George Zhou}
\affiliation{University of Southern Queensland, Centre for Astrophysics, West Street, Toowoomba, QLD 4350, Australia}
\author[0000-0001-9911-7388]{David~W.~Latham}
\affiliation{Center for Astrophysics \textbar \ Harvard \& Smithsonian, 60 Garden St, Cambridge, MA 02138, USA}
\author[0000-0002-8964-8377]{Samuel~N.~Quinn}
\affiliation{Center for Astrophysics \textbar \ Harvard \& Smithsonian, 60 Garden St, Cambridge, MA 02138, USA}
\author[0000-0001-6637-5401]{Allyson~Bieryla}
\affiliation{Center for Astrophysics \textbar \ Harvard \& Smithsonian, 60 Garden St, Cambridge, MA 02138, USA}
\author[0000-0001-6588-9574]{Karen~A.~Collins}
\affiliation{Center for Astrophysics \textbar \ Harvard \& Smithsonian, 60 Garden St, Cambridge, MA 02138, USA}
\author[0000-0003-3773-5142]{Jason~D.~Eastman}
\affiliation{Center for Astrophysics \textbar \ Harvard \& Smithsonian, 60 Garden St, Cambridge, MA 02138, USA}
% ...


\begin{abstract}
We present the discovery of \number\ntois\xspace transiting
giant planets that were initially detected using data from NASA's Transiting Exoplanet Survey Satellite (TESS) mission.
These new planets orbit relatively bright ($G \leq 12.5$) FGK host stars with orbital
periods between 1.6 and 8.2 days, and have radii between 0.9 and 1.7 Jupiter radii.
We performed follow-up ground-based photometry, high angular-resolution imaging, high-
resolution spectroscopy and radial velocity monitoring for each of these objects to confirm that they are planets
and determine their masses and other system parameters.
The planets' masses span more than an order of magnitude ($0.17\,\Mjup < \Mp < 3.3\,\Mjup$).
For two planets, 
\expandTOIs[\,b]{\eccentricTOIs}, we measured significant non-zero eccentricities of $0.11^{+0.05}_{-0.03}$ and $0.18^{+0.04}_{-0.05}$ respectively,
while for the other planets, the data typically provide a 1-$\sigma$ upper bound of 0.15 on the eccentricity.
These discoveries represent a major step toward assembling
a complete, magnitude-limited sample of transiting hot Jupiters around FGK stars.
\end{abstract}

\section{Introduction} \label{sec:intro}

Over the past seven years, NASA's Transiting Exoplanet Survey Satellite (TESS)
mission \citep{TESS_Ricker15} has discovered more than 7{,}000 transiting
planet candidates (known as TESS Objects of Interest, or TOIs).\footnote{\url{https://tess.mit.edu/toi-releases/}}
A significant fraction of the TOIs are
candidate giant planets on short-period orbits.
Such planets are always over-represented in transit surveys because of the strong selection bias favoring large planets and short periods. In fact, these ``hot Jupiters'' are a relatively
rare outcome of planet formation, occurring around $\lesssim 1\%$ of Sun-like
stars, and their origin is still unclear \citep[e.g.,][]{Dawson2018,Fortney2021}.

As an all-sky survey capable of detecting transiting hot Jupiters around
millions of stars, TESS presents the opportunity to deepen our understanding of these planets
at a demographic level -- for example, to study their occurrence rates as a
function of stellar mass \citep[e.g.,][]{Zhou2019a,Beleznay2022,Gan2022,Bryant2023}.
By combining the discoveries of ground-based transit surveys such as WASP
\citep{WASP_Pollacco2006}, HAT and HAT-South \citep{HAT_Bakos2004,HATS_Bakos2013}, KELT
\citep{KELT_Pepper2007,KELT_Pepper2012} and NGTS \citep{NGTS_Wheatley2018} with new
discoveries, and by providing a relatively homogeneous and high-quality dataset from which to quantify the selection function, TESS is enabling the construction of a large and statistically well understood 
sample of hot Jupiters \citep{Yee2021b}.

The planet candidates discovered by TESS need to survive additional observational tests for them
to be considered bona fide planets. Because of the relatively coarse angular resolution of the TESS cameras,
transits that appear to arise from planet-sized companions may in fact be due
to an unresolved blend of a bright star and a fainter eclipsing binary star,
causing the fractional amplitude of the eclipse signal to be ``diluted'' to planet-like proportions. Furthermore, the transit signal allows the candidate's radius to be calculated, but does not specify the mass.
Objects with Jupiter-like radii can have
masses ranging from below that
of Saturn to brown
dwarfs and low-mass stars; therefore, measuring the mass of the transiting companion
through radial velocity monitoring is essential to ensure a clean sample of
planetary-mass objects. 

We and several other groups have been observing hundreds of hot Jupiter
candidates from TESS in order to confirm and characterize them 
\citep[e.g.,][]{Rodriguez2019,Rodriguez2021,Rodriguez2023,Nielsen2019,Brahm2020,Psaridi2022,Psaridi2023,Knudstrup2022,Schulte2024,Ehrhardt2024}.
Much of this activity is coordinated by the \TESS Follow-up Observing
Program (TFOP; \citealt{TFOP_Collins2018}), which helps to organize
observations and avoid duplication of effort, with observations and observing
notes uploaded to the ExoFOP webpage \citep{NEA_ExoFoP_Christiansen2025}.\footnote{\url{https://exofop.ipac.caltech.edu/tess/}\label{footnote:exofop_url}}

Our contribution to this effort, the TESS Grand Unified Hot Jupiter Survey, is directed at
assembling a magnitude-limited ($G \leq 12.5)$ sample of transiting hot Jupiters
($P \leq 10$~days, $8\,\Rearth < \Rp < 24\,\Rearth$) orbiting FGK stars
($4500\,\mathrm{K} \leq \Teff \leq 6500\,\mathrm{K}$).
Paper \citetalias{Yee2022} in this
series presented 10 new planets \citep{Yee2022},
and Paper \citetalias{Yee2023} presented
20 new planets \citep{Yee2023}.
Here, we describe an additional 30 planets.
Their properties are summarized in Table \ref{sf:target_summary}.

\startlongtable
\begin{deluxetable*}{cccccccc}
\tablecaption{Summary of New Planetary Systems \label{tab:target_summary}}
\addtocounter{table}{1}
\pdfbookmark[2]{Table \thetable: Summary of New Planetary Systems}{target_summary}%
\addtocounter{table}{-1}
\tablehead{
    \colhead{TOI} & \colhead{TIC} & \colhead{$G$} & \colhead{Stellar \Teff} & \colhead{Stellar Radius} & \colhead{Orbital Period\tablenotemark{a}} & \colhead{Planet Radius} & \colhead{Planet Mass} \\
    & & \colhead{(mag)} & \colhead{(K)} & \colhead{(\Rsun)} & \colhead{(days)} & \colhead{(\Rjup)} & \colhead{(\Mjup)}
}
\startdata
TOI-2031A\,b & 470127886 & 11.11 & $6490^{+170}_{-130}$ & $1.241 \pm 0.022$ & $5.71548654(76)$ & $1.267 \pm 0.024$ & $0.80 \pm 0.23$ \\
TOI-2169A\,b & 8516795 & 10.96 & $6270 \pm 130$ & $1.820^{+0.047}_{-0.046}$ & $8.2148280(80)$ & $1.116^{+0.037}_{-0.036}$ & $1.037^{+0.077}_{-0.076}$ \\
TOI-2346\,b & 317483660 & 11.92 & $6480^{+200}_{-180}$ & $1.380^{+0.027}_{-0.026}$ & $3.3312029(24)$ & $1.438^{+0.043}_{-0.041}$ & $1.24 \pm 0.22$ \\
TOI-2382\,b/NGTS-37\,b & 2760219 & 11.85 & $6033^{+81}_{-83}$ & $1.417^{+0.041}_{-0.038}$ & $4.6612061(32)$ & $1.093^{+0.076}_{-0.051}$ & $0.74 \pm 0.25$ \\
TOI-2876\,b & 10827386 & 12.25 & $5270^{+110}_{-100}$ & $0.873^{+0.021}_{-0.020}$ & $6.2996430(70)$ & $0.860^{+0.030}_{-0.028}$ & $0.170 \pm 0.052$ \\
TOI-2886\,b & 318796593 & 12.19 & $6240 \pm 150$ & $1.241^{+0.024}_{-0.022}$ & $1.60200105(44)$ & $1.663^{+0.041}_{-0.035}$ & $1.40 \pm 0.23$ \\
TOI-2986\,b & 148497855 & 12.47 & $5906^{+66}_{-69}$ & $1.458^{+0.033}_{-0.031}$ & $3.2783662(44)$ & $0.829^{+0.023}_{-0.021}$ & $0.304^{+0.067}_{-0.086}$ \\
TOI-2992\,b & 49045066 & 12.38 & $6000^{+100}_{-110}$ & $1.545^{+0.046}_{-0.047}$ & $3.0078120(19)$ & $1.227 \pm 0.045$ & $3.33 \pm 0.39$ \\
TOI-3135\,b & 448098793 & 11.58 & $5890^{+150}_{-140}$ & $0.985 \pm 0.019$ & $3.7065585(13)$ & $1.220 \pm 0.030$ & $1.03^{+0.30}_{-0.27}$ \\
TOI-3160A\,b & 440872576 & 12.37 & $6190 \pm 170$ & $1.346^{+0.042}_{-0.041}$ & $3.9712810(25)$ & $1.228^{+0.046}_{-0.045}$ & $1.19^{+0.16}_{-0.15}$ \\
TOI-3464\,b & 290504044 & 12.50 & $6370^{+290}_{-250}$ & $1.560^{+0.046}_{-0.045}$ & $3.6515570(60)$ & $1.228^{+0.057}_{-0.053}$ & $1.25^{+0.27}_{-0.26}$ \\
TOI-3474\,b & 274367763 & 12.43 & $6020^{+170}_{-160}$ & $1.256 \pm 0.025$ & $3.8795105(45)$ & $1.309^{+0.033}_{-0.032}$ & $0.61^{+0.14}_{-0.15}$ \\
TOI-3486\,b & 221861843 & 12.11 & $4930^{+120}_{-110}$ & $0.788 \pm 0.017$ & $2.21778105(88)$ & $0.943^{+0.030}_{-0.029}$ & $0.312^{+0.057}_{-0.056}$ \\
TOI-3523A\,b & 417047499 & 12.50 & $6450 \pm 110$ & $1.448 \pm 0.035$ & $2.30458952(82)$ & $1.425^{+0.037}_{-0.035}$ & $1.37^{+0.12}_{-0.11}$ \\
TOI-3593\,b & 162289289 & 12.03 & $5550^{+100}_{-120}$ & $0.918^{+0.024}_{-0.022}$ & $3.8212867(16)$ & $0.984^{+0.028}_{-0.026}$ & $1.78^{+0.09}_{-0.13}$ \\
TOI-3682\,b & 28961316 & 11.87 & $5700 \pm 100$ & $1.735 \pm 0.045$ & $3.3462406(23)$ & $1.142^{+0.036}_{-0.035}$ & $0.218^{+0.075}_{-0.080}$ \\
TOI-3856\,b & 125552076 & 12.12 & $5622 \pm 50$ & $1.055 \pm 0.017$ & $2.04360352(94)$ & $1.213^{+0.025}_{-0.024}$ & $0.54 \pm 0.10$ \\
TOI-3877\,b & 144310492 & 12.36 & $5731^{+82}_{-81}$ & $1.445^{+0.043}_{-0.042}$ & $4.1235960(67)$ & $1.042^{+0.044}_{-0.043}$ & $0.314^{+0.069}_{-0.067}$ \\
TOI-3980\,b & 312548829 & 12.36 & $5840 \pm 140$ & $1.626^{+0.055}_{-0.050}$ & $3.6089250(30)$ & $1.29^{+0.20}_{-0.10}$ & $0.59^{+0.19}_{-0.17}$ \\
TOI-4214\,b & 409594381 & 11.45 & $6370^{+220}_{-210}$ & $1.434^{+0.040}_{-0.038}$ & $3.4913885(32)$ & $0.953^{+0.032}_{-0.030}$ & $0.520^{+0.053}_{-0.054}$ \\
TOI-4487A\,b & 193754373 & 11.92 & $6206^{+90}_{-88}$ & $1.634 \pm 0.046$ & $3.9540709(21)$ & $1.303^{+0.069}_{-0.064}$ & $1.25 \pm 0.16$ \\
TOI-4734\,b & 97714451 & 12.37 & $6070^{+160}_{-150}$ & $1.857^{+0.038}_{-0.036}$ & $6.235633(11)$ & $0.974^{+0.025}_{-0.023}$ & $0.194^{+0.033}_{-0.032}$ \\
TOI-4794\,b & 95589845 & 12.44 & $6390^{+160}_{-150}$ & $1.451^{+0.036}_{-0.030}$ & $3.5658116(63)$ & $1.281^{+0.044}_{-0.039}$ & $0.97^{+0.24}_{-0.23}$ \\
TOI-4961\,b & 420202798 & 12.26 & $5380^{+180}_{-170}$ & $0.912^{+0.032}_{-0.030}$ & $7.4791820(82)$ & $1.101^{+0.041}_{-0.039}$ & $1.80^{+0.21}_{-0.23}$ \\
TOI-5181A\,b & 346667887 & 12.21 & $6060^{+120}_{-140}$ & $1.530 \pm 0.041$ & $3.8922295(23)$ & $1.191^{+0.040}_{-0.038}$ & $1.04 \pm 0.17$ \\
TOI-5210\,b & 265979849 & 11.93 & $5720^{+100}_{-120}$ & $1.209^{+0.032}_{-0.026}$ & $4.5661061(24)$ & $0.957^{+0.032}_{-0.024}$ & $0.258 \pm 0.041$ \\
TOI-5322\,b & 336691874 & 12.47 & $5692^{+87}_{-83}$ & $1.462^{+0.028}_{-0.025}$ & $5.4233207(39)$ & $1.482^{+0.031}_{-0.027}$ & $0.579 \pm 0.081$ \\
TOI-5340\,b & 14156936 & 12.20 & $5990^{+130}_{-150}$ & $1.842^{+0.055}_{-0.054}$ & $4.9394392(72)$ & $1.345^{+0.051}_{-0.049}$ & $0.85 \pm 0.25$ \\
TOI-5386A\,b & 202425357 & 11.76 & $6094^{+67}_{-65}$ & $1.245 \pm 0.022$ & $3.62156552(83)$ & $1.312^{+0.029}_{-0.028}$ & $0.488^{+0.047}_{-0.045}$ \\
TOI-5592\,b & 88529975 & 12.49 & $6240 \pm 100$ & $1.393 \pm 0.038$ & $2.6085846(18)$ & $1.53^{+0.32}_{-0.16}$ & $0.899^{+0.070}_{-0.069}$%

\enddata
\tablenotetext{a}{The numbers in parentheses represent the uncertainty in the last two decimal digits.}
\tablecomments{This table summarizes the key stellar and planetary properties for the new hot Jupiter systems described in this paper. These parameters are derived from our global modelling of each system, described in \S\ref{M-sec:planet_char}.
The full set of planet and stellar parameters from these global fits can be found in Table \ref{tab:fitted_props}.}
\end{deluxetable*}
 \label{sf:target_summary}%

\section{Observations and Data} \label{sec:obs}

\subsection{TESS Photometry} \label{ssec:tess}

Each of the planets described here was initially detected by TESS as a periodic series of dips in apparent brightness, resulting from the passages of the planet directly in front of its host star.
All were first observed by TESS in the long-cadence Full Frame Image (FFI) mode,
either at 1800-second cadence during the Prime Mission (PM) or
at 600-second cadence during the Extended Mission (EM1).
Following their identification as planet candidates, some of the
targets were subsequently reobserved by \TESS with 120-second cadence.
For all of the systems analyzed here, the transits of the hot Jupiter
were the only transit signals detected in their respective TESS light curves.

The FFIs were downloaded from \TESS and calibrated using the
\texttt{tica} software \citep{TESS_QLP_Fausnaugh2020}.
Light-curves were then produced by the MIT Quick Look Pipeline (QLP), which
uses a difference-imaging algorithm to produce light curves for a
magnitude-limited set of stars in the TESS FFIs. The QLP is described
more fully in \citet{TESS_QLP_Huang2020a,TESS_QLP_Huang2020b,
TESS_QLP_Fausnaugh2020,TESS_QLP_Kunimoto2021,TESS_QLP_Kunimoto2022b}.
More recently, the TESS Science Processing Operations Center (SPOC) \citep{TESS_SPOC_Jenkins2016},
located at NASA Ames Research Center, has begun producing light curves for a subset
of targets in the FFIs \citep{TESS_SPOC_Caldwell2020}.

Almost all of the transit signals corresponding to the planet candidates
described in this paper were first detected by the ``faint-star'' search of the QLP light curves \citep{TESS_Faint_Kunimoto2022a}. Potential signals
were identified using a box-least squares (BLS) search algorithm
\citep{BLS_Kovacs2002,VARTOOLS_Hartman2016} and were subsequently triaged
and vetted through an automated pipeline before a final round of
manual vetting \citep{AstroNetTriage_Yu2019,TESS_Faint_Kunimoto2022a}. 

Initial detections for \countTOIs{\SPOCdetectedTOIs} of the planet candidates
(\expandTOIs[.01]{\SPOCdetectedTOIs}) were made 
by the transit search pipeline operated by SPOC \citep{TESS_SPOC_Jenkins2016},
which identifies Threshold Crossing Events (TCEs) and performs
a series of diagnostic tests and manual vetting to filter out
false positives \citep{TESS_DV_Twicken2018,TESS_DV_Li2019}.
The planet candidates from the QLP and SPOC searches were vetted
by the \TESS Science Office and announced to the community as
\TESS Objects of Interest (TOIs; \citealt{TESS_TOIs_Guerrero2021}).
We also inspected the Data Validation (DV) reports produced by
the SPOC pipeline, verifying that there were no offsets
between the in-transit and out-of-transit centroid locations,
which could indicate a nearby eclipsing binary false positive scenario.

One object, TOI-2346.01, was first flagged as a ``community TOI''
(cTOI) by \citet{DIAMANTE_Montalto2020}. They performed an independent BLS
transit search on FFI light curves extracted by the DIAmante pipeline
using a difference imaging method. Following additional vetting
from the \TESS Science Office, TOI-2346.01 was later also flagged
as a TOI \citep{TESS_cTOIs_Mireles2021}.

For our analysis, we used the \texttt{lightkurve} Python package
\citep{Lightkurve18} to download all of the available \TESS photometry from the
Mikulski Archive for Space Telescopes (MAST). We chose the
shortest cadence data available for each target in a given sector,
down to 120-second cadence.
When available, we used
the Presearch Data Conditioning (PDC; \citealt{TESS_PDC_Stumpe2012,TESS_PDC_Smith2012,TESS_PDC_Stumpe2014})
light curves produced by the SPOC pipeline.
We subsequently detrended the SPOC light curves by masking out the
transit events and iteratively fitting the out-of-transit continuum to
a basis spline, using the  \texttt{Keplerspline}\footnote{\url{https://github.com/avanderburg/keplersplinev2}} code \citep{Keplerspline_Vanderburg2014,Keplerspline_Shallue2018}.
For the remaining long-cadence data, we relied on the spline-detrended 
\texttt{KSPSAP} light curves from the QLP. Table \ref{tab:tess_summary}
summarizes the TESS data used in our analysis for each target.

\startlongtable
\begin{deluxetable}{cccr}
\tablecolumns{4}
\tablecaption{Summary of TESS Observations \label{tab:tess_summary}}
\addtocounter{table}{1}
\pdfbookmark[3]{Table \thetable: Summary of TESS Observations}{tess_summary}%
\addtocounter{table}{-1}
\tablehead{
    \colhead{Target} & \colhead{Sector} & \colhead{Source\tablenotemark{a}} & \colhead{Cadence (s)}
}
\startdata
TOI-2031A & 18,19,24--26 & SPOC & 1800 \\
$\cdots$ & 52,53,58--60,73,78,79,85 & SPOC & 120 \\
TOI-2169A & 26 & SPOC & 1800 \\
$\cdots$ & 40,53,54,80 & SPOC & 120 \\
TOI-2346 & 6 & SPOC & 1800 \\
$\cdots$ & 33 & SPOC & 600 \\
TOI-2382 & 2 & SPOC & 1800 \\
$\cdots$ & 29 & SPOC & 600 \\
$\cdots$ & 69 & SPOC & 120 \\
TOI-2876 & 7 & SPOC & 1800 \\
$\cdots$ & 33 & QLP & 600 \\
$\cdots$ & 34 & SPOC & 600 \\
TOI-2886 & 7 & QLP & 1800 \\
$\cdots$ & 33,34 & QLP & 600 \\
TOI-2986 & 9 & SPOC & 1800 \\
$\cdots$ & 36 & SPOC & 600 \\
$\cdots$ & 62 & SPOC & 120 \\
TOI-2992 & 9 & SPOC & 1800 \\
$\cdots$ & 36 & SPOC & 600 \\
$\cdots$ & 63 & SPOC & 120 \\
TOI-3135 & 11,12 & QLP & 1800 \\
$\cdots$ & 38 & QLP & 600 \\
$\cdots$ & 65 & SPOC & 120 \\
TOI-3160A & 11 & SPOC & 1800 \\
$\cdots$ & 38 & SPOC & 600 \\
$\cdots$ & 65 & SPOC & 120 \\
TOI-3464 & 12 & QLP & 1800 \\
$\cdots$ & 38 & QLP & 600 \\
$\cdots$ & 65 & SPOC & 120 \\
TOI-3474 & 12 & QLP & 1800 \\
$\cdots$ & 38 & QLP & 600 \\
$\cdots$ & 65 & SPOC & 120 \\
TOI-3486 & 12 & QLP & 1800 \\
$\cdots$ & 39 & QLP & 600 \\
$\cdots$ & 66 & SPOC & 120 \\
TOI-3523A & 14,15 & QLP & 1800 \\
$\cdots$ & 41 & QLP & 600 \\
$\cdots$ & 75,81,82 & SPOC & 120 \\
TOI-3593 & 15,16 & QLP & 1800 \\
$\cdots$ & 56,83 & SPOC & 120 \\
TOI-3682 & 18 & SPOC & 1800 \\
$\cdots$ & 42--44,70,71 & SPOC & 120 \\
TOI-3856 & 21 & SPOC & 1800 \\
$\cdots$ & 48 & SPOC & 120 \\
TOI-3877 & 22 & SPOC & 1800 \\
TOI-3980 & 17,18,24 & SPOC & 1800 \\
$\cdots$ & 58,85 & SPOC & 120 \\
TOI-4214 & 7 & SPOC & 1800 \\
$\cdots$ & 33,34 & SPOC & 600 \\
TOI-4487A & 14,15 & QLP & 1800 \\
$\cdots$ & 41,55,56,75,76,82,83 & SPOC & 120 \\
TOI-4734 & 6 & SPOC & 1800 \\
$\cdots$ & 33,44,45 & SPOC & 600 \\
$\cdots$ & 43 & QLP & 600 \\
$\cdots$ & 72 & SPOC & 120 \\
TOI-4794 & 7 & QLP & 1800 \\
$\cdots$ & 34 & QLP & 600 \\
TOI-4961 & 11 & QLP & 1800 \\
$\cdots$ & 37,38 & QLP & 600 \\
$\cdots$ & 64,65 & SPOC & 120 \\
TOI-5181A & 26 & SPOC & 1800 \\
$\cdots$ & 40,53,54 & SPOC & 600 \\
$\cdots$ & 80 & SPOC & 120 \\
TOI-5210 & 15 & SPOC & 1800 \\
$\cdots$ & 41,55 & SPOC & 600 \\
$\cdots$ & 82 & SPOC & 120 \\
TOI-5322 & 42,43 & SPOC & 600 \\
$\cdots$ & 57,70,84 & SPOC & 120 \\
TOI-5340 & 42--44 & SPOC & 600 \\
$\cdots$ & 71 & SPOC & 120 \\
TOI-5386A & 15,16,22,23 & SPOC & 1800 \\
$\cdots$ & 48--50,75,76 & SPOC & 120 \\
TOI-5592 & 20 & SPOC & 1800 \\
$\cdots$ & 47 & SPOC & 600 \\
$\cdots$ & 60,74 & SPOC & 120%
\enddata
\tablenotetextmod{a}{The source column indicates the High-Level Science Product (HLSP) source of the TESS light-curves used in the analysis.}
\tablecommentsmodline{The raw TESS data are available on MAST, while the flattened and normalized TESS photometry used in our analysis are provided as online supplementary material (Data behind the Figure for Figure \subref*{fig:toi2031_multiplot}).}
\end{deluxetable}
 \label{sf:tess_summary}%

\subsection{NGTS Photometry} \label{ssec:ngts_photometry}%
TOI-2382/NGTS-37 was also observed by the Next Generation Transit Survey~\citep[NGTS,][]{NGTS_Wheatley2018}, based at the ESO Paranal Observatory, Chile. NGTS is an automated array of twelve 20cm optical telescopes, which are fitted with a custom filter spanning $520-890~$nm. Each telescope is independently mounted and covers a field-of-view of 8 square degrees, allowing NGTS to observe up to twelve different fields on a given night and find sufficient comparison stars for even the brightest of its targets. This capability has been used to conduct a wide-field transit survey for the majority of operation time since September 2015. A single telescope is capable of achieving a photometric precision corresponding to a root-mean-squared flux of 400~ppm in 30 minutes for stars brighter than 12th magnitude in the TESS bandpass; this is approximately the level of fluctuations expected from scintillation~\citep{ngtsppm,ngtsscintillation}. 

Survey data are reduced using standard aperture photometry routines and detrended for systematics, then passed through \texttt{ORION} (BLS algorithm), as described in \cite{NGTS_Wheatley2018}. 
Prior to the issuance of the TOI alert for TOI-2382/NGTS-37, NGTS made an indepedendent detection of a transit signal with a period of 4.661285~days and depth 3.6~parts per thousand (ppt), based on survey data acquired between Apr 2019 and Jan 2020.
NGTS returned to observe this target between Dec 2021 and Aug 2022, with a total of ten epochs of observation containing full or partial transits.
We included the detrended and normalized data in our global analysis of this system.

\subsection{Follow-up Ground-Based Photometry} \label{ssec:sg1}

In order to confirm the planet candidates from \TESS, we obtained ground-based follow-up observations from a variety of facilities.
One important component of these observations is seeing-limited time-series photometry spanning the expected times of transits.
Because the ground-based observations have a typical angular resolution of a few arcseconds, much better than that of \TESS (which has a pixel scale of 21$^{\prime\prime}$), these observations can confirm which star is actually undergoing the fading events detected by \TESS. This helps to evaluate false positive scenarios.
The additional transit observations also help to refine the transit ephemeris and thereby facilitate future studies of the system. 
For some of our targets, we observed the transit events in multiple photometric bands to test for chromatic depth variations that are indicative of a stellar companion rather than a planet.

The ground-based time-series photometric observations were coordinated through the TFOP Sub-Group 1 (SG1) for Seeing-limited Photometry \citep{TFOP_Collins2018}, and are summarized in Table \ref{tab:sg1_summary}.
The \texttt{TAPIR} software \citep{TAPIR_Jensen2013} was used for scheduling of the transit observations.
\texttt{AstroImageJ} \citep{AstroImageJ_Collins17} was used to perform data reduction for the majority of the observations. Exposures were flat-fielded and calibrated using bias, dark, and flat frames.
Aperture photometry was performed for the target star and a selection of comparison stars, which were then
used to generate a differential photometry light curve, together with auxiliary PSF shape and position
parameters that were used to detrend the light curve.

The complete set of observations are summarized in Table \ref{tab:sg1_summary}, with full facility names, locations, and references included in the note to the table.
Here, we provide additional details only for the observations which used different data reductions procedures from the standard TFOP SG1 process described above.

\paragraph{LCOGT} \countTOIs{\LCOTOIs} targets were observed using telescopes in the Las Cumbres Observatory Global Telescope network (LCOGT; \citealt{LCOGT_Brown2013}). The data were calibrated using the standard LCOGT \textit{BANZAI} pipeline \citep{LCOGT_BANZAI_McCully2018} and photometric observations extracted using \texttt{AstroImageJ} \citep{AstroImageJ_Collins17}.

\paragraph{MuSCAT2} \expandTOIs{\MuscatTOIs} were observed by the MuSCAT2 multi-color imager \citep{MuSCAT2_Narita2018} on the 1.5 m Telescopio Carlos S\'{a}nchez (TCS) at Teide Observatory in Spain, which is capable of obtaining photometry in up to four bands simultaneously. These data were reduced using the custom pipeline described in \citet{MuSCAT2_Parviainen2020}.

\paragraph{TRAPPIST} We obtained follow-up photometric data for \expandTOIs{\TRAPPISTSouthTOIs} using the robotic TRAPPIST-South telescope at La Silla Observatory in Chile, and for \expandTOIs{\TRAPPISTNorthTOIs} from TRAPPIST-North at Oukaimeden Observatory in Morocco respectively.
Data calibration and photometric extraction for the TRAPPIST observations were performed using the {\tt PROSE} pipeline \citep{TRAPPIST_PROSE_Garcia2022}.

\paragraph{Dragonfly} TOI-3856 was observed by the Dragonfly Telephoto Array \citep{Dragonfly_Danieli2020} at the New Mexico Skies telescope hosting facility. Dragonfly is a remote telescope consisting of an array of small telephoto lenses roughly equivalent to a 1.0 m refractor. Observations were made simultaneously in $g^\prime$ and $r^\prime$ bands and the data were reduced by a custom differential aperture photometry pipeline.

\paragraph{Feder} TOI-2031 was observed from the Feder Observatory at Minnesota State University Moorhead. Images were reduced using the Python package \texttt{ccdproc} \citep{CCDProc_Craig2019} and aperture photometry performed with \texttt{stellarphot} \citep{Stellarphot_Craig2021} following the conventions used in \texttt{AstroImageJ}.

\paragraph{PEST} The Perth Exoplanet Survey Telescope is a private observatory in Perth, Australia. The PEST observations for \expandTOIs{\PESTTOIs} were also reduced using a custom pipeline.\footnote{\url{https://pestobservatory.com/the-pest-pipeline/}}

We used the ground-based photometry in our global modelling of each system (\S\ref{sec:planet_char}). We fitted the planet transit model while simultaneously detrending against the parameters listed in Table \ref{tab:sg1_summary}, using an additive detrending model.
Some of the light curves were excluded from the fit because no transit events were detected due to uncertainty in the transit ephemerides, poor weather, or larger than expected photometric scatter in the light curves.
These excluded light curves are indicated in Table \ref{tab:sg1_summary}.

\begin{rotatepage}
\movetabledown=0.25in
\begin{longrotatetable}
\begin{deluxetable*}{cccccrcrc}
\tablecolumns{9}
\tablecaption{Summary of Ground-Based Photometric Follow-Up Observations \label{tab:sg1_summary}}
\addtocounter{table}{1}
\pdfbookmark[3]{Table \thetable: Summary of Ground-Based Photometric Follow-Up}{sg1_summary}%
\addtocounter{table}{-1}
\tablehead{
    \colhead{Target} & \colhead{Facility/Instrument} & \colhead{Aperture} & \colhead{Filter} &
    \colhead{Date} & \colhead{Cadence} & \colhead{Used in Fit} & \colhead{Precision\tablenotemark{a}} & \colhead{Detrending Vectors} \\
    & & \colhead{(m)} & & \colhead{(UT)} & \colhead{(s)} & & \colhead{(mmag)} &
}
\startdata
TOI-2031A & Feder Observatory & 0.4 & $r^\prime$ & 2020 Jun 27 & 60 & Y & 1.4 & Airmass \\
$\cdots$ & LCO Teide/SBIG-6303 & 0.4 & $g^\prime$ & 2020 Jun 28 & 50 & Y & 1.6 & Total Counts \\
$\cdots$ & LCO Teide/SBIG-6303 & 0.4 & $i^\prime$ & 2020 Jun 28 & 50 & Y & 1.6 & Airmass \\
$\cdots$ & BYU OPO & 0.3 & $R$ & 2020 Jul 15 & 80 & Y & 1.4 & Airmass \\
$\cdots$ & LCO Haleakala/SBIG-6303 & 0.4 & $g^\prime$ & 2020 Aug 01 & 45 & Y & 1.1 & -- \\
$\cdots$ & LCO Haleakala/SBIG-6303 & 0.4 & $i^\prime$ & 2020 Aug 01 & 50 & Y & 1.0 & -- \\
$\cdots$ & Feder Observatory & 0.4 & $r^\prime$ & 2020 Aug 06 & 60 & Y & 1.5 & Airmass, Width \\
$\cdots$ & Feder Observatory & 0.4 & $B$ & 2020 Aug 06 & 60 & Y & 2.8 & Airmass, Width \\
$\cdots$ & CRCAO & 0.61 & $B$ & 2020 Oct 09 & 240 & Y & 1.2 & Airmass, Y (T1) \\
$\cdots$ & CRCAO & 0.61 & $I$ & 2020 Oct 09 & 120 & Y & 0.7 & Airmass \\
$\cdots$ & LCO McDonald/SBIG-6303 & 0.4 & $i^\prime$ & 2023 Jul 06 & 76 & Y & 1.4 & Total Counts \\
TOI-2169A & LCO SAAO/Sinistro & 1.0 & $z^\prime$ & 2020 Aug 13 & 43 & Y & 0.5 & Width, Sky/Pixel \\
$\cdots$ & OPM/RC8 & 0.2 & $I$ & 2020 Aug 13 & 170 & Y & 1.8 & Airmass, Y (T1) \\
$\cdots$ & Waffelow Creek Observatory & 0.36 & $r^\prime$ & 2020 Aug 22 & 60 & Y & 1.0 & Width, BJD$_\mathrm{TDB}$, Airmass \\
$\cdots$ & LCO SSO/Sinistro & 1.0 & $z^\prime$ & 2020 Aug 30 & 43 & Y & 0.5 & Width, BJD$_\mathrm{TDB}$ \\
$\cdots$ & LCO SAAO/Sinistro & 1.0 & $B$ & 2020 Sep 15 & 28 & Y & 0.8 & Width, Total Counts, X (T1) \\
$\cdots$ & LCO Teide/Sinistro & 1.0 & $z^\prime$ & 2024 May 24 & 135 & Y & 0.6 & Airmass \\
$\cdots$ & LCO Teide/Sinistro & 1.0 & $B$ & 2024 May 24 & 135 & Y & 0.6 & Sky/Pixel \\
TOI-2346 & Hazelwood & 0.318 & $R$ & 2020 Dec 13 & 240 & N & -- & -- \\
$\cdots$ & PEST & 0.3 & $R$ & 2021 Dec 08 & 120 & Y & 2.1 & -- \\
$\cdots$ & LCO SAAO/SBIG-6303 & 0.4 & $g^\prime$ & 2021 Dec 11 & 100 & N & -- & -- \\
$\cdots$ & LCO Teide/SBIG-6303 & 0.4 & $g^\prime$ & 2022 Oct 21 & 100 & Y & 1.6 & BJD$_\mathrm{TDB}$ \\
$\cdots$ & SUTO/OTIVAR & 0.3 & $B$ & 2022 Nov 09 & 180 & Y & 3.6 & Meridian Flip, Airmass \\
TOI-2382/NGTS-37 & NGTS & 0.2 & $NGTS$ & 2019 Nov 13 & 13 & Y & 1.1 & -- \\
$\cdots$ & LCO SAAO/Sinistro & 1.0 & $g^\prime$ & 2021 Jul 10 & 26 & Y & 0.3 & Airmass, Y (T1) \\
$\cdots$ & LCO SAAO/Sinistro & 1.0 & $i^\prime$ & 2021 Jul 10 & 19 & Y & 0.3 & Total Counts \\
TOI-2876 & PEST & 0.3 & $R$ & 2022 Jan 02 & 120 & Y & 1.3 & Airmass \\
$\cdots$ & Hazelwood & 0.318 & $g^\prime$ & 2022 Dec 02 & 240 & Y & 1.7 & Airmass, Meridian Flip \\
TOI-2886 & Brierfield & 0.36 & $R$ & 2021 Sep 21 & 120 & Y & 3.5 & Airmass \\
$\cdots$ & LCO Haleakala/SBIG-6303 & 0.4 & $i^\prime$ & 2021 Oct 27 & 110 & Y & 1.5 & Airmass, FWHM \\
$\cdots$ & El Sauce & 0.51 & $B$ & 2021 Dec 09 & 180 & Y & 1.1 & Airmass, Total Counts \\
$\cdots$ & TRAPPIST-North & 0.6 & $z^\prime$ & 2024 Jan 21 & 30 & Y & 0.6 & Width, Meridian Flip \\
TOI-2986 & FLWO/KeplerCam & 1.2 & $i^\prime$ & 2022 Feb 12 & 40 & Y & 1.1 & FWHM \\
TOI-2992 & El Sauce & 0.51 & $R$ & 2023 Jan 16 & 90 & Y & 1.0 & Airmass \\
TOI-3135 & Hazelwood & 0.318 & $R$ & 2021 Jun 29 & 90 & Y & 1.2 & Width, BJD$_\mathrm{TDB}$, Y (T1) \\
$\cdots$ & Brierfield & 0.36 & $B$ & 2022 Apr 03 & 240 & Y & 2.2 & Airmass \\
$\cdots$ & El Sauce & 0.51 & $R$ & 2022 May 03 & 240 & Y & 2.6 & Airmass, Sky/Pixel \\
TOI-3160A & Brierfield & 0.36 & $R$ & 2022 Jun 08 & 240 & Y & 2.5 & Airmass, Sky/Pixel \\
$\cdots$ & Brierfield & 0.36 & $B$ & 2022 Jun 22 & 240 & Y & 2.0 & Airmass \\
$\cdots$ & LCO SAAO/Sinistro & 1.0 & $g^\prime$ & 2023 Jun 06 & 140 & Y & 0.7 & BJD$_\mathrm{TDB}$, Width \\
$\cdots$ & LCO SAAO/Sinistro & 1.0 & $i^\prime$ & 2023 Jun 06 & 140 & Y & 0.7 & BJD$_\mathrm{TDB}$, Total Counts \\
TOI-3464 & TRAPPIST-South & 0.6 & $z^\prime$ & 2021 Jul 29 & 35 & N & -- & -- \\
$\cdots$ & LCO CTIO/Sinistro & 1.0 & $i^\prime$ & 2021 Jul 30 & 37 & N & -- & -- \\
$\cdots$ & PEST & 0.3 & $r^\prime$ & 2022 May 29 & 120 & Y & 1.5 & SKY \\
TOI-3474 & El Sauce & 0.51 & $R$ & 2022 Jun 08 & 60 & Y & 0.9 & Sky/Pixel, Y (T1) \\
$\cdots$ & PEST & 0.3 & $g^\prime$ & 2022 Jul 01 & 120 & Y & 1.6 & Airmass \\
$\cdots$ & PEST & 0.3 & $i^\prime$ & 2022 Jul 01 & 120 & Y & 2.3 & Airmass \\
TOI-3486 & Hazelwood & 0.318 & $R$ & 2021 Jun 21 & 120 & Y & 1.0 & BJD$_\mathrm{TDB}$, Meridian Flip, Sky/Pixel \\
$\cdots$ & LCO SSO/SBIG-6303 & 0.4 & $g^\prime$ & 2021 Jul 02 & 130 & Y & 1.5 & Airmass, Width \\
TOI-3523A & TRAPPIST-North & 0.6 & $z^\prime$ & 2021 Jul 07 & 60 & Y & 0.8 & Airmass, Sky/Pixel \\
$\cdots$ & Phillips Academy Observatory & 0.5 & $R$ & 2022 Apr 30 & 240 & Y & 1.1 & Airmass \\
$\cdots$ & SUTO/OTIVAR & 0.3 & $B$ & 2022 Jun 13 & 300 & Y & 4.6 & Airmass \\
$\cdots$ & T100 & 1.0 & $r^\prime$ & 2022 Jul 13 & 160 & N & -- & -- \\
$\cdots$ & LCO Teide/SBIG-6303 & 0.4 & $g^\prime$ & 2022 Aug 05 & 150 & Y & 1.7 & Airmass \\
$\cdots$ & LCO Teide/SBIG-6303 & 0.4 & $i^\prime$ & 2022 Aug 05 & 150 & Y & 1.6 & Y (T1) \\
$\cdots$ & TCS/MuSCAT2 & 1.52 & $g^\prime$ & 2024 Sep 22 & 11 & Y & 0.4 & Airmass, Sky Entropy, X Shift, Y Shift \\
$\cdots$ & TCS/MuSCAT2 & 1.52 & $r^\prime$ & 2024 Sep 22 & 11 & Y & 0.5 & Airmass, Sky Entropy, X Shift, Y Shift \\
$\cdots$ & TCS/MuSCAT2 & 1.52 & $i^\prime$ & 2024 Sep 22 & 11 & Y & 0.5 & Airmass, Sky Entropy, X Shift, Y Shift \\
$\cdots$ & TCS/MuSCAT2 & 1.52 & $z^\prime$ & 2024 Sep 22 & 11 & Y & 0.6 & Airmass, Sky Entropy, X Shift, Y Shift \\
TOI-3593 & TCS/MuSCAT2 & 1.52 & $g^\prime$ & 2021 Sep 09 & 11 & N & -- & -- \\
$\cdots$ & TCS/MuSCAT2 & 1.52 & $i^\prime$ & 2021 Sep 09 & 16 & N & -- & -- \\
$\cdots$ & TCS/MuSCAT2 & 1.52 & $z^\prime$ & 2021 Sep 09 & 11 & N & -- & -- \\
$\cdots$ & OAUV/TURIA2 & 0.3 & $R$ & 2021 Sep 28 & 75 & Y & 2.6 & X (T1), Total Counts \\
$\cdots$ & GMU/SBIG-16803 & 0.8 & $R$ & 2021 Nov 05 & 65 & Y & 1.7 & Airmass, Total Counts \\
$\cdots$ & LCO Teide/Sinistro & 1.0 & $g^\prime$ & 2022 Aug 11 & 42 & Y & 0.4 & -- \\
$\cdots$ & LCO Teide/Sinistro & 1.0 & $i^\prime$ & 2022 Aug 11 & 22 & Y & 0.3 & -- \\
$\cdots$ & Wendelstein Observatory & 0.43 & $r^\prime$ & 2022 Sep 03 & 150 & Y & 0.9 & Airmass \\
TOI-3682 & FLWO/KeplerCam & 1.2 & $i^\prime$ & 2022 Sep 17 & 30 & Y & 0.5 & Airmass \\
$\cdots$ & LCO Haleakala/SBIG-6303 & 0.4 & $g^\prime$ & 2022 Dec 03 & 80 & Y & 0.9 & Airmass \\
TOI-3856 & TRAPPIST-North & 0.6 & $z^\prime$ & 2022 Feb 02 & 20 & Y & 0.7 & Airmass, Total Counts \\
$\cdots$ & Whitin/CDK700 & 0.7 & $R$ & 2022 Feb 11 & 25 & Y & 0.7 & Airmass, Sky/Pixel \\
$\cdots$ & Adams Observatory & 0.61 & $I$ & 2022 Feb 13 & 180 & Y & 1.0 & Airmass \\
$\cdots$ & Dragonfly Telephoto Array & - & $g^\prime$ & 2022 Feb 15 & 64 & Y & 1.8 &  FWHM,  SKY,  AIRMASS \\
$\cdots$ & Dragonfly Telephoto Array & - & $r^\prime$ & 2022 Feb 15 & 64 & Y & 1.5 &  FWHM,  SKY,  AIRMASS \\
$\cdots$ & TCS/MuSCAT2 & 1.52 & $g^\prime$ & 2022 May 03 & 16 & Y & 0.4 & Airmass, Sky Entropy, X Shift, Y Shift \\
$\cdots$ & TCS/MuSCAT2 & 1.52 & $r^\prime$ & 2022 May 03 & 11 & Y & 0.4 & Airmass, Sky Entropy, X Shift, Y Shift \\
$\cdots$ & TCS/MuSCAT2 & 1.52 & $i^\prime$ & 2022 May 03 & 6 & Y & 0.4 & Airmass, Sky Entropy, X Shift, Y Shift \\
$\cdots$ & TCS/MuSCAT2 & 1.52 & $z^\prime$ & 2022 May 03 & 6 & Y & 0.5 & Airmass, Sky Entropy, X Shift, Y Shift \\
TOI-3877 & GMU/SBIG-16803 & 0.8 & $R$ & 2022 Feb 10 & 65 & N & -- & Airmass, Width, Y (T1) \\
$\cdots$ & Waffelow Creek Observatory & 0.36 & $r^\prime$ & 2022 Feb 11 & 90 & N & -- & -- \\
$\cdots$ & FLWO/KeplerCam & 1.2 & $i^\prime$ & 2022 Feb 11 & 38 & N & -- & -- \\
$\cdots$ & LCO CTIO/Sinistro & 1.0 & $i^\prime$ & 2023 Feb 21 & 60 & N & -- & -- \\
$\cdots$ & LCO Teide/Sinistro & 1.0 & $i^\prime$ & 2023 Mar 01 & 30 & Y & 0.4 & Width \\
$\cdots$ & LCO Teide/Sinistro & 1.0 & $i^\prime$ & 2023 Mar 05 & 60 & Y & 2.2 & Airmass, Width, Total Counts \\
$\cdots$ & LCO Teide/Sinistro & 1.0 & $g^\prime$ & 2023 Apr 03 & 40 & Y & 1.5 & Width \\
$\cdots$ & CMO/RC600 & 0.6 & $R$ & 2023 Apr 03 & 36 & Y & 1.1 & Airmass \\
TOI-3980 & TCS/MuSCAT2 & 1.52 & $g^\prime$ & 2021 Aug 29 & 10 & Y & 0.4 & Airmass, Sky Entropy, X Shift, Y Shift \\
$\cdots$ & TCS/MuSCAT2 & 1.52 & $i^\prime$ & 2021 Aug 29 & 10 & Y & 0.5 & Airmass, Sky Entropy, X Shift, Y Shift \\
$\cdots$ & TCS/MuSCAT2 & 1.52 & $z^\prime$ & 2021 Aug 29 & 10 & Y & 0.4 & Airmass, Sky Entropy, X Shift, Y Shift \\
TOI-4214 & GMU/SBIG-16803 & 0.8 & $R$ & 2021 Dec 13 & 60 & Y & 4.1 & Total Counts, Width, Y (T1) \\
$\cdots$ & PEST & 0.3 & $R$ & 2021 Dec 24 & 120 & Y & 1.0 & Airmass, SKY \\
$\cdots$ & El Sauce & 0.51 & $B$ & 2023 Jan 23 & 180 & Y & 1.0 & Airmass \\
TOI-4487A & OAUV/TURIA1 & 0.14 & $R$ & 2022 Apr 01 & 75 & Y & 3.2 & Airmass \\
TOI-4734 & LCO McDonald/Sinistro & 1.0 & $g^\prime$ & 2022 Nov 01 & 40 & Y & 0.3 & Total Counts \\
$\cdots$ & LCO CTIO/Sinistro & 1.0 & $i^\prime$ & 2022 Nov 26 & 30 & Y & 0.5 & X (T1) \\
$\cdots$ & LCO CTIO/Sinistro & 1.0 & $g^\prime$ & 2022 Dec 21 & 40 & Y & 0.5 & Airmass \\
$\cdots$ & LCO Teide/Sinistro & 1.0 & $g^\prime$ & 2022 Dec 21 & 40 & Y & 0.4 & Total Counts \\
$\cdots$ & LCO McDonald/Sinistro & 1.0 & $i^\prime$ & 2022 Dec 27 & 30 & Y & 0.4 & Airmass, Total Counts \\
TOI-4794 & Hazelwood & 0.318 & $R$ & 2022 Feb 06 & 240 & Y & 1.7 & BJD$_\mathrm{TDB}$ \\
TOI-4961 & Hazelwood & 0.318 & $R$ & 2022 Feb 12 & 180 & Y & 1.7 & Total Counts, BJD$_\mathrm{TDB}$ \\
$\cdots$ & TRAPPIST-South & 0.6 & $z^\prime$ & 2023 Apr 21 & 90 & Y & 1.8 & Airmass, X (T1), FWHM \\
TOI-5181A & FLWO/KeplerCam & 1.2 & $i^\prime$ & 2022 Jun 04 & 35 & Y & 0.8 & X (T1), Y (T1) \\
$\cdots$ & TCS/MuSCAT2 & 1.52 & $g^\prime$ & 2024 Aug 21 & 36 & Y & 0.6 & Airmass, X Shift, Y Shift, Sky Entropy \\
$\cdots$ & TCS/MuSCAT2 & 1.52 & $r^\prime$ & 2024 Aug 21 & 45 & Y & 0.6 & Airmass, X Shift, Y Shift, Sky Entropy \\
$\cdots$ & TCS/MuSCAT2 & 1.52 & $i^\prime$ & 2024 Aug 21 & 16 & Y & 0.5 & Airmass, X Shift, Y Shift, Sky Entropy \\
$\cdots$ & TCS/MuSCAT2 & 1.52 & $z^\prime$ & 2024 Aug 21 & 36 & Y & 0.6 & Airmass, X Shift, Y Shift, Sky Entropy \\
TOI-5210 & OAUV/T50 & 0.5 & $R$ & 2022 May 31 & 90 & Y & 1.4 & Total Counts, Y (T1) \\
$\cdots$ & OAA & 0.4 & $I$ & 2022 Aug 03 & 120 & Y & 1.1 & Airmass \\
TOI-5322 & SUTO/OTIVAR & 0.3 & $B$ & 2022 Oct 30 & 300 & Y & 3.8 & Meridian Flip \\
TOI-5340 & MLO & 0.356 & $I$ & 2022 Sep 23 & 300 & N & -- & Airmass, Meridian Flip \\
$\cdots$ & FLWO/KeplerCam & 1.2 & $i^\prime$ & 2022 Oct 03 & 40 & Y & 0.8 & Sky/Pixel \\
TOI-5386A & SUTO/OTIVAR & 0.3 & $R$ & 2022 Apr 16 & 180 & Y & 2.3 & Airmass, Y (T1) \\
$\cdots$ & LCO Teide/SBIG-6303 & 0.4 & $g^\prime$ & 2022 May 04 & 70 & Y & 1.0 & Width, BJD$_\mathrm{TDB}$, Meridian Flip \\
$\cdots$ & LCO McDonald/SBIG-6303 & 0.4 & $g^\prime$ & 2023 Mar 23 & 60 & Y & 0.8 & -- \\
$\cdots$ & Whitin/CDK700 & 0.7 & $B$ & 2023 Apr 14 & 30 & Y & 1.2 & Airmass \\
$\cdots$ & TCS/MuSCAT2 & 1.52 & $g^\prime$ & 2023 Jun 21 & 16 & Y & 0.5 & X Shift, Y Shift, Sky Entropy, Airmass \\
$\cdots$ & TCS/MuSCAT2 & 1.52 & $r^\prime$ & 2023 Jun 21 & 16 & Y & 0.4 & X Shift, Y Shift, Sky Entropy, Airmass \\
$\cdots$ & TCS/MuSCAT2 & 1.52 & $i^\prime$ & 2023 Jun 21 & 16 & Y & 0.5 & X Shift, Y Shift, Sky Entropy, Airmass \\
$\cdots$ & TCS/MuSCAT2 & 1.52 & $z^\prime$ & 2023 Jun 21 & 16 & Y & 0.5 & X Shift, Y Shift, Sky Entropy, Airmass \\
TOI-5592 & CMO/RC600 & 0.6 & $R$ & 2022 Nov 01 & 80 & Y & 0.5 & Airmass, FWHM \\
$\cdots$ & Acton Sky Portal & 0.36 & $r^\prime$ & 2022 Nov 20 & 32 & Y & 1.1 & Airmass, J.D.-2400000 \\
$\cdots$ & TRAPPIST-North & 0.6 & $I$ & 2022 Nov 27 & 40 & Y & 1.4 & Airmass \\
$\cdots$ & LCO Haleakala/SBIG-6303 & 0.4 & $g^\prime$ & 2023 Mar 25 & 140 & Y & 0.7 & Airmass%
\enddata
\tablenotetext{a}{Precision is computed as the rms of the residuals when the observed data points are subtracted from the best-fit transit and detrending model, scaled to a fixed 600s timescale.}
\tablecomments{
The ground-based follow-up photometry are publicly avilable via ExoFoP\footnote{\url{https://exofop.ipac.caltech.edu/tess/}\label{footnote:exofop_url}} and are also provided as online supplementary material (Data behind the Figure for Figure \subref*{fig:toi2031_multiplot}).\\
The following facilities were used for ground-based photometric observations: 0.4m and 1.0m telescopes of the
Las Cumbres Observatory Global Telescope (LCOGT; \citealt{LCOGT_Brown2013}) using sites at
Cerro Tololo Inter-American Observation (CTIO), Haleakalā Observatory, McDonald Observatory, South African Astronomical Observatory (SAAO), Siding Spring Observatory (SSO), and Teide Observatory;
Feder Observatory in Glyndon, Minnesota;
Brigham Young University (BYU) Orson Pratt Observatory (OPO);
C. R. Chambliss Astronomical Observatory (CRCAO) at Kutztown University in Kutztown, Pennsylvania;
the Private observatory of the Mount at Saint-Pierre-du-Mont, France (OPM);
Waffelow Creek Observatory (WCO) in Nacogdoches, Texas;
Hazelwood Observatory in Churchill, Australia;
the Perth Exoplanet Survey Telescope (PEST);
the 0.3m OTIVAR telescopes at Silesian University of Technology Observatories (SUTO);
the Next Generation Transit Survey \citep[NGTS;][]{NGTS_Wheatley2018} at ESO Paranal Observatory;
Brierfield Observatory in Bowral, Australia;
the 0.36m and 0.51m telescopes at the El Sauce Observatory, Chile;
TRAPPIST-South at La Silla Observatory and TRAPPIST-North at Oukaimeden Observatory \citep{TRAPPIST_Jehin2011,TRAPPIST_Gillon2011,TRAPPIST-North_Barkaoui2019};
KeplerCam on the Fred Lawrence Whipple Observatory (FLWO) 1.2m telescope;
Phillips Academy Observatory in Andover, Massachusetts;
the T100 telescope at T\"urkiye National Observatories in Antalya, T\"urkiye;
the MuSCAT2 four-color imager at the Telescopio Carlos S\'{a}nchez (TCS) at Teide Observatory \citet{MuSCAT2_Narita2018};
the TURIA1 0.14m, TURIA2 0.3m, and T50 0.5m telescopes of the Observatori Astronòmic de la Universitat de València (OAUV) in Aras de los Olmos, Valencia, Spain;
the 0.8m telescope at George Mason University (GMU) in Fairfax, Virginia, with automation described in \citet{GMU_Reefe2022};
Wendelstein Observatory in Munich, Germany;
Wellesley College Whitin Observatory in Wellesley, Massachusetts;
Adams Observatory at Austin College in Sherman, Texas;
the Dragonfly Telephoto Array at the New Mexico Skies telescope hosting facility \citep{Dragonfly_Danieli2020};
Caucasian Mountain Observatory (CMO) in Kislovodsk, Russia;
Observatori Astron\`{a}mic Albany\`{a} (OAA) in Albany\`{a}, Spain;
the Maury Lewin Astronomical Observatory (MLO) in Glendora, California;
and the Acton Sky Portal private observatory in Acton, Massachusetts.

}
\end{deluxetable*}
\end{longrotatetable}
\end{rotatepage}

\ifSubfilesClassLoaded{%
\bibliography{../instruments}
}{}

\end{document} \label{sf:sg1_summary}%

\subsection{High Angular Resolution Imaging} \label{ssec:imaging}

In order to check for stellar companions closer to the planet candidate host stars than can be resolved in seeing-limited images, we obtained images with specialized cameras and techniques with higher angular resolution for each target.
Such observations are key for identifying close blended eclipsing binaries, as well as to improve estimates of flux dilution when computing the planet candidates' transit depths.
These observations were coordinated through the TFOP High-Resolution Imaging Sub-Group 3 (SG3).

\paragraph{Gemini 'Alopeke/Zorro}
\expandTOIs{\GeminiNorthTOIs} were observed with Gemini-N/'Alopeke, while \expandTOIs{\GeminiSouthTOIs} were observed with Gemini-S/Zorro. 'Alopeke and Zorro are twin speckle imagers on the Gemini telescopes \citep{Gemini_Zorro_Alopeke_Scott2021}. Observations were made in two bands centered at 562~nm and 832~nm and reduced with the standard pipeline described in  \citet{Speckle_Reduction_Howell2011}.
Final data products provide details for any close companion detected as well as 5$\sigma$ magnitude
contrast curves.

\paragraph{SAI Speckle Polarimeter}
We obtained $I$-band observations of \countTOIs{\SAISpeckleTOIs} targets with the Speckle Polarimeter \citep{SAI_Safonov2017,SAI_Strakhov2023}. This is an optical speckle imaging instrument on the 2.5 m telescope at the Caucasian Mountain Observatory (CMO) of Sternberg Astronomical Institute (SAI) of Lomonosov Moscow State University.

\paragraph{Lick Shane/ShARCS}
\expandTOIs{\ShaneTOIs} were observed in $J$ and $K_s$ bands with the ShARCS adaptive optics camera on the Shane 3m telescope at Lick observatory \citep{ShaneAO_Kupke2012,ShaneAO_Gavel2014,ShaneAO_McGurk2014}. These observations were reduced with the \texttt{SImMER} pipeline described in \citet{SIMmER_Savel2020}. Further details of the Shane/ShARCS AO follow-up of TESS Objects of Interest are included in Dressing et al. (2025, submitted).

\paragraph{SOAR HRCam}
\countTOIs{\SOARTOIs} of the targets were observed with the High-Resolution Camera (HRCam; \citealt{SOAR_Tokovinin2008}) on the Southern Astrophysical Research (SOAR) 4.1m telescope.
The data were reduced according to the procedures described in \citet{SOAR_Tokovinin2018}, while the overall observation strategy for \TESS planet candidate follow-up is described in \citet{SOAR_TESS_Ziegler2019,SOAR_TESS_Ziegler2021}.

\paragraph{Palomar PHARO}
We observed \expandTOIs{\PalomarTOIs} with the Palomar High Angular Resolution Observer (PHARO; \citealt{PHARO_Hayward2001}), a near-infrared adaptive optics camera on the 200-in Hale telescope at Palomar Observatory.
Observations were made in the Br$\gamma$ filter, $H$cont, and $K$cont filters.
The data were reduced with the pipeline described in \citet{PHARO_Furlan2017}.

\paragraph{WIYN NESSI}
Finally, we observed \countTOIs{\NESSITOIs} objects with the NN-Explore Exoplanet Stellar Speckle Imager (NESSI; \citealt{NESSI_Scott2018}) on the WIYN 3.5m telescope at Kitt Peak National Observatory (KPNO).
Observations were made using the 832~nm filter for all targets, with additional observations using the 562~nm filter for some of them.
\citet{NESSI_Strategy_Howell2021} describes the observing strategy and data reduction procedures for \TESS planet candidate host stars in detail.

We summarize all the high angular-resolution imaging observations in Table \ref{tab:imaging_obs}.
We discuss the systems with detected companions below (\S\ref{sssec:ao_detected_companions}) and provide the reconstructed images and sensitivity limits in Figure \subref*{fig:toi3160_imaging}.
For all other targets, no stellar companions were identified down to the respective observations' detection limits.
Figures showing the contrast curves and reconstructed images for these non-detections are provided in Figure \ref{fig:high_res_imaging_nocomp} in Appendix \ref{sec:ao_no_comp}, and are available on ExoFOP.\textsuperscript{\ref{footnote:exofop_url}}

\subsubsection{Detected Close Companions} \label{sssec:ao_detected_companions}

Of the thirty targets described in this work, high angular-resolution imaging detected close ($< 3^{\prime\prime}$) stellar companions in five cases.
Figure \subref*{fig:toi3160_imaging} shows the observations for these five targets.
The observed properties of the companions, along with wider companions detected by \Gaia (\S\ref{ssec:gaia_companions}), are collected in Table \ref{tab:stellar_comps}.

\paragraph{TOI-3160} This target was imaged by SOAR HRCam in $I_c$ band and found to have a close companion with $\Delta I_c = 3.3$~mag, at a separation of $0\farcs33$\ifjournal\else{} (Fig. \subref{fig:toi3160_imaging})\fi.
Gemini-South Zorro observations of the same target, made in a narrow bandpass centered at 832~nm, also detected the companion and placed a stricter lower limit on the contrast ratio between the two components ($\Delta$ mag $\gtrsim 4$).
The two observations of the companion are inconsistent with each other and may indicate that it is a variable source.
In our subsequent analyses, we adopted the smaller contrast ratio, as a brighter companion would have the most impact on the inferred planetary properties.
Given the two stars' close proximity, we assumed the pair is gravitationally bound, as previous work \citep[e.g.,][]{Matson2018} suggests that stars with separations  $\lesssim 1^{\prime\prime}$ are more likely to be a bound pair than chance alignments.

\paragraph{TOI-3464} The SOAR HRCam observations of  this target revealed a faint ($\Delta I_c = 5.5$~mag) companion at an angular separation of $2\farcs7$\ifjournal\else{} (Fig. \subref{fig:toi3464_imaging})\fi.
This stellar neighbor was also detected by the \Gaia spacecraft, which is typically able to resolve close stellar companions down to $\sim 1\farcs0$.
The companion in the \Gaia DR3 catalog \citep{GaiaEDR3_Brown2021} has an angular separation of $2\farcs64$ and $\Delta G = 5.47$~mag.
The neighbor appears to be an unrelated background star, with a measured parallax and proper motion $>5\sigma$ discrepant from those of TOI-3464 (\S\ref{ssec:gaia_companions}).

\paragraph{TOI-3523} We observed this target with Palomar/PHARO and the SAI Speckle Polarimeter, detecting a secondary star with a separation of $0\farcs67$, and a contrast ratio with the primary of 2.058~mag in $H$cont, 2.105~mag in Br$\gamma$, and 3.5~mag in $I$\ifjournal\else{} (Fig. \subref{fig:toi3464_imaging})\fi.
This companion is too close to have allowed for an independent detection by \Gaia, but it is probably the reason why the \Gaia catalog
reports an anomalously high value for the excess noise in the astrometric measurements of the primary; specifically, the Renormalized Unit Weight Error (RUWE) is 2.828.
In addition, the \Gaia mission identified a nearby star ($8\farcs5$ separation) that has a  parallax and proper motion nearly matching those of the primary star (Table \ref{tab:stellar_comps}).
In the catalog of stellar binaries by \citet{GaiaEDR3_Binaries_El-Badry2021}, the pair has a chance-alignment ratio (an approximation for the chance-alignment probability) of $\mathcal{R}_\mathrm{chance} = 1.24\times10^{-2}$.
If all three stars are indeed bound, TOI-3523 would be an unusual case of a hierarchical system featuring three stars and a hot Jupiter.

\paragraph{TOI-5181} Observations of this target identified a stellar companion with an angular separation of $\approx 1\farcs5$ that is roughly 4.4~mag fainter in $I$ from SOAR HRCam and 3.2~mag fainter in $H$cont from Palomar/PHARO imaging\ifjournal\else{} (Fig. \subref{fig:toi5181_imaging})\fi.
Additional differential magnitudes are listed in Table \ref{tab:stellar_comps}.
This object was also detected by \Gaia, and has parallax and proper motion measurements consistent with being a bound companion (Table \ref{tab:stellar_comps}, \S\ref{ssec:gaia_companions}).

\paragraph{TOI-5386} This target was found to be a binary stellar system by WIYN/NESSI and Palomar/PHARO observations.
The Palomar/PHARO observations measured the companion to be 1.8 and 2.1 magnitudes fainter in Br$\gamma$ and $H$cont bands respectively, and to have a separation of $0\farcs24$\ifjournal\else{} (Fig \subref{fig:toi5386_imaging})\fi.
The WIYN/NESSI observations suggested a much fainter companion, with $\Delta$~mag $\approx 4$ at 832~nm, but the fit was uncertain due to the small separation between the two stars.
As with TOI-3160, we adopted the brighter properties for the companion.

The companion to TOI-3464 is too faint to affect the interpretation of our follow-up observations. However, in the other four cases, the light from the companion cannot be neglected.  We describe how we model their effects in Section \ref{ssec:companion_modelling}.




\startlongtable
\begin{deluxetable*}{ccccccc}
\tablecolumns{7}
\tablecaption{Summary of High-Resolution Imaging Observations \label{tab:imaging_obs}}
\addtocounter{table}{1}
\pdfbookmark[3]{Table \thetable: Summary of High-Resolution Imaging}{imaging_obs}%
\addtocounter{table}{-1}
\tablehead{
    \colhead{Target} & \colhead{Telescope} & \colhead{Instrument} & \colhead{Filter} &
    \colhead{Date} & \colhead{Image Type} & \colhead{Comp. Detected\tablenotemark{a}}
}
\startdata
TOI-2031A & Gemini-N (8 m) & 'Alopeke & 562 nm & 2020 Aug 04 & Speckle & N \\
$\cdots$ & Gemini-N (8 m) & 'Alopeke & 832 nm & 2020 Aug 04 & Speckle & N \\
TOI-2169A & SAI-2.5m (2.5 m) & Speckle Polarimeter & $I_c$ & 2021 May 06 & Speckle & N \\
$\cdots$ & Shane (3 m) & ShARCS & $J$ & 2021 May 31 & AO & N \\
$\cdots$ & Shane (3 m) & ShARCS & $K_s$ & 2021 May 31 & AO & N \\
$\cdots$ & SOAR (4.1 m) & HRCam & $I_c$ & 2022 Apr 15 & Speckle & N \\
$\cdots$ & Palomar (5 m) & PHARO & Br$\gamma$ & 2023 Jun 29 & AO & N \\
$\cdots$ & Palomar (5 m) & PHARO & $H$cont & 2023 Jun 29 & AO & N \\
TOI-2346 & SOAR (4.1 m) & HRCam & $I_c$ & 2020 Dec 03 & Speckle & N \\
TOI-2382 & SOAR (4.1 m) & HRCam & $I_c$ & 2020 Dec 03 & Speckle & N \\
TOI-2876 & SOAR (4.1 m) & HRCam & $I_c$ & 2021 Nov 20 & Speckle & N \\
TOI-2886 & SOAR (4.1 m) & HRCam & $I_c$ & 2021 Nov 20 & Speckle & N \\
TOI-2986 & Shane (3 m) & ShARCS & $K_s$ & 2021 Dec 18 & AO & N \\
$\cdots$ & Shane (3 m) & ShARCS & $J$ & 2021 Dec 18 & AO & N \\
$\cdots$ & SOAR (4.1 m) & HRCam & $I_c$ & 2022 Apr 15 & Speckle & N \\
TOI-2992 & SOAR (4.1 m) & HRCam & $I_c$ & 2022 Apr 15 & Speckle & N \\
TOI-3135 & SOAR (4.1 m) & HRCam & $I_c$ & 2021 Jul 14 & Speckle & N \\
TOI-3160 & Gemini-S (8 m) & Zorro & 832 nm & 2021 Jul 20 & Speckle & Y \\
$\cdots$ & SOAR (4.1 m) & HRCam & $I_c$ & 2022 Apr 15 & Speckle & Y \\
TOI-3464 & SOAR (4.1 m) & HRCam & $I_c$ & 2022 Apr 15 & Speckle & Y \\
TOI-3474 & SOAR (4.1 m) & HRCam & $I_c$ & 2021 Oct 01 & Speckle & N \\
TOI-3486 & SOAR (4.1 m) & HRCam & $I_c$ & 2021 Jul 14 & Speckle & N \\
TOI-3523 & Palomar (5 m) & PHARO & Br$\gamma$ & 2023 Jun 07 & AO & Y \\
$\cdots$ & Palomar (5 m) & PHARO & $H$cont & 2023 Jun 07 & AO & Y \\
$\cdots$ & SAI-2.5m (2.5 m) & Speckle Polarimeter & $I_c$ & 2023 Aug 02 & Speckle & Y \\
$\cdots$ & SAI-2.5m (2.5 m) & Speckle Polarimeter & $I_c$ & 2023 Aug 27 & Speckle & Y \\
TOI-3593 & Palomar (5 m) & PHARO & Br$\gamma$ & 2021 Aug 24 & AO & N \\
TOI-3682 & SOAR (4.1 m) & HRCam & $I_c$ & 2021 Oct 18 & Speckle & N \\
$\cdots$ & WIYN (3.5 m) & NESSI & 832 nm & 2021 Oct 28 & Speckle & N \\
$\cdots$ & WIYN (3.5 m) & NESSI & 562 nm & 2021 Oct 28 & Speckle & N \\
$\cdots$ & SAI-2.5m (2.5 m) & Speckle Polarimeter & $I_c$ & 2021 Oct 29 & Speckle & N \\
$\cdots$ & Palomar (5 m) & PHARO & Kcont & 2024 Sep 22 & AO & N \\
TOI-3856 & SAI-2.5m (2.5 m) & Speckle Polarimeter & $I_c$ & 2022 Dec 02 & Speckle & N \\
$\cdots$ & WIYN (3.5 m) & NESSI & 562 nm & 2023 Feb 04 & Speckle & N \\
$\cdots$ & WIYN (3.5 m) & NESSI & 832 nm & 2023 Feb 04 & Speckle & N \\
$\cdots$ & Palomar (5 m) & PHARO & Kcont & 2023 Nov 27 & AO & N \\
TOI-3877 & Shane (3 m) & ShARCS & $K_s$ & 2022 Jan 16 & AO & N \\
$\cdots$ & Shane (3 m) & ShARCS & $J$ & 2022 Jan 16 & AO & N \\
$\cdots$ & SAI-2.5m (2.5 m) & Speckle Polarimeter & $I_c$ & 2022 Mar 21 & Speckle & N \\
$\cdots$ & WIYN (3.5 m) & NESSI & 832 nm & 2022 Apr 21 & Speckle & N \\
TOI-3980 & SAI-2.5m (2.5 m) & Speckle Polarimeter & $I_c$ & 2021 Oct 30 & Speckle & N \\
$\cdots$ & Gemini-N (8 m) & 'Alopeke & 562 nm & 2022 Sep 14 & Speckle & N \\
$\cdots$ & Gemini-N (8 m) & 'Alopeke & 832 nm & 2022 Sep 14 & Speckle & N \\
TOI-4214 & SOAR (4.1 m) & HRCam & $I_c$ & 2021 Nov 20 & Speckle & N \\
$\cdots$ & Shane (3 m) & ShARCS & $K_s$ & 2021 Dec 18 & AO & N \\
$\cdots$ & Shane (3 m) & ShARCS & $J$ & 2021 Dec 18 & AO & N \\
TOI-4487 & SAI-2.5m (2.5 m) & Speckle Polarimeter & $I_c$ & 2021 Oct 30 & Speckle & N \\
$\cdots$ & WIYN (3.5 m) & NESSI & 832 nm & 2022 May 05 & Speckle & N \\
TOI-4734 & SAI-2.5m (2.5 m) & Speckle Polarimeter & $I_c$ & 2022 Dec 18 & Speckle & N \\
$\cdots$ & Gemini-S (8 m) & Zorro & 562 nm & 2023 Jan 09 & Speckle & N \\
$\cdots$ & Gemini-S (8 m) & Zorro & 832 nm & 2023 Jan 09 & Speckle & N \\
$\cdots$ & WIYN (3.5 m) & NESSI & 562 nm & 2023 Jan 27 & Speckle & N \\
$\cdots$ & WIYN (3.5 m) & NESSI & 832 nm & 2023 Jan 27 & Speckle & N \\
$\cdots$ & WIYN (3.5 m) & NESSI & 562 nm & 2023 Feb 05 & Speckle & N \\
$\cdots$ & WIYN (3.5 m) & NESSI & 832 nm & 2023 Feb 05 & Speckle & N \\
TOI-4794 & SOAR (4.1 m) & HRCam & $I_c$ & 2022 Apr 15 & Speckle & N \\
TOI-4961 & SOAR (4.1 m) & HRCam & $I_c$ & 2022 Mar 20 & Speckle & N \\
TOI-5181 & WIYN (3.5 m) & NESSI & 832 nm & 2022 May 05 & Speckle & N \\
$\cdots$ & SOAR (4.1 m) & HRCam & $I_c$ & 2022 Jun 10 & Speckle & Y \\
$\cdots$ & Palomar (5 m) & PHARO & Br$\gamma$ & 2023 Jun 30 & AO & Y \\
$\cdots$ & Palomar (5 m) & PHARO & $H$cont & 2023 Jun 30 & AO & Y \\
$\cdots$ & SAI-2.5m (2.5 m) & Speckle Polarimeter & $I_c$ & 2023 Oct 23 & Speckle & Y \\
TOI-5210 & WIYN (3.5 m) & NESSI & 832 nm & 2022 May 05 & Speckle & N \\
$\cdots$ & SOAR (4.1 m) & HRCam & $I_c$ & 2022 Jun 10 & Speckle & N \\
$\cdots$ & SAI-2.5m (2.5 m) & Speckle Polarimeter & $I_c$ & 2023 Aug 28 & Speckle & N \\
TOI-5322 & SAI-2.5m (2.5 m) & Speckle Polarimeter & $I_c$ & 2022 Dec 01 & Speckle & N \\
$\cdots$ & WIYN (3.5 m) & NESSI & 562 nm & 2023 Jan 28 & Speckle & N \\
$\cdots$ & WIYN (3.5 m) & NESSI & 832 nm & 2023 Jan 28 & Speckle & N \\
TOI-5340 & SAI-2.5m (2.5 m) & Speckle Polarimeter & $I_c$ & 2022 Dec 10 & Speckle & N \\
$\cdots$ & WIYN (3.5 m) & NESSI & 562 nm & 2023 Feb 05 & Speckle & N \\
$\cdots$ & WIYN (3.5 m) & NESSI & 832 nm & 2023 Feb 05 & Speckle & N \\
TOI-5386 & WIYN (3.5 m) & NESSI & 832 nm & 2022 Apr 20 & Speckle & Y \\
$\cdots$ & Palomar (5 m) & PHARO & Br$\gamma$ & 2023 Jul 02 & AO & Y \\
$\cdots$ & Palomar (5 m) & PHARO & $H$cont & 2023 Jul 02 & AO & Y \\
TOI-5592 & SAI-2.5m (2.5 m) & Speckle Polarimeter & $I_c$ & 2022 Dec 21 & Speckle & N%

\enddata
\tablenotetext{a}{Whether a secondary star was detected in these observations. Those observations are shown in Figure \subref*{fig:toi3160_imaging}. The remaining observations are shown in Figure Set \ref{fig:high_res_imaging_nocomp}.}
\end{deluxetable*}
 \label{sf:sg3_summary}%

\ifjournal
\figsetstart
\figsetnum{1}
\figsettitle{High-resolution imaging of hot Jupiter hosts, with detected companions.}

\figsetgrpstart
\figsetgrpnum{1.1}
\figsetgrptitle{High-resolution imaging observations of TOI-3160}
\figsetplot{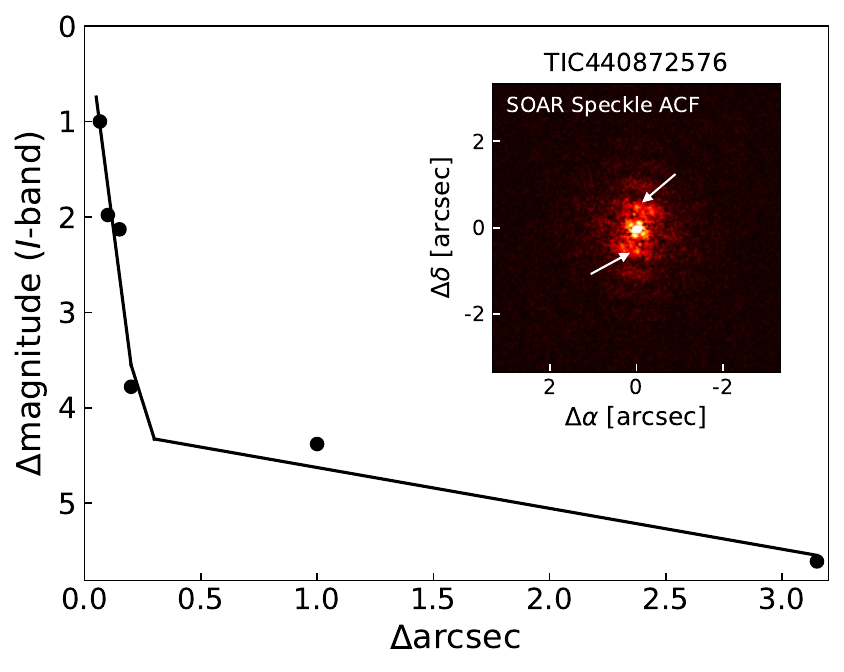}
\figsetplot{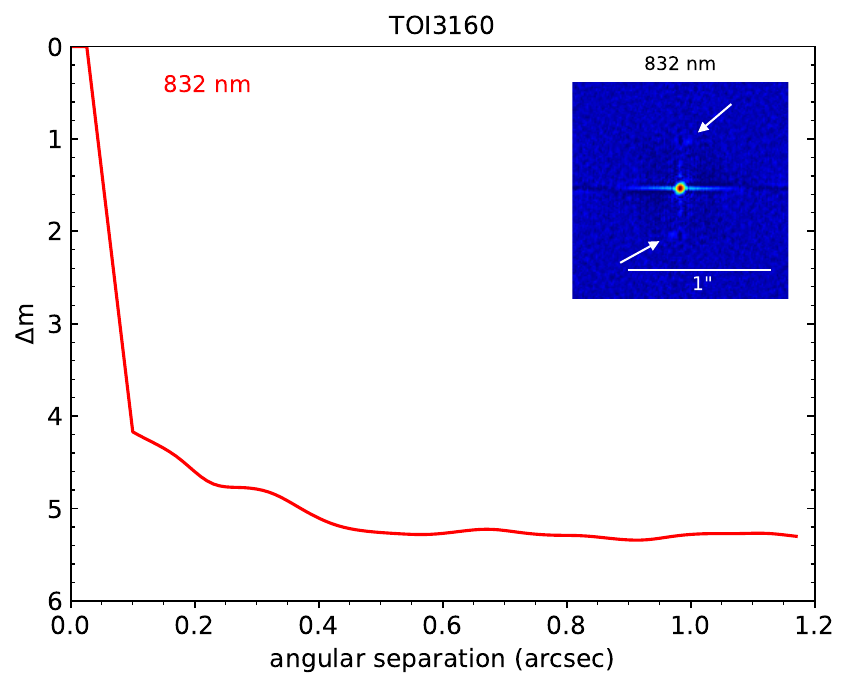}
\figsetgrpnote{%
Panel 1: Speckle sensitivity curve (solid line) and auto-correlation function (ACF, inset image) from the SOAR HRCam observation of TOI-3160.
A companion with $\Delta I = 3.3$~mag and a separation of $0\farcs33$ was detected, and can be seen in the
ACF image just above and below the central star.
Panel 2: Speckle 5$\sigma$ magnitude sensitivity curve (solid line) and reconstructed image (inset) from the Gemini-South/Zorro observation of TOI-3160 at 832 nm.
In both panels, the companion detection is marked with a white arrow. Note that there is a 180$^\circ$ degeneracy in the companion's position angle from the speckle imaging observations.%
}
\figsetgrpend

\figsetgrpstart
\figsetgrpnum{1.2}
\figsetgrptitle{High-resolution imaging observations of TOI-3464}
\figsetplot{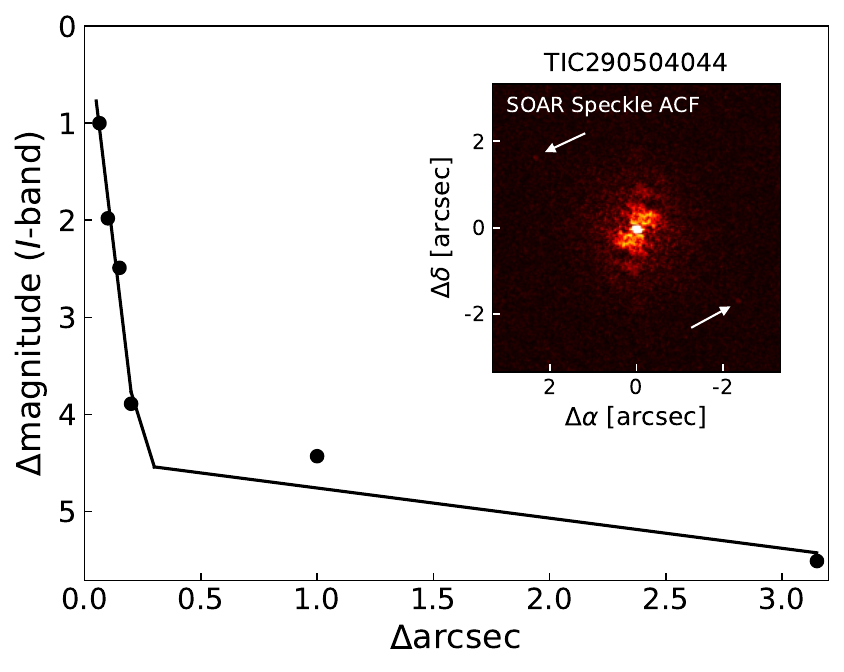}
\figsetgrpnote{%
Speckle sensitivity curve (solid line) and auto-correlation function (ACF, inset image) from the SOAR HRCam observations of TOI-3464.
The faint ($\Delta I = 5.5$~mag) companion can be seen in the top left and bottom right corners of the ACF image, marked with a white arrow.%
}
\figsetgrpend

\figsetgrpstart
\figsetgrpnum{1.3}
\figsetgrptitle{High-resolution imaging observations of TOI-3523}
\figsetplot{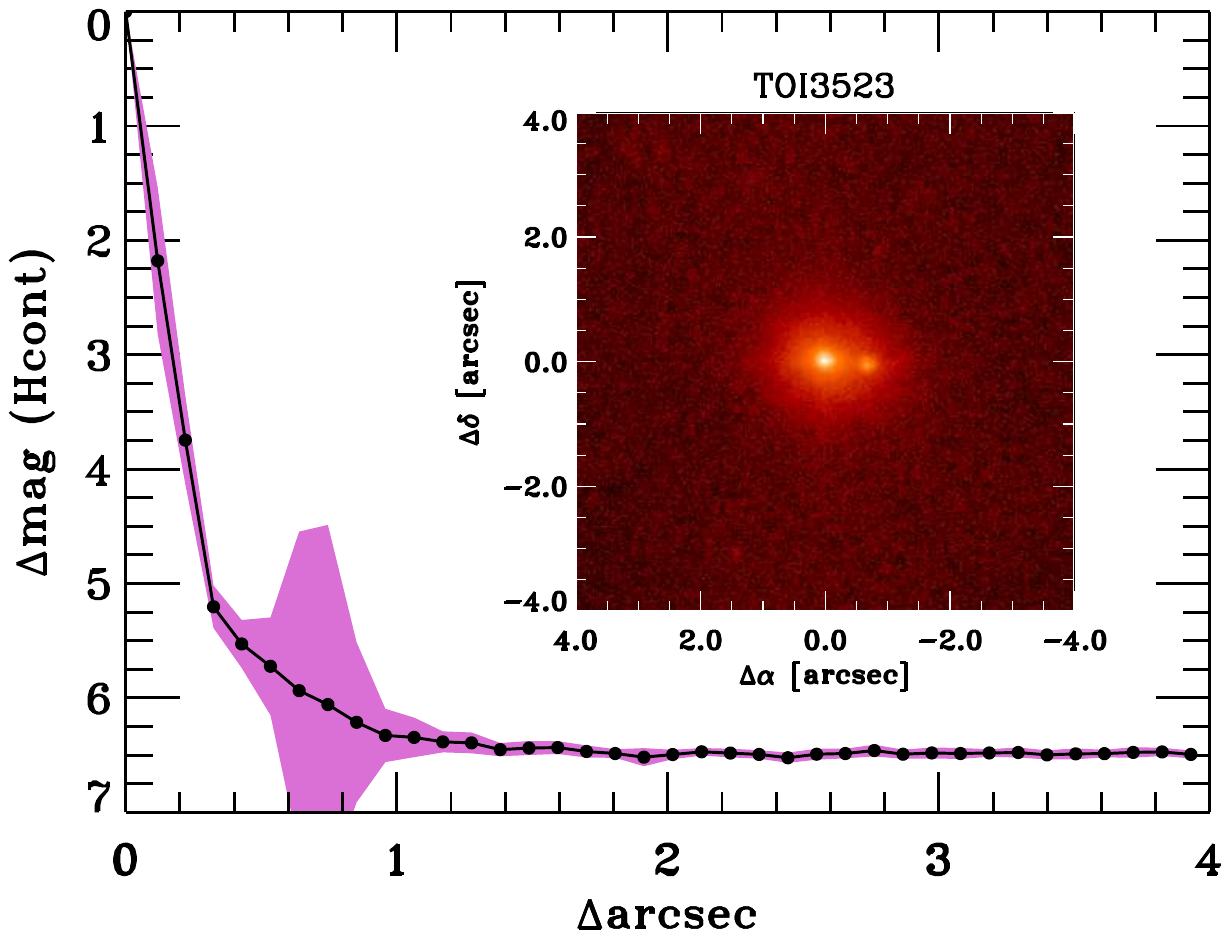}
\figsetplot{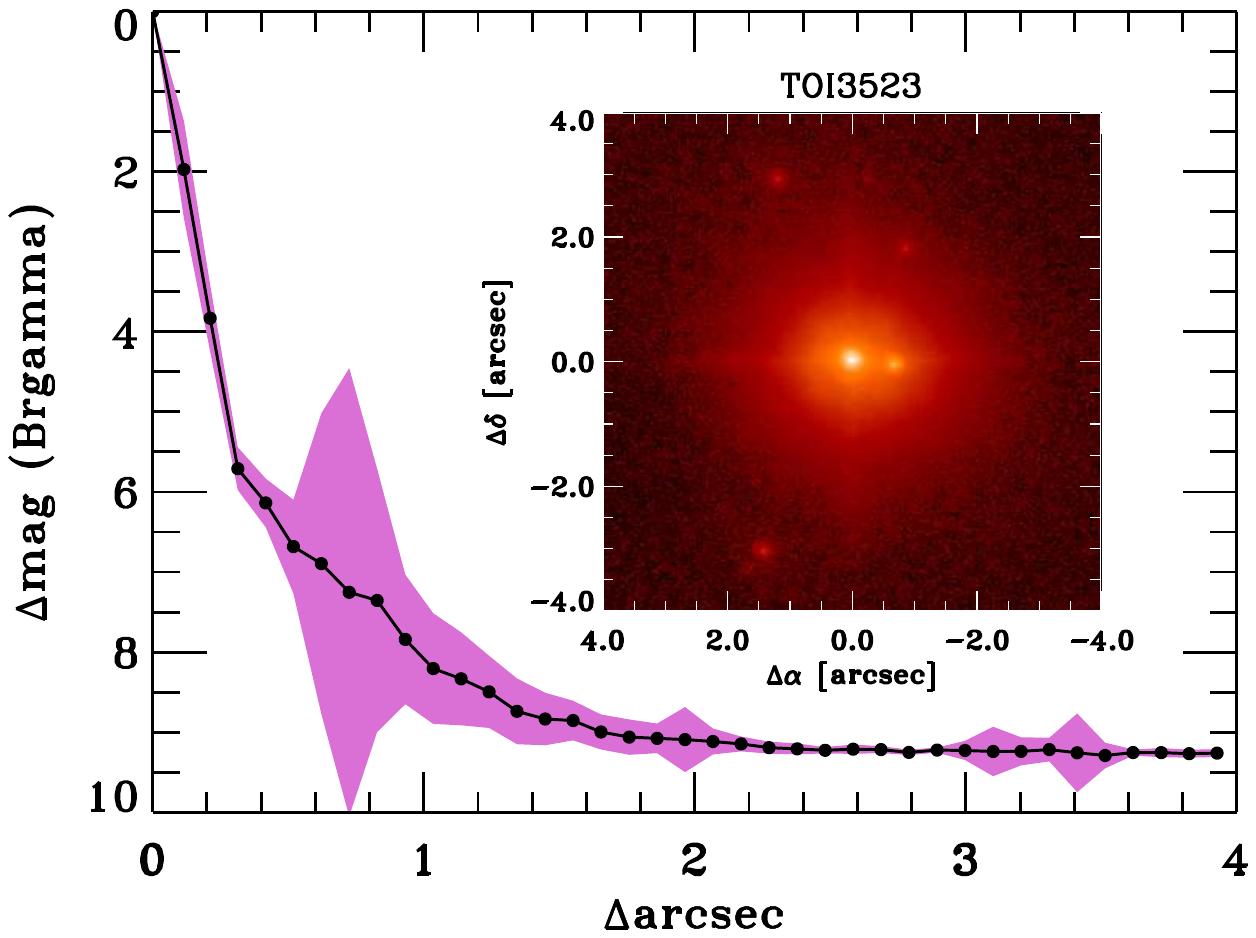}
\figsetplot{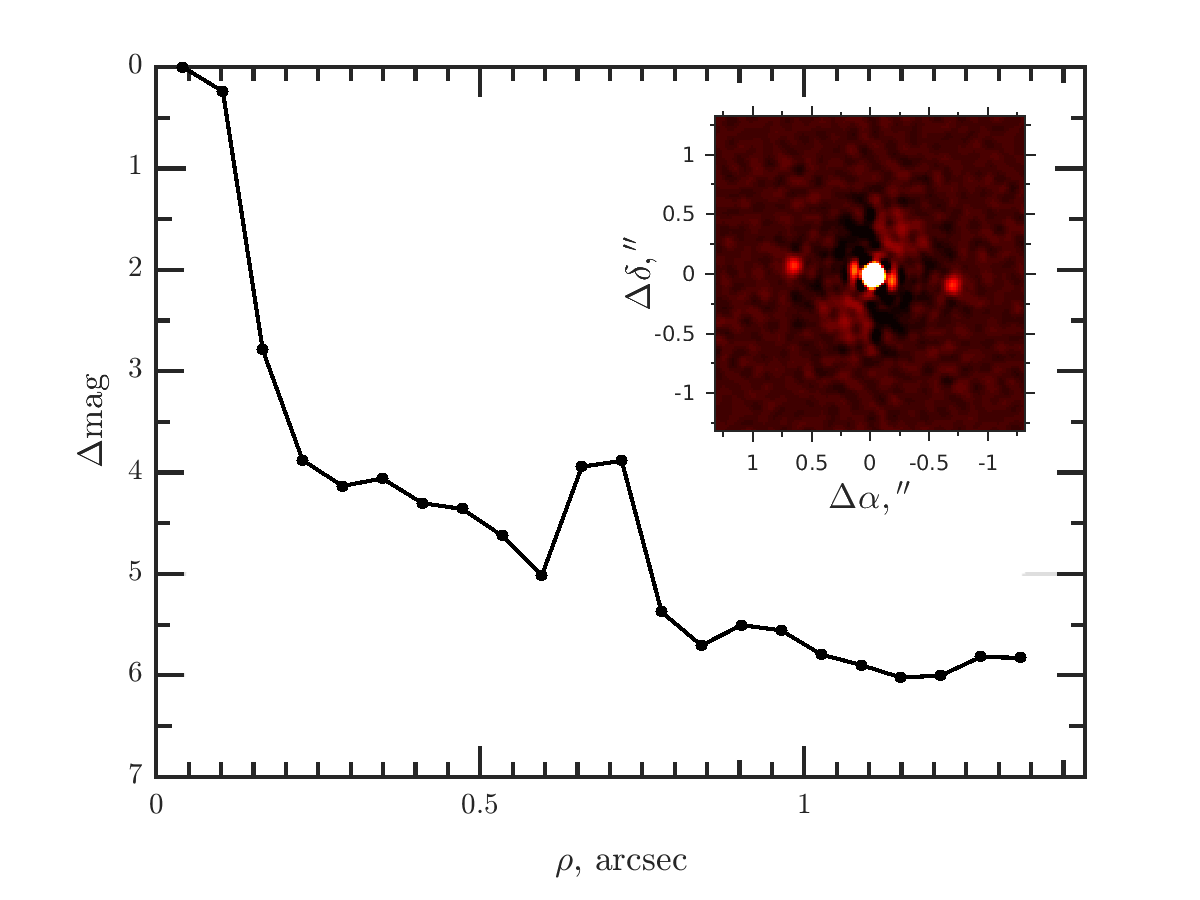}
\figsetgrpnote{%
Panel 1: Palomar PHARO adaptive optics image of TOI-3523 (inset) and sensitivity curve (solid line),
in the $H$cont band.
Panel 2: Same as above, but for observation in the narrow Br$\gamma$ band.
Panel 3: SAI speckle polarimeter $I$-band reconstructed image and sensitivity curve for observation of TOI-3523.
A companion is clearly detected $0\farcs67$ due east of the primary in all observations.
}
\figsetgrpend

\figsetgrpstart
\figsetgrpnum{1.4}
\figsetgrptitle{High-resolution imaging observations of TOI-5181}
\figsetplot{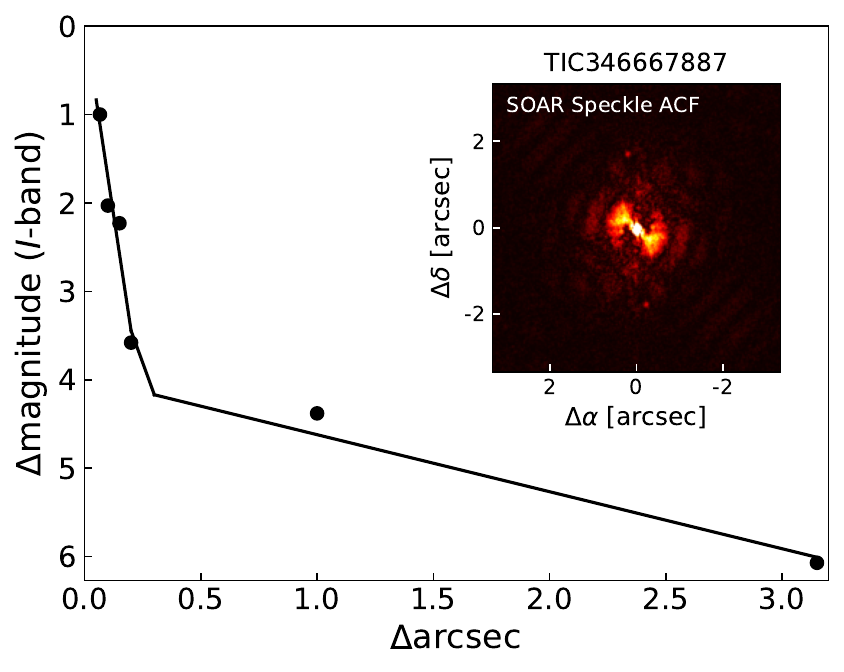}
\figsetplot{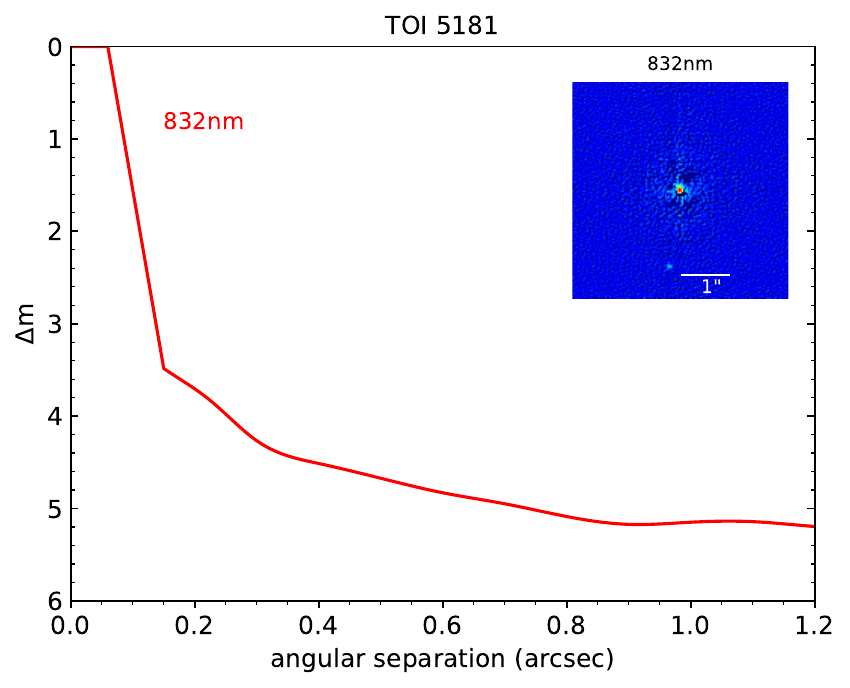}
\figsetplot{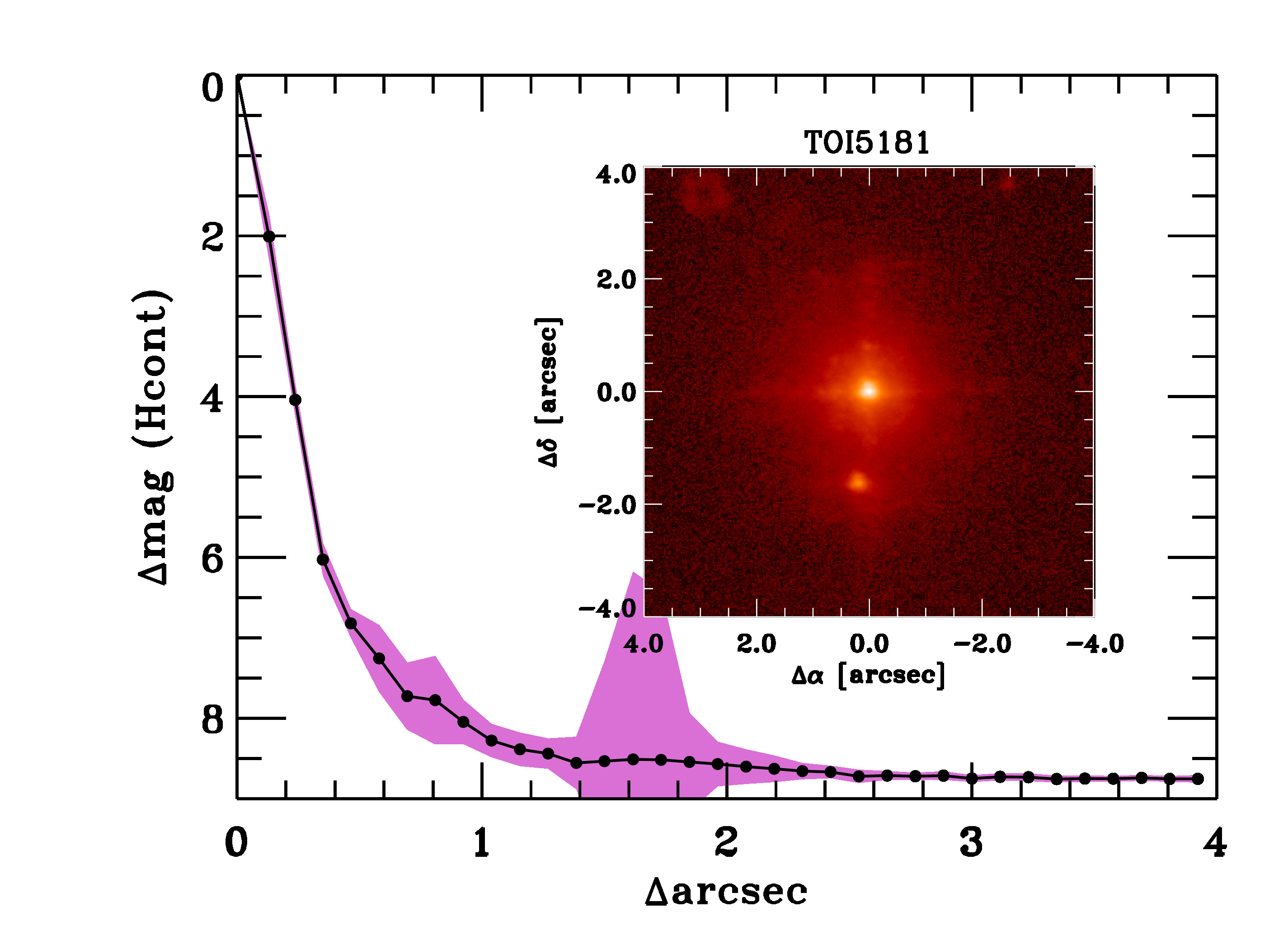}
\figsetplot{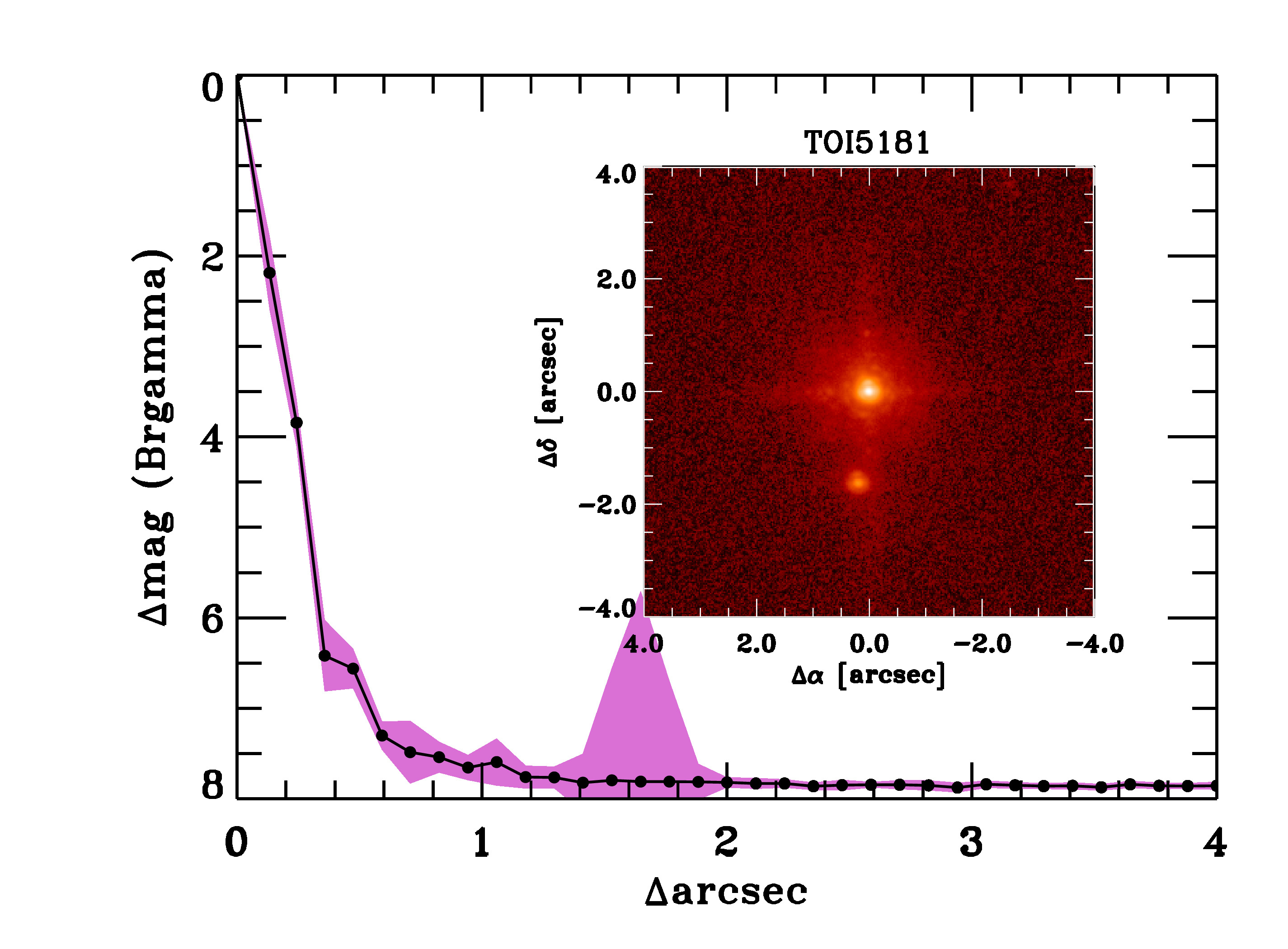}
\figsetplot{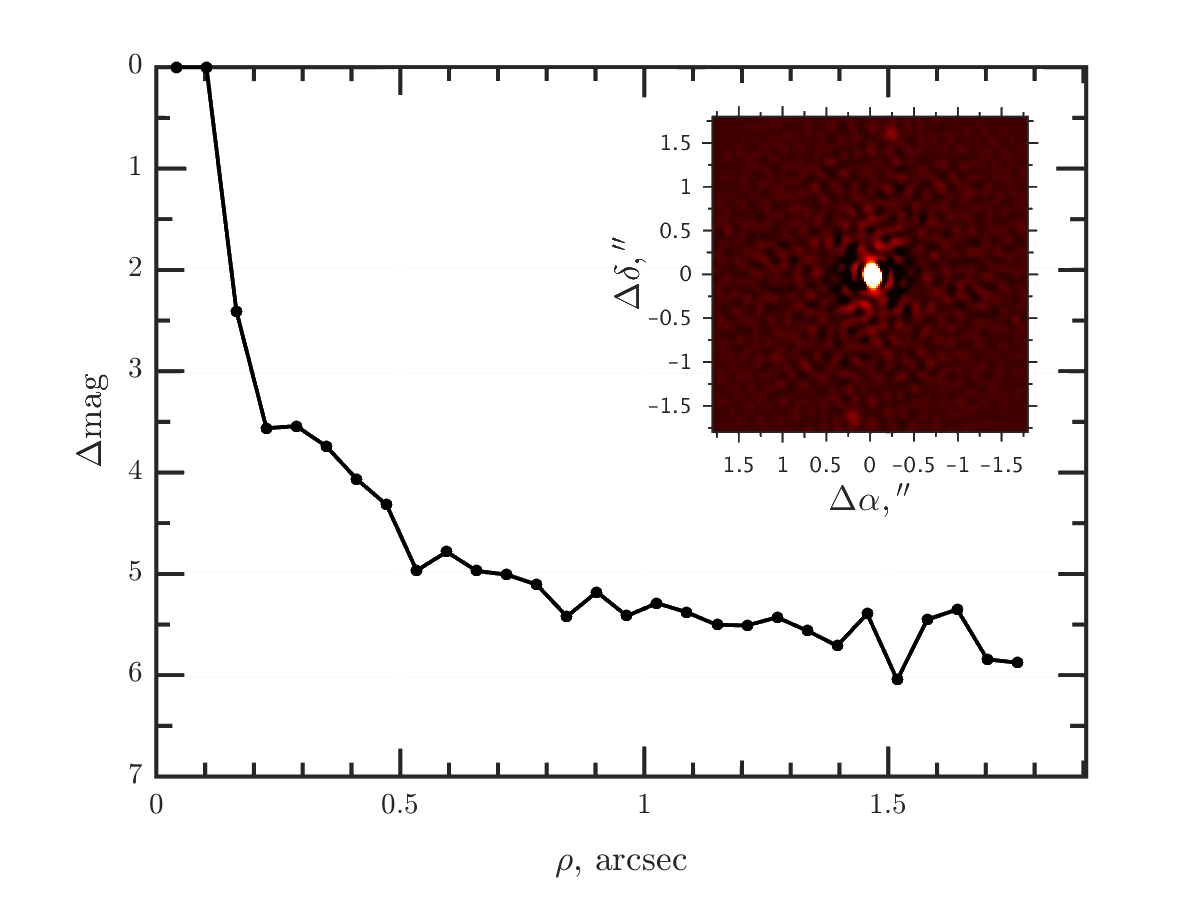}
\figsetgrpnote{%
SOAR HRCam, WIYN/NESSI, Palomar/PHARO $H$cont, Palomar/PHARO Br$\gamma$, and SAI Speckle Polarimeter observations of TOI-5181. Evidence of the companion is south of the central star.
In each figure, the solid lines show the detection sensitivity curve of the respective observation.%
}
\figsetgrpend

\figsetgrpstart
\figsetgrpnum{1.5}
\figsetgrptitle{High-resolution imaging observations of TOI-5386}
\figsetplot{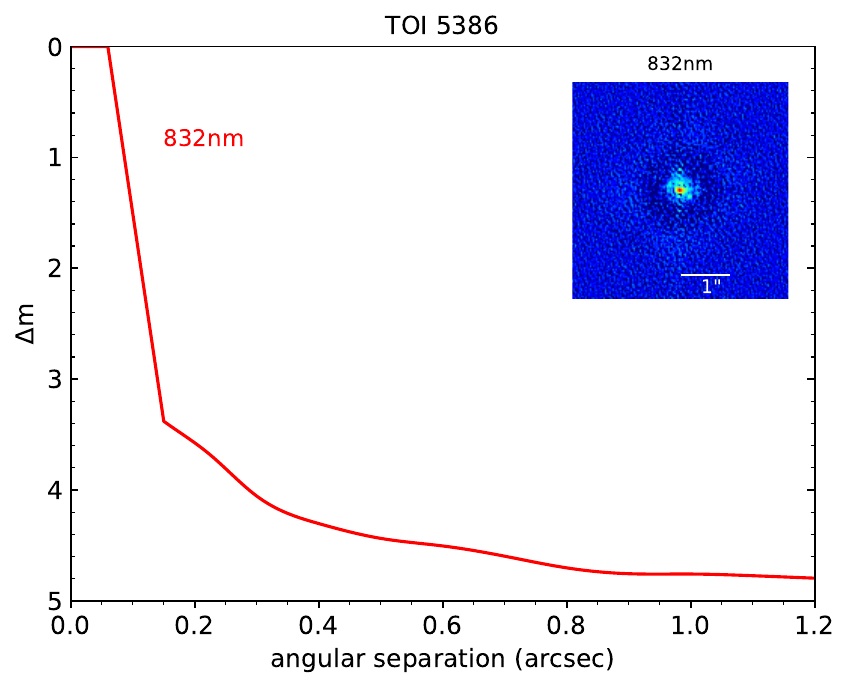}
\figsetplot{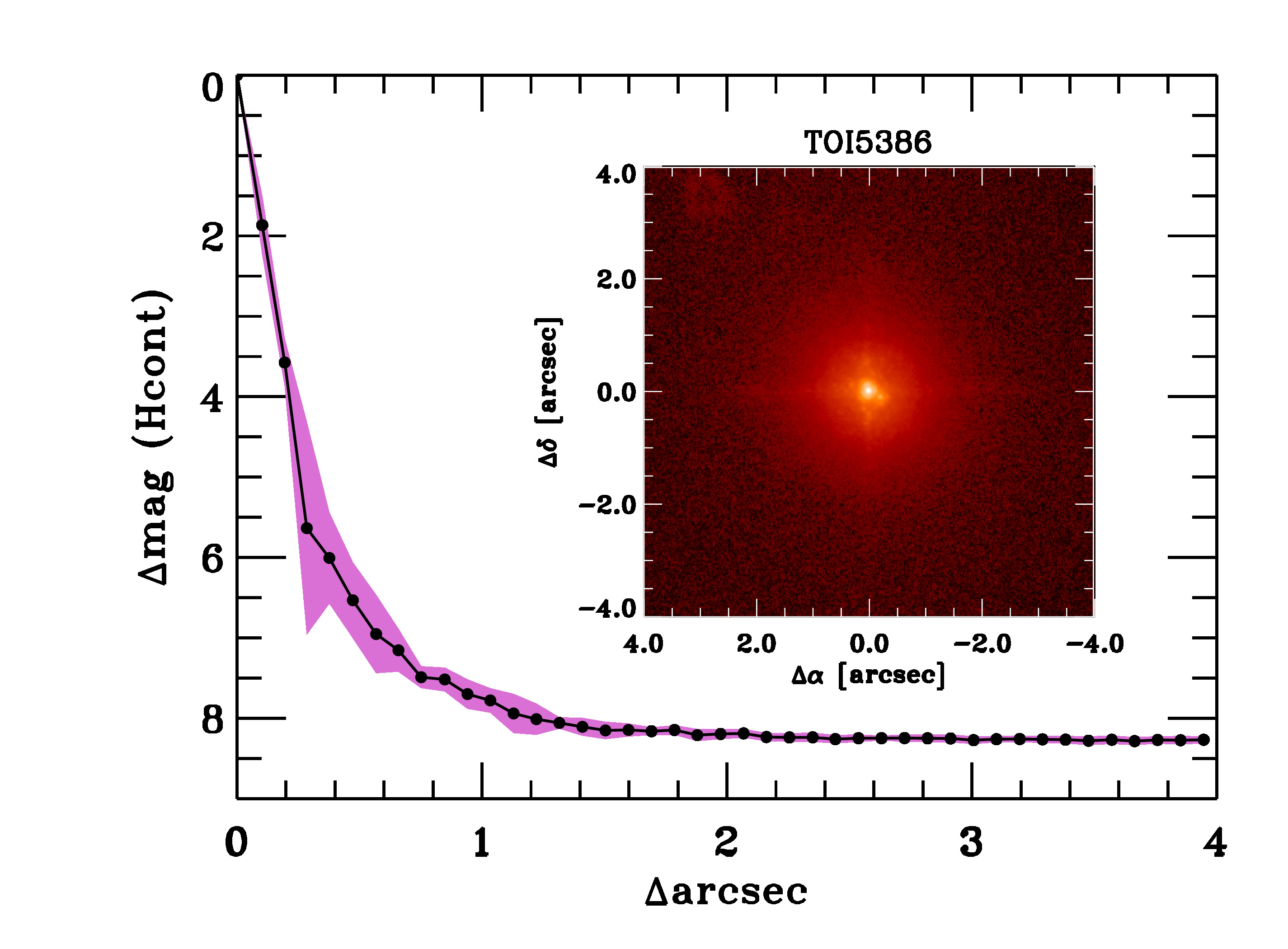}
\figsetplot{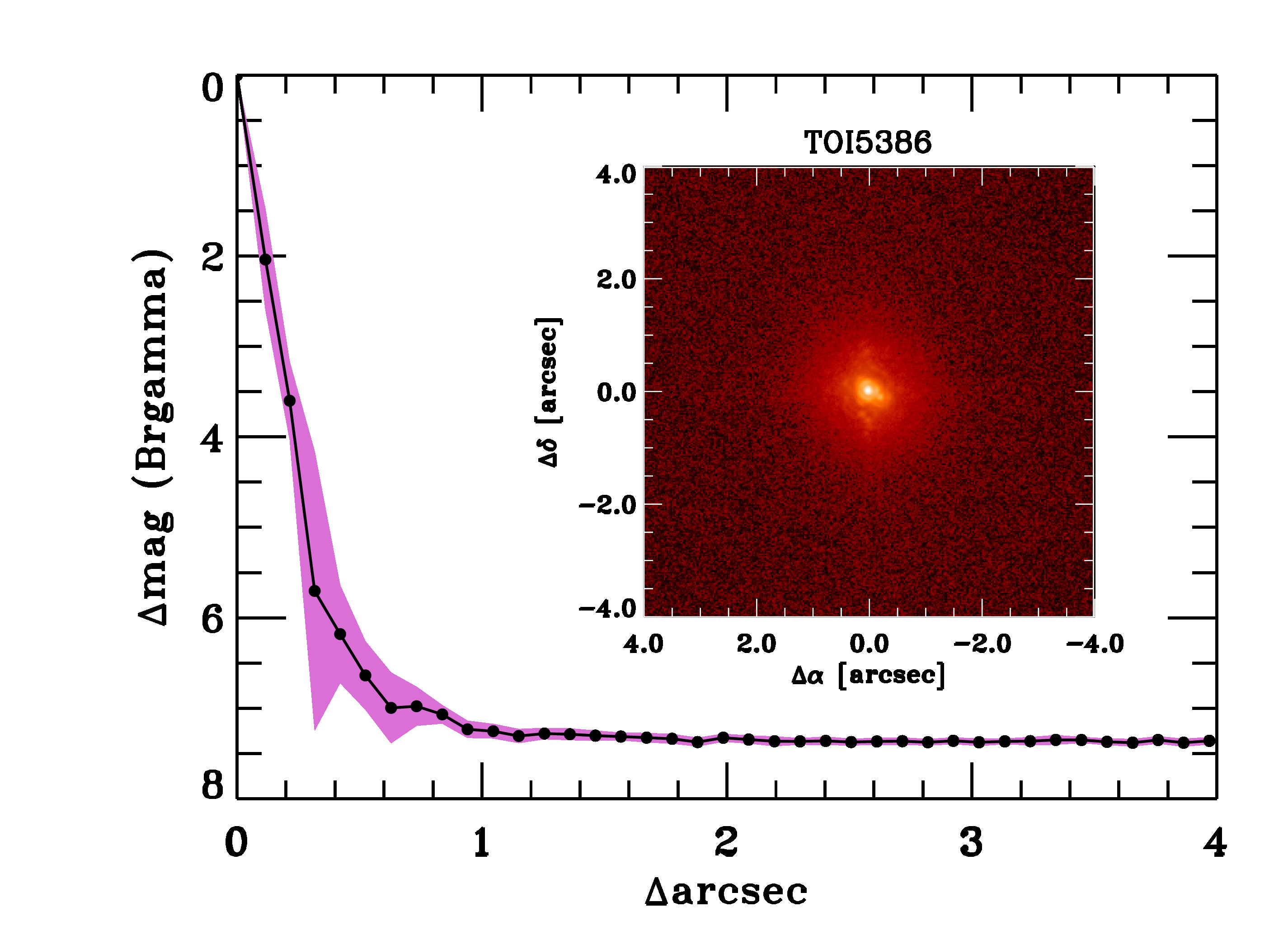}
\figsetgrpnote{%
WIYN/NESSI, Palomar/PHARO $H$cont, Palomar/PHARO Br$\gamma$ observations
of TOI-5386. A close companion is detected at $\approx0\farcs3$ from the primary.%
}
\figsetgrpend

\figsetend
\fi
\pauselinenumbers
\begin{figure}
\subfloat[Top: Speckle sensitivity curve (solid line) and auto-correlation function (ACF, inset image) from the SOAR HRCam observation of TOI-3160.
A companion with $\Delta I = 3.3$~mag and a separation of $0\farcs33$ was detected, and can be seen in the
ACF image just above and below the central star.
Bottom: Speckle 5$\sigma$ magnitude sensitivity curve (solid line) and reconstructed image (inset) from the Gemini-South/Zorro observation of TOI-3160 at 832 nm.
In both panels, the companion detection is marked with a white arrow. Note that there is a 180$^\circ$ degeneracy in the companion's position angle from the speckle imaging observations.
\label{fig:toi3160_imaging}]{
\begin{minipage}{\linewidth}
\centering
\includegraphics[width=0.9\linewidth]{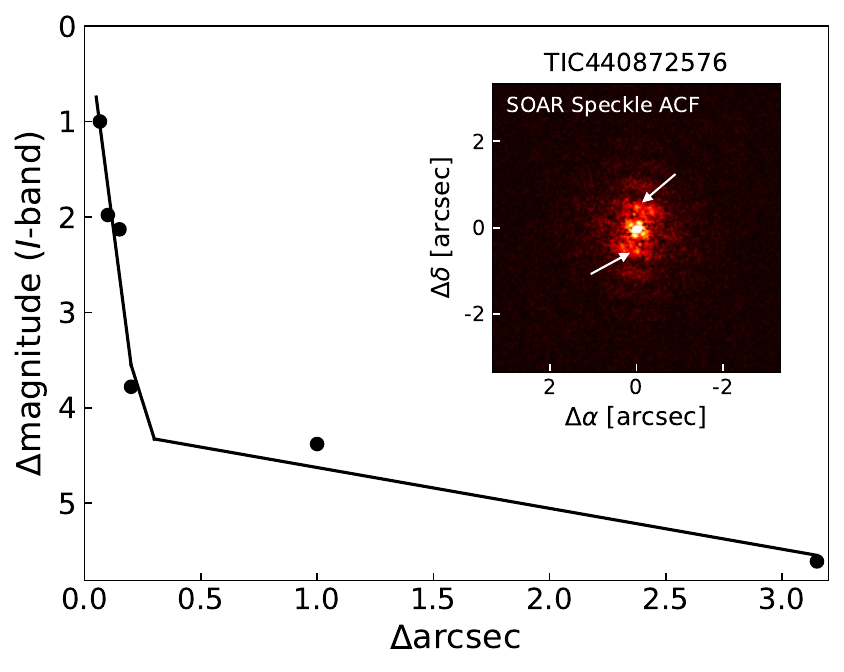}
\includegraphics[width=0.9\linewidth]{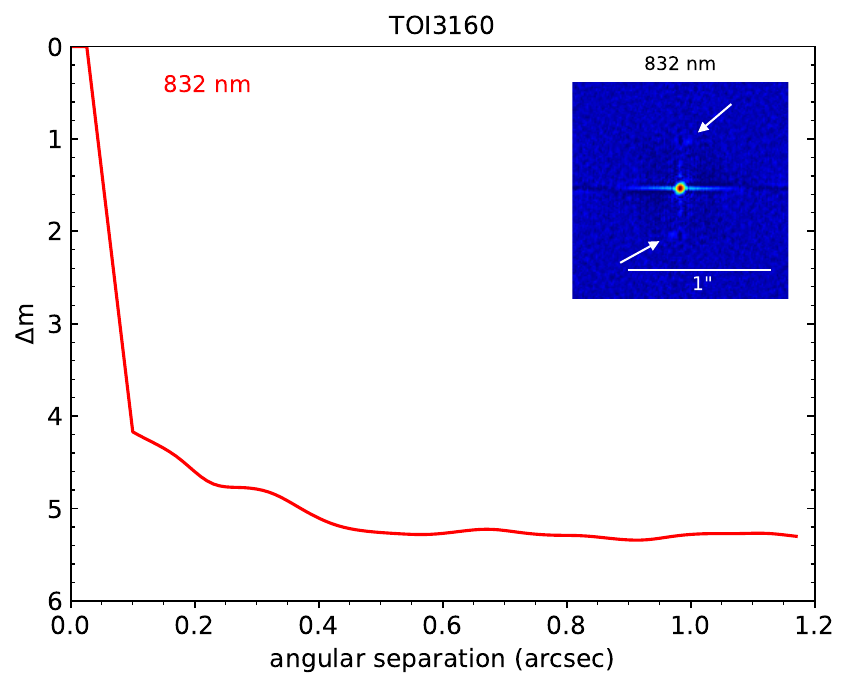}
\end{minipage}
}
\addtocounter{figure}{1}
\end{figure}
\resumelinenumbers

\ifjournal
\else
\pauselinenumbers
\begin{figure}
\centering
\ContinuedFloat
\subfloat[Speckle sensitivity curve (solid line) and auto-correlation function (ACF, inset image) from the SOAR HRCam observations of TOI-3464.
The faint ($\Delta I = 5.5$~mag) companion can be seen in the top left and bottom right corners of the ACF image, marked with a white arrow.
\label{fig:toi3464_imaging}]{
\includegraphics[width=0.9\linewidth]{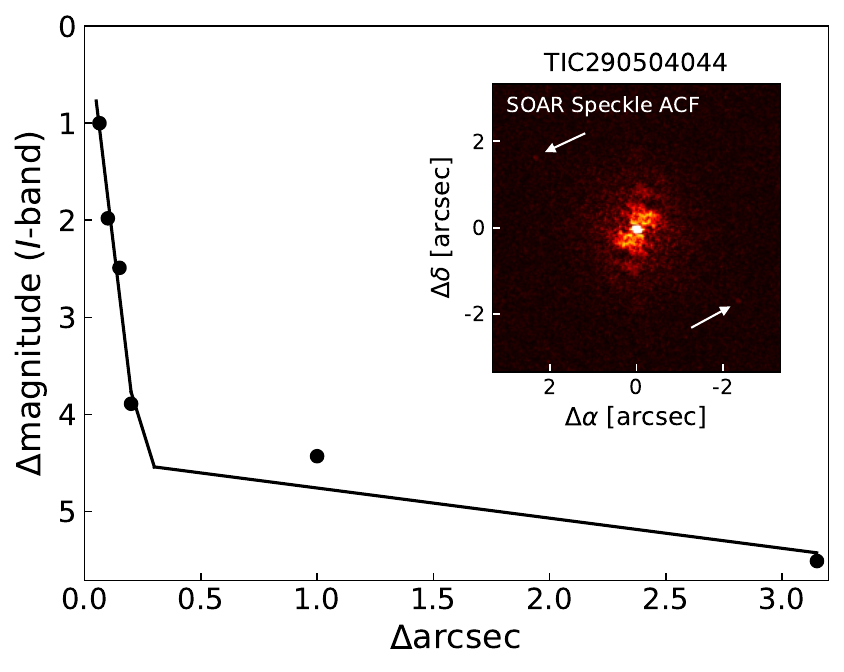}}
\addtocounter{figure}{1}
\end{figure}

\begin{figure}
\ContinuedFloat
\subfloat[Top: Palomar PHARO adaptive optics image of TOI-3523 (inset) and sensitivity curve (solid line),
in the $H$cont band.
Middle: Same as above, but for observation in the narrow Br$\gamma$ band.
Bottom: SAI speckle polarimeter $I$-band reconstructed image and sensitivity curve for observation of TOI-3523.
A companion is clearly detected $0\farcs67$ due east of the primary in all observations.
\label{fig:toi3523_imaging}]{
\begin{minipage}{\linewidth}
\centering
\includegraphics[width=0.9\linewidth]{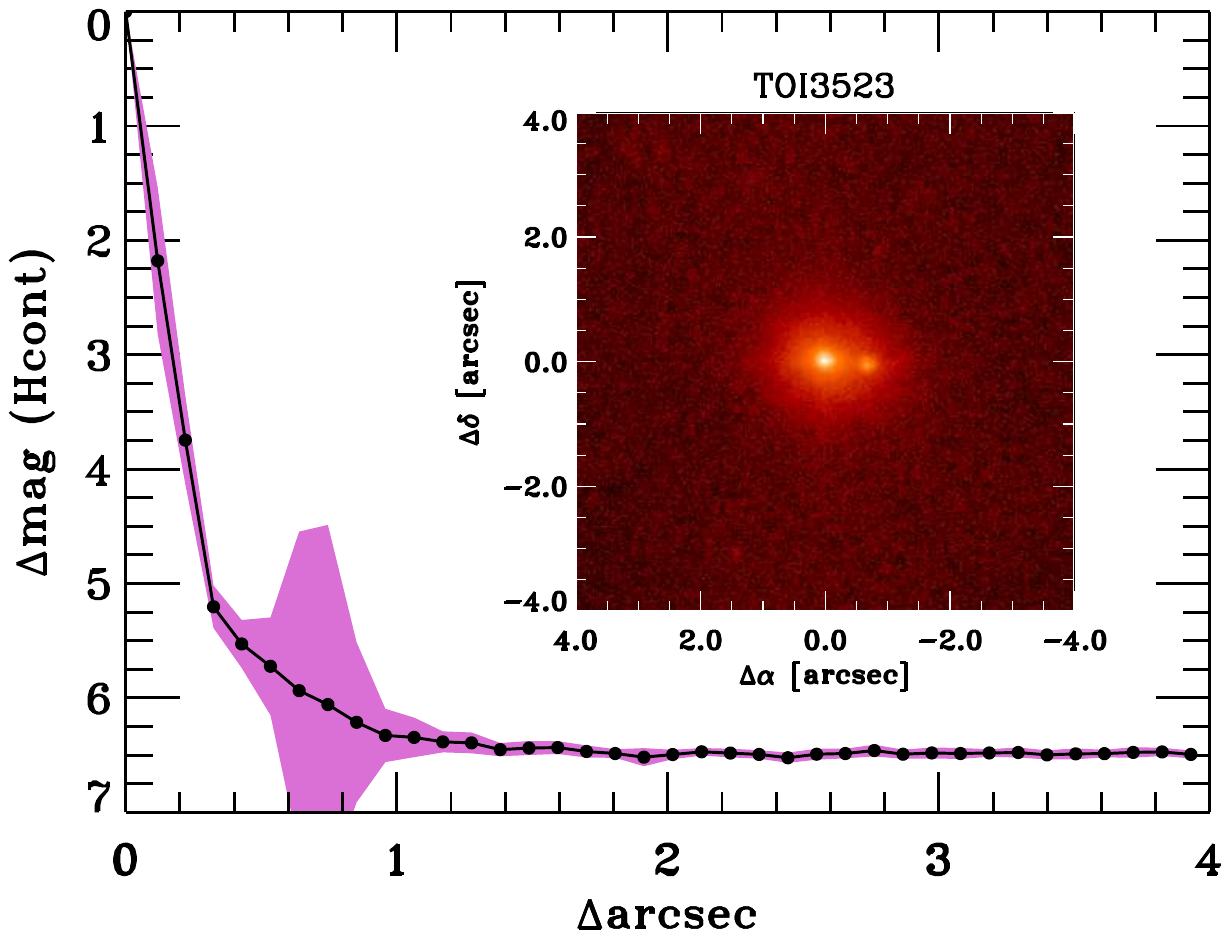}
\includegraphics[width=0.9\linewidth]{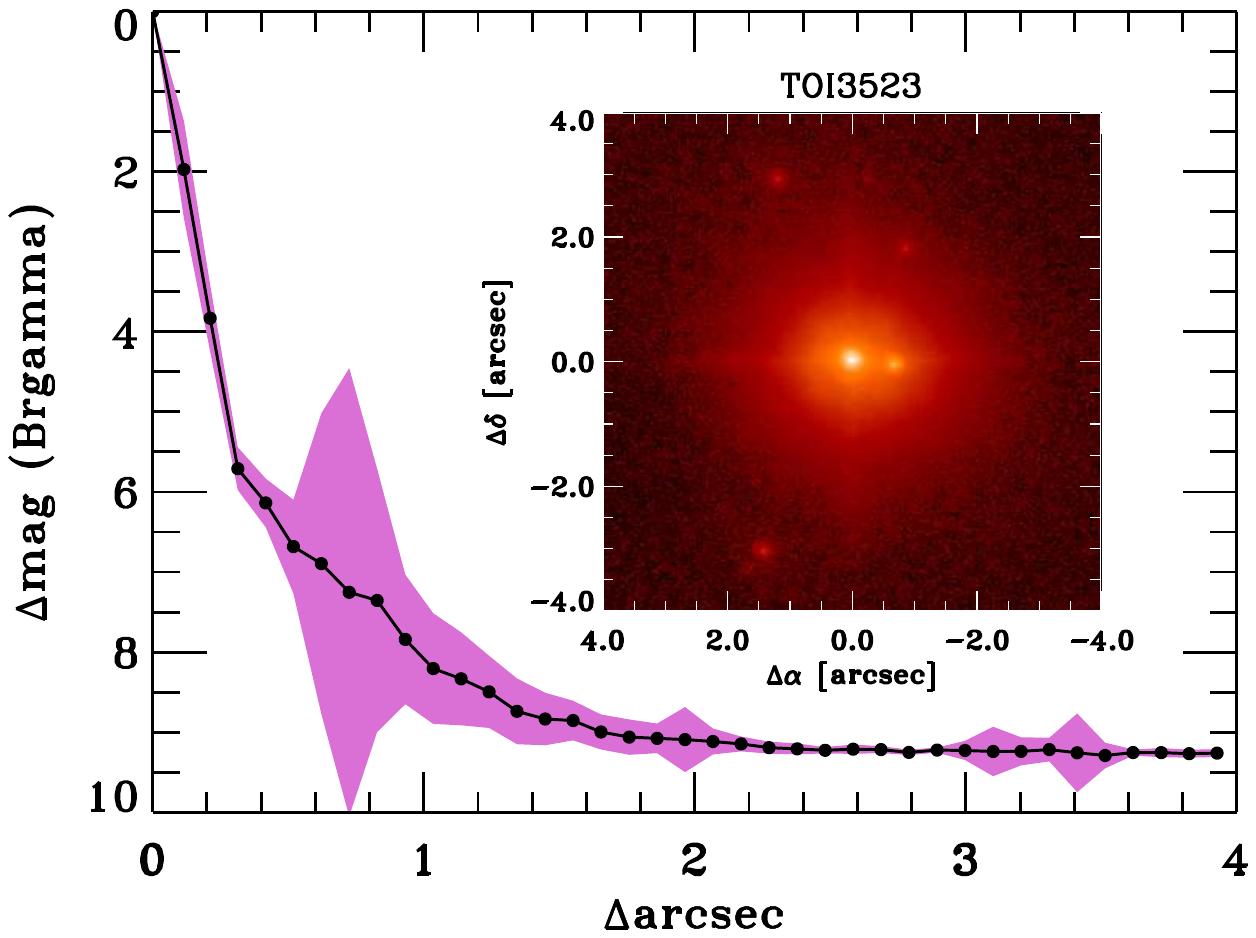}
\includegraphics[width=0.85\linewidth]{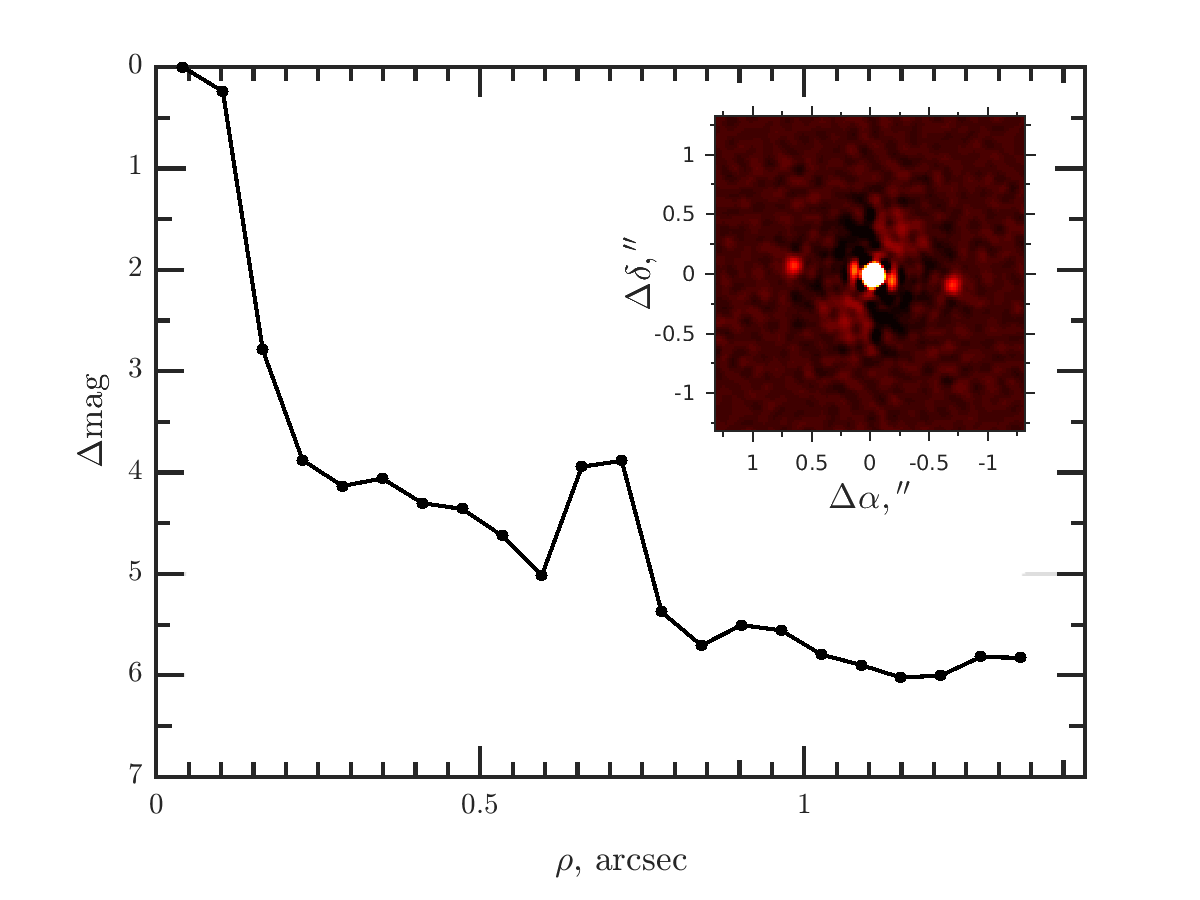}
\end{minipage}
}
\addtocounter{figure}{1}
\end{figure}

\begin{figure*}
\centering
\ContinuedFloat
\subfloat[From left to right, top to bottom: SOAR HRCam, WIYN/NESSI, Palomar/PHARO $H$cont, Palomar/PHARO Br$\gamma$, and SAI Speckle Polarimeter observations of TOI-5181. Evidence of the companion is south of the central star.
In each figure, the solid lines show the detection sensitivity curve of the respective observation.
]{
\label{fig:toi5181_imaging}
\begin{minipage}{\linewidth}
\centering
\includegraphics[width=0.33\linewidth]{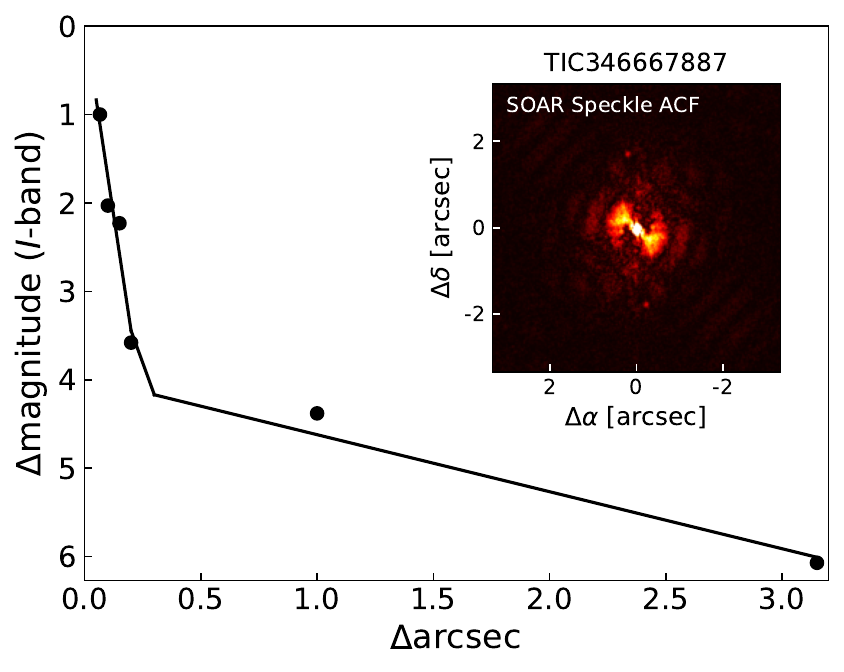}
\includegraphics[width=0.33\linewidth]{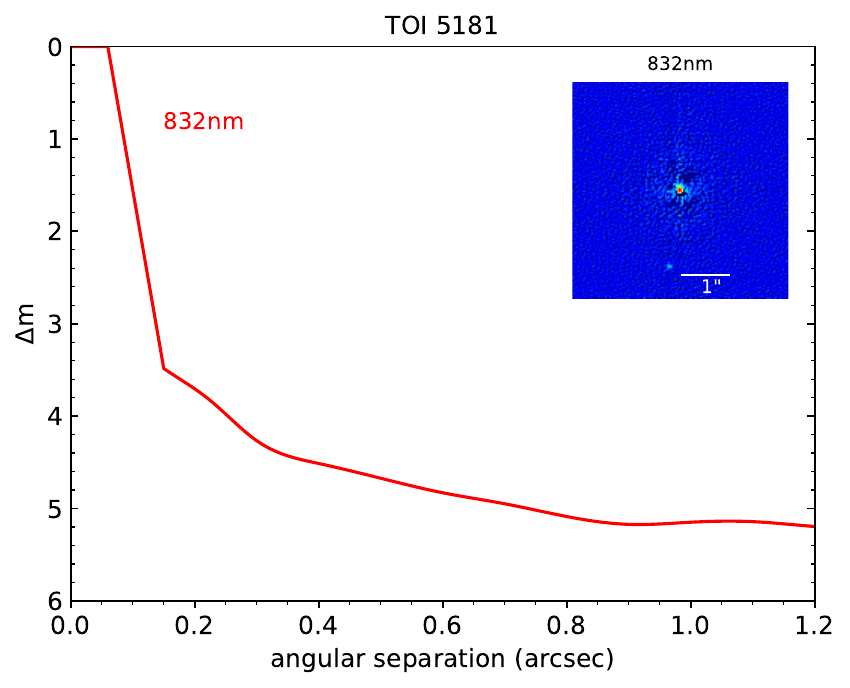}\\
\includegraphics[width=0.33\linewidth]{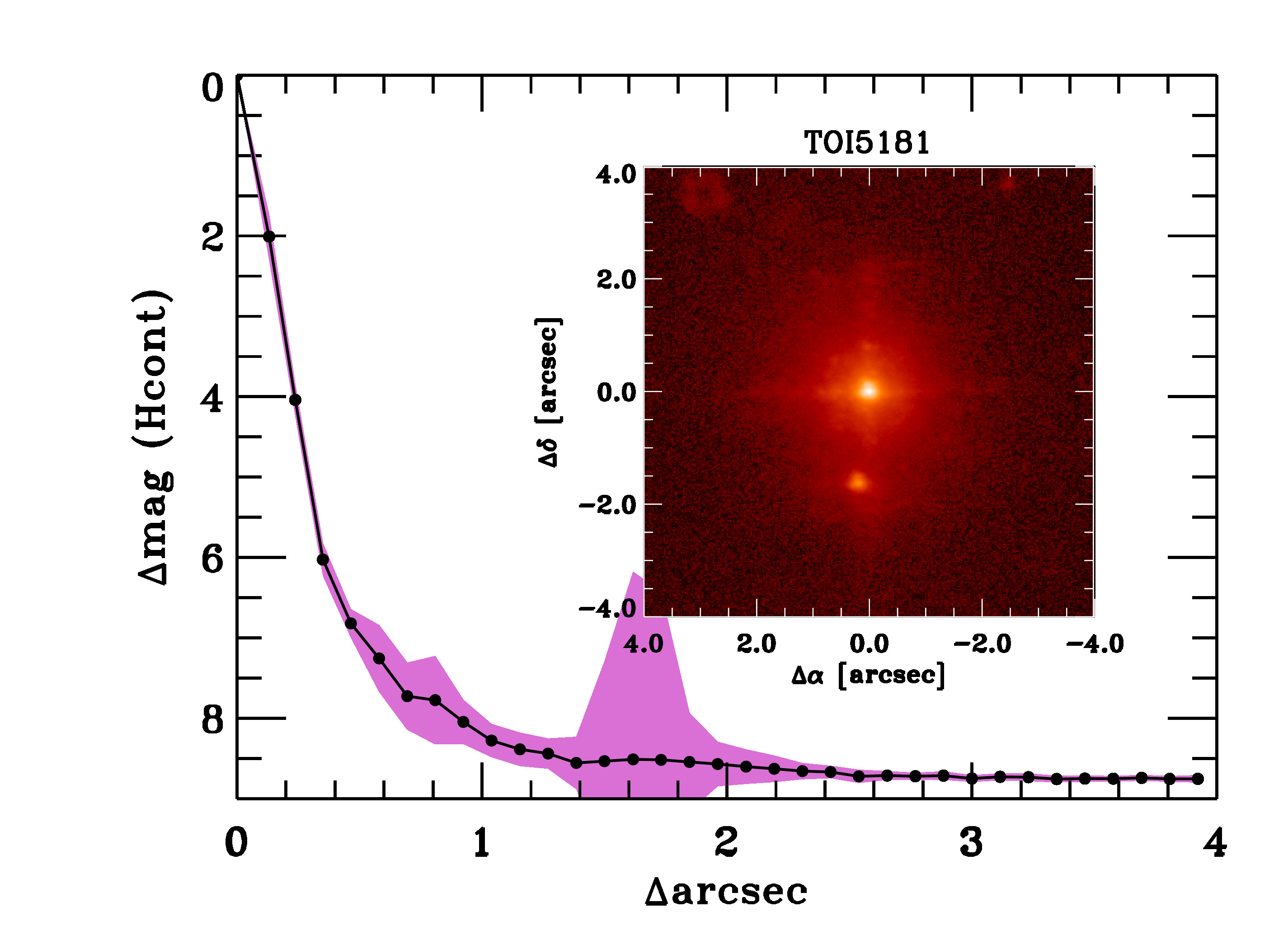}
\includegraphics[width=0.33\linewidth]{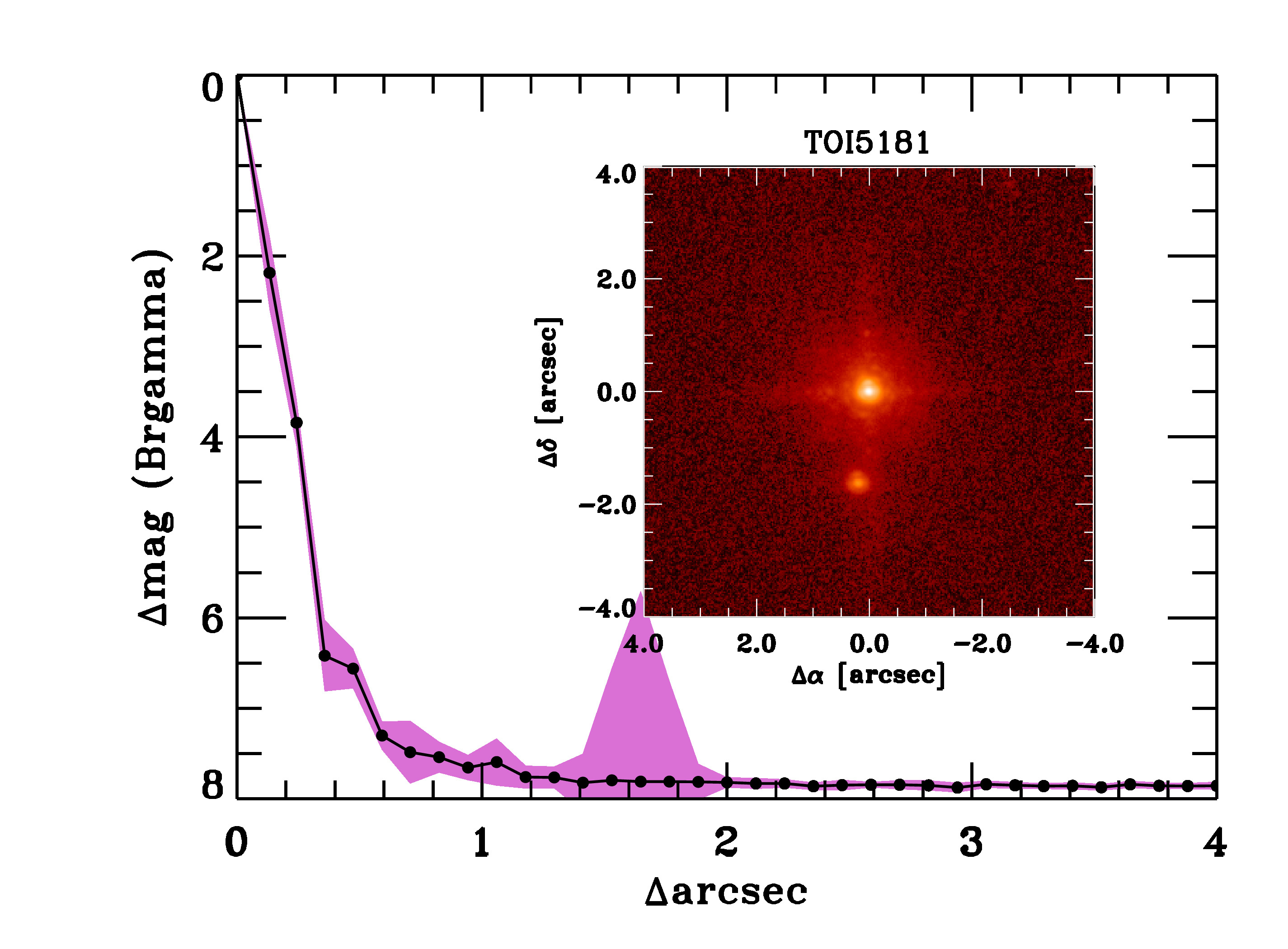}
\includegraphics[width=0.32\linewidth]{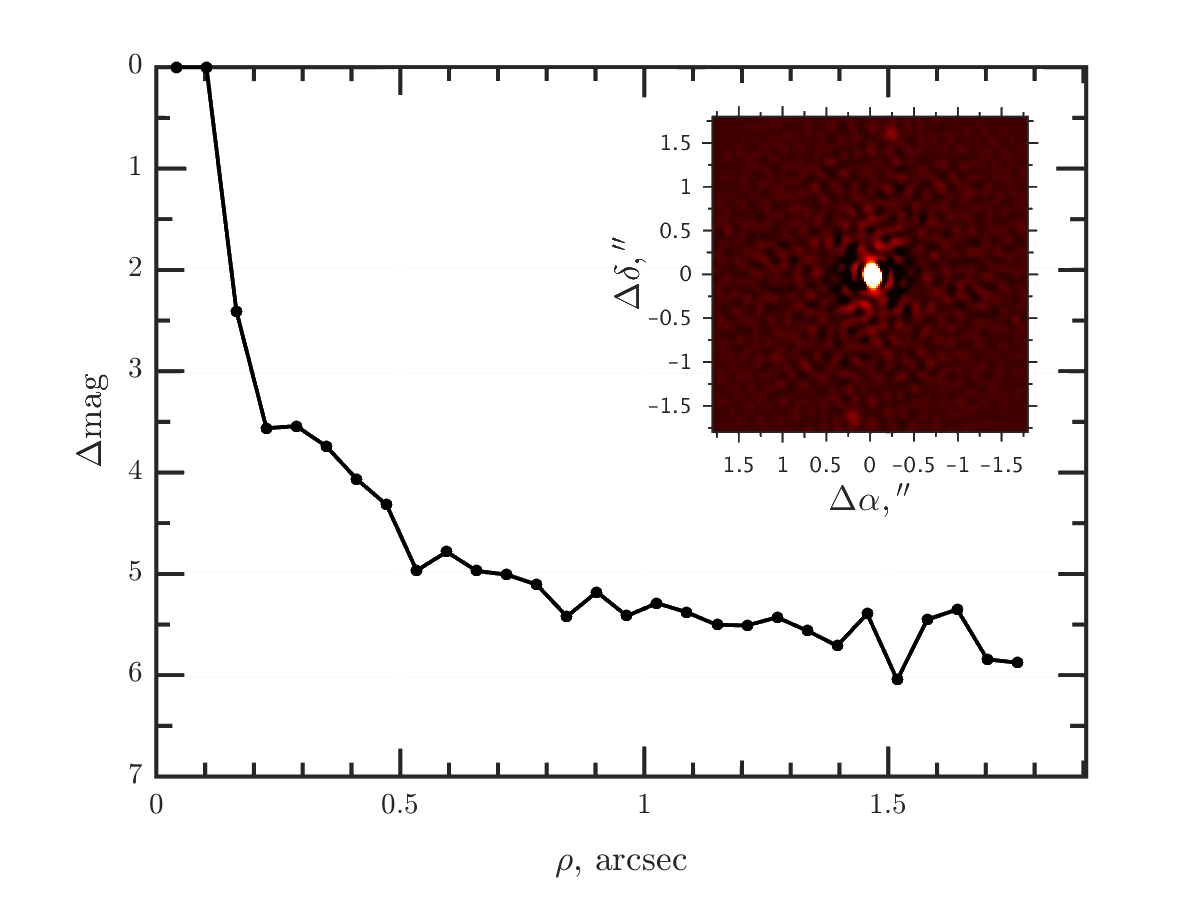}
\end{minipage}}
\addtocounter{figure}{1}
\end{figure*}

\begin{figure*}
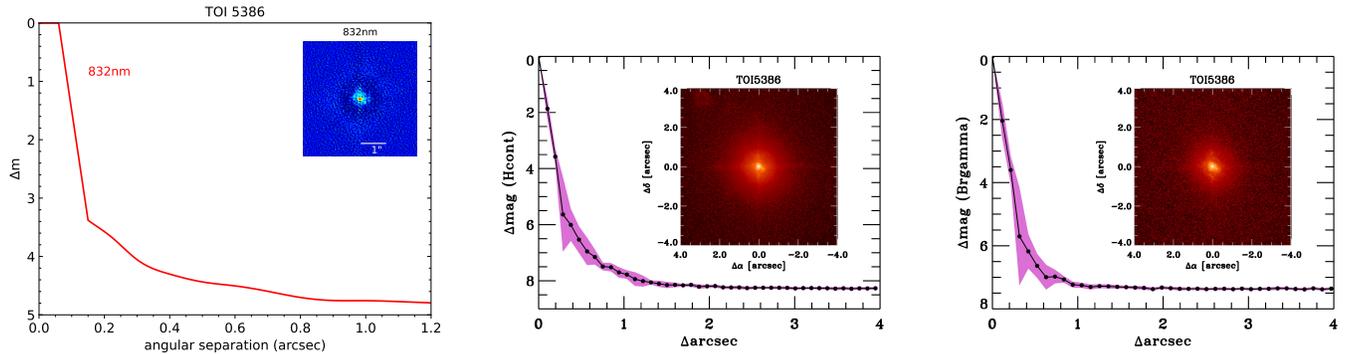

\centering
\ContinuedFloat
\subfloat[Left to right: WIYN/NESSI, Palomar/PHARO $H$cont, Palomar/PHARO Br$\gamma$ observations
of TOI-5386. A close companion is detected at $\approx0\farcs3$ from the primary.]{
\label{fig:toi5386_imaging}
\begin{minipage}{\linewidth}
\includegraphics[width=0.33\linewidth]{toi5386_imaging_WIYN_832nm_20220420.pdf}
\includegraphics[width=0.33\linewidth]{toi5386_imaging_Palomar_Hcont_20230702.jpg}
\includegraphics[width=0.33\linewidth]{toi5386_imaging_Palomar_Brgamma_20230702.jpg}
\end{minipage}}
\addtocounter{figure}{1}
\end{figure*}
\resumelinenumbers
\fi

\subsection{High-Resolution Spectroscopy} \label{ssec:spec}

We obtained high-resolution spectroscopy for each of the candidate host stars,
in order to measure precise relative radial velocities (RVs),
enabling the confirmation of the planets and the determination of their masses. The spectroscopic data were also used to characterize the host stars,
as described in \S\ref{sec:stellar_char}. We summarize the spectroscopic
observations in Table \ref{tab:rvs}. The measured RVs and uncertainties
are provided as a machine-readable table accompanying Figure
 \subref*{fig:toi2031_multiplot}.

\paragraph{TRES Spectroscopy}
We observed \countTOIs{\TRESTOIs} targets with the Tillinghast Reflector Echelle
Spectrograph (TRES; \citealt{TRES_Furesz2008}) on the FLWO 1.5m Tillinghast
Reflector. For most of these objects, we obtained two ``reconnaissance spectra'' at
opposite orbital quadratures. Our goals
were to rule out spectroscopic binaries, perform spectral typing, and obtain
an initial mass estimate of the orbiting object.
The spectra were reduced following the procedures
described by \citet{TRES_Buchhave2010} and \citet{TRES_Quinn2012}.
\added{We derived relative radial-velocities by cross-correlating each spectrum
against a template constructed by summing all the observations for the same target.}
For \expandTOIs{\TRESorbitTOIs}, we obtained additional
high signal-to-noise (S/N) measurements to measure the spectroscopic orbit of the planet.
We fitted the TRES data for TOI-2886 and TOI-5340 in our global analysis;
however in the case of TOI-2031, the scatter in the TRES RV measurements was
comparable to the planetary RV semi-amplitude as revealed by other observations,
so we did not include the TRES data in the final analysis.

\paragraph{HIRES Spectroscopy}
\countTOIs{\HIRESTOIs} of our targets were observed with the High Resolution Echelle 
Spectrometer (HIRES; \citealt{HIRES_Vogt94}) on the Keck-I 10m telescope on
Mauna Kea. We obtained 6--12 observations of each object through the iodine
cell, using the standard observing mode and informal queue system organized
by the California Planet Search (CPS; \citealt{CPS_Howard2010}).
We reduced the spectra using the standard CPS pipelines,
and radial velocities were extracted with the matched template technique
developed by \citet{CPS_Dalba2020a}. This technique substitutes a closely-matched
archival HIRES template spectrum to extract the relative radial velocities, avoiding the need to obtain an expensive high S/N iodine-free
template spectrum for each target. As recommended by
\citet{CPS_Dalba2020a}, we added 4.7~m/s in quadrature to the instrumental
error bars to account for the additional scatter introduced by this technique.

\paragraph{NEID Spectroscopy}
We observed \countTOIs{\NEIDTOIs} targets (\expandTOIs{\NEIDTOIs}) with the NEID
spectrograph on the WIYN telescope at KPNO. These data were obtained in high-resolution (HR) mode, providing
a spectral resolution of $R \sim 110{,}000$. The echelle spectra were
reduced with v1.2 of the standard NEID Data Reduction Pipeline (NEID-DRP)\footnote{\url{https://neid.ipac.caltech.edu/docs/NEID-DRP}}
developed by the NEID instrument team.
To extract RVs, the NEID-DRP computes a cross-correlation between the
observed spectrum and a stellar line mask appropriate to the target
star's spectral type \citep{Baranne1996,Pepe2002}.
\added{A G2 line mask was used for TOI-2196, TOI-2986, TOI-4734, and TOI-5386,
while a G8 line mask was used for TOI-3682, TOI-3980 and TOI-5340.}

The NEID observations were affected by the Contreras Fire in
June 2022 at KPNO, which required the safe shutdown and warming of
the NEID spectrograph.
Following the restart of observations, the spectrograph was used in ``low-precision mode'' between 2023 Nov 11 and Nov 23
while the instrument was still thermally settling. After analysis
of our observations and those of standard stars, we found that even with the
extra noise, the precision was acceptable,
and we did not
treat those data differently.
However, the thermal cycling of the instrument resulted in a small
shift of the optical components leading to an relative RV offset;
thus, for \expandTOIs{\NEIDoffsetTOIs}, we allowed for an additive offset between the RV scales of the data obtained before and after the interruption due to the fire.
The pre-fire and post-fire datasets are labelled as ``NEID1'' and ``NEID2'' in
Table \ref{tab:rvs} and Figure \subref*{fig:toi2031_multiplot}.

\paragraph{PFS Spectroscopy}
\expandTOIs{\PFSTOIs} were observed with the Planet Finder Spectrograph (PFS;
\citealt{PFS_Crane2006,PFS_Crane2008,PFS_Crane2010}) on the Magellan II
Clay 6.5m telescope at Las Campanas Observatory in Chile. We observed these
targets using the 0.3x2.5$"$ slit in 3x3 binning mode, achieving a spectral
resolution of $R\approx110{,}000$. Data were collected with an iodine cell
in the optical path. We additionally
obtained a high signal-to-noise ratio, iodine-free template \replaced{spectrum}{observation} for each target, which
was used to extract relative radial velocities using the custom IDL
pipeline from \citet{PFS_Butler1996}.

\paragraph{CHIRON Spectroscopy}
We observed \countTOIs{\CHIRONTOIs} targets with the CTIO High Resolution Spectrometer
(CHIRON; \citealt{CHIRON_Tokovinin2013,CHIRON_Paredes2021}) on the 1.5m
telescope at CTIO. For \expandTOIs{\CHIRONreconTOIs}, we only obtained
3, 3 and 1 observations respectively with CHIRON, as reconnaissance spectra, 
before continuing observations with PFS.
All observations were taken with the image slicer, with a spectral resolution of $\sim 80{,}000$.
Wavelength solutions were derived using bracketing calibration observations of a ThAr lamp.
The spectra were extracted from the 2D images using the standard CHIRON pipeline, 
\citep{CHIRON_Paredes2021}.
We measured radial velocities from the CHIRON spectra using the procedure described
in \citep{CHIRON_Zhou2020}. In brief, we performed a least-squares deconvolution
between the observed spectra and \replaced{non-rotating ATLAS9 synthetic spectral templates \citep{ATLAS_Castelli2003}}{a non-rotating synthetic spectral template. The template was chosen from the ATLAS9 spectral library \citep{ATLAS_Castelli2003} by selecting the synthetic spectrum that best matched the observed spectrum}. The line profiles are modeled via a rotation broadening kernel
to determine the rotational broadening velocity and stellar radial velocity.

\paragraph{Additional Spectroscopic Analysis}
We assessed the data quality for each observation and excluded any observations
that had an unusually low signal-to-noise ratio
\added{compared with the expected flux given the star's brightness, reflecting
poor conditions during the observation,} or that showed contamination from scattered moonlight.
As part of our planet confirmation procedure, we also checked for line shape
changes that may arise from stellar activity or blended background
eclipsing binaries, which could lead to spurious RV variations. We used 
the procedures described in \citet{Hartman2019} to compute the CCF widths
and bisector inverse slopes (BIS) for each spectrum, using only orders
free from iodine lines. These measurements are provided in the data tables
accompanying this manuscript.

We found no significant variations in the CCF widths for any of our targets.
For all but one of the targets discussed in this paper,
we did not find any correlations between the RVs and BIS measurements
more significant than the $p < 0.05$ level.
The only case where we detected a significant RV-BIS correlation was for the NEID
observations of TOI-4734, where the bisector spans varied by a few tens of m/s,
comparable in magnitude to the RV variations. Performing linear regression between
the RV and BIS measurements resulted in a Pearson product-moment correlation coefficient
of $R^2 = 0.66$, with a $p$-value of 0.02. Such variations could indicate a false positive
scenario such as a blended stellar eclipsing binary \citep[e.g.,][]{Torres2004,Hartman2011}.
However, we found no correlation between the RVs and BIS measured from the HIRES data for the
same system. Furthermore, the bisector span measurements in the NEID data were also
correlated with the overall signal-to-noise of each spectrum ($p = 0.04$), suggesting
that the apparent correlation with the RVs may have been spurious. Given the lack of other
indicators for stellar binarity from transit depth chromaticity, high angular resolution
imaging, or \Gaia RUWE measurements ($\mathrm{RUWE} = 0.981$), we do not further consider the
stellar blend scenario for this system.

\startlongtable
\begin{deluxetable*}{ccrrcc}
\tablecolumns{6}
\tablecaption{Summary of Radial Velocity Measurements \label{tab:rvs}}
\addtocounter{table}{1}
\pdfbookmark[3]{Table \thetable: Summary of RV Data}{rv_summary}%
\addtocounter{table}{-1}
\tablehead{
    \colhead{Target} & \colhead{Instrument} & \colhead{N$_\mathrm{obs}$} & \colhead{Median $\sigma_\mathrm{RV}$ } & \colhead{First Observation Date} & \colhead{Last Observation Date} \\
    &  &  & \colhead{(m/s)$^{a}$} & \colhead{(UT)} & \colhead{(UT)}
}
\startdata
TOI-2031A & Keck-I/HIRES & 7 & 7.4 & 2022 Aug 11 & 2022 Sep 12 \\
$\cdots$ & FLWO/TRES & 8 & 52.0 & 2020 Jul 08 & 2020 Aug 14 \\
TOI-2169A & WIYN/NEID & 13 & 12.8 & 2021 Nov 19 & 2023 Jul 17 \\
$\cdots$ & FLWO/TRES & 2 & 130.1 & 2020 Aug 20 & 2020 Sep 10 \\
TOI-2346 & Keck-I/HIRES & 9 & 12.4 & 2022 Sep 07 & 2022 Nov 14 \\
$\cdots$ & FLWO/TRES & 3 & 197.8 & 2020 Nov 19 & 2020 Dec 01 \\
TOI-2382 & CTIO-1.5m/CHIRON & 11 & 39.0 & 2021 Aug 30 & 2021 Dec 22 \\
$\cdots$ & FLWO/TRES & 2 & 45.4 & 2020 Nov 19 & 2020 Dec 19 \\
TOI-2876 & CTIO-1.5m/CHIRON & 3 & 30.0 & 2023 Jan 14 & 2023 Jan 20 \\
$\cdots$ & Magellan-Clay/PFS & 6 & 3.0 & 2022 Nov 14 & 2023 May 08 \\
$\cdots$ & FLWO/TRES & 2 & 30.7 & 2021 Nov 14 & 2021 Nov 18 \\
TOI-2886 & FLWO/TRES & 12 & 35.5 & 2021 Oct 30 & 2023 Feb 26 \\
TOI-2986 & WIYN/NEID & 12 & 9.3 & 2022 May 10 & 2023 Apr 08 \\
$\cdots$ & FLWO/TRES & 2 & 43.0 & 2021 Dec 09 & 2022 Jan 06 \\
TOI-2992 & CTIO-1.5m/CHIRON & 6 & 58.0 & 2022 Apr 12 & 2022 Jun 09 \\
TOI-3135 & CTIO-1.5m/CHIRON & 7 & 66.0 & 2022 Mar 22 & 2022 Aug 09 \\
TOI-3160A & Magellan-Clay/PFS & 6 & 13.0 & 2022 Mar 14 & 2022 Aug 04 \\
TOI-3464 & CTIO-1.5m/CHIRON & 9 & 56.0 & 2022 Aug 06 & 2023 Apr 11 \\
TOI-3474 & CTIO-1.5m/CHIRON & 11 & 38.0 & 2022 Aug 02 & 2023 Apr 23 \\
TOI-3486 & CTIO-1.5m/CHIRON & 3 & 57.0 & 2021 Aug 23 & 2021 Sep 07 \\
$\cdots$ & Magellan-Clay/PFS & 6 & 3.0 & 2022 Jun 19 & 2022 Sep 05 \\
TOI-3523A & Keck-I/HIRES & 7 & 7.8 & 2022 Sep 01 & 2022 Sep 16 \\
$\cdots$ & FLWO/TRES & 2 & 61.0 & 2021 Aug 05 & 2021 Sep 03 \\
TOI-3593 & Keck-I/HIRES & 8 & 5.3 & 2022 Jul 31 & 2022 Sep 12 \\
$\cdots$ & FLWO/TRES & 2 & 29.8 & 2021 Sep 03 & 2021 Sep 09 \\
TOI-3682 & WIYN/NEID & 8 & 9.8 & 2021 Oct 30 & 2022 Jan 10 \\
$\cdots$ & FLWO/TRES & 2 & 40.9 & 2021 Sep 19 & 2021 Oct 01 \\
TOI-3856 & Keck-I/HIRES & 10 & 5.4 & 2022 Feb 21 & 2023 Jun 07 \\
$\cdots$ & FLWO/TRES & 3 & 35.0 & 2021 Dec 18 & 2022 Jan 25 \\
TOI-3877 & Keck-I/HIRES & 9 & 5.6 & 2022 Feb 15 & 2022 Jun 12 \\
$\cdots$ & FLWO/TRES & 2 & 35.1 & 2022 Jan 02 & 2022 Jan 04 \\
TOI-3980 & WIYN/NEID & 6 & 11.0 & 2021 Nov 20 & 2022 Feb 15 \\
$\cdots$ & FLWO/TRES & 2 & 24.8 & 2021 Sep 03 & 2021 Sep 16 \\
TOI-4214 & CTIO-1.5m/CHIRON & 16 & 30.5 & 2021 Oct 23 & 2022 Feb 08 \\
$\cdots$ & Magellan-Clay/PFS & 6 & 3.0 & 2023 Feb 08 & 2023 May 08 \\
$\cdots$ & FLWO/TRES & 2 & 37.2 & 2021 Oct 30 & 2021 Nov 08 \\
TOI-4487A & Keck-I/HIRES & 6 & 7.8 & 2022 Jul 15 & 2022 Aug 31 \\
TOI-4734 & Keck-I/HIRES & 12 & 5.9 & 2022 Aug 31 & 2023 May 02 \\
$\cdots$ & WIYN/NEID & 14 & 8.7 & 2022 Nov 24 & 2024 Jan 31 \\
$\cdots$ & FLWO/TRES & 3 & 36.7 & 2022 Jan 07 & 2022 Dec 22 \\
TOI-4794 & CTIO-1.5m/CHIRON & 12 & 67.5 & 2023 Jan 19 & 2023 Apr 26 \\
$\cdots$ & FLWO/TRES & 2 & 50.5 & 2022 Jan 13 & 2022 Jan 22 \\
TOI-4961 & CTIO-1.5m/CHIRON & 1 & 46.0 & 2022 Apr 06 & 2022 Apr 06 \\
$\cdots$ & Magellan-Clay/PFS & 7 & 2.9 & 2023 Feb 09 & 2023 May 08 \\
TOI-5181A & Keck-I/HIRES & 7 & 6.6 & 2022 Jun 08 & 2022 Jul 29 \\
$\cdots$ & FLWO/TRES & 2 & 38.7 & 2022 Mar 19 & 2022 Mar 25 \\
TOI-5210 & Keck-I/HIRES & 9 & 5.7 & 2022 Jun 26 & 2022 Jul 25 \\
$\cdots$ & FLWO/TRES & 2 & 35.1 & 2022 Apr 22 & 2022 May 03 \\
TOI-5322 & Keck-I/HIRES & 6 & 5.7 & 2022 Sep 01 & 2022 Sep 12 \\
$\cdots$ & FLWO/TRES & 2 & 34.5 & 2022 Jun 29 & 2022 Jul 07 \\
TOI-5340 & WIYN/NEID & 5 & 22.3 & 2022 Dec 20 & 2023 Mar 08 \\
$\cdots$ & FLWO/TRES & 12 & 33.6 & 2022 Mar 06 & 2022 Sep 17 \\
TOI-5386A & Keck-I/HIRES & 7 & 5.4 & 2022 Jun 10 & 2022 Jul 15 \\
$\cdots$ & WIYN/NEID & 5 & 6.6 & 2023 Mar 28 & 2023 Jun 01 \\
$\cdots$ & FLWO/TRES & 2 & 46.0 & 2022 Mar 27 & 2022 Apr 05 \\
TOI-5592 & Keck-I/HIRES & 7 & 6.6 & 2022 Oct 18 & 2023 Jan 09 \\
$\cdots$ & FLWO/TRES & 2 & 42.2 & 2022 Oct 03 & 2022 Oct 12%
\enddata
\tablenotetext{a}{Median instrumental RV uncertainty for each target and instrument.}
\tablecomments{The complete table of RV measurements is available in machine-readable form as Data behind the Figure for Figure Set \subref*{fig:toi2031_multiplot}, provided as relative RVs with an arbitrary target- and instrument-specific offset subtracted.}
\end{deluxetable*} 
 \label{sf:rv_summary}%

\subsection{Catalog Photometry and Astrometry} \label{ssec:catalog}

We queried various catalogs to obtain archival data for each target.
These were used as inputs to our global analysis, where they constrain the
host star properties through their spectral energy distributions.
We obtained the photometric and astrometric data from the \Gaia Data Release 3 \citep{GaiaEDR3_Brown2021,GaiaEDR3_Riello2021,GaiaEDR3_Lindegren2021,GaiaDR3_Vallenari2022},
as well as the photometry from the \TESS Input Catalog (TIC; \citealp{TIC_Stassun2018,TIC_Stassun2019}), the AAVSO Photometric All Sky Survey (APASS) DR6 as compiled in the UCAC4 catalog \citep{UCAC4_Zacharias2013}, \textit{2MASS} \citep{TMASS_Cutri2003,TMASS_Skrutskie2006}, and \textit{WISE} \citep{WISE_Cutri2012} catalogs.
Our use of $BVgri$ photometry from APASS DR6 is one point of difference from Papers \citetalias{Yee2022} and \citetalias{Yee2023}, which only fitted $B$, $V$-band photometry from the Tycho-2 catalog.
We used the \texttt{astroquery} package \citep{Astroquery_Ginsburg2019} to perform our catalog queries.
The \Gaia astrometry was corrected for the known parallax zero-point error \citep{GaiaEDR3_Lindegren2021} using the code provided by the \Gaia team.\footnote{\url{https://gitlab.com/icc-ub/public/gaiadr3_zeropoint}}

\begin{rotatepage}

\foreach \n [count=\ni] in {0,1,...,\numTables} {%
\movetabledown=1.5in
\begin{rotatetable}
\begin{deluxetable*}{lcccccc}
\ifnum\ni=1%
\tablecaption{Catalog Photometry and Astrometry of Planet Host Stars \label{tab:catalog_props}}
\addtocounter{table}{1}
\pdfbookmark[3]{Table \thetable: Catalog Stellar Properties}{catalog_props}%
\addtocounter{table}{-1}
\else
\tablecaption{\textit{(Continued)}}
\fi
\tabletypesize{\small}
\input{catalog_props_\n}

\ifnum\ni=\lastTableNum%
\tablerefsmod{\textbf{Sources:} (1) - \Gaia DR3 \citep{GaiaEDR3_Brown2021};
(2) - \TESS Input Catalog \citep{TIC_Stassun2019};
(3) - APASS DR6 as contained in UCAC4 \citep{UCAC4_Zacharias2013};
(4) - 2MASS \citep{TMASS_Cutri2003,TMASS_Skrutskie2006};
(5) - WISE \citep{WISE_Cutri2012}.}
\tablecommentsmod{The catalog photometry presented here has not been corrected for contamination by nearby stellar companions (\S\ref{M-ssec:companion_modelling}).\\
The data in this table are available in machine-readable form.}
\fi
\end{deluxetable*}
\end{rotatetable}
\addtocounter{table}{-1}
}
\addtocounter{table}{1}
\end{rotatepage}
\pdfpageattr{}

\ifSubfilesClassLoaded{%
\bibliography{catalogs}
}{}
\end{document} \label{sf:catalog_props}%
\pdfpageattr{}

\subsubsection{\Gaia-detected Companions} \label{ssec:gaia_companions}


\foreach \n [count=\ni] in {0,1,2} {%
\centering
\begin{deluxetable*}{lcccccc}
\ifnum\ni=1%
\tablecaption{Observed Properties of Stellar Companions \label{tab:stellar_comps}}
\addtocounter{table}{1}
\pdfbookmark[4]{Table \thetable: Observed Properties of Stellar Companions}{stellar_comps}%
\addtocounter{table}{-1}
\else
\tablecaption{\textit{(Continued)}}
\fi
\tabletypesize{\scriptsize}
\input{stellar_companions_\n}

\ifnum\ni=3%
\tablecommentsmod{
This table contains observed properties of planet hosts and stellar companions detected either through
high angular-resolution imaging (\S\ref{sssec:ao_detected_companions}) or as a bound companion in \Gaia (\S\ref{ssec:gaia_companions}).
If the companion star was detected by \Gaia, $\mathcal{R}_\mathrm{chance}$ is the ``chance alignment ratio'' defined in \citet{GaiaEDR3_Binaries_El-Badry2021}; objects with $\mathcal{R} < 0.1$ have a high likelihood of being a bound pair.
$T$ magnitude is from the TESS Input Catalog \citep{TIC_Stassun2018,TIC_Stassun2019}, while all remaining parameters, including systemic RVs, are drawn from \Gaia DR3 \citep{GaiaDR3_Vallenari2022,GaiaDR3_RVs_Katz2022}.
If the companion was detected in high angular-resolution imaging, we provide the magnitude differences measured from those observations.
}
\fi
\end{deluxetable*}
\addtocounter{table}{-1}
}
\addtocounter{table}{1}

\ifSubfilesClassLoaded{%
\bibliography{../catalogs}
}{}

\end{document} \label{sf:all_companions}%

The precise parallax and proper motion measurements from \Gaia also provide the opportunity to check for wide stellar companions to our planet host stars.
\citet{GaiaEDR3_Binaries_El-Badry2021} compiled a catalog of stellar binaries derived from \Gaia DR3, requiring potential wide binary candidates to have parallaxes, projected separations, and proper motions consistent with a Keplerian orbit.
\Gaia is able to identify binaries even with relatively faint secondaries down to angular separations of $\gtrsim 1"$, making it complementary to the high-angular resolution imaging we obtained for these objects, which probe closer separations.

We checked our list of planet candidate hosts against the \citet{GaiaEDR3_Binaries_El-Badry2021} stellar binary catalog.
Five of the targets --- TOI-2031, TOI-2169, TOI-3523, TOI-4487, and TOI-5181 --- have nearby stars with similar parallaxes and proper motions consistent with being a bound companion, and their properties are described in Table \ref{tab:stellar_comps}.
For TOI-2169, both components also have RV measurements from the \Gaia Radial Velocity Spectrograph (RVS) instrument that are consistent within the uncertainties, providing further evidence that the pair is bound.
As described in \S\ref{sssec:ao_detected_companions}, the companion to TOI-5181 was also detected in high angular-resolution imaging.

These pairs have angular separations ranging from $1\farcs6$ to 46$^{\prime\prime}$, corresponding to projected physical separations of 800-12000~AU.
In each case, the star that exhibits the planetary transits is the brighter component of the stellar system.
\citet{Ngo2016} found that systems that host hot Jupiters are more likely than field stars to have stellar companions between 50-2000~AU; the ability of \Gaia to identify stellar companions over a wide range of separations will soon enable studies extending that result.

For completeness, we also include the \Gaia astrometry and photometry for a star that is $2\farcs6$ away from TOI-3464 that was also detected by speckle interferometry by SOAR HRCam.
In this case, the \Gaia parallax measurements indicate that the pair are not associated; they are a chance alignment of stars at different distances.


\section{Spectroscopic Stellar Characterization} \label{sec:stellar_char}

We characterized each planet host star using the high-resolution spectra
obtained as part of our radial velocity monitoring. The resulting stellar
properties are reported in Table \ref{tab:spec_props}.

For stars with observations from PFS, HIRES, NEID, or CHIRON, we used the \texttt{SpecMatch-Emp} code
\citep{SpecMatchEmp_Yee2017}\footnote{\url{https://github.com/samuelyeewl/specmatch-emp}}
to match the target spectrum to a library of empirically observed spectra from
stars with well-determined properties and thereby derive \Teff, \Rstar, and \feh.
For PFS and HIRES observations, we used the iodine-free template spectra.
For the NEID and CHIRON observations, we
stacked all observations to obtain a higher S/N spectrum for this analysis
(we note that in Papers \citetalias{Yee2022} and \citetalias{Yee2023}, we used
only a single observation for this purpose).
Because the empirical library spectra have different intrinsic levels
of broadening from stellar rotation, $\vsini$ measurements from this technique are unreliable.
Instead, we measured \vsini with the \texttt{SpecMatch-Synth} code
\citep{SpecMatchSynth_Petigura2015},
which uses the \citet{Kurucz1993} synthetic spectral library to match the target spectrum
and accounts for both rotational broadening and macroturbulent broadening
using the  $v_\mathrm{mac}-\Teff$ relation from \citet{Valenti2005}.

For stars with observations from TRES or CHIRON, we also obtained stellar
parameters from the Stellar Parameter Classification code (SPC;
\citealt{SPC_Buchhave2012}), which uses the \citet{Kurucz1993} model grid.
We classified each TRES observation of the same target separately, and we adopted the
median value from all observations and their standard deviation as the
uncertainty for each parameter, with an error floor of 50\,K in \Teff,
0.1~dex in \logg, 0.08~dex in \feh, and 0.5~\kms in \vsini.
\added{The adopted error floors are based on the results from \citet{SPC_Buchhave2012},
who performed an empirical uncertainty analysis using stars with multiple observations
to determine the internal uncertainty as a function of observation signal-to-noise.}

The CHIRON spectra were analyzed by matching against a library of
$\sim$10,000 observed spectra that were previously classified using SPC,
as described by \citet{CHIRON_Zhou2020}. For these results, we adopted
an error floor of 50\,K in \Teff, 0.1~dex in \logg, 0.08~dex in \feh, and
0.5~\kms in \vsini, \added{the same as those used for the SPC analysis of
the TRES spectra.}

All spectroscopic properties are recorded in Table \ref{tab:spec_props}.
When multiple derivations were available, we found
the parameters from different methods to be broadly consistent.
We used only the spectroscopically-derived \feh as a prior for our global fits (\S\ref{sec:planet_char}),
adopting the \texttt{SpecMatch}-derived value when available for consistency
with our previous work.
\added{We note that the quoted uncertainties for the spectroscopic measurements
in this section do not account for the systematic uncertainties between
model grids or fundamental limits in the calibration of physical parameter scales
\citep[e.g.,][]{Tayar2022}. We refer the reader to the stellar properties tabulated
in Table \ref{tab:fitted_props} for values and uncertainties derived from a full
global modeling of the system and accounting for these limitations.}


\startlongtable
\begin{deluxetable*}{cccccccccc}
\tablecaption{Spectroscopic Stellar Properties \label{tab:spec_props}}
\addtocounter{table}{1}
\pdfbookmark[2]{Table \thetable: Spectroscopic Stellar Properties}{spec_props}%
\addtocounter{table}{-1}
\tablecolumns{9}
\tablehead{
    \colhead{Target} & \colhead{Code} & \colhead{Instrument} &
    \colhead{\Teff} & \colhead{\Rstar} & \colhead{\logg} & \colhead{\feh} &
    \colhead{\vsini} & \colhead{\vmac} &
    \colhead{Adopted}\tablenotemark{a} \\
     & & & \colhead{(K)} & \colhead{(\Rsun)} & \colhead{(dex)} & \colhead{(dex)} & \colhead{(\kms)} & \colhead{(\kms)} &
}
\startdata
TOI-2031A & SpecMatch & HIRES & 6135 $\pm$ 110 & 1.24 $\pm$ 0.22 & -- & -0.28 $\pm$ 0.09 & 5.8 $\pm$ 1.0 & 4.6 $\pm$ 0.2 & Y \\
$\cdots$ & SPC & TRES & 6027 $\pm$ 99 & -- & 4.22 $\pm$ 0.16 & -0.04 $\pm$ 0.09 & 7.8 $\pm$ 0.7 & -- & N \\
TOI-2169A & SpecMatch & NEID & 6072 $\pm$ 110 & 1.76 $\pm$ 0.32 & -- & 0.10 $\pm$ 0.09 & 8.5 $\pm$ 1.0 & 4.7 $\pm$ 0.2 & Y \\
$\cdots$ & SPC & TRES & 6114 $\pm$ 84 & -- & 4.06 $\pm$ 0.14 & 0.29 $\pm$ 0.08 & 11.0 $\pm$ 0.5 & -- & N \\
TOI-2346 & SpecMatch & HIRES & 6187 $\pm$ 110 & 1.67 $\pm$ 0.30 & -- & -0.19 $\pm$ 0.09 & 11.1 $\pm$ 1.0 & 5.0 $\pm$ 0.2 & Y \\
$\cdots$ & SPC & TRES & 6300 $\pm$ 95 & -- & 4.25 $\pm$ 0.16 & 0.18 $\pm$ 0.09 & 14.8 $\pm$ 0.6 & -- & N \\
TOI-2382 & SpecMatch & CHIRON & 5947 $\pm$ 110 & 1.24 $\pm$ 0.22 & -- & 0.23 $\pm$ 0.09 & 3.3 $\pm$ 1.0 & 4.2 $\pm$ 0.2 & Y \\
$\cdots$ & SPC & CHIRON & 5867 $\pm$ 98 & -- & 4.08 $\pm$ 0.12 & 0.14 $\pm$ 0.10 & 5.7 $\pm$ 0.5 & -- & N \\
$\cdots$ & SPC & TRES & 5892 $\pm$ 50 & -- & 4.24 $\pm$ 0.10 & 0.34 $\pm$ 0.08 & 4.9 $\pm$ 0.5 & -- & N \\
TOI-2876 & SpecMatch & PFS & 5145 $\pm$ 110 & 0.95 $\pm$ 0.09 & -- & 0.32 $\pm$ 0.09 & 1.1 $\pm$ 1.0 & 3.2 $\pm$ 0.2 & Y \\
$\cdots$ & SPC & CHIRON & 5314 $\pm$ 110 & -- & 4.01 $\pm$ 0.46 & 0.18 $\pm$ 0.10 & 5.1 $\pm$ 0.5 & -- & N \\
$\cdots$ & SPC & TRES & 5274 $\pm$ 50 & -- & 4.44 $\pm$ 0.10 & 0.34 $\pm$ 0.08 & 3.1 $\pm$ 0.5 & -- & N \\
TOI-2886 & SpecMatch & TRES & 6011 $\pm$ 85 & -- & -- & 0.30 $\pm$ 0.08 & -- & -- & N \\
$\cdots$ & SPC & TRES & 6011 $\pm$ 85 & -- & 4.29 $\pm$ 0.14 & 0.30 $\pm$ 0.08 & 8.6 $\pm$ 0.5 & -- & Y \\
TOI-2986 & SpecMatch & NEID & 5749 $\pm$ 110 & 1.63 $\pm$ 0.29 & -- & 0.14 $\pm$ 0.09 & 2.7 $\pm$ 1.0 & 4.2 $\pm$ 0.2 & Y \\
$\cdots$ & SPC & TRES & 5907 $\pm$ 50 & -- & 4.13 $\pm$ 0.10 & 0.12 $\pm$ 0.08 & 4.7 $\pm$ 0.5 & -- & N \\
TOI-2992 & SpecMatch & CHIRON & 6019 $\pm$ 110 & 1.49 $\pm$ 0.27 & -- & 0.11 $\pm$ 0.09 & 7.7 $\pm$ 1.0 & 4.4 $\pm$ 0.2 & Y \\
$\cdots$ & SPC & CHIRON & 6030 $\pm$ 50 & -- & 4.18 $\pm$ 0.10 & 0.04 $\pm$ 0.10 & 8.6 $\pm$ 0.5 & -- & N \\
TOI-3135 & SpecMatch & CHIRON & 5844 $\pm$ 110 & 1.07 $\pm$ 0.19 & -- & 0.10 $\pm$ 0.09 & 11.6 $\pm$ 1.0 & 4.3 $\pm$ 0.2 & Y \\
$\cdots$ & SPC & CHIRON & 5844 $\pm$ 61 & -- & 4.46 $\pm$ 0.10 & -0.05 $\pm$ 0.10 & 12.4 $\pm$ 0.5 & -- & N \\
TOI-3160A & SpecMatch & PFS & 6163 $\pm$ 110 & 1.45 $\pm$ 0.26 & -- & -0.16 $\pm$ 0.09 & 9.3 $\pm$ 1.0 & 4.7 $\pm$ 0.2 & Y \\
$\cdots$ & SPC & TRES & 5776 $\pm$ 95 & -- & 3.49 $\pm$ 0.16 & 0.41 $\pm$ 0.09 & 13.7 $\pm$ 0.6 & -- & N \\
TOI-3464 & SpecMatch & CHIRON & 6051 $\pm$ 110 & 1.72 $\pm$ 0.31 & -- & 0.02 $\pm$ 0.09 & 5.9 $\pm$ 1.0 & 4.4 $\pm$ 0.2 & Y \\
$\cdots$ & SPC & CHIRON & 6096 $\pm$ 52 & -- & 4.13 $\pm$ 0.10 & -0.02 $\pm$ 0.12 & 7.2 $\pm$ 0.5 & -- & N \\
TOI-3474 & SpecMatch & CHIRON & 5755 $\pm$ 110 & 1.32 $\pm$ 0.24 & -- & 0.18 $\pm$ 0.09 & 3.3 $\pm$ 1.0 & 4.0 $\pm$ 0.2 & Y \\
$\cdots$ & SPC & CHIRON & 5736 $\pm$ 166 & -- & 4.10 $\pm$ 0.19 & 0.03 $\pm$ 0.19 & 5.6 $\pm$ 0.5 & -- & N \\
TOI-3486 & SpecMatch & PFS & 4933 $\pm$ 110 & 0.80 $\pm$ 0.08 & -- & 0.23 $\pm$ 0.09 & 2.7 $\pm$ 1.0 & 2.8 $\pm$ 0.2 & Y \\
$\cdots$ & SPC & CHIRON & 5134 $\pm$ 121 & -- & 4.57 $\pm$ 0.10 & -0.07 $\pm$ 0.10 & 11.0 $\pm$ 0.5 & -- & N \\
TOI-3523A & SpecMatch & HIRES & 6208 $\pm$ 110 & 1.69 $\pm$ 0.30 & -- & -0.29 $\pm$ 0.09 & 8.0 $\pm$ 1.0 & 4.9 $\pm$ 0.2 & Y \\
$\cdots$ & SPC & TRES & 6378 $\pm$ 81 & -- & 4.20 $\pm$ 0.14 & 0.25 $\pm$ 0.08 & 10.3 $\pm$ 0.5 & -- & N \\
TOI-3593 & SpecMatch & HIRES & 5318 $\pm$ 110 & 1.01 $\pm$ 0.18 & -- & 0.37 $\pm$ 0.09 & 1.9 $\pm$ 1.0 & 3.6 $\pm$ 0.2 & Y \\
$\cdots$ & SPC & TRES & 5510 $\pm$ 50 & -- & 4.40 $\pm$ 0.10 & 0.46 $\pm$ 0.08 & 3.8 $\pm$ 0.5 & -- & N \\
TOI-3682 & SpecMatch & NEID & 5614 $\pm$ 110 & 1.72 $\pm$ 0.31 & -- & 0.40 $\pm$ 0.09 & 3.5 $\pm$ 1.0 & 3.9 $\pm$ 0.2 & Y \\
$\cdots$ & SPC & TRES & 5820 $\pm$ 50 & -- & 4.18 $\pm$ 0.10 & 0.47 $\pm$ 0.08 & 5.3 $\pm$ 0.5 & -- & N \\
TOI-3856 & SpecMatch & HIRES & 5463 $\pm$ 110 & 1.16 $\pm$ 0.21 & -- & 0.42 $\pm$ 0.09 & 2.7 $\pm$ 1.0 & 3.7 $\pm$ 0.2 & Y \\
$\cdots$ & SPC & TRES & 5653 $\pm$ 50 & -- & 4.43 $\pm$ 0.10 & 0.40 $\pm$ 0.08 & 3.7 $\pm$ 0.5 & -- & N \\
TOI-3877 & SpecMatch & HIRES & 5674 $\pm$ 110 & 1.37 $\pm$ 0.25 & -- & 0.25 $\pm$ 0.09 & 2.6 $\pm$ 1.0 & 3.9 $\pm$ 0.2 & Y \\
$\cdots$ & SPC & TRES & 5589 $\pm$ 60 & -- & 3.98 $\pm$ 0.10 & 0.36 $\pm$ 0.08 & 5.1 $\pm$ 0.5 & -- & N \\
TOI-3980 & SpecMatch & NEID & 5634 $\pm$ 110 & 1.76 $\pm$ 0.32 & -- & 0.14 $\pm$ 0.09 & 3.3 $\pm$ 1.0 & 4.0 $\pm$ 0.2 & Y \\
$\cdots$ & SPC & TRES & 5817 $\pm$ 50 & -- & 4.14 $\pm$ 0.10 & 0.16 $\pm$ 0.08 & 4.9 $\pm$ 0.5 & -- & N \\
TOI-4214 & SpecMatch & PFS & 5830 $\pm$ 110 & 1.50 $\pm$ 0.27 & -- & 0.23 $\pm$ 0.09 & 4.0 $\pm$ 1.0 & 4.4 $\pm$ 0.2 & Y \\
$\cdots$ & SPC & CHIRON & 5993 $\pm$ 91 & -- & 4.06 $\pm$ 0.15 & 0.13 $\pm$ 0.10 & 6.2 $\pm$ 0.5 & -- & N \\
$\cdots$ & SPC & TRES & 6013 $\pm$ 59 & -- & 4.24 $\pm$ 0.10 & 0.44 $\pm$ 0.08 & 6.2 $\pm$ 0.5 & -- & N \\
TOI-4487A & SpecMatch & HIRES & 6080 $\pm$ 110 & 1.50 $\pm$ 0.27 & -- & 0.16 $\pm$ 0.09 & 8.5 $\pm$ 1.0 & 4.5 $\pm$ 0.2 & Y \\
TOI-4734 & SpecMatch & HIRES & 5763 $\pm$ 110 & 1.88 $\pm$ 0.34 & -- & 0.34 $\pm$ 0.09 & 4.8 $\pm$ 1.0 & 4.2 $\pm$ 0.2 & Y \\
$\cdots$ & SPC & TRES & 5923 $\pm$ 50 & -- & 4.23 $\pm$ 0.10 & 0.44 $\pm$ 0.08 & 6.0 $\pm$ 0.5 & -- & N \\
TOI-4794 & SpecMatch & CHIRON & 6201 $\pm$ 110 & 1.49 $\pm$ 0.27 & -- & -0.21 $\pm$ 0.09 & 7.9 $\pm$ 1.0 & 4.6 $\pm$ 0.2 & Y \\
$\cdots$ & SPC & CHIRON & 6253 $\pm$ 86 & -- & 4.10 $\pm$ 0.10 & -0.06 $\pm$ 0.10 & 8.5 $\pm$ 0.5 & -- & N \\
$\cdots$ & SPC & TRES & 6205 $\pm$ 60 & -- & 4.25 $\pm$ 0.11 & 0.07 $\pm$ 0.08 & 9.5 $\pm$ 0.5 & -- & N \\
TOI-4961 & SpecMatch & PFS & 5348 $\pm$ 110 & 1.02 $\pm$ 0.18 & -- & 0.21 $\pm$ 0.09 & 1.8 $\pm$ 1.0 & 3.5 $\pm$ 0.2 & Y \\
$\cdots$ & SPC & CHIRON & 5370 $\pm$ 50 & -- & 3.83 $\pm$ 0.10 & 0.22 $\pm$ 0.10 & 5.2 $\pm$ 0.5 & -- & N \\
TOI-5181A & SpecMatch & HIRES & 5924 $\pm$ 110 & 1.62 $\pm$ 0.29 & -- & 0.27 $\pm$ 0.09 & 7.4 $\pm$ 1.0 & 4.5 $\pm$ 0.2 & Y \\
$\cdots$ & SPC & TRES & 6070 $\pm$ 59 & -- & 4.28 $\pm$ 0.10 & 0.41 $\pm$ 0.08 & 8.6 $\pm$ 0.5 & -- & N \\
TOI-5210 & SpecMatch & HIRES & 5781 $\pm$ 110 & 1.15 $\pm$ 0.21 & -- & 0.28 $\pm$ 0.09 & 3.0 $\pm$ 1.0 & 4.2 $\pm$ 0.2 & Y \\
$\cdots$ & SPC & TRES & 5952 $\pm$ 50 & -- & 4.35 $\pm$ 0.10 & 0.32 $\pm$ 0.08 & 4.8 $\pm$ 0.5 & -- & N \\
TOI-5322 & SpecMatch & HIRES & 5667 $\pm$ 110 & 1.20 $\pm$ 0.22 & -- & -0.37 $\pm$ 0.09 & 2.2 $\pm$ 1.0 & 3.9 $\pm$ 0.2 & Y \\
$\cdots$ & SPC & TRES & 5750 $\pm$ 62 & -- & 4.30 $\pm$ 0.11 & -0.08 $\pm$ 0.08 & 3.4 $\pm$ 0.5 & -- & N \\
TOI-5340 & SpecMatch & NEID & 6114 $\pm$ 110 & 1.92 $\pm$ 0.35 & -- & 0.18 $\pm$ 0.09 & 8.5 $\pm$ 1.0 & 4.8 $\pm$ 0.2 & Y \\
$\cdots$ & SPC & TRES & 6221 $\pm$ 92 & -- & 4.09 $\pm$ 0.15 & 0.23 $\pm$ 0.09 & 10.2 $\pm$ 0.6 & -- & N \\
TOI-5386A & SpecMatch & HIRES & 5828 $\pm$ 110 & 1.42 $\pm$ 0.26 & -- & 0.22 $\pm$ 0.09 & 2.3 $\pm$ 1.0 & 4.3 $\pm$ 0.2 & Y \\
$\cdots$ & SPC & TRES & 5927 $\pm$ 50 & -- & 4.27 $\pm$ 0.10 & 0.25 $\pm$ 0.08 & 4.2 $\pm$ 0.5 & -- & N \\
TOI-5592 & SpecMatch & HIRES & 6099 $\pm$ 110 & 1.28 $\pm$ 0.23 & -- & -0.20 $\pm$ 0.09 & 4.6 $\pm$ 1.0 & 4.7 $\pm$ 0.2 & Y \\
$\cdots$ & SPC & TRES & 6218 $\pm$ 69 & -- & 4.13 $\pm$ 0.12 & 0.03 $\pm$ 0.08 & 6.8 $\pm$ 0.5 & -- & N%

\enddata
\tablenotetext{a}{Denotes whether the spectroscopic metallicity \feh is used as a prior to the global fits (\S\ref{sec:planet_char}).}
\end{deluxetable*}
 \label{sf:spec_props}%

\section{Planetary System Characterization} \label{sec:planet_char}

For each of the 30 objects presented in this paper, we jointly modelled the
broad-band photometry, TESS and ground-based light curves, and RV time-series
using the \Exofast code \citep{ExoFAST_Eastman2013,ExoFASTv2_Eastman19}. This
allowed us to comprehensively characterize the stellar and planetary properties
of each system in a self-consistent manner.
We followed the same general fitting procedure used in Papers
\citetalias{Yee2022} and \citetalias{Yee2023}; we first outline the key points,
before describing any target-specific modifications.

We used as input data the flattened \TESS light curves clipped to
three transit durations around each transit midpoint, as well as the ground-based
follow-up light curves that had a clear transit detection, as marked in
Table \ref{tab:sg1_summary}. \Exofast models the light curves with a quadratic
limb-darkened transit model \citep{Mandel02,Agol2020}, using separate limb-darkening
coefficients for each filter. A prior was placed on these coefficients using the
values from \citet{Claret2011,Claret2017}, interpolated to the parameters of the
stellar model, ensuring self-consistency.
For the long-cadence \TESS data, the transit model was computed at 120 second intervals
and integrated over the 600 second or 1800 second duration of each exposure,
accounting for the smearing of the transit shape \citep{Kipping2010}.
To allow for imperfections in the dilution
corrections used when producing the \TESS light curves, we fitted for an
additional dilution term for those transits. We imposed a Gaussian prior with
width equal to 10\% of the factors already used in the SPOC and QLP light curves.
For the ground-based time-series photometry, we simultaneously fitted the transit
model together with a linear additive detrending model in the variables
listed in Table \ref{tab:sg1_summary}.
We also fitted for small offsets in the flux baseline $F_0$ and added variance
$\sigma^2$ for each sector of \TESS data and each ground-based light curve to
account for errors in continuum normalization and uncertainty estimates.

We fitted the radial velocity data for each target in \Exofast with a separate
per-instrument offset and jitter term $\sigma_J^2$ added in quadrature with the
instrumental uncertainties. We exclude observations from a given instrument
in the fit when a target only had three or fewer measurements for that instrument;
however we found that those data were consistent with the other RV measurements
of the same target.
For \expandTOIs{\fitslopeTOIs}, we additionally fitted for a linear drift in the RVs.
These long-term accelerations may be indicative of additional long-period objects in these
systems, and further monitoring could help determine the nature
of these outer companions.
In the remaining cases, there was no evidence for trends in the RVs, although most 
of these systems only have relatively short RV observation baselines.

Together with the planetary properties, \Exofast simultaneously fits a model
for the star based on the MIST stellar evolutionary models \citep{MIST0_Dotter2016,MISTI_Choi2016}.
The stellar model is constrained by the \Gaia DR3 parallax and broadband
catalog photometry (Table \ref{tab:catalog_props}), where we imposed minimum
uncertainties of 0.02~mag on the \textit{Gaia}, APASS, and 2MASS photometry,
and 0.03~mag on the WISE photometry. We used a prior on the stellar
metallicity based on our spectroscopic characterization (Table \ref{tab:spec_props}),
as well as an upper limit on the line-of-sight extinction $A_V$ based on the
dust maps from \citet{Schlegel1998,Schlafly2011}. Fitting for the parameters of both the planet
and the star ensures a self-consistent solution.
In particular, the planet's transit duration places an empirical constraint on the
stellar density that can improve the precision of the other stellar properties
beyond the systematic uncertainties from stellar modelling \citep{Tayar2022,Eastman2023}.

In addition to deriving the best-fit stellar and planetary parameters,
\Exofast derives uncertainties on each parameter using a differential
evolution Markov Chain Monte Carlo (DE-MCMC) algorithm to explore the posterior
probability distribution. We deemed the MCMC to be converged according to
the following criteria: Gelman-Rubin statistic \citep{GelmanRubin}
$\mathrm{GR} < 1.05$ and $>1000$ statistically independent MCMC draws.
We also inspected the posterior probability distributions and MCMC chains to
ensure they were well-mixed.

We performed two fits for each system, one in which the orbital eccentricity was held fixed at zero, and a second in which the eccentricity
parameters were allowed to vary. In most
cases, the best-fit eccentricity is not significantly different from zero,
and the circular model is preferred according to the Bayesian Information
Criterion (BIC; \citealt{BIC_Schwarz1978}), and we adopt those fits as our
fiducial model. We report the 1-$\sigma$ upper limits on the eccentricities
in Table \ref{tab:fitted_props} for those systems, and both sets of fit
results are provided in the machine-readable tables. For \expandTOIs{\eccentricTOIs},
the data led to a two-sided constraint on the eccentricity, and we adopt the eccentric model.

We report the best-fit results and
1-$\sigma$ uncertainties for the key stellar and planet properties in Table
\ref{tab:fitted_props}, with the remaining fit parameters, such as the
RV slope $\dot{\gamma}$ and transit dilution factors, are reported in Table
\ref{tab:additional_fit_params}. The best-fit models and data are plotted in
Figure \subref*{fig:toi2031_multiplot}. All data and fit results are
provided as machine-readable tables accompanying this paper.

\subsection{Grazing Transits} \label{ssec:grazing_transits}
\expandTOIs{\grazingTOIs} were found to have V-shaped transits suggesting a
grazing transit geometry. This gives rise to a degeneracy between the impact
parameter $b$ and the planet-star radius ratio, $R_p/R_\star$. As a result, the
posterior distributions have long tails out to large and unphysical values for $\Rp$. To enable the MCMC exploration to converge, we placed an upper limit of
$\Rp < 2.5\Rjup$. Because the exact choice of this upper limit affects the posterior
median, we instead report the mode and 67\% highest-density interval (HDI) of the
posterior distribution for $b$, \Rp, and other parameters that depend directly
on these quantities.
The mode was derived by computing the maximum of a Gaussian kernel-density estimation (KDE)
smoothing of the posterior probability distribution.

Such V-shaped transits may also be caused by stellar eclipsing binary (EB)
systems, particularly if the EB is blended with the light of a nearby star,
resulting in relatively shallow transits with a depth similar to that of a
planetary transit \citep[e.g.,][]{Torres2004}.
For \expandTOIs{\grazingTOIs}, the achromaticity of the transit events as seen
from multi-color ground-based transit photometry and lack of significant variation
in the spectral line bisectors make such a scenario unlikely.

To make this conclusion quantitative, we performed a detailed blend analysis
of these two systems, using the procedures described in \citet{Hartman2019},
who built on earlier work including that of \citet{Torres2004}.
We jointly modelled the light curves, broadband photometry, and
stellar atmospheric parameters for each system, as either a single star orbited
by a giant planet, or as a system containing a background eclipsing
binary whose light is blended with the primary star. In both cases,
the planetary scenario provides a better explanation for the data as determined
by the Bayesian Information Criterion (BIC), with $\Delta\mathrm{BIC} = 22.9$ for
TOI-3980 and $\Delta\mathrm{BIC} = 33.6$ for TOI-5592.
We also simulated radial velocities and bisector span variations that would be
expected from the blended EB scenario. For both systems, the predicted variation in 
bisector spans given the measured radial velocity variation is
incompatible with the observations, allowing us to rule out this false positive
scenario with high confidence.

\subsection{Close Companions} \label{ssec:companion_modelling}

As noted in \S\ref{sssec:ao_detected_companions}, five of our targets had close
companions detected in high angular-resolution imaging. For TOI-3464, the nearby
star is sufficiently distant ($2\farcs6$) and faint ($\Delta I_c$ = 5.5~mag) that
it would not have significantly affected the interpretation of our follow-up observations. Furthermore,
given that the star was also identified in the TIC, its light was already
included in the blending corrections used for creating the light curve for TOI-3464.

For the other four targets, we accounted for the secondary stars in our joint fits using
an updated version of \Exofast that can simultaneously model multiple stars.
Given the close separation of the companions (and in the case of TOI-5181AB, their
consistent parallax measurements from \Gaia), we assumed that all four systems
comprise bound pairs of coeval stars. As such, when modelling the two stars using
the MIST evolutionary models, we required that they have the same initial \feh,
age,\footnote{We placed a Gaussian penalty on the age difference between the two
stars, with a width of 0.1 Gyr, to account for potential systematic uncertainties
in the MIST models.} distance, and extinction. To constrain the secondary stars,
we fit the contrast ratios measured from our imaging observations. With the exception
of TOI-5181AB, we assumed that all other broadband photometric measurements are
of the blended light of both components. Because \Gaia resolved the TOI-5181AB
pair, we fit their \Gaia $G$-band magnitudes separately, and assumed that the
$G_\mathrm{BP}$, $G_\mathrm{RP}$ magnitudes, which are only available for the primary,
represent the flux from that star alone.

We also modelled the dilution of the transit light curves due to the secondary stars'
flux. \Exofast uses the model SED of the two stars to compute the relative flux in
each band and thereby correct for the dilution from the blended secondary.
We did this for all ground-based light curves for these systems, as well as the TESS
data for TOI-3160, TOI-3523, and TOI-5386. For TOI-5181, the TESS light curve
is already corrected for blending due to TOI-5181B. We report the best-fit stellar
parameters for the secondary components of these four systems in Table \ref{tab:comp_fitted_props}.

The presence of close secondary stars in these systems introduces the possibility
that the secondary component is an eclipsing binary star, and the observed transit signals are not due to a hot Jupiter orbiting the brighter star,
but are instead the diluted eclipses of the secondary component. Because of the components' close separation, our ground-based
follow-up observations and centroid analysis of the TESS data are unable to definitely
show that the primary star is responsible for the transit signals.
For TOI-3160, TOI-3523, and TOI-5386, the spectroscopic observations with Magellan/PFS and Keck/HIRES 
encompassed the light from both components.
We detected no clear evidence for secondary lines in these spectra,
although if these systems are indeed bound, the low velocity separation would
make it difficult to separate multiple spectral components at the expected contrast ratios.

We performed a blend analysis
similar to that described in the previous section \citep{Hartman2019}, considering the following scenarios:
(1) there is a planet around the brighter star; (2) the fainter star is a blended eclipsing binary that is gravitationally bound to the primary star; 
(3) the fainter star is a blended eclipsing binary that is unassociated with the primary star.
In each case, the first scenario resulted in the best fit to the data, as measured by the $\chi^2$ of the fit.
The evidence favoring the planet explanation for these systems include
the lack of chromaticity in the measured depths of the transits
observed from the ground, the in-phase variation of the RVs with
the expected orbital ephemeris; and the lack of associated line shape variations
as measured through the bisector slopes.

Even in scenario (1), the observed spectra for TOI-3160, TOI-3523, and TOI-5386
are blends of the two stars, which may bias the measured RVs. We simulated spectra
of the two stars, assuming properties of the primary and secondary stars drawn from
the posterior distributions of our global system characterization. We then injected
Keplerian motion into models of the blended spectra and measured the ratio of the
true RV semi-amplitude to the measured value. We found that the presence of the
nearby star would dilute the measured $K$ for each system, and that the true value
of $K$ can be determined by multiplying the measured (diluted) value by $1.00506 \pm 0.00051$, $1.0545 \pm 0.0094$, and
$1.00581 \pm 0.00031$ respectively. These scaling factors were applied to the RV semi-amplitude $K$,
planet mass \Mp, and all other dependent variables reported in Table \ref{tab:fitted_props}.

\subsection{TOI-4961 Transit Depth}

TOI-4961 was observed by TESS in Sectors 11, 37, 38, 64 and 65, with light curves
produced by QLP for all five sectors and by SPOC for Sectors 64 and 65.
We noted that the transit depths measured from the QLP light curves, while consistent
between sectors, were inconsistent with the depths in the SPOC light curves, even
for the sectors where both were available.
We suspect that this inconsistency is caused by differences in the way dilution from nearby stars is corrected for
by the two pipelines.
Because the SPOC data produced transit depths more consistent with the ground-based
follow-up photometry, we favored those results.
We fit independent additional dilution factors for the SPOC and QLP light curves,
where instead of imposing a Gaussian prior, we allowed the dilution to vary freely
for the QLP light curves.
We found that an additional $27.5\pm1.8\%$ dilution is necessary to bring the QLP
data into consistency with the other transit light curves.

\subsection{TOI-2031 Literature Comparison}

During the preparation of this manuscript, \citet{Wang2024} independently confirmed TOI-2031\,b as a planet using NEID RV data. We did not include their published NEID RVs in our analysis, as those data were primarily taken during the planet's transit for the purpose of measuring the stellar obliquity using the Rossiter-McLaughlin effect. \citet{Wang2024} did not report a mass for the planet from their analysis; otherwise all of their other measured stellar and planet parameters were consistent with those presented in this work.

\section{Discussion} \label{sec:discussion}

We have presented the discovery and characterization of thirty new giant
planets from \TESS, with orbital periods ranging between 1.6 and 8.2 days.
Our global modelling of the space- and ground-based observations with
\Exofast provides us with precise stellar and planet properties for each
system. In this section, we compared the planets' properties to those of other
known planets in the NASA Exoplanet Archive \citep{ExoplanetArchive_Akeson2013,NEA_ExoFoP_Christiansen2025,ExoplanetArchive_PSCompPars},\footnote{\texttt{Planetary Systems Composite Parameters} table, accessed 12 May 2025 at
\url{https://exoplanetarchive.ipac.caltech.edu/}}
and highlight some notable systems.

\begin{figure}
\plotone{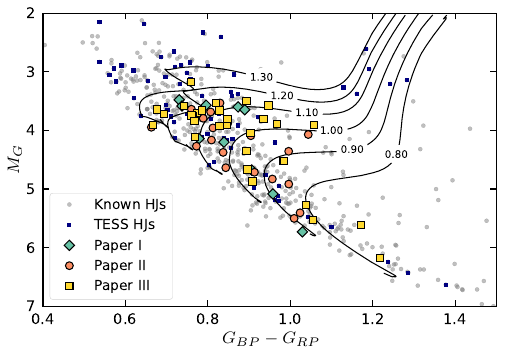}
\caption{\Gaia color-magnitude diagram for transiting hot Jupiter hosts
($P < 10$~days, $8 \,\Rearth < \Rp < 24\,\Rearth$ ) from the NASA Exoplanet Archive.
Yellow squares represent the stellar hosts of the newly-confirmed planets in
this paper; green diamonds and orange circles show the systems from previous
works in this survey (Papers \citetalias{Yee2022} and \citetalias{Yee2023}).
Navy blue squares show other discoveries from \TESS, while all previously-known
hot Jupiter hosts are plotted in gray.
We also plot MIST stellar evolutionary tracks for a range of stellar masses
corresponding to a metallicity of $\feh = +0.20$~dex, close to the median
metallicity of our sample.
The absolute magnitude and colors plotted here have not been corrected for the
effects of extinction and reddening.
\label{fig:hj_pop_cmd}
}
\end{figure}

Figure
\ref{fig:hj_pop_cmd} is a color-magnitude diagram of the planet host stars in our sample,
alongside those of previously-known transiting hot Jupiter hosts. By construction, the planets described in this paper
orbit FGK stars ($0.8\,\Msun < \Mstar < 1.5\,\Msun$), with the majority
being on the main sequence, while a handful are somewhat evolved subgiants.
The stellar sample has a median metallicity of $\feh = +0.20$~dex, consistent
with the general trend toward metal enrichment of hot Jupiter hosts
\citep[e.g.,][]{Gonzalez1997,Santos2004,Fischer2005}. 

\begin{figure*}
\begin{interactive}{js}{population_interactive_figure.html}
\epsscale{1.1}
\plotone{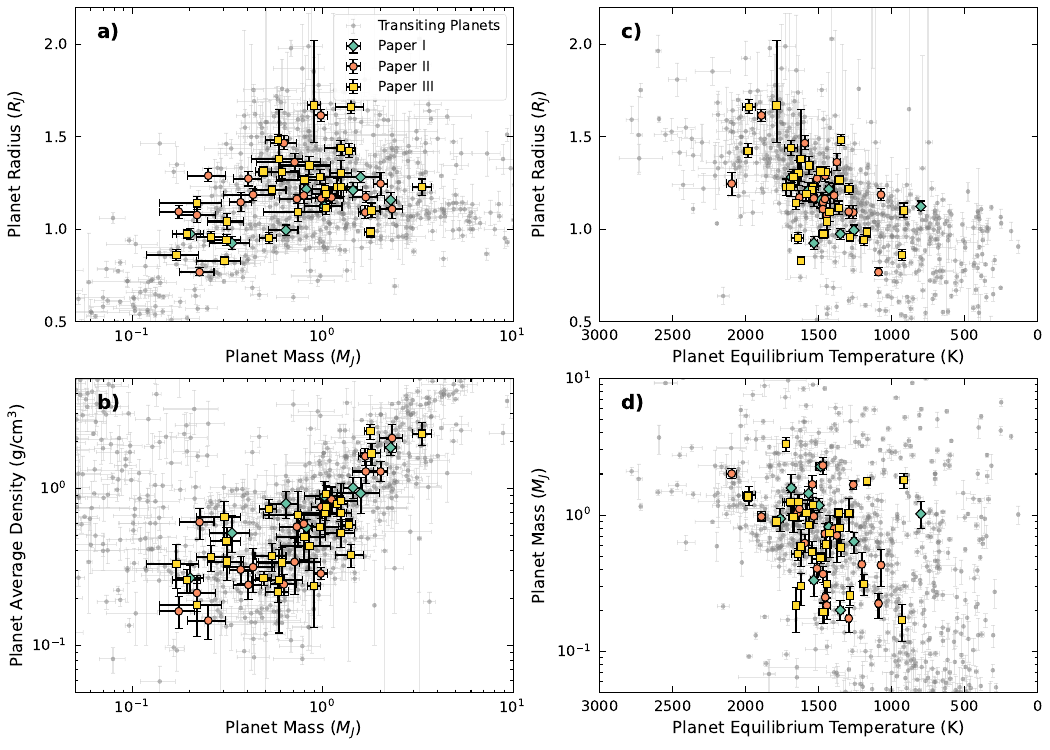}
\end{interactive}
\caption{The new planets confirmed by our survey in context. Gray points show previously known transiting planets from the NASA Exoplanet Archive with masses and radii measured to better than 2.5-$\sigma$.
\textbf{a):} Planet masses and radii. The new planets have masses that span more than an
order of magnitude, ranging from sub-Saturns
to super-Jupiters, while having roughly similar radii.
\textbf{b):} Planet bulk densities as a function of mass. The new planets show a similar spread in density compared with previously known planets, with the least dense object, TOI-3682\,b, having a bulk density $\rhop < 0.2\,\gcc$.
\textbf{c) and d):} Planet radii and masses as a function of planet
equilibrium temperature. The well-known trend of increasing radius inflation
with stellar insolation is clearly seen; the two most strongly irradiated planets
in the sample, TOI-2886\,b and TOI-3523\,b, are both significantly inflated. The dearth
of planets with intermediate size on close orbits -- the ``hot Neptune desert''
-- is an intriguing feature in the exoplanet distribution; in our sample,
TOI-2986\,b and TOI-3682\,b appear to be on the edge of this desert and may be
interesting targets for future study.
An interactive version of this figure that allows zooming and panning\added{, as well as the highlighting of specific sources,} 
is available in the online journal.
\label{fig:population_multipanel}}
\end{figure*}


The thirty new planets have radii of $0.84\,\Rjup < \Rp < 1.8\,\Rjup$,
while their masses and bulk densities 
span more than an order of magnitude ($0.17\,\Mjup < \Mp <
3.35\,\Mjup$, $0.18\,\gcc < \rhop < 2.3\,\gcc$), as plotted in
Figures \ref{fig:population_multipanel}a and \ref{fig:population_multipanel}b.
Six of the planets in this paper have masses comparable to or lower than that of Saturn
($\Mp \lesssim 0.3\,\Mjup$), with the two least massive objects (TOI-2876\,b,
$\Mp = 0.17\pm0.05\,\Mjup$; TOI-4734\,b, $\Mp = 0.18\pm0.04\,\Mjup$) being
remarkably similar in mass and orbital period ($P \approx 6.3$~days).
The latter two planets reside close to the transition between the gas giants
dominated by thick H/He atmospheres and the smaller sub-Saturns. If giant
planets form through core accretion followed by runaway gas accretion, the wide
range of their masses may indicate the diversity of conditions in which these
processes take place, including cases where gas accretion may slow or halt
before the planet reaches $\sim 1\,\Mjup$.

Figure \ref{fig:population_multipanel}c and \ref{fig:population_multipanel}d
show the planet radii and masses as a
function of their equilibrium temperature, assuming zero albedo and a uniform temperature across the entire planetary photosphere. The planets in our sample are all highly irradiated and
are consistent with the previously-known trend of increasing radius inflation
with incident stellar flux \citep[e.g.,][]{Kovacs2010,Laughlin2011}. A few of
the planets are also perched on the edge of the ``hot Neptune desert'', an
observed lack of planets between the sizes of Neptune and Jupiter at short
orbital periods \citep{Szabo2011,Mazeh2016,Castro-Gonzalez2024}. One proposed explanation for
this feature is photoevaporation at high stellar irradiation; if this is the case,
these planets may be good targets for observations of atmospheric outflows
\citep[e.g.,][]{Vissapragada2022,Guilluy2023}.
Indeed, one of the objects on the edge of the ``hot Neptune desert'', TOI-3682\,b,
is also the least dense planet confirmed in this paper, with a bulk density of
$\rhop = 0.18\pm0.07\,\gcc$ and may be especially susceptible to mass loss.


For most of the systems described in this paper, the precision and relatively
sparse sampling of our RV observations do not allow us to put strong
constraints on the planets' orbital eccentricities, with median 1- and 2-$\sigma$
upper limits on the eccentricities of 0.14 and 0.27. In all but two cases, the
circular fit is preferred over the eccentric models according to the Bayesian Information
Criterion (BIC; \citealt{BIC_Schwarz1978}). For \expandTOIs[\,b]{\eccentricTOIs},
we measured eccentricities of $e =0.106^{+0.053}_{-0.029}$
and $0.182^{+0.040}_{-0.048}$ respectively.

TOI-4961\,b has the second-longest orbital period
($P = 7.48$~days) and largest scaled semimajor axis ($a/\Rstar = 17.4$) in the sample of thirty planets presented in this paper, giving
rise to a long tidal circularization timescale of $\tcirc = 21\pm6$~Gyr, as
computed from Equation (3) of \citet{Adams2006} and assuming a planetary tidal quality factor $Q_P = 10^6$.
Thus, the moderate eccentricity
of this object may be residual eccentricity from a high-eccentricity formation pathway.

We also measured a small but significantly ($3.8\sigma$) non-zero eccentricity
for TOI-3593\,b, which has a relatively short orbital period of just $P = 3.82$
days. Interestingly, our joint modelling points to a relatively young solar-mass
stellar host with an age of $1.8_{-1.3}^{+2.9}$~Gyr, comparable to the computed
tidal circularization timescale of $\tcirc = 1.7\pm0.4$~Gyr, so this may be a
planet whose orbit is still circularizing. Still, we stress that this estimate
does not account for uncertainties in our knowledge of the tidal quality factor $Q_P$.
Additional observations of the other long-period planets in our sample
to refine their eccentricity measurements or a population-level analysis of
the hot Jupiter eccentricity distribution could help us better understand the
formation history of these planets.

Many of the planets described in this manuscript will be suitable for detailed
atmospheric characterization by JWST and \textit{Ariel}.
In particular, \textit{Ariel} is expected to probe the atmospheres of $\sim 1000$ giant planets,
providing the first demographic census of exoplanetary atmospheres and compositions \citep{ARIEL_Tinetti2016}.
The ARIEL Science Consortium released a list of unconfirmed TESS planet candidates that could potentially
be suitable for observation \citep{Edwards2022};\footnote{\url{https://exofop.ipac.caltech.edu/tess/view_toi_ariel.php}}
this list includes five of the planets newly confirmed in this paper:
TOI-2031\,b, TOI-2886\,b, TOI-3523\,b, TOI-4214\,b, and TOI-5322\,b. The confirmations
and mass measurements presented here will enable more refined target selection and
the interpretation of the atmospheric spectroscopy observations.

\section{Conclusions} \label{sec:conclusion}

We present in this paper the discovery of thirty giant planets that were initially
detected using data from NASA's \TESS mission, and subsequently confirmed through
ground-based follow-up photometric, imaging, and spectroscopic observations.
This is the third batch of planets to be confirmed by our survey, which
has the goal of assembling a homogeneous sample of $\approx 400$ hot Jupiters
orbiting FGK stars brighter than $G < 12.5$ \citep{Yee2021b,Yee2022,Yee2023}.
Together with other efforts to confirm new hot Jupiters from \TESS in the
broader community \citep[e.g.,][]{Rodriguez2019,Rodriguez2021,Rodriguez2023,Nielsen2019,Zhou2019a,Ikwut-Ukwa2021,Brahm2020,Wong2021,Kabath2022,Knudstrup2022,Psaridi2022,Subjak2022,Wittenmyer2022,Schulte2024,Ehrhardt2024,Thomas2025}, we will soon achieve the largest ever sample of hot
Jupiters from a uniform transit survey.
Such a catalog of planets, when combined with a detailed characterization
of its completeness and reliability, will enable demographic study
of the hot Jupiter population; future works in this series will address these issues.

The continued follow-up and confirmation of planet candidates from \TESS has
already expanded our knowledge of giant planet demographics, providing new
clues as to their formation. \citet{Zhou2019a} and \citet{Beleznay2022} used
samples containing $\approx40$ and $\approx100$ hot Jupiters to study their
occurrence rates around AFG stars; while \citet{Gan2022} and \citet{Bryant2023} examined the frequency of hot Jupiters around M dwarfs. These works appear to suggest that the occurrence of
close-in giant planets peaks around roughly solar-type stars. \TESS is also
providing insights into giant planets around evolved stars \citep[e.g.,][]{Saunders2021,Grunblatt2022}.
\citet{Yee2023b} examined the observed period distribution of giant
planets and found no evidence for a dependence of the period distribution
on host star metallicity, placing constraints on the relative fraction of
hot Jupiters that might arise from different formation pathways.

The discovery of new transiting hot Jupiters also presents new opportunities
for more detailed characterization of these planets, including atmospheric
characterization,
detection of gravitational interactions with stellar tides or other planets
in the same system, and the measurement of stellar obliquities through the
Rossiter-McLaughlin effect. A complete sample of hot
Jupiters will provide new targets for these investigations, as well as an
understanding of the underlying population from which these targets are drawn.
Thus, \TESS is enabling new efforts to seek out a global picture of hot Jupiter
populations and their properties across the HR diagram.

\begin{rotatepage}
\foreach \n [count=\ni] in {0,1,...,\numTables} {%
\movetableright=-1in
\movetabledown=2.3in
\begin{rotatetable}
\begin{deluxetable*}{l>{\centering}cccccc}
\ifnum\ni=1%
\tablecaption{Median Values and 68\% Confidence Intervals for Fitted Stellar and Planetary Parameters \label{tab:fitted_props}}
\addtocounter{table}{1}
\pdfbookmark[1]{Table \thetable: Fitted Parameters}{fitted_props}%
\addtocounter{table}{-1}
\else
\tablecaption{\textit{(Continued)}}
\fi
\tabletypesize{\footnotesize}
\input{fit_results_\n}

\ifnum\ni=\lastTableNum%
\tablecommentsmod{
This table contains the fit results from the preferred fit for each target.\\
Table 3 in \citet{ExoFASTv2_Eastman19} provides a detailed description of all derived and fitted parameters.\\ 
$^a$\,This is 68\% upper limit on eccentricity derived from the eccentric fits. \\
$^b$\,The tidal circularization timescale is computed with Equation (3) of \citet{Adams2006}, assuming a tidal quality factor $Q_S = 10^6$. \\
$^c$\,The stellar metallicity when the star was formed, that define the grid points for the MIST stellar evolutionary tracks.\\ 
$^d$\,The equal evolutionary phase (EEP) corresponds to specific points in the stellar evolutionary tracks, as described in \citet{MIST0_Dotter2016}. \\ 
Table \ref{M-tab:fitted_props} is published in its entirety in the electronic
edition of the journal. This version only shows the
results from the preferred fit for each target. The full version includes
these results and fits where the eccentricity was allowed to float. Note that
the full version also includes the results from the additional fit parameters outlined in Table \ref{tab:additional_fit_params}.}
\fi
\end{deluxetable*}
\end{rotatetable}
\addtocounter{table}{-1}
}
\addtocounter{table}{1}
\end{rotatepage}

\ifSubfilesClassLoaded{%
\bibliography{../software}
}{}

\end{document} \label{sf:fit_results}%

\providecommand{\main}{..}
\documentclass[\main/main.tex]{subfiles}

\ifSubfilesClassLoaded{\journalfalse
\addtocounter{figure}{3}}{}

\begin{document}

\ifjournal
\figsetstart
\figsetnum{4}
\figsettitle{Data and \Exofast Fit Results}

\figsetgrpstart
\figsetgrpnum{4.1}
\figsetgrptitle{Data and \Exofast fit results for TOI-2031\,b}
\figsetplot{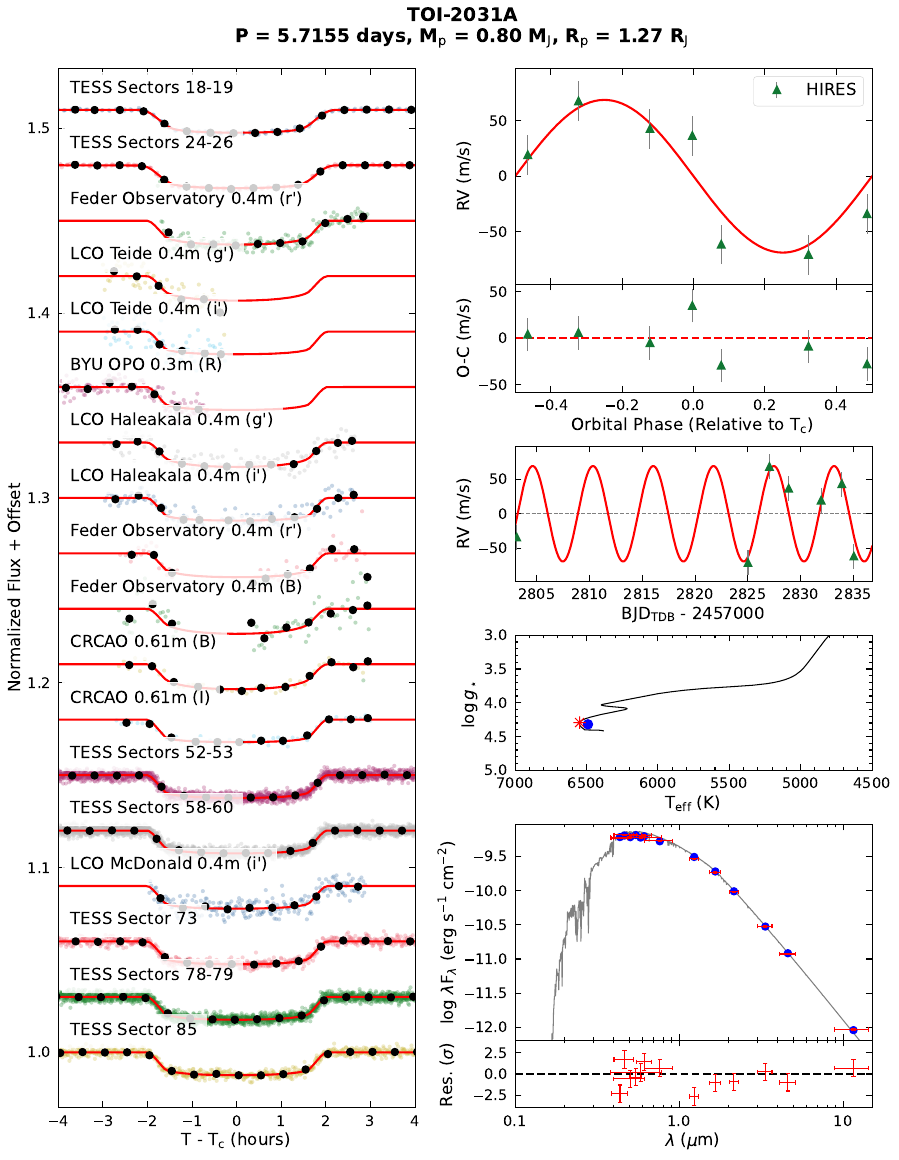}
\figsetgrpnote{Data and \Exofast fit results for TOI-2031\,b.
\textbf{Left:} \TESS and ground-based light-curves, phase-folded onto the best-fit period and time of conjunction.
Faint colored points represent the unbinned data, while large black circles show the time-series data binned to 30-min cadence.
The best-fit transit model in each band is shown as the red line.
\textbf{Top right:} RV observations, also phased onto the best-fit orbital period.
Error bars represent the fitted per-instrument jitter term $\sigma_\mathrm{jit}$ added in quadrature to the instrumental uncertainties.
The red line shows the best-fit RV model.
We plot the residuals after subtracting the model in the middle subpanel, and the unphased RV data and model time-series in the lower subpanel.
\textbf{Middle right:} The best-fit MIST stellar evolution track (black line), with a red asterisk marking the position along the track corresponding to the best-fit stellar age.
The blue point represents the best-fit stellar \Teff and \logg.
The discrepancy between the blue point and red asterisk are well within the fitted uncertainties in each parameter, indicating no tension between the different constraints on the stellar properties.
\textbf{Bottom right:} The observed stellar fluxes from the \Gaia, UCAC, 2MASS and WISE catalogs are plotted in red, with horizontal error bars corresponding to the width of the photometric bandpass.
The blue points show the best-fit model flux derived from the stellar properties and MIST bolometric correction grid.
We plot in gray an atmospheric model from \citet{Kurucz1993} corresponding to the best-fit stellar parameters for illustrative purposes only, as the fit is performed directly to the MIST grid.
The TESS and ground-based time-series photometry, as well as the RV measurements, are available as Data behind the Figure.}
\figsetgrpend

\figsetgrpstart
\figsetgrpnum{4.2}
\figsetgrptitle{Data and \Exofast fit results for TOI-2169\,b}
\figsetplot{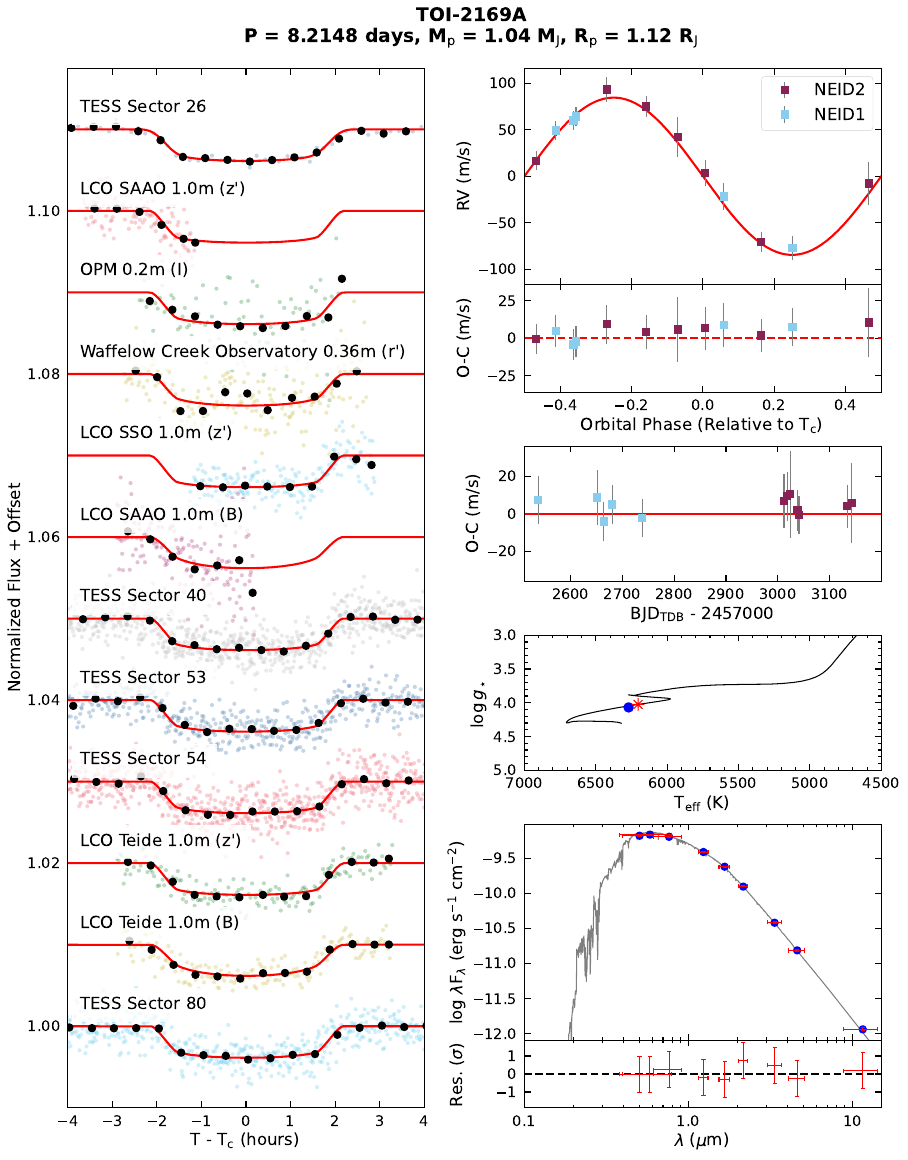}
\figsetgrpnote{Data and \Exofast fit results for TOI-2169\,b.
\textbf{Left:} \TESS and ground-based light-curves, phase-folded onto the best-fit period and time of conjunction.
Faint colored points represent the unbinned data, while large black circles show the time-series data binned to 30-min cadence.
The best-fit transit model in each band is shown as the red line.
\textbf{Top right:} RV observations, also phased onto the best-fit orbital period.
Error bars represent the fitted per-instrument jitter term $\sigma_\mathrm{jit}$ added in quadrature to the instrumental uncertainties.
The red line shows the best-fit RV model.
We plot the residuals after subtracting the model in the middle subpanel, and the unphased RV data and model time-series in the lower subpanel.
\textbf{Middle right:} The best-fit MIST stellar evolution track (black line), with a red asterisk marking the position along the track corresponding to the best-fit stellar age.
The blue point represents the best-fit stellar \Teff and \logg.
The discrepancy between the blue point and red asterisk are well within the fitted uncertainties in each parameter, indicating no tension between the different constraints on the stellar properties.
\textbf{Bottom right:} The observed stellar fluxes from the \Gaia, UCAC, 2MASS and WISE catalogs are plotted in red, with horizontal error bars corresponding to the width of the photometric bandpass.
The blue points show the best-fit model flux derived from the stellar properties and MIST bolometric correction grid.
We plot in gray an atmospheric model from \citet{Kurucz1993} corresponding to the best-fit stellar parameters for illustrative purposes only, as the fit is performed directly to the MIST grid.
The TESS and ground-based time-series photometry, as well as the RV measurements, are available as Data behind the Figure.}
\figsetgrpend

\figsetgrpstart
\figsetgrpnum{4.3}
\figsetgrptitle{Data and \Exofast fit results for TOI-2346\,b}
\figsetplot{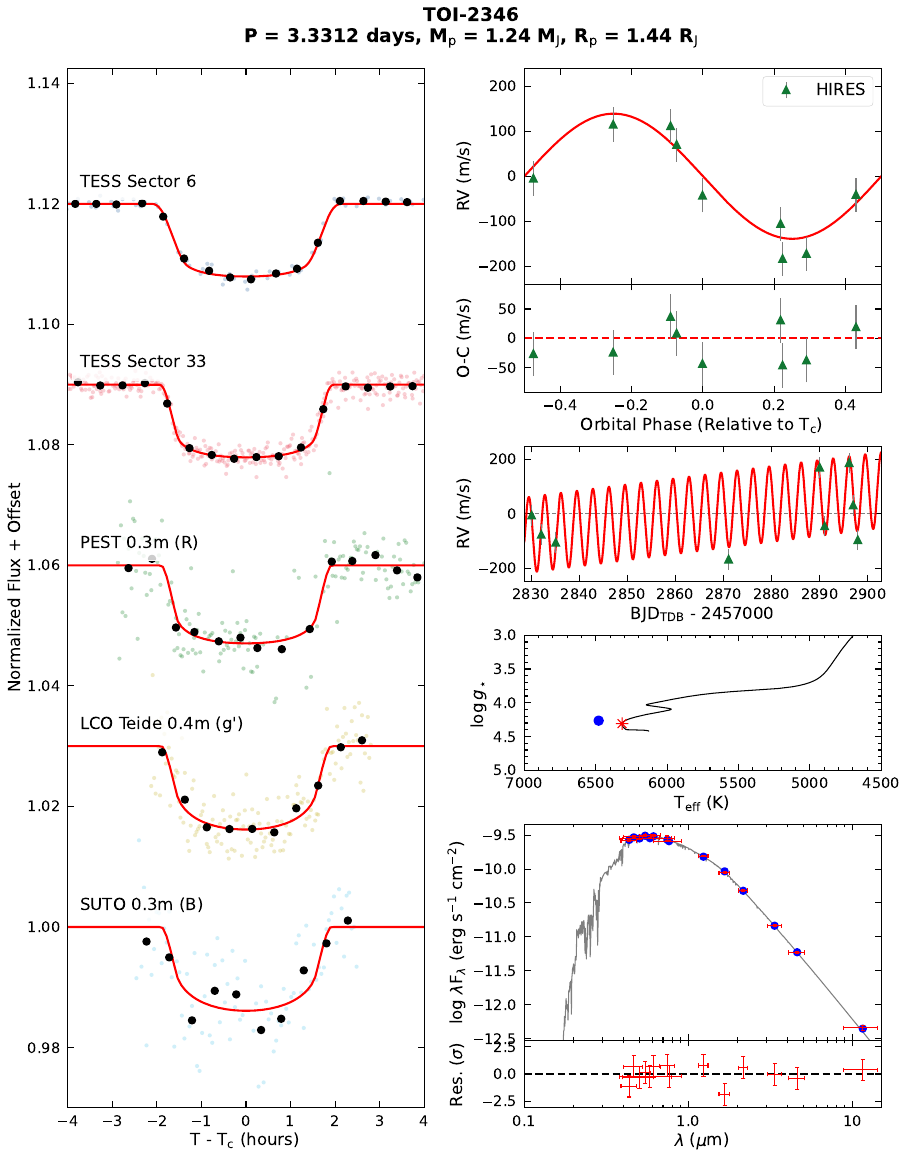}
\figsetgrpnote{Data and \Exofast fit results for TOI-2346\,b.
\textbf{Left:} \TESS and ground-based light-curves, phase-folded onto the best-fit period and time of conjunction.
Faint colored points represent the unbinned data, while large black circles show the time-series data binned to 30-min cadence.
The best-fit transit model in each band is shown as the red line.
\textbf{Top right:} RV observations, also phased onto the best-fit orbital period.
Error bars represent the fitted per-instrument jitter term $\sigma_\mathrm{jit}$ added in quadrature to the instrumental uncertainties.
The red line shows the best-fit RV model.
We plot the residuals after subtracting the model in the middle subpanel, and the unphased RV data and model time-series in the lower subpanel.
\textbf{Middle right:} The best-fit MIST stellar evolution track (black line), with a red asterisk marking the position along the track corresponding to the best-fit stellar age.
The blue point represents the best-fit stellar \Teff and \logg.
The discrepancy between the blue point and red asterisk are well within the fitted uncertainties in each parameter, indicating no tension between the different constraints on the stellar properties.
\textbf{Bottom right:} The observed stellar fluxes from the \Gaia, UCAC, 2MASS and WISE catalogs are plotted in red, with horizontal error bars corresponding to the width of the photometric bandpass.
The blue points show the best-fit model flux derived from the stellar properties and MIST bolometric correction grid.
We plot in gray an atmospheric model from \citet{Kurucz1993} corresponding to the best-fit stellar parameters for illustrative purposes only, as the fit is performed directly to the MIST grid.
The TESS and ground-based time-series photometry, as well as the RV measurements, are available as Data behind the Figure.}
\figsetgrpend

\figsetgrpstart
\figsetgrpnum{4.4}
\figsetgrptitle{Data and \Exofast fit results for TOI-2382\,b}
\figsetplot{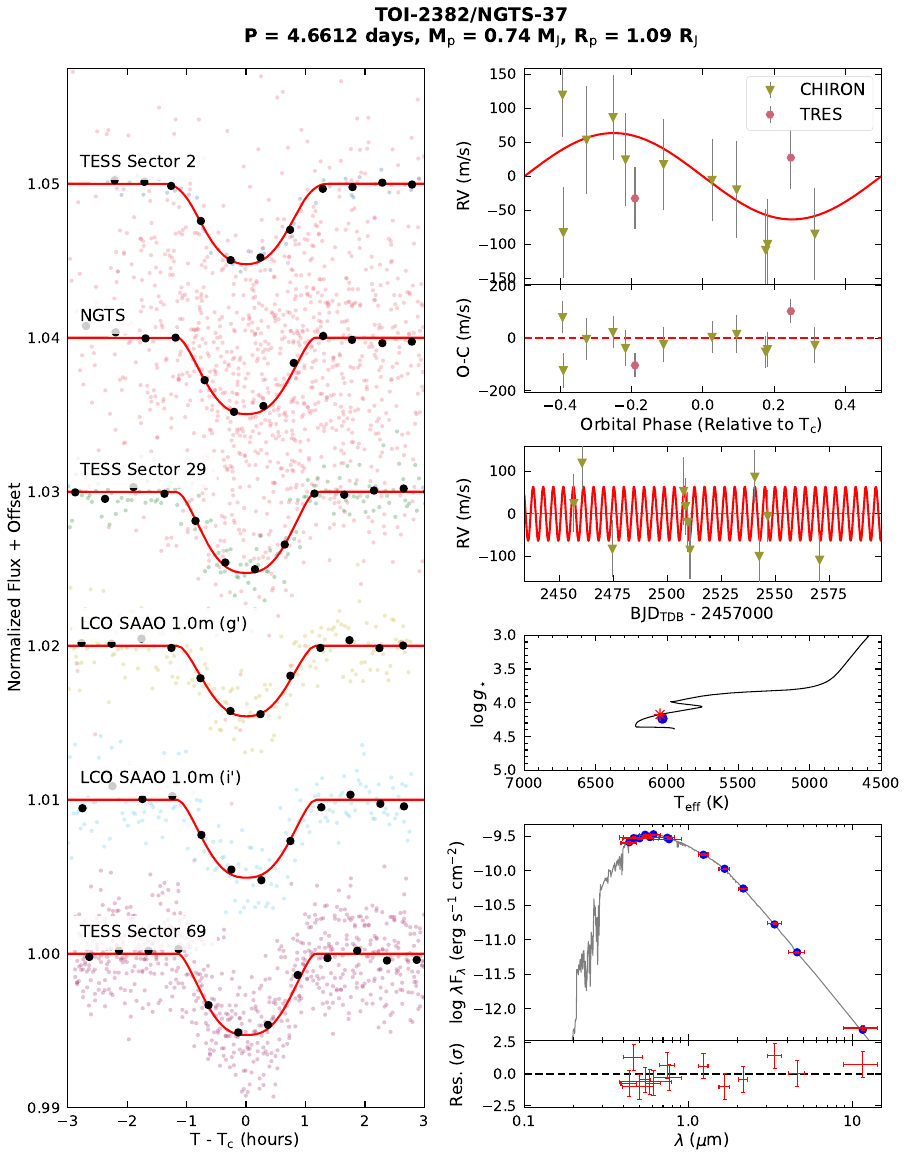}
\figsetgrpnote{Data and \Exofast fit results for TOI-2382\,b.
\textbf{Left:} \TESS and ground-based light-curves, phase-folded onto the best-fit period and time of conjunction.
Faint colored points represent the unbinned data, while large black circles show the time-series data binned to 30-min cadence.
The best-fit transit model in each band is shown as the red line.
\textbf{Top right:} RV observations, also phased onto the best-fit orbital period.
Error bars represent the fitted per-instrument jitter term $\sigma_\mathrm{jit}$ added in quadrature to the instrumental uncertainties.
The red line shows the best-fit RV model.
We plot the residuals after subtracting the model in the middle subpanel, and the unphased RV data and model time-series in the lower subpanel.
\textbf{Middle right:} The best-fit MIST stellar evolution track (black line), with a red asterisk marking the position along the track corresponding to the best-fit stellar age.
The blue point represents the best-fit stellar \Teff and \logg.
The discrepancy between the blue point and red asterisk are well within the fitted uncertainties in each parameter, indicating no tension between the different constraints on the stellar properties.
\textbf{Bottom right:} The observed stellar fluxes from the \Gaia, UCAC, 2MASS and WISE catalogs are plotted in red, with horizontal error bars corresponding to the width of the photometric bandpass.
The blue points show the best-fit model flux derived from the stellar properties and MIST bolometric correction grid.
We plot in gray an atmospheric model from \citet{Kurucz1993} corresponding to the best-fit stellar parameters for illustrative purposes only, as the fit is performed directly to the MIST grid.
The TESS and ground-based time-series photometry, as well as the RV measurements, are available as Data behind the Figure.}
\figsetgrpend

\figsetgrpstart
\figsetgrpnum{4.5}
\figsetgrptitle{Data and \Exofast fit results for TOI-2876\,b}
\figsetplot{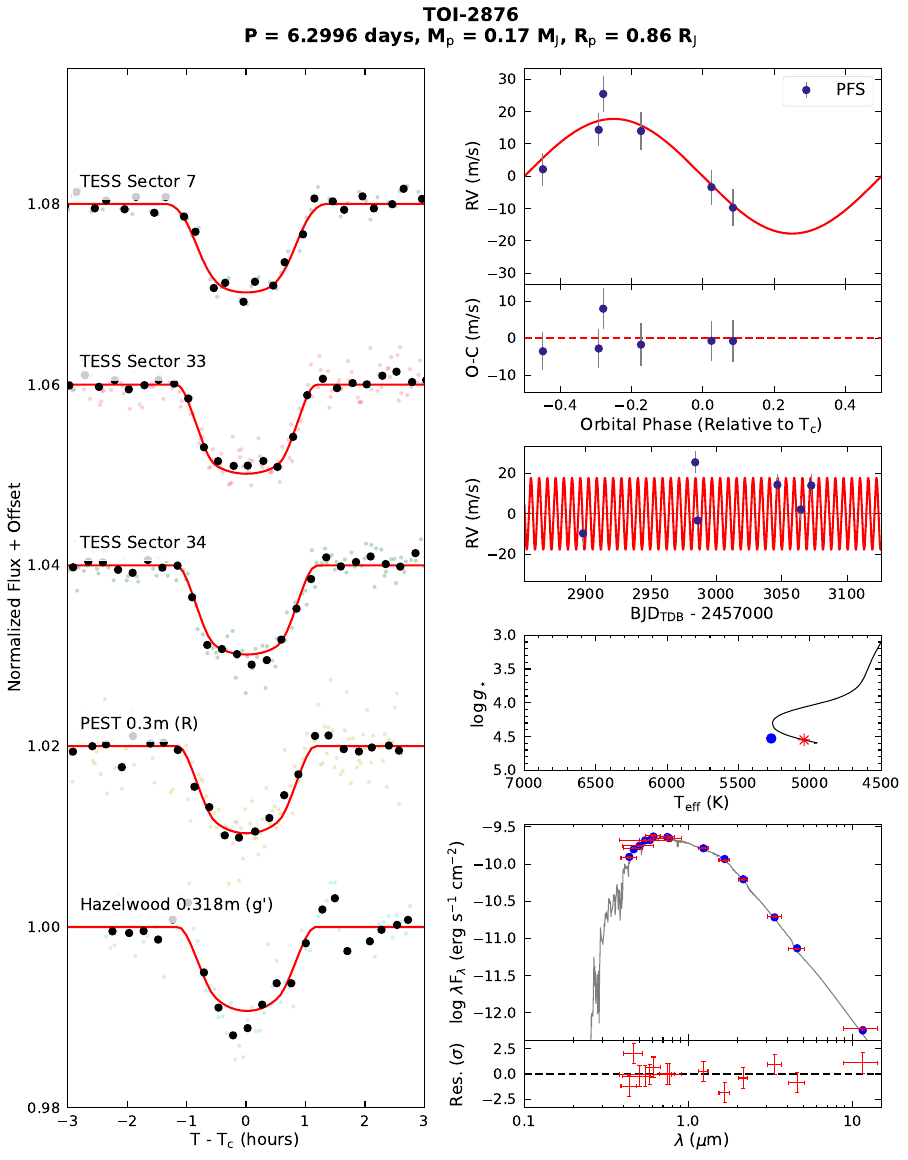}
\figsetgrpnote{Data and \Exofast fit results for TOI-2876\,b.
\textbf{Left:} \TESS and ground-based light-curves, phase-folded onto the best-fit period and time of conjunction.
Faint colored points represent the unbinned data, while large black circles show the time-series data binned to 30-min cadence.
The best-fit transit model in each band is shown as the red line.
\textbf{Top right:} RV observations, also phased onto the best-fit orbital period.
Error bars represent the fitted per-instrument jitter term $\sigma_\mathrm{jit}$ added in quadrature to the instrumental uncertainties.
The red line shows the best-fit RV model.
We plot the residuals after subtracting the model in the middle subpanel, and the unphased RV data and model time-series in the lower subpanel.
\textbf{Middle right:} The best-fit MIST stellar evolution track (black line), with a red asterisk marking the position along the track corresponding to the best-fit stellar age.
The blue point represents the best-fit stellar \Teff and \logg.
The discrepancy between the blue point and red asterisk are well within the fitted uncertainties in each parameter, indicating no tension between the different constraints on the stellar properties.
\textbf{Bottom right:} The observed stellar fluxes from the \Gaia, UCAC, 2MASS and WISE catalogs are plotted in red, with horizontal error bars corresponding to the width of the photometric bandpass.
The blue points show the best-fit model flux derived from the stellar properties and MIST bolometric correction grid.
We plot in gray an atmospheric model from \citet{Kurucz1993} corresponding to the best-fit stellar parameters for illustrative purposes only, as the fit is performed directly to the MIST grid.
The TESS and ground-based time-series photometry, as well as the RV measurements, are available as Data behind the Figure.}
\figsetgrpend

\figsetgrpstart
\figsetgrpnum{4.6}
\figsetgrptitle{Data and \Exofast fit results for TOI-2886\,b}
\figsetplot{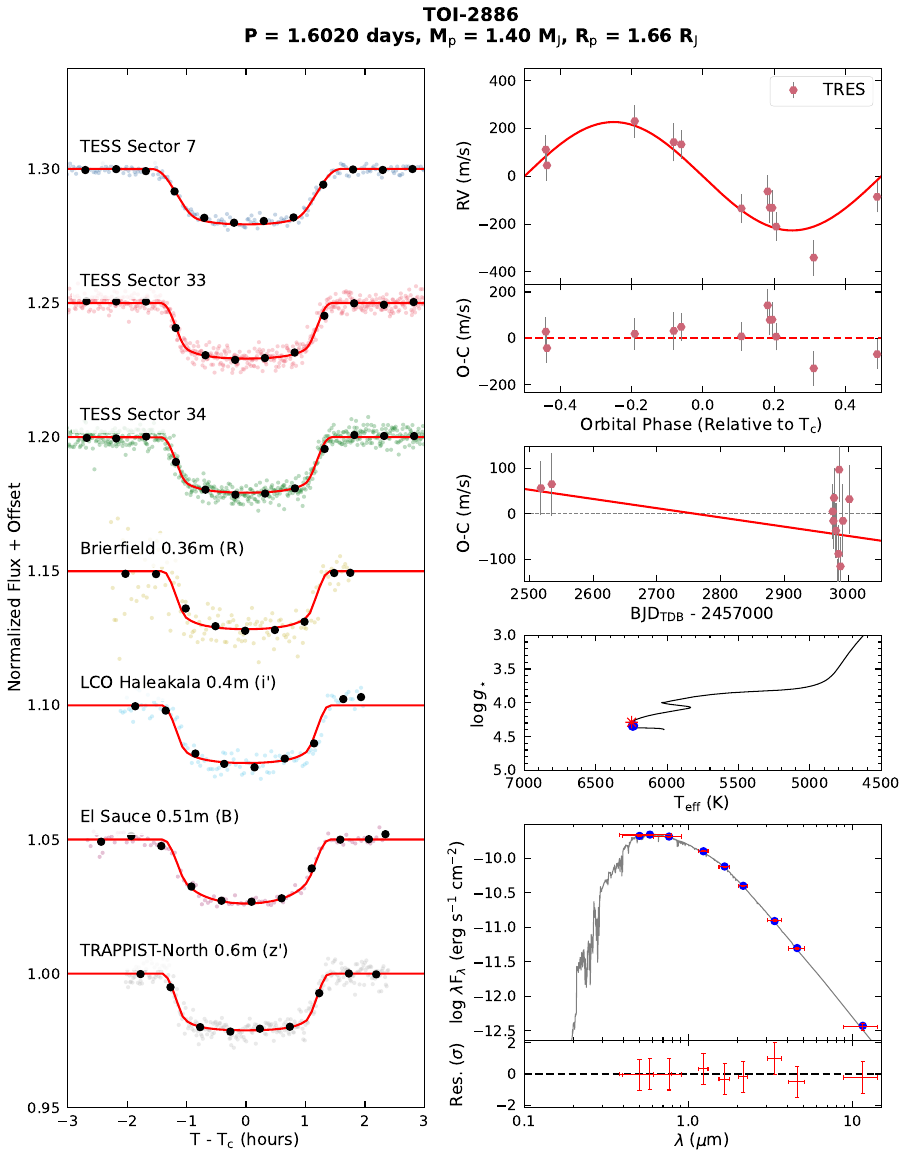}
\figsetgrpnote{Data and \Exofast fit results for TOI-2886\,b.
\textbf{Left:} \TESS and ground-based light-curves, phase-folded onto the best-fit period and time of conjunction.
Faint colored points represent the unbinned data, while large black circles show the time-series data binned to 30-min cadence.
The best-fit transit model in each band is shown as the red line.
\textbf{Top right:} RV observations, also phased onto the best-fit orbital period.
Error bars represent the fitted per-instrument jitter term $\sigma_\mathrm{jit}$ added in quadrature to the instrumental uncertainties.
The red line shows the best-fit RV model.
We plot the residuals after subtracting the model in the middle subpanel, and the unphased RV data and model time-series in the lower subpanel.
\textbf{Middle right:} The best-fit MIST stellar evolution track (black line), with a red asterisk marking the position along the track corresponding to the best-fit stellar age.
The blue point represents the best-fit stellar \Teff and \logg.
The discrepancy between the blue point and red asterisk are well within the fitted uncertainties in each parameter, indicating no tension between the different constraints on the stellar properties.
\textbf{Bottom right:} The observed stellar fluxes from the \Gaia, UCAC, 2MASS and WISE catalogs are plotted in red, with horizontal error bars corresponding to the width of the photometric bandpass.
The blue points show the best-fit model flux derived from the stellar properties and MIST bolometric correction grid.
We plot in gray an atmospheric model from \citet{Kurucz1993} corresponding to the best-fit stellar parameters for illustrative purposes only, as the fit is performed directly to the MIST grid.
The TESS and ground-based time-series photometry, as well as the RV measurements, are available as Data behind the Figure.}
\figsetgrpend

\figsetgrpstart
\figsetgrpnum{4.7}
\figsetgrptitle{Data and \Exofast fit results for TOI-2986\,b}
\figsetplot{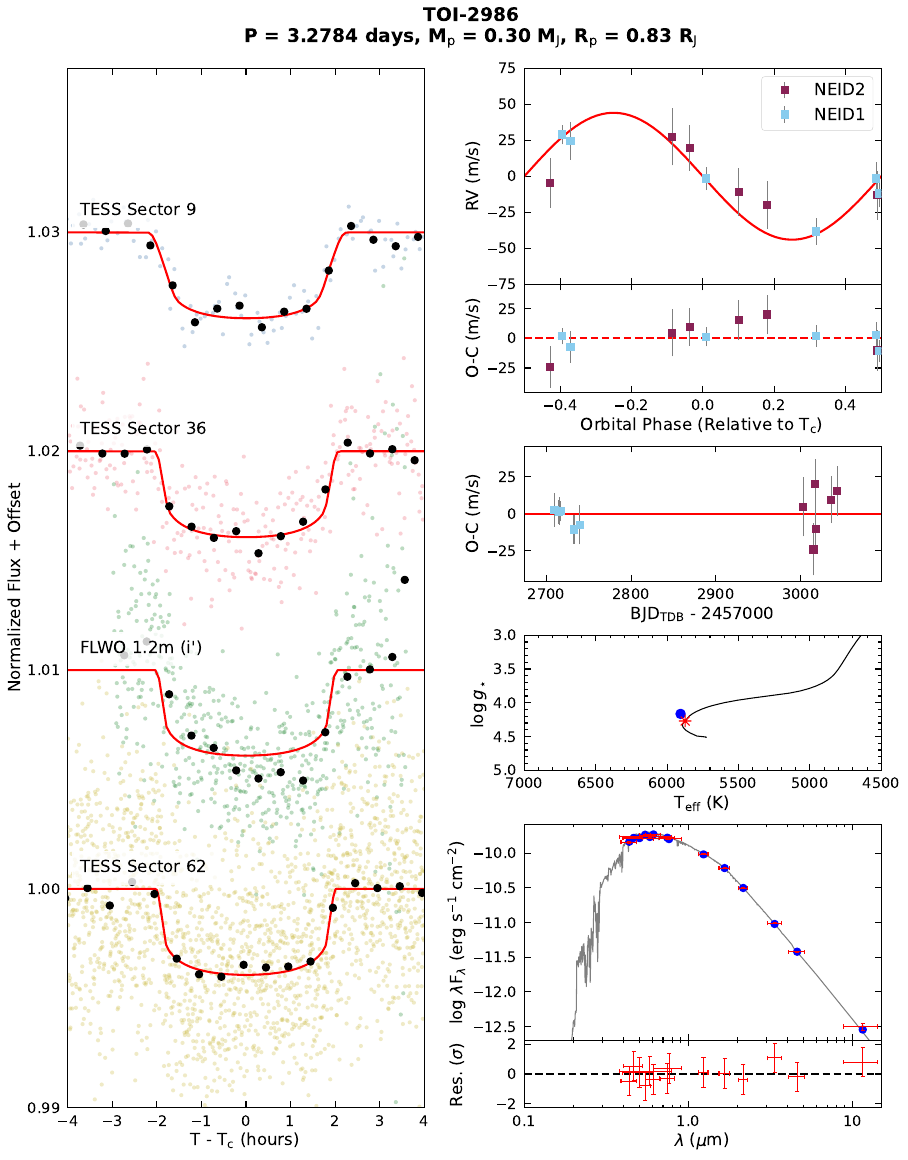}
\figsetgrpnote{Data and \Exofast fit results for TOI-2986\,b.
\textbf{Left:} \TESS and ground-based light-curves, phase-folded onto the best-fit period and time of conjunction.
Faint colored points represent the unbinned data, while large black circles show the time-series data binned to 30-min cadence.
The best-fit transit model in each band is shown as the red line.
\textbf{Top right:} RV observations, also phased onto the best-fit orbital period.
Error bars represent the fitted per-instrument jitter term $\sigma_\mathrm{jit}$ added in quadrature to the instrumental uncertainties.
The red line shows the best-fit RV model.
We plot the residuals after subtracting the model in the middle subpanel, and the unphased RV data and model time-series in the lower subpanel.
\textbf{Middle right:} The best-fit MIST stellar evolution track (black line), with a red asterisk marking the position along the track corresponding to the best-fit stellar age.
The blue point represents the best-fit stellar \Teff and \logg.
The discrepancy between the blue point and red asterisk are well within the fitted uncertainties in each parameter, indicating no tension between the different constraints on the stellar properties.
\textbf{Bottom right:} The observed stellar fluxes from the \Gaia, UCAC, 2MASS and WISE catalogs are plotted in red, with horizontal error bars corresponding to the width of the photometric bandpass.
The blue points show the best-fit model flux derived from the stellar properties and MIST bolometric correction grid.
We plot in gray an atmospheric model from \citet{Kurucz1993} corresponding to the best-fit stellar parameters for illustrative purposes only, as the fit is performed directly to the MIST grid.
The TESS and ground-based time-series photometry, as well as the RV measurements, are available as Data behind the Figure.}
\figsetgrpend

\figsetgrpstart
\figsetgrpnum{4.8}
\figsetgrptitle{Data and \Exofast fit results for TOI-2992\,b}
\figsetplot{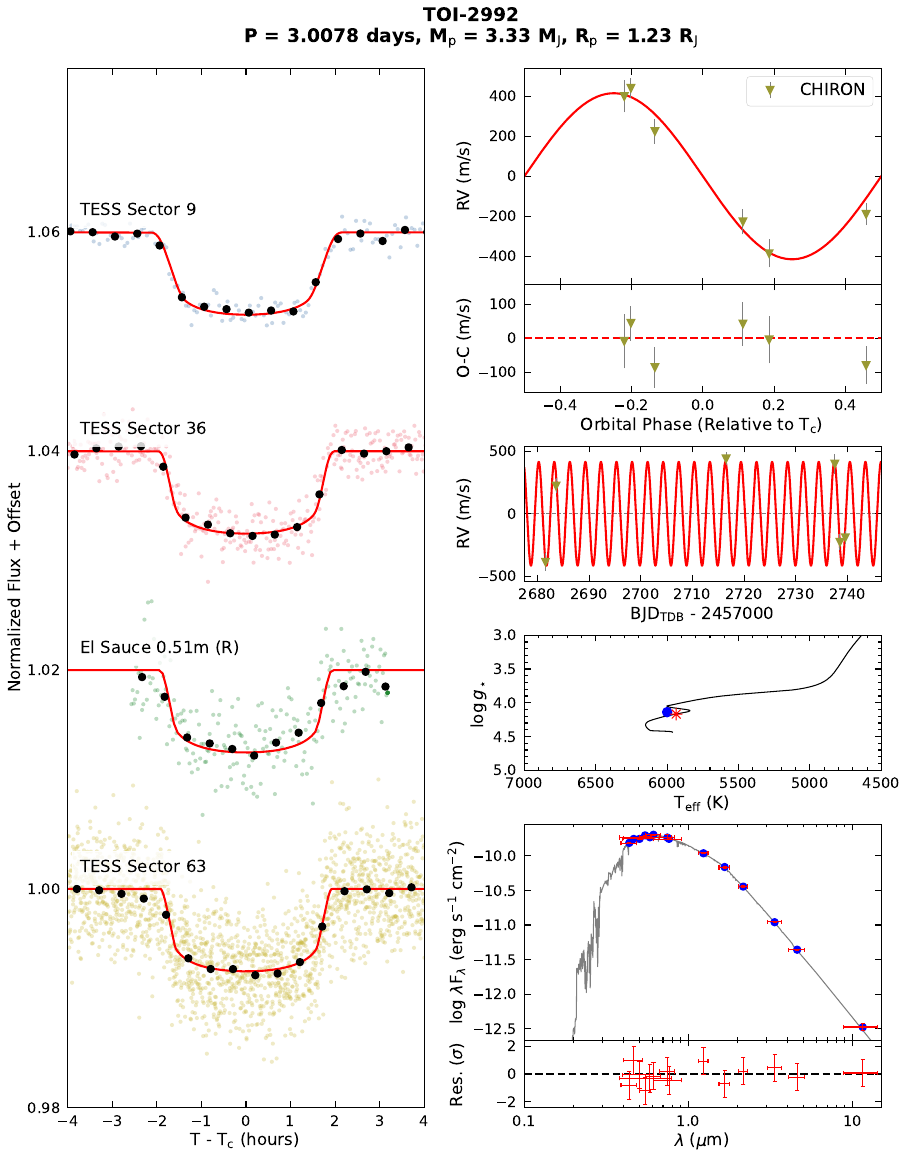}
\figsetgrpnote{Data and \Exofast fit results for TOI-2992\,b.
\textbf{Left:} \TESS and ground-based light-curves, phase-folded onto the best-fit period and time of conjunction.
Faint colored points represent the unbinned data, while large black circles show the time-series data binned to 30-min cadence.
The best-fit transit model in each band is shown as the red line.
\textbf{Top right:} RV observations, also phased onto the best-fit orbital period.
Error bars represent the fitted per-instrument jitter term $\sigma_\mathrm{jit}$ added in quadrature to the instrumental uncertainties.
The red line shows the best-fit RV model.
We plot the residuals after subtracting the model in the middle subpanel, and the unphased RV data and model time-series in the lower subpanel.
\textbf{Middle right:} The best-fit MIST stellar evolution track (black line), with a red asterisk marking the position along the track corresponding to the best-fit stellar age.
The blue point represents the best-fit stellar \Teff and \logg.
The discrepancy between the blue point and red asterisk are well within the fitted uncertainties in each parameter, indicating no tension between the different constraints on the stellar properties.
\textbf{Bottom right:} The observed stellar fluxes from the \Gaia, UCAC, 2MASS and WISE catalogs are plotted in red, with horizontal error bars corresponding to the width of the photometric bandpass.
The blue points show the best-fit model flux derived from the stellar properties and MIST bolometric correction grid.
We plot in gray an atmospheric model from \citet{Kurucz1993} corresponding to the best-fit stellar parameters for illustrative purposes only, as the fit is performed directly to the MIST grid.
The TESS and ground-based time-series photometry, as well as the RV measurements, are available as Data behind the Figure.}
\figsetgrpend

\figsetgrpstart
\figsetgrpnum{4.9}
\figsetgrptitle{Data and \Exofast fit results for TOI-3135\,b}
\figsetplot{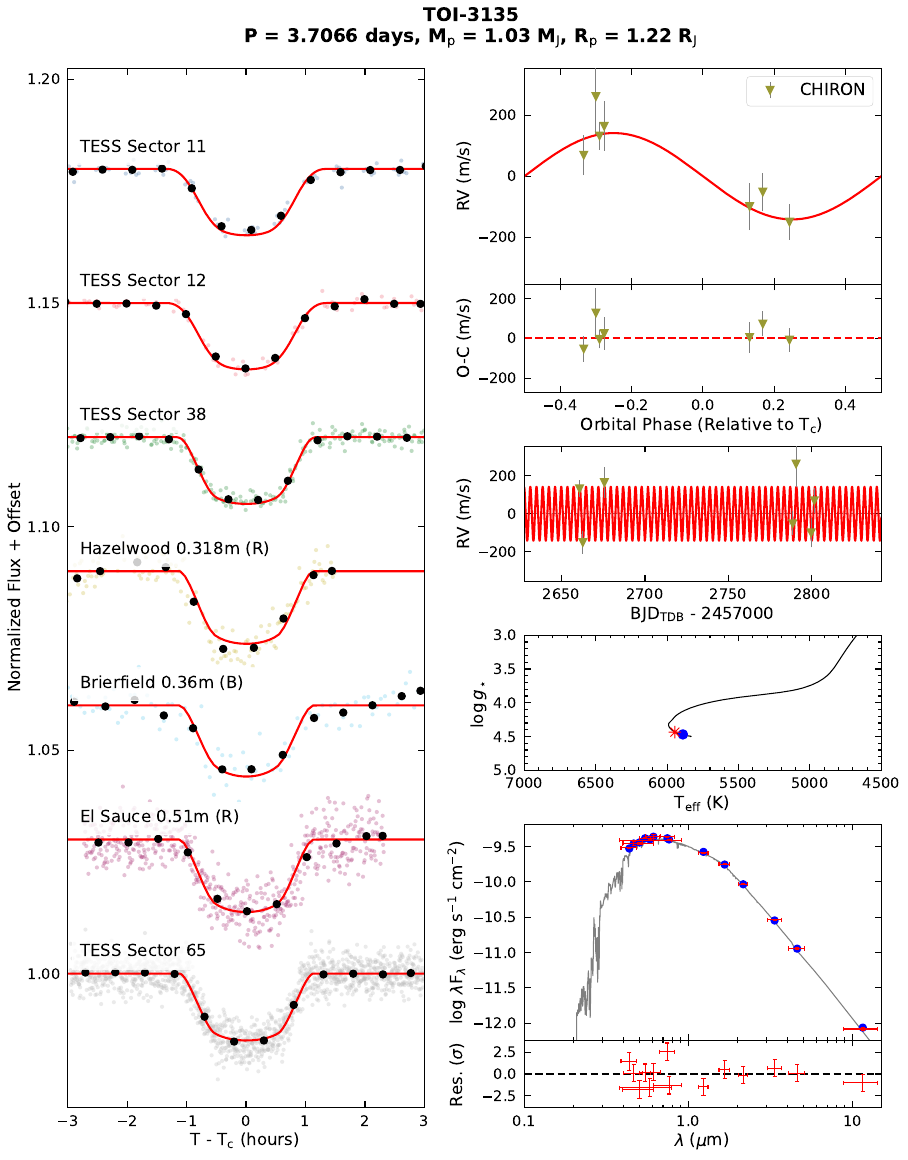}
\figsetgrpnote{Data and \Exofast fit results for TOI-3135\,b.
\textbf{Left:} \TESS and ground-based light-curves, phase-folded onto the best-fit period and time of conjunction.
Faint colored points represent the unbinned data, while large black circles show the time-series data binned to 30-min cadence.
The best-fit transit model in each band is shown as the red line.
\textbf{Top right:} RV observations, also phased onto the best-fit orbital period.
Error bars represent the fitted per-instrument jitter term $\sigma_\mathrm{jit}$ added in quadrature to the instrumental uncertainties.
The red line shows the best-fit RV model.
We plot the residuals after subtracting the model in the middle subpanel, and the unphased RV data and model time-series in the lower subpanel.
\textbf{Middle right:} The best-fit MIST stellar evolution track (black line), with a red asterisk marking the position along the track corresponding to the best-fit stellar age.
The blue point represents the best-fit stellar \Teff and \logg.
The discrepancy between the blue point and red asterisk are well within the fitted uncertainties in each parameter, indicating no tension between the different constraints on the stellar properties.
\textbf{Bottom right:} The observed stellar fluxes from the \Gaia, UCAC, 2MASS and WISE catalogs are plotted in red, with horizontal error bars corresponding to the width of the photometric bandpass.
The blue points show the best-fit model flux derived from the stellar properties and MIST bolometric correction grid.
We plot in gray an atmospheric model from \citet{Kurucz1993} corresponding to the best-fit stellar parameters for illustrative purposes only, as the fit is performed directly to the MIST grid.
The TESS and ground-based time-series photometry, as well as the RV measurements, are available as Data behind the Figure.}
\figsetgrpend

\figsetgrpstart
\figsetgrpnum{4.10}
\figsetgrptitle{Data and \Exofast fit results for TOI-3160\,b}
\figsetplot{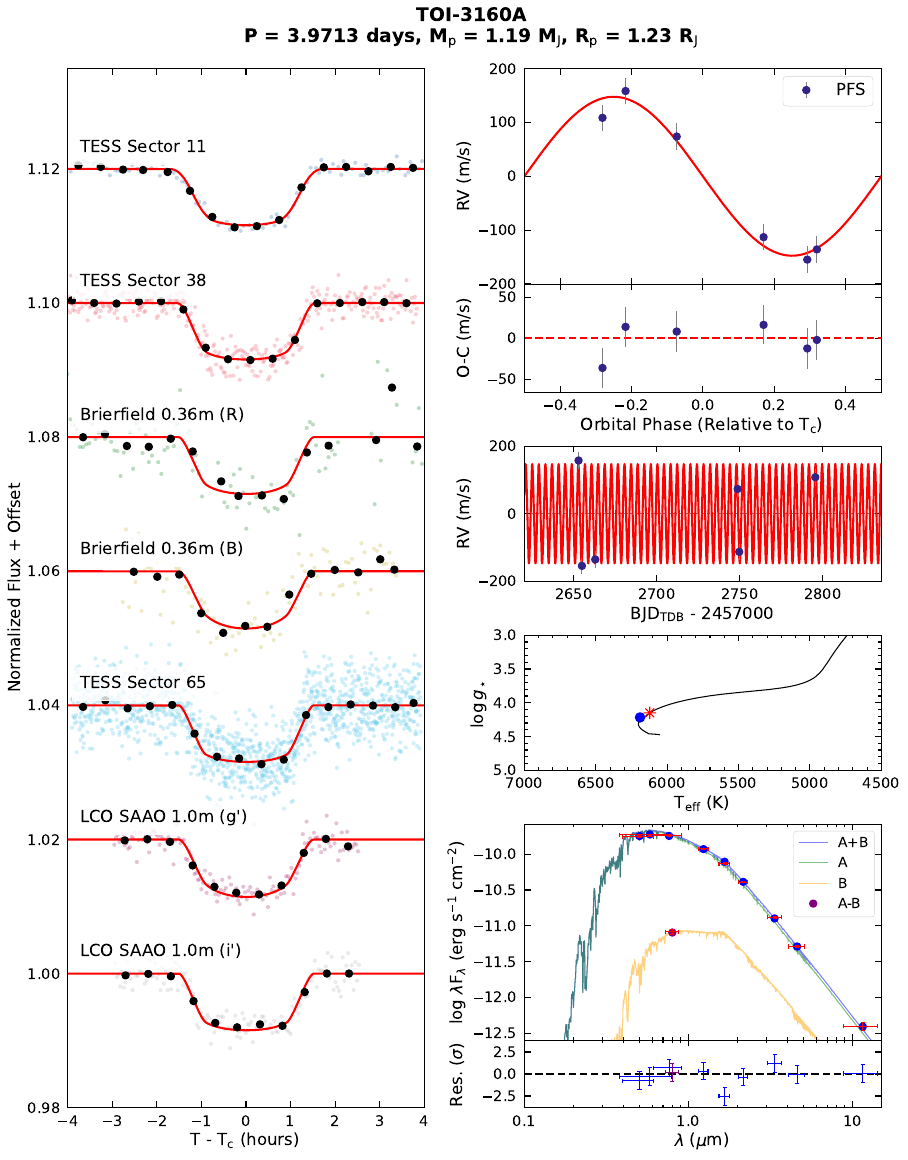}
\figsetgrpnote{Data and \Exofast fit results for TOI-3160\,b.
\textbf{Left:} \TESS and ground-based light-curves, phase-folded onto the best-fit period and time of conjunction.
Faint colored points represent the unbinned data, while large black circles show the time-series data binned to 30-min cadence.
The best-fit transit model in each band is shown as the red line.
\textbf{Top right:} RV observations, also phased onto the best-fit orbital period.
Error bars represent the fitted per-instrument jitter term $\sigma_\mathrm{jit}$ added in quadrature to the instrumental uncertainties.
The red line shows the best-fit RV model.
We plot the residuals after subtracting the model in the middle subpanel, and the unphased RV data and model time-series in the lower subpanel.
\textbf{Middle right:} The best-fit MIST stellar evolution track (black line), with a red asterisk marking the position along the track corresponding to the best-fit stellar age.
The blue point represents the best-fit stellar \Teff and \logg.
The discrepancy between the blue point and red asterisk are well within the fitted uncertainties in each parameter, indicating no tension between the different constraints on the stellar properties.
\textbf{Bottom right:} The observed stellar fluxes from the \Gaia, UCAC, 2MASS and WISE catalogs are plotted in red, with horizontal error bars corresponding to the width of the photometric bandpass.
The blue points show the best-fit model flux derived from the stellar properties and MIST bolometric correction grid.
We plot in gray an atmospheric model from \citet{Kurucz1993} corresponding to the best-fit stellar parameters for illustrative purposes only, as the fit is performed directly to the MIST grid.
The TESS and ground-based time-series photometry, as well as the RV measurements, are available as Data behind the Figure.}
\figsetgrpend

\figsetgrpstart
\figsetgrpnum{4.11}
\figsetgrptitle{Data and \Exofast fit results for TOI-3464\,b}
\figsetplot{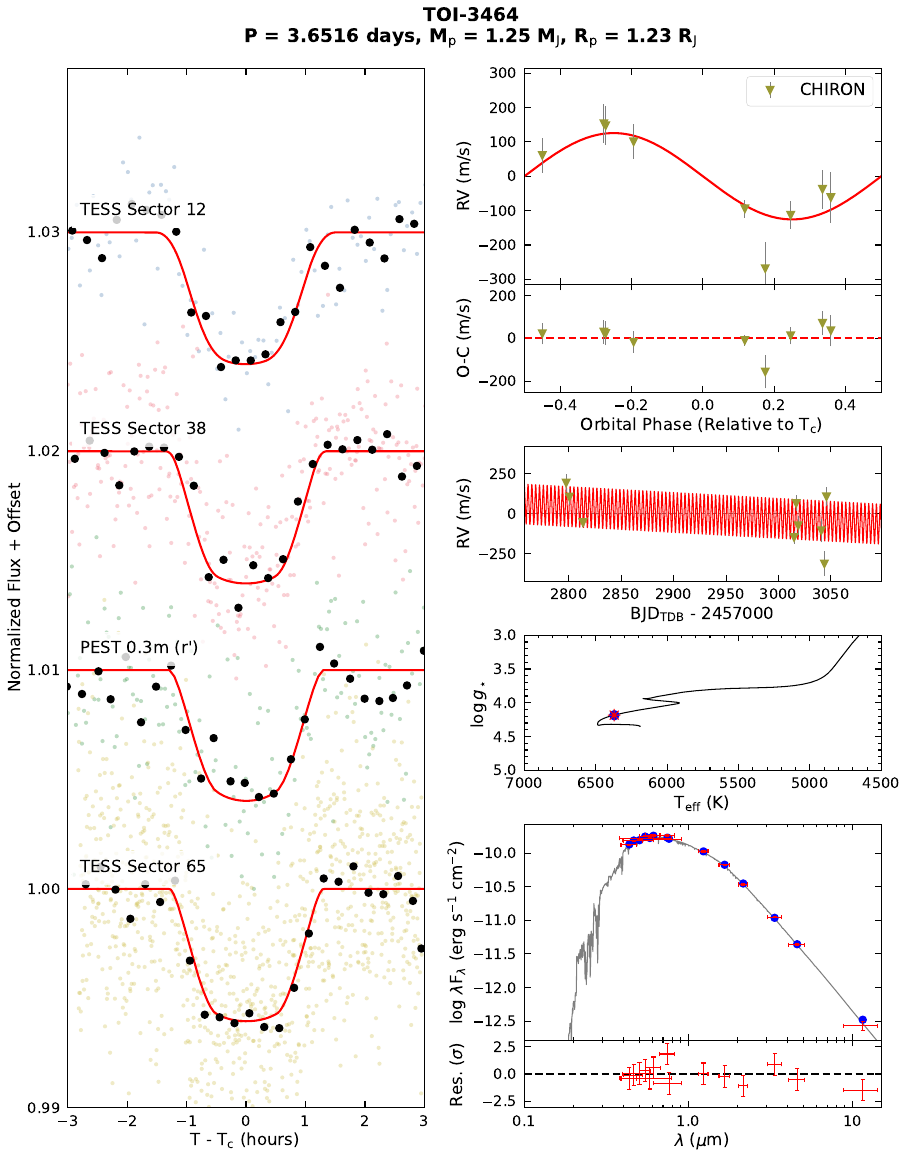}
\figsetgrpnote{Data and \Exofast fit results for TOI-3464\,b.
\textbf{Left:} \TESS and ground-based light-curves, phase-folded onto the best-fit period and time of conjunction.
Faint colored points represent the unbinned data, while large black circles show the time-series data binned to 30-min cadence.
The best-fit transit model in each band is shown as the red line.
\textbf{Top right:} RV observations, also phased onto the best-fit orbital period.
Error bars represent the fitted per-instrument jitter term $\sigma_\mathrm{jit}$ added in quadrature to the instrumental uncertainties.
The red line shows the best-fit RV model.
We plot the residuals after subtracting the model in the middle subpanel, and the unphased RV data and model time-series in the lower subpanel.
\textbf{Middle right:} The best-fit MIST stellar evolution track (black line), with a red asterisk marking the position along the track corresponding to the best-fit stellar age.
The blue point represents the best-fit stellar \Teff and \logg.
The discrepancy between the blue point and red asterisk are well within the fitted uncertainties in each parameter, indicating no tension between the different constraints on the stellar properties.
\textbf{Bottom right:} The observed stellar fluxes from the \Gaia, UCAC, 2MASS and WISE catalogs are plotted in red, with horizontal error bars corresponding to the width of the photometric bandpass.
The blue points show the best-fit model flux derived from the stellar properties and MIST bolometric correction grid.
We plot in gray an atmospheric model from \citet{Kurucz1993} corresponding to the best-fit stellar parameters for illustrative purposes only, as the fit is performed directly to the MIST grid.
The TESS and ground-based time-series photometry, as well as the RV measurements, are available as Data behind the Figure.}
\figsetgrpend

\figsetgrpstart
\figsetgrpnum{4.12}
\figsetgrptitle{Data and \Exofast fit results for TOI-3474\,b}
\figsetplot{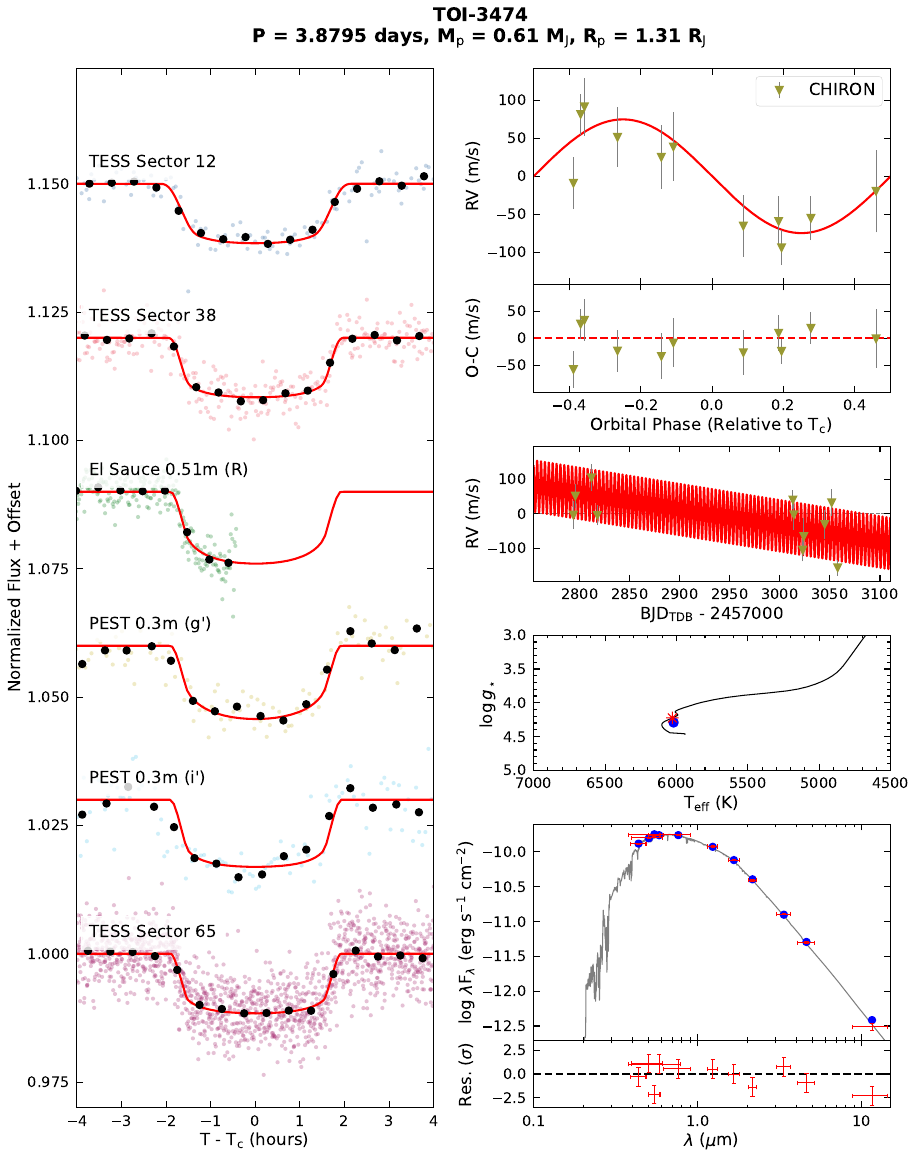}
\figsetgrpnote{Data and \Exofast fit results for TOI-3474\,b.
\textbf{Left:} \TESS and ground-based light-curves, phase-folded onto the best-fit period and time of conjunction.
Faint colored points represent the unbinned data, while large black circles show the time-series data binned to 30-min cadence.
The best-fit transit model in each band is shown as the red line.
\textbf{Top right:} RV observations, also phased onto the best-fit orbital period.
Error bars represent the fitted per-instrument jitter term $\sigma_\mathrm{jit}$ added in quadrature to the instrumental uncertainties.
The red line shows the best-fit RV model.
We plot the residuals after subtracting the model in the middle subpanel, and the unphased RV data and model time-series in the lower subpanel.
\textbf{Middle right:} The best-fit MIST stellar evolution track (black line), with a red asterisk marking the position along the track corresponding to the best-fit stellar age.
The blue point represents the best-fit stellar \Teff and \logg.
The discrepancy between the blue point and red asterisk are well within the fitted uncertainties in each parameter, indicating no tension between the different constraints on the stellar properties.
\textbf{Bottom right:} The observed stellar fluxes from the \Gaia, UCAC, 2MASS and WISE catalogs are plotted in red, with horizontal error bars corresponding to the width of the photometric bandpass.
The blue points show the best-fit model flux derived from the stellar properties and MIST bolometric correction grid.
We plot in gray an atmospheric model from \citet{Kurucz1993} corresponding to the best-fit stellar parameters for illustrative purposes only, as the fit is performed directly to the MIST grid.
The TESS and ground-based time-series photometry, as well as the RV measurements, are available as Data behind the Figure.}
\figsetgrpend

\figsetgrpstart
\figsetgrpnum{4.13}
\figsetgrptitle{Data and \Exofast fit results for TOI-3486\,b}
\figsetplot{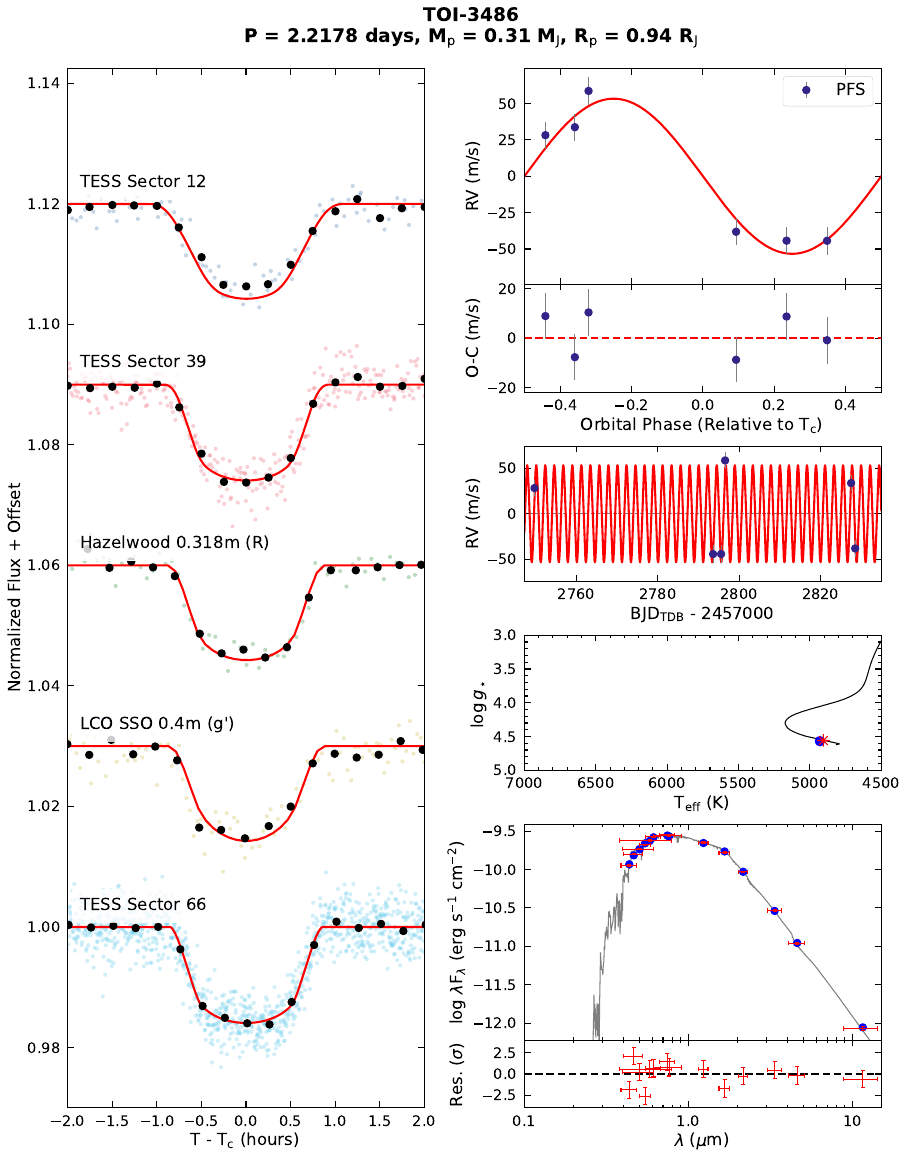}
\figsetgrpnote{Data and \Exofast fit results for TOI-3486\,b.
\textbf{Left:} \TESS and ground-based light-curves, phase-folded onto the best-fit period and time of conjunction.
Faint colored points represent the unbinned data, while large black circles show the time-series data binned to 30-min cadence.
The best-fit transit model in each band is shown as the red line.
\textbf{Top right:} RV observations, also phased onto the best-fit orbital period.
Error bars represent the fitted per-instrument jitter term $\sigma_\mathrm{jit}$ added in quadrature to the instrumental uncertainties.
The red line shows the best-fit RV model.
We plot the residuals after subtracting the model in the middle subpanel, and the unphased RV data and model time-series in the lower subpanel.
\textbf{Middle right:} The best-fit MIST stellar evolution track (black line), with a red asterisk marking the position along the track corresponding to the best-fit stellar age.
The blue point represents the best-fit stellar \Teff and \logg.
The discrepancy between the blue point and red asterisk are well within the fitted uncertainties in each parameter, indicating no tension between the different constraints on the stellar properties.
\textbf{Bottom right:} The observed stellar fluxes from the \Gaia, UCAC, 2MASS and WISE catalogs are plotted in red, with horizontal error bars corresponding to the width of the photometric bandpass.
The blue points show the best-fit model flux derived from the stellar properties and MIST bolometric correction grid.
We plot in gray an atmospheric model from \citet{Kurucz1993} corresponding to the best-fit stellar parameters for illustrative purposes only, as the fit is performed directly to the MIST grid.
The TESS and ground-based time-series photometry, as well as the RV measurements, are available as Data behind the Figure.}
\figsetgrpend

\figsetgrpstart
\figsetgrpnum{4.14}
\figsetgrptitle{Data and \Exofast fit results for TOI-3523\,b}
\figsetplot{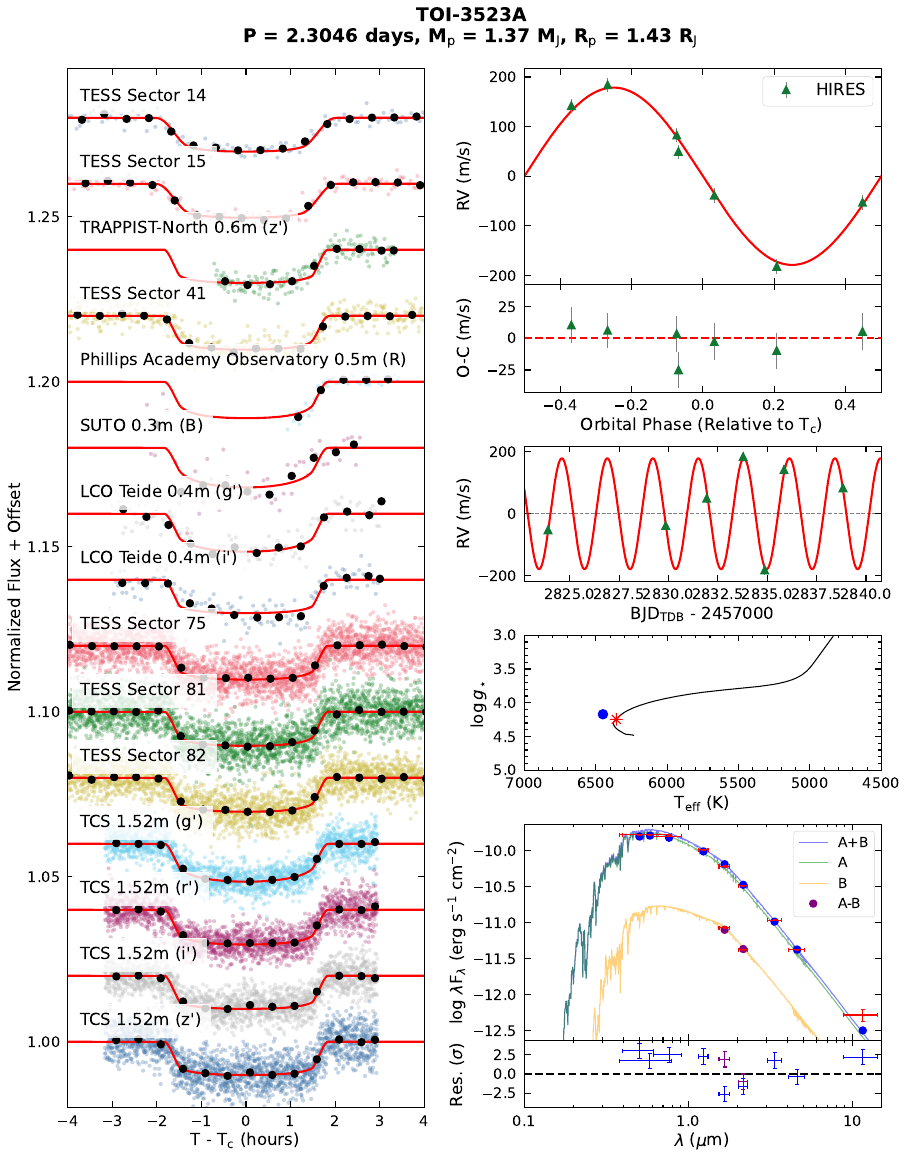}
\figsetgrpnote{Data and \Exofast fit results for TOI-3523\,b.
\textbf{Left:} \TESS and ground-based light-curves, phase-folded onto the best-fit period and time of conjunction.
Faint colored points represent the unbinned data, while large black circles show the time-series data binned to 30-min cadence.
The best-fit transit model in each band is shown as the red line.
\textbf{Top right:} RV observations, also phased onto the best-fit orbital period.
Error bars represent the fitted per-instrument jitter term $\sigma_\mathrm{jit}$ added in quadrature to the instrumental uncertainties.
The red line shows the best-fit RV model.
We plot the residuals after subtracting the model in the middle subpanel, and the unphased RV data and model time-series in the lower subpanel.
\textbf{Middle right:} The best-fit MIST stellar evolution track (black line), with a red asterisk marking the position along the track corresponding to the best-fit stellar age.
The blue point represents the best-fit stellar \Teff and \logg.
The discrepancy between the blue point and red asterisk are well within the fitted uncertainties in each parameter, indicating no tension between the different constraints on the stellar properties.
\textbf{Bottom right:} The observed stellar fluxes from the \Gaia, UCAC, 2MASS and WISE catalogs are plotted in red, with horizontal error bars corresponding to the width of the photometric bandpass.
The blue points show the best-fit model flux derived from the stellar properties and MIST bolometric correction grid.
We plot in gray an atmospheric model from \citet{Kurucz1993} corresponding to the best-fit stellar parameters for illustrative purposes only, as the fit is performed directly to the MIST grid.
The TESS and ground-based time-series photometry, as well as the RV measurements, are available as Data behind the Figure.}
\figsetgrpend

\figsetgrpstart
\figsetgrpnum{4.15}
\figsetgrptitle{Data and \Exofast fit results for TOI-3593\,b}
\figsetplot{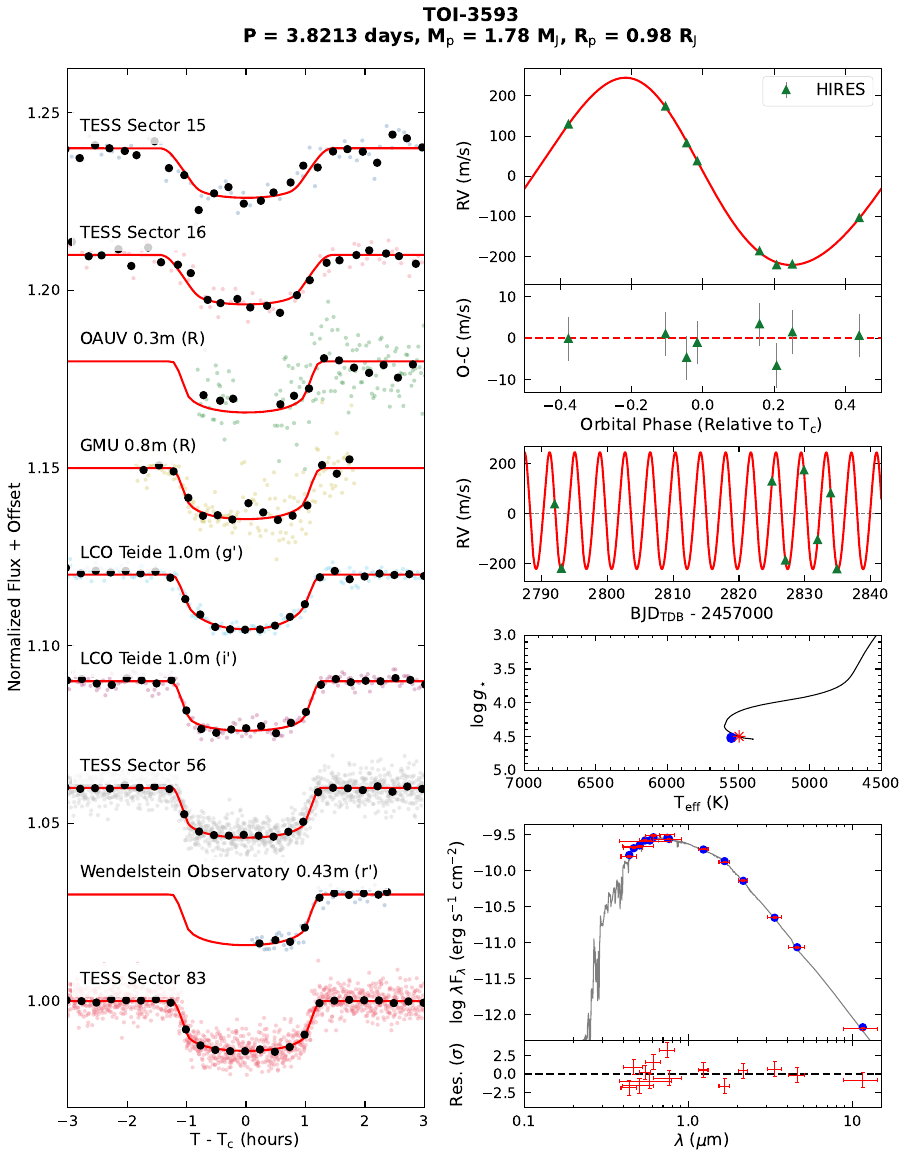}
\figsetgrpnote{Data and \Exofast fit results for TOI-3593\,b.
\textbf{Left:} \TESS and ground-based light-curves, phase-folded onto the best-fit period and time of conjunction.
Faint colored points represent the unbinned data, while large black circles show the time-series data binned to 30-min cadence.
The best-fit transit model in each band is shown as the red line.
\textbf{Top right:} RV observations, also phased onto the best-fit orbital period.
Error bars represent the fitted per-instrument jitter term $\sigma_\mathrm{jit}$ added in quadrature to the instrumental uncertainties.
The red line shows the best-fit RV model.
We plot the residuals after subtracting the model in the middle subpanel, and the unphased RV data and model time-series in the lower subpanel.
\textbf{Middle right:} The best-fit MIST stellar evolution track (black line), with a red asterisk marking the position along the track corresponding to the best-fit stellar age.
The blue point represents the best-fit stellar \Teff and \logg.
The discrepancy between the blue point and red asterisk are well within the fitted uncertainties in each parameter, indicating no tension between the different constraints on the stellar properties.
\textbf{Bottom right:} The observed stellar fluxes from the \Gaia, UCAC, 2MASS and WISE catalogs are plotted in red, with horizontal error bars corresponding to the width of the photometric bandpass.
The blue points show the best-fit model flux derived from the stellar properties and MIST bolometric correction grid.
We plot in gray an atmospheric model from \citet{Kurucz1993} corresponding to the best-fit stellar parameters for illustrative purposes only, as the fit is performed directly to the MIST grid.
The TESS and ground-based time-series photometry, as well as the RV measurements, are available as Data behind the Figure.}
\figsetgrpend

\figsetgrpstart
\figsetgrpnum{4.16}
\figsetgrptitle{Data and \Exofast fit results for TOI-3682\,b}
\figsetplot{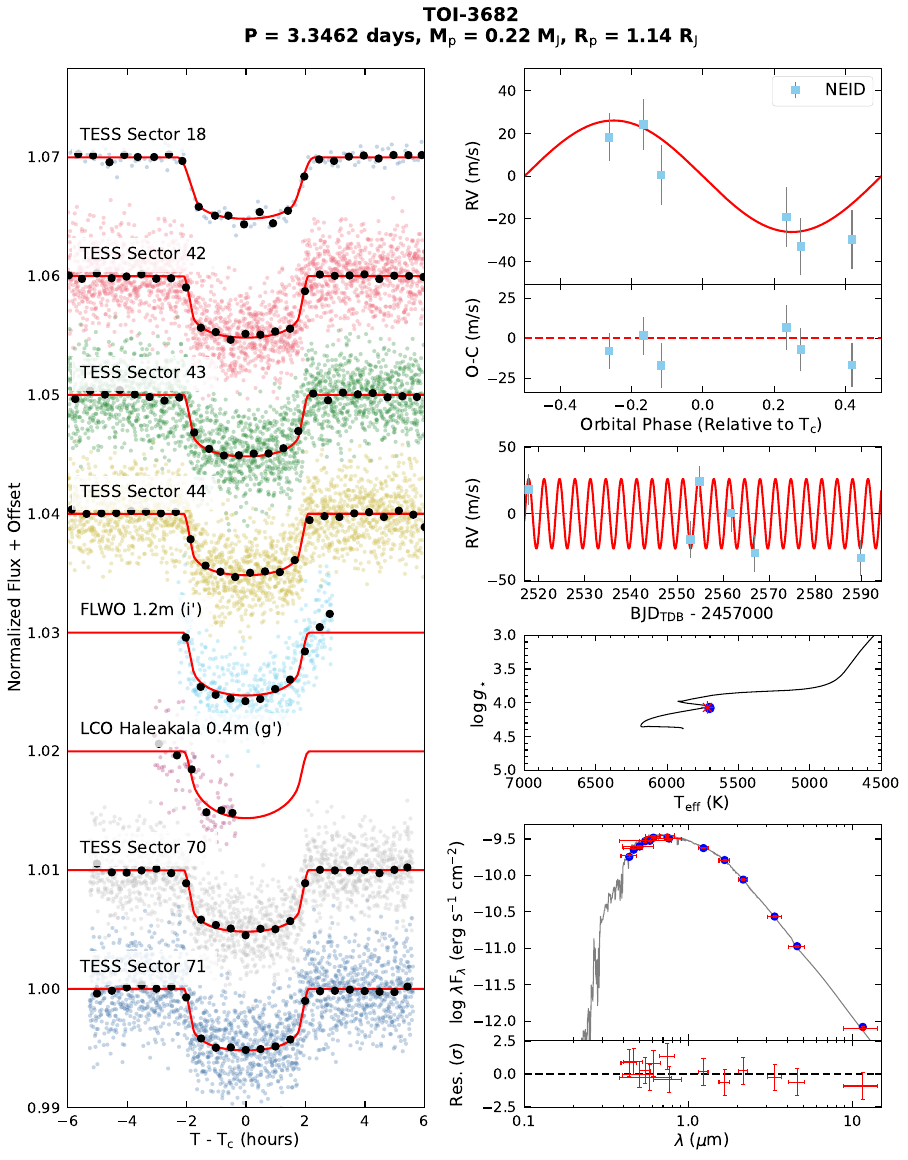}
\figsetgrpnote{Data and \Exofast fit results for TOI-3682\,b.
\textbf{Left:} \TESS and ground-based light-curves, phase-folded onto the best-fit period and time of conjunction.
Faint colored points represent the unbinned data, while large black circles show the time-series data binned to 30-min cadence.
The best-fit transit model in each band is shown as the red line.
\textbf{Top right:} RV observations, also phased onto the best-fit orbital period.
Error bars represent the fitted per-instrument jitter term $\sigma_\mathrm{jit}$ added in quadrature to the instrumental uncertainties.
The red line shows the best-fit RV model.
We plot the residuals after subtracting the model in the middle subpanel, and the unphased RV data and model time-series in the lower subpanel.
\textbf{Middle right:} The best-fit MIST stellar evolution track (black line), with a red asterisk marking the position along the track corresponding to the best-fit stellar age.
The blue point represents the best-fit stellar \Teff and \logg.
The discrepancy between the blue point and red asterisk are well within the fitted uncertainties in each parameter, indicating no tension between the different constraints on the stellar properties.
\textbf{Bottom right:} The observed stellar fluxes from the \Gaia, UCAC, 2MASS and WISE catalogs are plotted in red, with horizontal error bars corresponding to the width of the photometric bandpass.
The blue points show the best-fit model flux derived from the stellar properties and MIST bolometric correction grid.
We plot in gray an atmospheric model from \citet{Kurucz1993} corresponding to the best-fit stellar parameters for illustrative purposes only, as the fit is performed directly to the MIST grid.
The TESS and ground-based time-series photometry, as well as the RV measurements, are available as Data behind the Figure.}
\figsetgrpend

\figsetgrpstart
\figsetgrpnum{4.17}
\figsetgrptitle{Data and \Exofast fit results for TOI-3856\,b}
\figsetplot{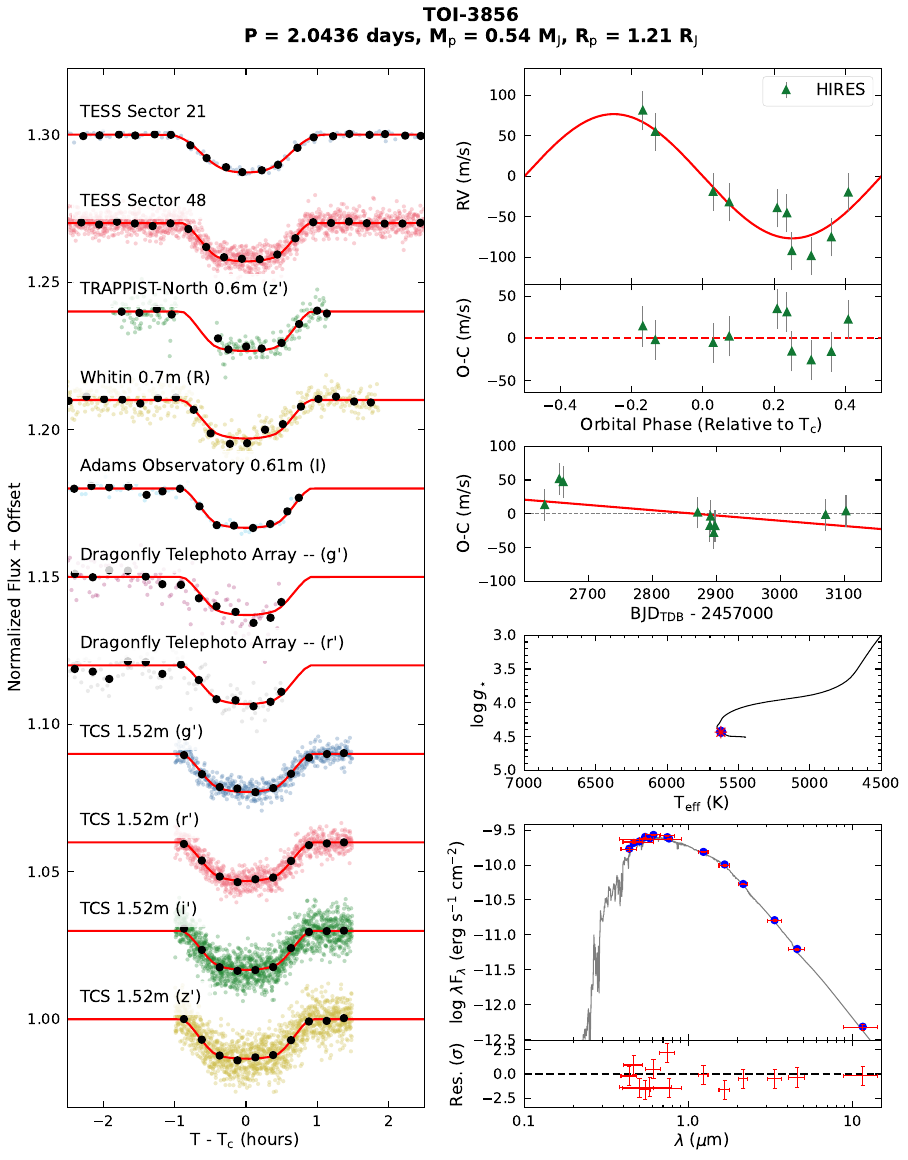}
\figsetgrpnote{Data and \Exofast fit results for TOI-3856\,b.
\textbf{Left:} \TESS and ground-based light-curves, phase-folded onto the best-fit period and time of conjunction.
Faint colored points represent the unbinned data, while large black circles show the time-series data binned to 30-min cadence.
The best-fit transit model in each band is shown as the red line.
\textbf{Top right:} RV observations, also phased onto the best-fit orbital period.
Error bars represent the fitted per-instrument jitter term $\sigma_\mathrm{jit}$ added in quadrature to the instrumental uncertainties.
The red line shows the best-fit RV model.
We plot the residuals after subtracting the model in the middle subpanel, and the unphased RV data and model time-series in the lower subpanel.
\textbf{Middle right:} The best-fit MIST stellar evolution track (black line), with a red asterisk marking the position along the track corresponding to the best-fit stellar age.
The blue point represents the best-fit stellar \Teff and \logg.
The discrepancy between the blue point and red asterisk are well within the fitted uncertainties in each parameter, indicating no tension between the different constraints on the stellar properties.
\textbf{Bottom right:} The observed stellar fluxes from the \Gaia, UCAC, 2MASS and WISE catalogs are plotted in red, with horizontal error bars corresponding to the width of the photometric bandpass.
The blue points show the best-fit model flux derived from the stellar properties and MIST bolometric correction grid.
We plot in gray an atmospheric model from \citet{Kurucz1993} corresponding to the best-fit stellar parameters for illustrative purposes only, as the fit is performed directly to the MIST grid.
The TESS and ground-based time-series photometry, as well as the RV measurements, are available as Data behind the Figure.}
\figsetgrpend

\figsetgrpstart
\figsetgrpnum{4.18}
\figsetgrptitle{Data and \Exofast fit results for TOI-3877\,b}
\figsetplot{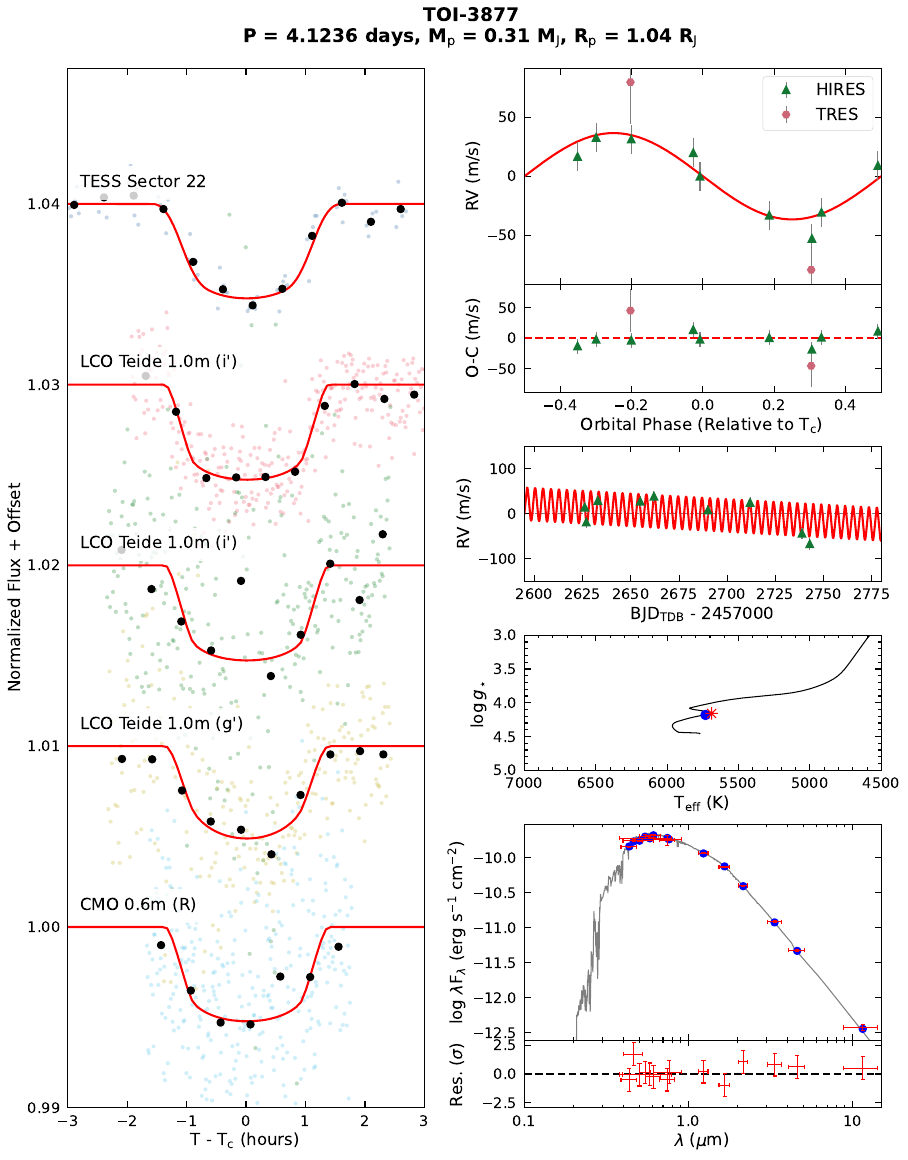}
\figsetgrpnote{Data and \Exofast fit results for TOI-3877\,b.
\textbf{Left:} \TESS and ground-based light-curves, phase-folded onto the best-fit period and time of conjunction.
Faint colored points represent the unbinned data, while large black circles show the time-series data binned to 30-min cadence.
The best-fit transit model in each band is shown as the red line.
\textbf{Top right:} RV observations, also phased onto the best-fit orbital period.
Error bars represent the fitted per-instrument jitter term $\sigma_\mathrm{jit}$ added in quadrature to the instrumental uncertainties.
The red line shows the best-fit RV model.
We plot the residuals after subtracting the model in the middle subpanel, and the unphased RV data and model time-series in the lower subpanel.
\textbf{Middle right:} The best-fit MIST stellar evolution track (black line), with a red asterisk marking the position along the track corresponding to the best-fit stellar age.
The blue point represents the best-fit stellar \Teff and \logg.
The discrepancy between the blue point and red asterisk are well within the fitted uncertainties in each parameter, indicating no tension between the different constraints on the stellar properties.
\textbf{Bottom right:} The observed stellar fluxes from the \Gaia, UCAC, 2MASS and WISE catalogs are plotted in red, with horizontal error bars corresponding to the width of the photometric bandpass.
The blue points show the best-fit model flux derived from the stellar properties and MIST bolometric correction grid.
We plot in gray an atmospheric model from \citet{Kurucz1993} corresponding to the best-fit stellar parameters for illustrative purposes only, as the fit is performed directly to the MIST grid.
The TESS and ground-based time-series photometry, as well as the RV measurements, are available as Data behind the Figure.}
\figsetgrpend

\figsetgrpstart
\figsetgrpnum{4.19}
\figsetgrptitle{Data and \Exofast fit results for TOI-3980\,b}
\figsetplot{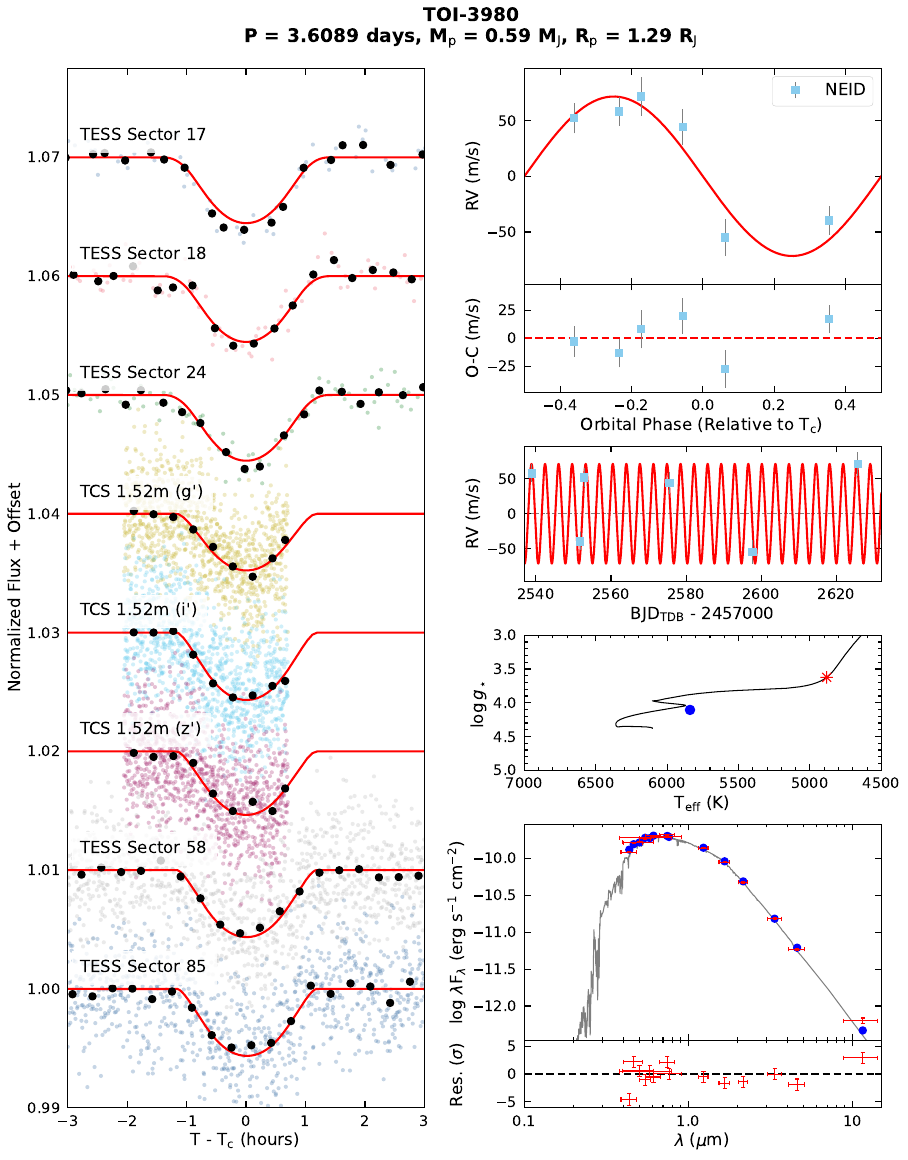}
\figsetgrpnote{Data and \Exofast fit results for TOI-3980\,b.
\textbf{Left:} \TESS and ground-based light-curves, phase-folded onto the best-fit period and time of conjunction.
Faint colored points represent the unbinned data, while large black circles show the time-series data binned to 30-min cadence.
The best-fit transit model in each band is shown as the red line.
\textbf{Top right:} RV observations, also phased onto the best-fit orbital period.
Error bars represent the fitted per-instrument jitter term $\sigma_\mathrm{jit}$ added in quadrature to the instrumental uncertainties.
The red line shows the best-fit RV model.
We plot the residuals after subtracting the model in the middle subpanel, and the unphased RV data and model time-series in the lower subpanel.
\textbf{Middle right:} The best-fit MIST stellar evolution track (black line), with a red asterisk marking the position along the track corresponding to the best-fit stellar age.
The blue point represents the best-fit stellar \Teff and \logg.
The discrepancy between the blue point and red asterisk are well within the fitted uncertainties in each parameter, indicating no tension between the different constraints on the stellar properties.
\textbf{Bottom right:} The observed stellar fluxes from the \Gaia, UCAC, 2MASS and WISE catalogs are plotted in red, with horizontal error bars corresponding to the width of the photometric bandpass.
The blue points show the best-fit model flux derived from the stellar properties and MIST bolometric correction grid.
We plot in gray an atmospheric model from \citet{Kurucz1993} corresponding to the best-fit stellar parameters for illustrative purposes only, as the fit is performed directly to the MIST grid.
The TESS and ground-based time-series photometry, as well as the RV measurements, are available as Data behind the Figure.}
\figsetgrpend

\figsetgrpstart
\figsetgrpnum{4.20}
\figsetgrptitle{Data and \Exofast fit results for TOI-4214\,b}
\figsetplot{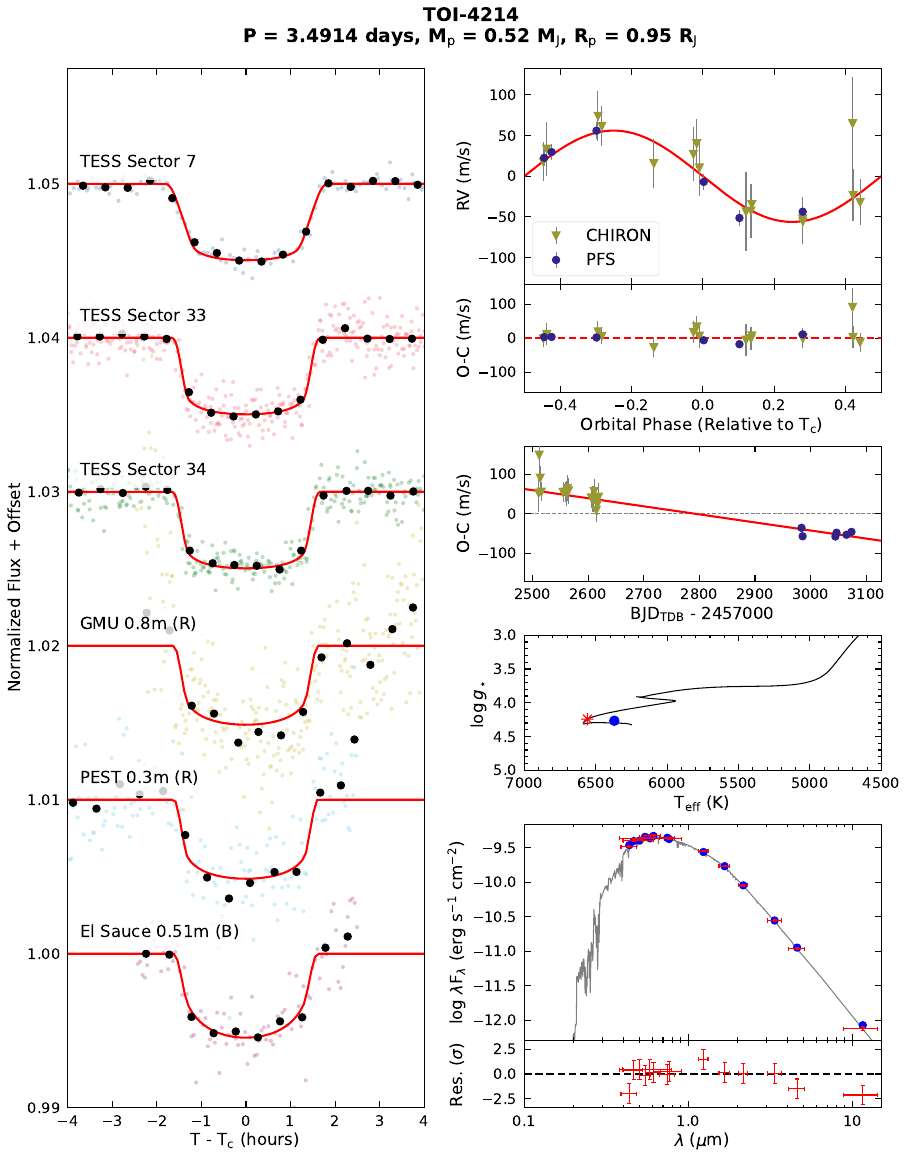}
\figsetgrpnote{Data and \Exofast fit results for TOI-4214\,b.
\textbf{Left:} \TESS and ground-based light-curves, phase-folded onto the best-fit period and time of conjunction.
Faint colored points represent the unbinned data, while large black circles show the time-series data binned to 30-min cadence.
The best-fit transit model in each band is shown as the red line.
\textbf{Top right:} RV observations, also phased onto the best-fit orbital period.
Error bars represent the fitted per-instrument jitter term $\sigma_\mathrm{jit}$ added in quadrature to the instrumental uncertainties.
The red line shows the best-fit RV model.
We plot the residuals after subtracting the model in the middle subpanel, and the unphased RV data and model time-series in the lower subpanel.
\textbf{Middle right:} The best-fit MIST stellar evolution track (black line), with a red asterisk marking the position along the track corresponding to the best-fit stellar age.
The blue point represents the best-fit stellar \Teff and \logg.
The discrepancy between the blue point and red asterisk are well within the fitted uncertainties in each parameter, indicating no tension between the different constraints on the stellar properties.
\textbf{Bottom right:} The observed stellar fluxes from the \Gaia, UCAC, 2MASS and WISE catalogs are plotted in red, with horizontal error bars corresponding to the width of the photometric bandpass.
The blue points show the best-fit model flux derived from the stellar properties and MIST bolometric correction grid.
We plot in gray an atmospheric model from \citet{Kurucz1993} corresponding to the best-fit stellar parameters for illustrative purposes only, as the fit is performed directly to the MIST grid.
The TESS and ground-based time-series photometry, as well as the RV measurements, are available as Data behind the Figure.}
\figsetgrpend

\figsetgrpstart
\figsetgrpnum{4.21}
\figsetgrptitle{Data and \Exofast fit results for TOI-4487\,b}
\figsetplot{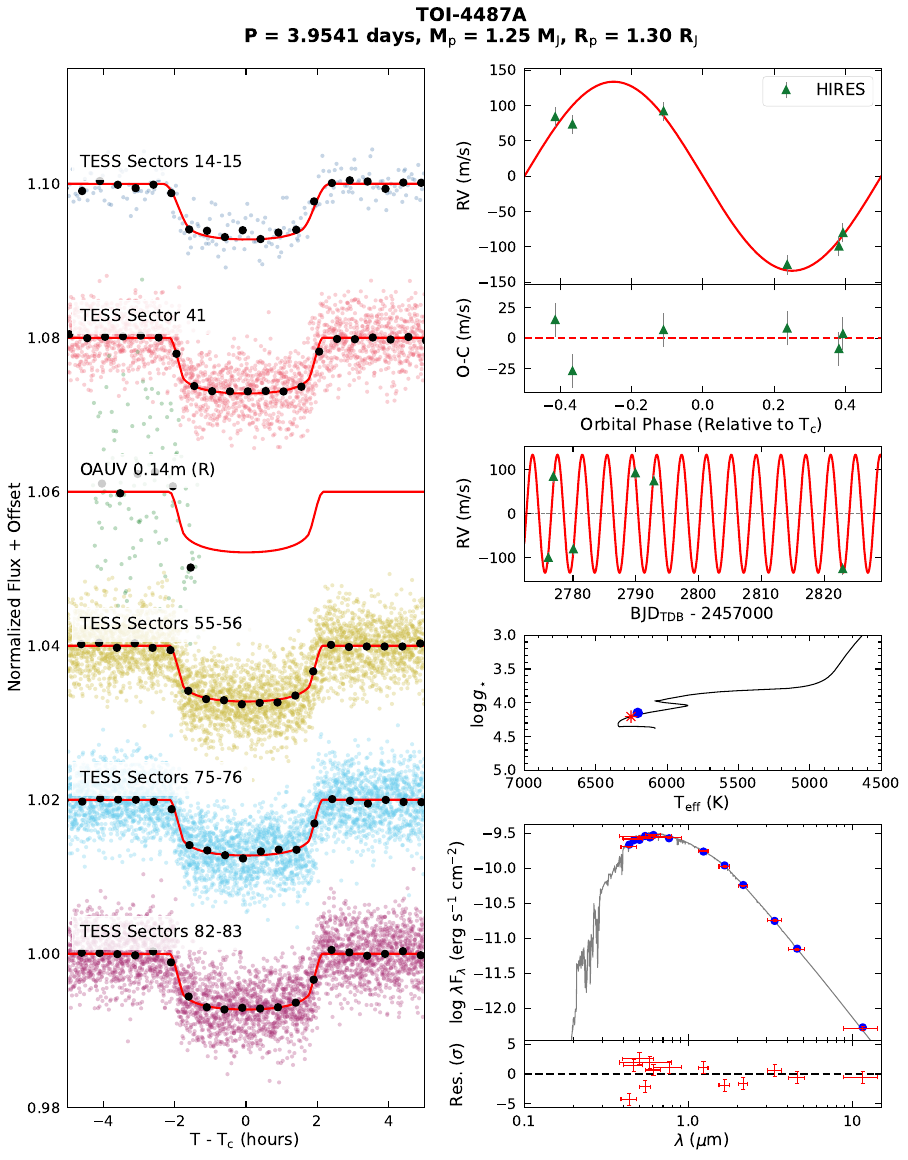}
\figsetgrpnote{Data and \Exofast fit results for TOI-4487\,b.
\textbf{Left:} \TESS and ground-based light-curves, phase-folded onto the best-fit period and time of conjunction.
Faint colored points represent the unbinned data, while large black circles show the time-series data binned to 30-min cadence.
The best-fit transit model in each band is shown as the red line.
\textbf{Top right:} RV observations, also phased onto the best-fit orbital period.
Error bars represent the fitted per-instrument jitter term $\sigma_\mathrm{jit}$ added in quadrature to the instrumental uncertainties.
The red line shows the best-fit RV model.
We plot the residuals after subtracting the model in the middle subpanel, and the unphased RV data and model time-series in the lower subpanel.
\textbf{Middle right:} The best-fit MIST stellar evolution track (black line), with a red asterisk marking the position along the track corresponding to the best-fit stellar age.
The blue point represents the best-fit stellar \Teff and \logg.
The discrepancy between the blue point and red asterisk are well within the fitted uncertainties in each parameter, indicating no tension between the different constraints on the stellar properties.
\textbf{Bottom right:} The observed stellar fluxes from the \Gaia, UCAC, 2MASS and WISE catalogs are plotted in red, with horizontal error bars corresponding to the width of the photometric bandpass.
The blue points show the best-fit model flux derived from the stellar properties and MIST bolometric correction grid.
We plot in gray an atmospheric model from \citet{Kurucz1993} corresponding to the best-fit stellar parameters for illustrative purposes only, as the fit is performed directly to the MIST grid.
The TESS and ground-based time-series photometry, as well as the RV measurements, are available as Data behind the Figure.}
\figsetgrpend

\figsetgrpstart
\figsetgrpnum{4.22}
\figsetgrptitle{Data and \Exofast fit results for TOI-4734\,b}
\figsetplot{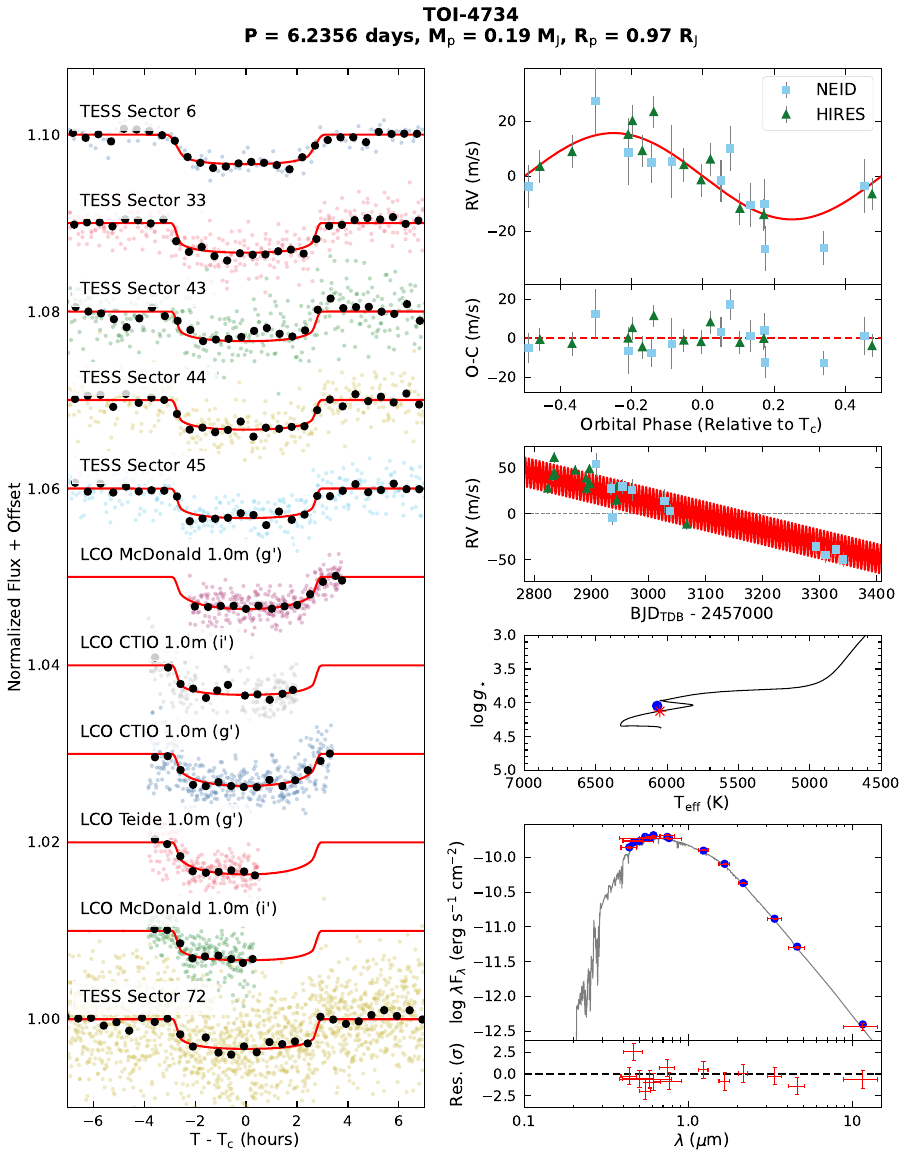}
\figsetgrpnote{Data and \Exofast fit results for TOI-4734\,b.
\textbf{Left:} \TESS and ground-based light-curves, phase-folded onto the best-fit period and time of conjunction.
Faint colored points represent the unbinned data, while large black circles show the time-series data binned to 30-min cadence.
The best-fit transit model in each band is shown as the red line.
\textbf{Top right:} RV observations, also phased onto the best-fit orbital period.
Error bars represent the fitted per-instrument jitter term $\sigma_\mathrm{jit}$ added in quadrature to the instrumental uncertainties.
The red line shows the best-fit RV model.
We plot the residuals after subtracting the model in the middle subpanel, and the unphased RV data and model time-series in the lower subpanel.
\textbf{Middle right:} The best-fit MIST stellar evolution track (black line), with a red asterisk marking the position along the track corresponding to the best-fit stellar age.
The blue point represents the best-fit stellar \Teff and \logg.
The discrepancy between the blue point and red asterisk are well within the fitted uncertainties in each parameter, indicating no tension between the different constraints on the stellar properties.
\textbf{Bottom right:} The observed stellar fluxes from the \Gaia, UCAC, 2MASS and WISE catalogs are plotted in red, with horizontal error bars corresponding to the width of the photometric bandpass.
The blue points show the best-fit model flux derived from the stellar properties and MIST bolometric correction grid.
We plot in gray an atmospheric model from \citet{Kurucz1993} corresponding to the best-fit stellar parameters for illustrative purposes only, as the fit is performed directly to the MIST grid.
The TESS and ground-based time-series photometry, as well as the RV measurements, are available as Data behind the Figure.}
\figsetgrpend

\figsetgrpstart
\figsetgrpnum{4.23}
\figsetgrptitle{Data and \Exofast fit results for TOI-4794\,b}
\figsetplot{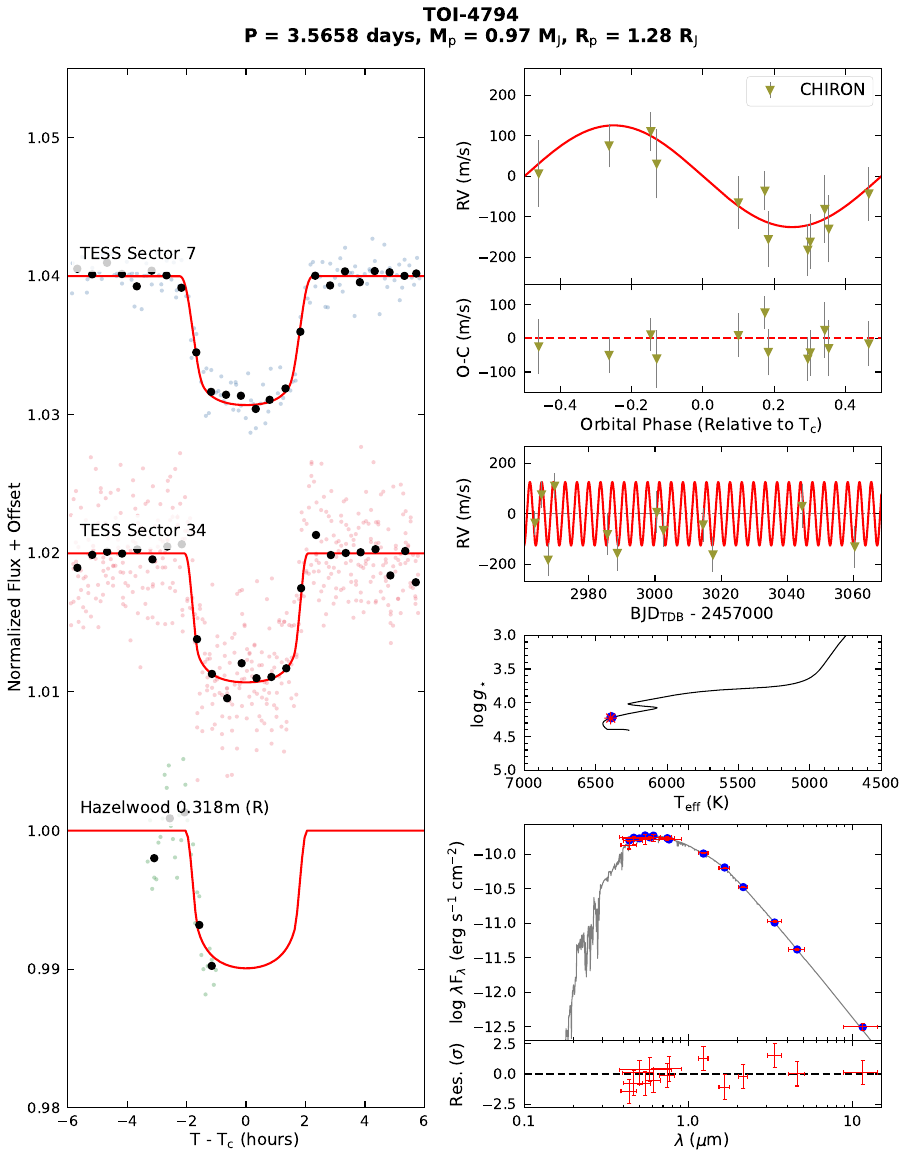}
\figsetgrpnote{Data and \Exofast fit results for TOI-4794\,b.
\textbf{Left:} \TESS and ground-based light-curves, phase-folded onto the best-fit period and time of conjunction.
Faint colored points represent the unbinned data, while large black circles show the time-series data binned to 30-min cadence.
The best-fit transit model in each band is shown as the red line.
\textbf{Top right:} RV observations, also phased onto the best-fit orbital period.
Error bars represent the fitted per-instrument jitter term $\sigma_\mathrm{jit}$ added in quadrature to the instrumental uncertainties.
The red line shows the best-fit RV model.
We plot the residuals after subtracting the model in the middle subpanel, and the unphased RV data and model time-series in the lower subpanel.
\textbf{Middle right:} The best-fit MIST stellar evolution track (black line), with a red asterisk marking the position along the track corresponding to the best-fit stellar age.
The blue point represents the best-fit stellar \Teff and \logg.
The discrepancy between the blue point and red asterisk are well within the fitted uncertainties in each parameter, indicating no tension between the different constraints on the stellar properties.
\textbf{Bottom right:} The observed stellar fluxes from the \Gaia, UCAC, 2MASS and WISE catalogs are plotted in red, with horizontal error bars corresponding to the width of the photometric bandpass.
The blue points show the best-fit model flux derived from the stellar properties and MIST bolometric correction grid.
We plot in gray an atmospheric model from \citet{Kurucz1993} corresponding to the best-fit stellar parameters for illustrative purposes only, as the fit is performed directly to the MIST grid.
The TESS and ground-based time-series photometry, as well as the RV measurements, are available as Data behind the Figure.}
\figsetgrpend

\figsetgrpstart
\figsetgrpnum{4.24}
\figsetgrptitle{Data and \Exofast fit results for TOI-4961\,b}
\figsetplot{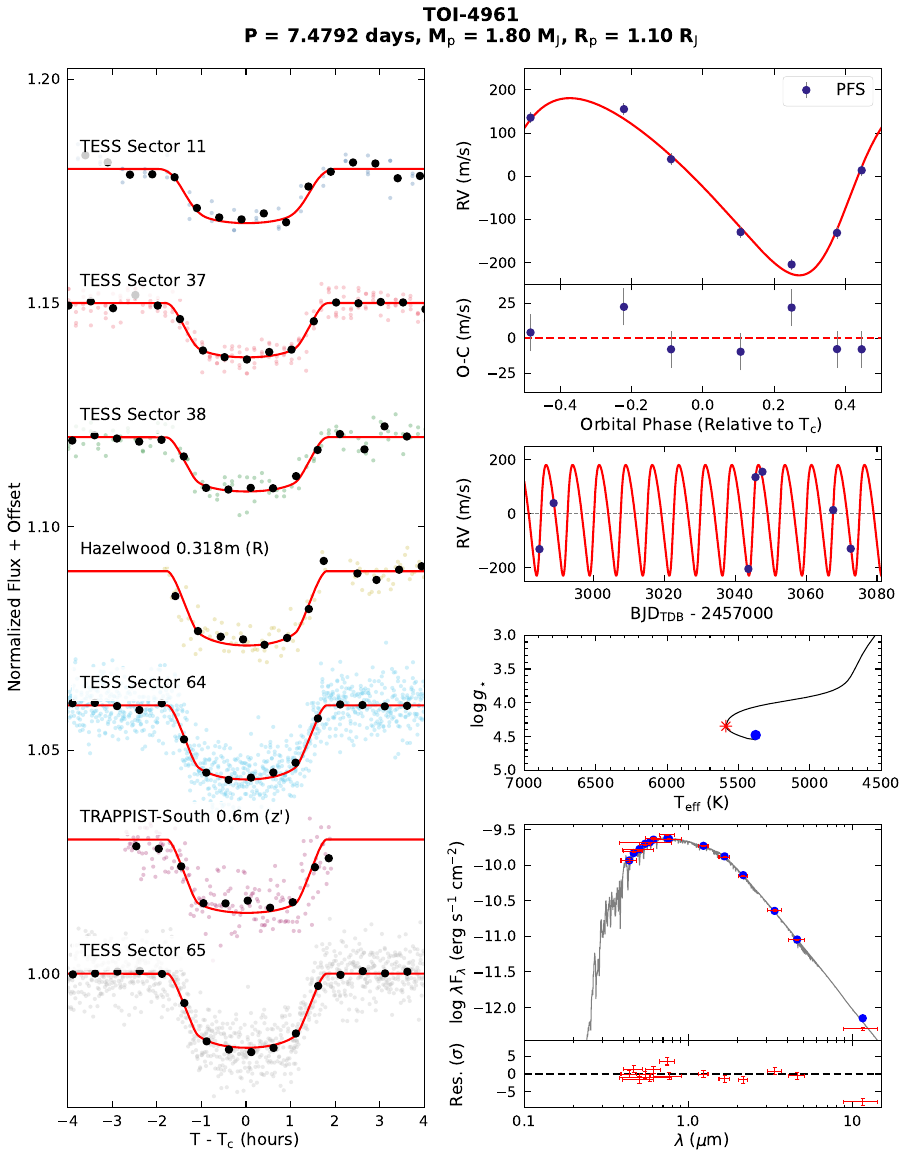}
\figsetgrpnote{Data and \Exofast fit results for TOI-4961\,b.
\textbf{Left:} \TESS and ground-based light-curves, phase-folded onto the best-fit period and time of conjunction.
Faint colored points represent the unbinned data, while large black circles show the time-series data binned to 30-min cadence.
The best-fit transit model in each band is shown as the red line.
\textbf{Top right:} RV observations, also phased onto the best-fit orbital period.
Error bars represent the fitted per-instrument jitter term $\sigma_\mathrm{jit}$ added in quadrature to the instrumental uncertainties.
The red line shows the best-fit RV model.
We plot the residuals after subtracting the model in the middle subpanel, and the unphased RV data and model time-series in the lower subpanel.
\textbf{Middle right:} The best-fit MIST stellar evolution track (black line), with a red asterisk marking the position along the track corresponding to the best-fit stellar age.
The blue point represents the best-fit stellar \Teff and \logg.
The discrepancy between the blue point and red asterisk are well within the fitted uncertainties in each parameter, indicating no tension between the different constraints on the stellar properties.
\textbf{Bottom right:} The observed stellar fluxes from the \Gaia, UCAC, 2MASS and WISE catalogs are plotted in red, with horizontal error bars corresponding to the width of the photometric bandpass.
The blue points show the best-fit model flux derived from the stellar properties and MIST bolometric correction grid.
We plot in gray an atmospheric model from \citet{Kurucz1993} corresponding to the best-fit stellar parameters for illustrative purposes only, as the fit is performed directly to the MIST grid.
The TESS and ground-based time-series photometry, as well as the RV measurements, are available as Data behind the Figure.}
\figsetgrpend

\figsetgrpstart
\figsetgrpnum{4.25}
\figsetgrptitle{Data and \Exofast fit results for TOI-5181\,b}
\figsetplot{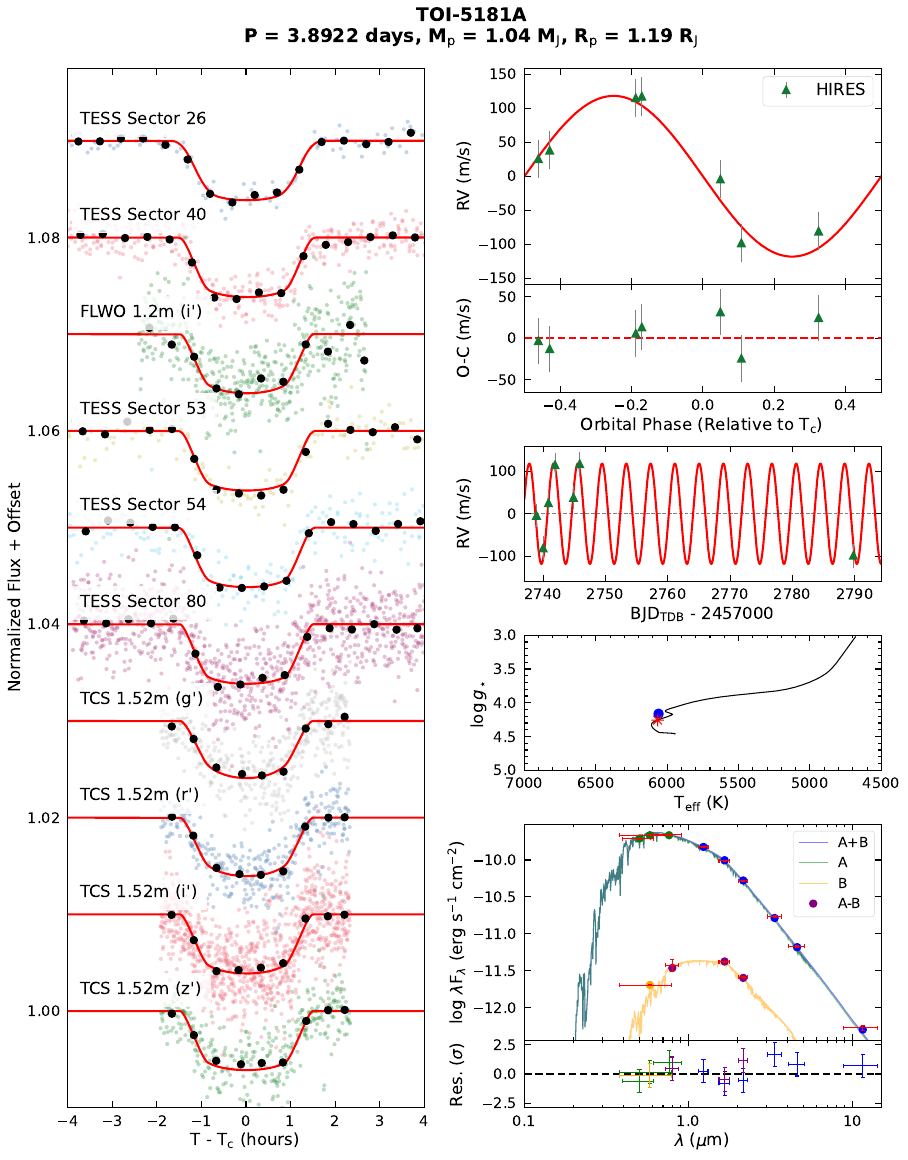}
\figsetgrpnote{Data and \Exofast fit results for TOI-5181\,b.
\textbf{Left:} \TESS and ground-based light-curves, phase-folded onto the best-fit period and time of conjunction.
Faint colored points represent the unbinned data, while large black circles show the time-series data binned to 30-min cadence.
The best-fit transit model in each band is shown as the red line.
\textbf{Top right:} RV observations, also phased onto the best-fit orbital period.
Error bars represent the fitted per-instrument jitter term $\sigma_\mathrm{jit}$ added in quadrature to the instrumental uncertainties.
The red line shows the best-fit RV model.
We plot the residuals after subtracting the model in the middle subpanel, and the unphased RV data and model time-series in the lower subpanel.
\textbf{Middle right:} The best-fit MIST stellar evolution track (black line), with a red asterisk marking the position along the track corresponding to the best-fit stellar age.
The blue point represents the best-fit stellar \Teff and \logg.
The discrepancy between the blue point and red asterisk are well within the fitted uncertainties in each parameter, indicating no tension between the different constraints on the stellar properties.
\textbf{Bottom right:} The observed stellar fluxes from the \Gaia, UCAC, 2MASS and WISE catalogs are plotted in red, with horizontal error bars corresponding to the width of the photometric bandpass.
The blue points show the best-fit model flux derived from the stellar properties and MIST bolometric correction grid.
We plot in gray an atmospheric model from \citet{Kurucz1993} corresponding to the best-fit stellar parameters for illustrative purposes only, as the fit is performed directly to the MIST grid.
The TESS and ground-based time-series photometry, as well as the RV measurements, are available as Data behind the Figure.}
\figsetgrpend

\figsetgrpstart
\figsetgrpnum{4.26}
\figsetgrptitle{Data and \Exofast fit results for TOI-5210\,b}
\figsetplot{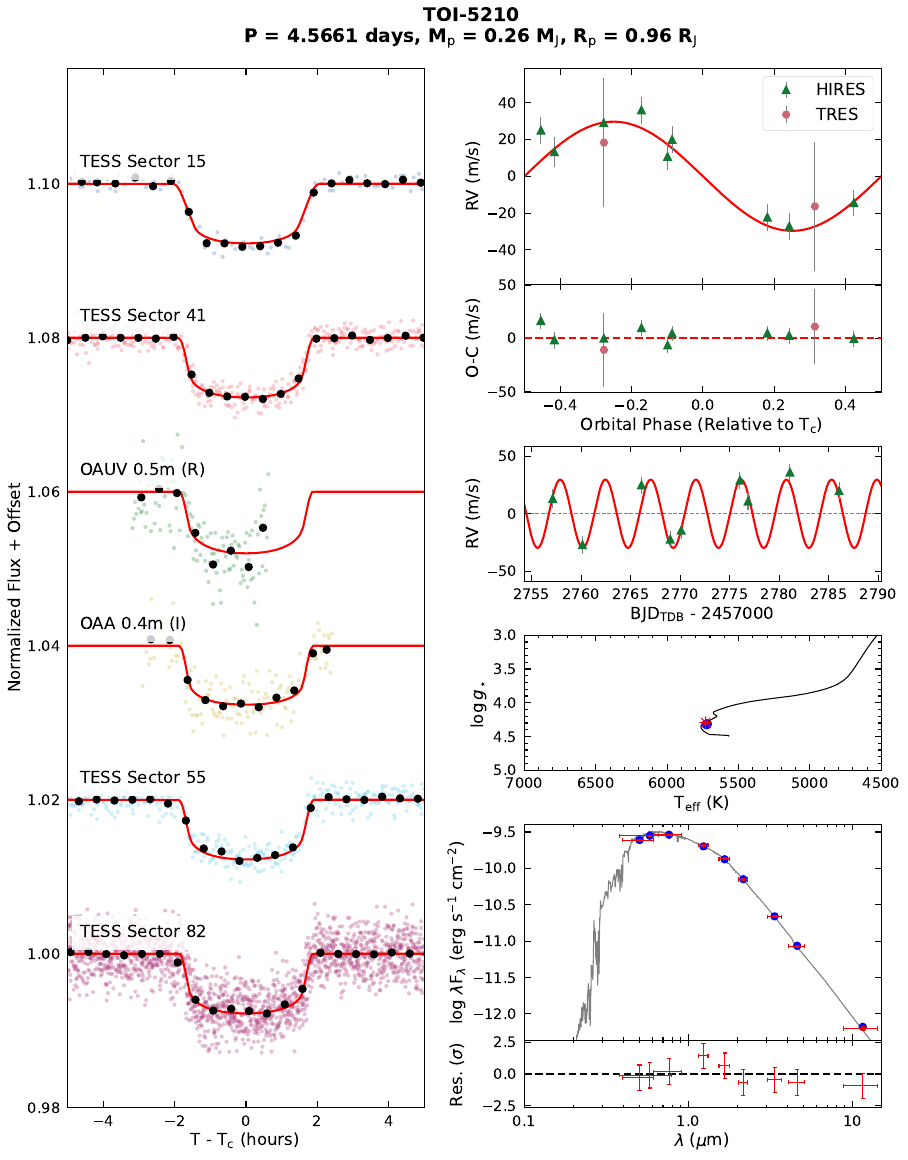}
\figsetgrpnote{Data and \Exofast fit results for TOI-5210\,b.
\textbf{Left:} \TESS and ground-based light-curves, phase-folded onto the best-fit period and time of conjunction.
Faint colored points represent the unbinned data, while large black circles show the time-series data binned to 30-min cadence.
The best-fit transit model in each band is shown as the red line.
\textbf{Top right:} RV observations, also phased onto the best-fit orbital period.
Error bars represent the fitted per-instrument jitter term $\sigma_\mathrm{jit}$ added in quadrature to the instrumental uncertainties.
The red line shows the best-fit RV model.
We plot the residuals after subtracting the model in the middle subpanel, and the unphased RV data and model time-series in the lower subpanel.
\textbf{Middle right:} The best-fit MIST stellar evolution track (black line), with a red asterisk marking the position along the track corresponding to the best-fit stellar age.
The blue point represents the best-fit stellar \Teff and \logg.
The discrepancy between the blue point and red asterisk are well within the fitted uncertainties in each parameter, indicating no tension between the different constraints on the stellar properties.
\textbf{Bottom right:} The observed stellar fluxes from the \Gaia, UCAC, 2MASS and WISE catalogs are plotted in red, with horizontal error bars corresponding to the width of the photometric bandpass.
The blue points show the best-fit model flux derived from the stellar properties and MIST bolometric correction grid.
We plot in gray an atmospheric model from \citet{Kurucz1993} corresponding to the best-fit stellar parameters for illustrative purposes only, as the fit is performed directly to the MIST grid.
The TESS and ground-based time-series photometry, as well as the RV measurements, are available as Data behind the Figure.}
\figsetgrpend

\figsetgrpstart
\figsetgrpnum{4.27}
\figsetgrptitle{Data and \Exofast fit results for TOI-5322\,b}
\figsetplot{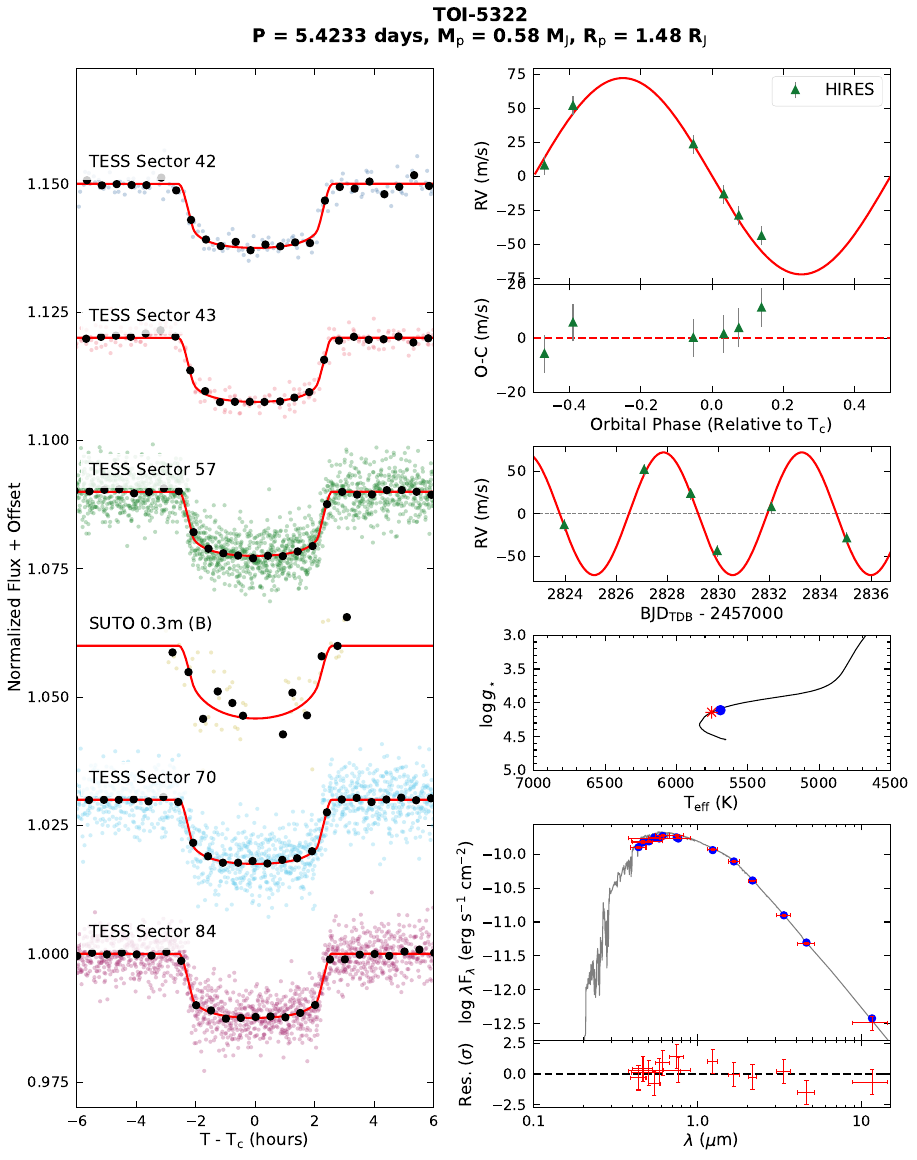}
\figsetgrpnote{Data and \Exofast fit results for TOI-5322\,b.
\textbf{Left:} \TESS and ground-based light-curves, phase-folded onto the best-fit period and time of conjunction.
Faint colored points represent the unbinned data, while large black circles show the time-series data binned to 30-min cadence.
The best-fit transit model in each band is shown as the red line.
\textbf{Top right:} RV observations, also phased onto the best-fit orbital period.
Error bars represent the fitted per-instrument jitter term $\sigma_\mathrm{jit}$ added in quadrature to the instrumental uncertainties.
The red line shows the best-fit RV model.
We plot the residuals after subtracting the model in the middle subpanel, and the unphased RV data and model time-series in the lower subpanel.
\textbf{Middle right:} The best-fit MIST stellar evolution track (black line), with a red asterisk marking the position along the track corresponding to the best-fit stellar age.
The blue point represents the best-fit stellar \Teff and \logg.
The discrepancy between the blue point and red asterisk are well within the fitted uncertainties in each parameter, indicating no tension between the different constraints on the stellar properties.
\textbf{Bottom right:} The observed stellar fluxes from the \Gaia, UCAC, 2MASS and WISE catalogs are plotted in red, with horizontal error bars corresponding to the width of the photometric bandpass.
The blue points show the best-fit model flux derived from the stellar properties and MIST bolometric correction grid.
We plot in gray an atmospheric model from \citet{Kurucz1993} corresponding to the best-fit stellar parameters for illustrative purposes only, as the fit is performed directly to the MIST grid.
The TESS and ground-based time-series photometry, as well as the RV measurements, are available as Data behind the Figure.}
\figsetgrpend

\figsetgrpstart
\figsetgrpnum{4.28}
\figsetgrptitle{Data and \Exofast fit results for TOI-5340\,b}
\figsetplot{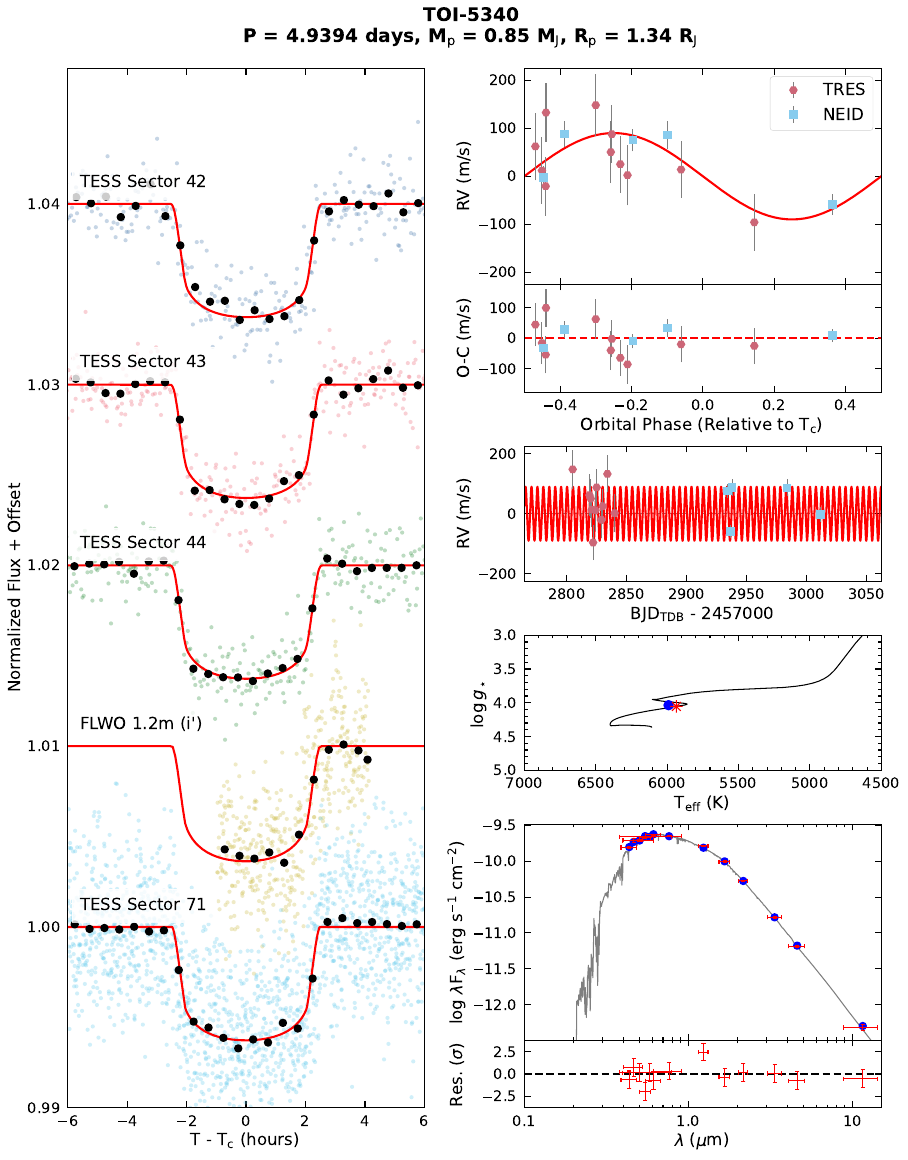}
\figsetgrpnote{Data and \Exofast fit results for TOI-5340\,b.
\textbf{Left:} \TESS and ground-based light-curves, phase-folded onto the best-fit period and time of conjunction.
Faint colored points represent the unbinned data, while large black circles show the time-series data binned to 30-min cadence.
The best-fit transit model in each band is shown as the red line.
\textbf{Top right:} RV observations, also phased onto the best-fit orbital period.
Error bars represent the fitted per-instrument jitter term $\sigma_\mathrm{jit}$ added in quadrature to the instrumental uncertainties.
The red line shows the best-fit RV model.
We plot the residuals after subtracting the model in the middle subpanel, and the unphased RV data and model time-series in the lower subpanel.
\textbf{Middle right:} The best-fit MIST stellar evolution track (black line), with a red asterisk marking the position along the track corresponding to the best-fit stellar age.
The blue point represents the best-fit stellar \Teff and \logg.
The discrepancy between the blue point and red asterisk are well within the fitted uncertainties in each parameter, indicating no tension between the different constraints on the stellar properties.
\textbf{Bottom right:} The observed stellar fluxes from the \Gaia, UCAC, 2MASS and WISE catalogs are plotted in red, with horizontal error bars corresponding to the width of the photometric bandpass.
The blue points show the best-fit model flux derived from the stellar properties and MIST bolometric correction grid.
We plot in gray an atmospheric model from \citet{Kurucz1993} corresponding to the best-fit stellar parameters for illustrative purposes only, as the fit is performed directly to the MIST grid.
The TESS and ground-based time-series photometry, as well as the RV measurements, are available as Data behind the Figure.}
\figsetgrpend

\figsetgrpstart
\figsetgrpnum{4.29}
\figsetgrptitle{Data and \Exofast fit results for TOI-5386\,b}
\figsetplot{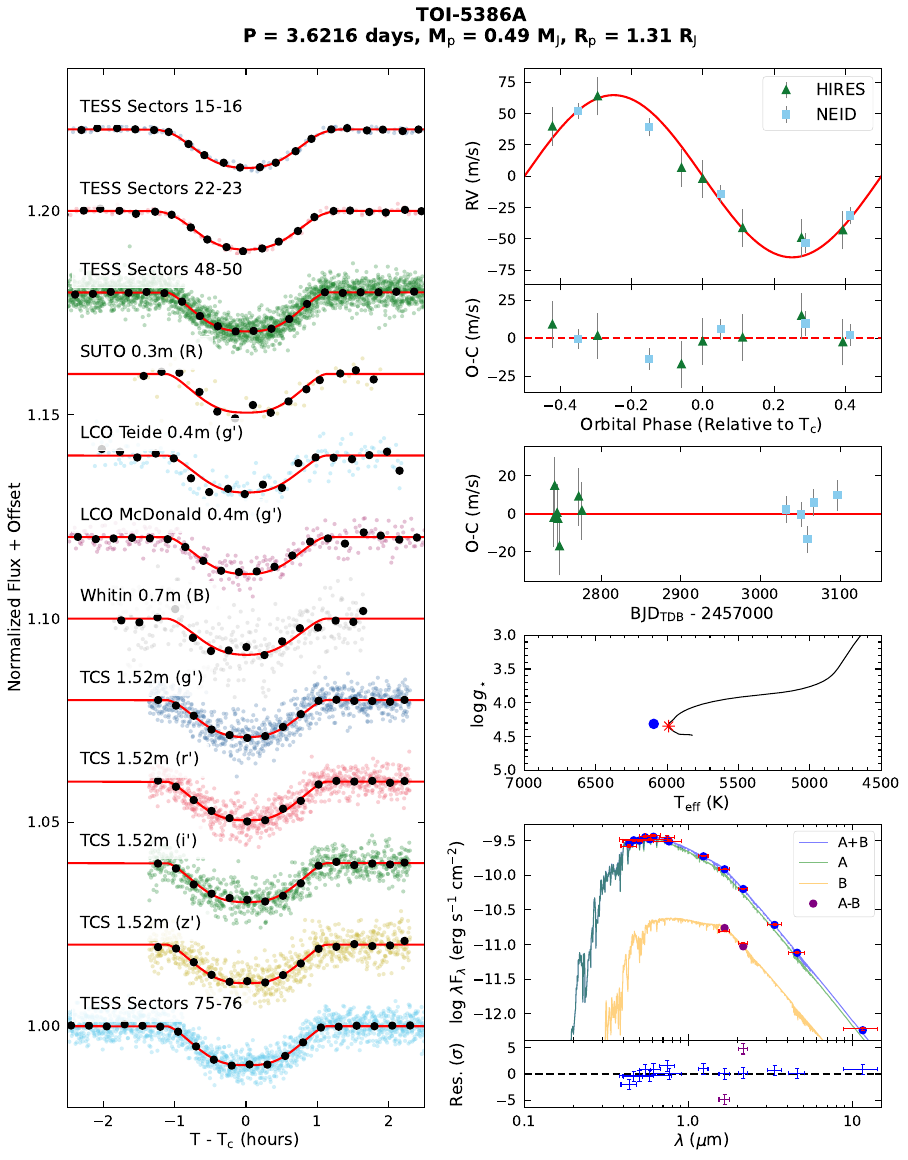}
\figsetgrpnote{Data and \Exofast fit results for TOI-5386\,b.
\textbf{Left:} \TESS and ground-based light-curves, phase-folded onto the best-fit period and time of conjunction.
Faint colored points represent the unbinned data, while large black circles show the time-series data binned to 30-min cadence.
The best-fit transit model in each band is shown as the red line.
\textbf{Top right:} RV observations, also phased onto the best-fit orbital period.
Error bars represent the fitted per-instrument jitter term $\sigma_\mathrm{jit}$ added in quadrature to the instrumental uncertainties.
The red line shows the best-fit RV model.
We plot the residuals after subtracting the model in the middle subpanel, and the unphased RV data and model time-series in the lower subpanel.
\textbf{Middle right:} The best-fit MIST stellar evolution track (black line), with a red asterisk marking the position along the track corresponding to the best-fit stellar age.
The blue point represents the best-fit stellar \Teff and \logg.
The discrepancy between the blue point and red asterisk are well within the fitted uncertainties in each parameter, indicating no tension between the different constraints on the stellar properties.
\textbf{Bottom right:} The observed stellar fluxes from the \Gaia, UCAC, 2MASS and WISE catalogs are plotted in red, with horizontal error bars corresponding to the width of the photometric bandpass.
The blue points show the best-fit model flux derived from the stellar properties and MIST bolometric correction grid.
We plot in gray an atmospheric model from \citet{Kurucz1993} corresponding to the best-fit stellar parameters for illustrative purposes only, as the fit is performed directly to the MIST grid.
The TESS and ground-based time-series photometry, as well as the RV measurements, are available as Data behind the Figure.}
\figsetgrpend

\figsetgrpstart
\figsetgrpnum{4.30}
\figsetgrptitle{Data and \Exofast fit results for TOI-5592\,b}
\figsetplot{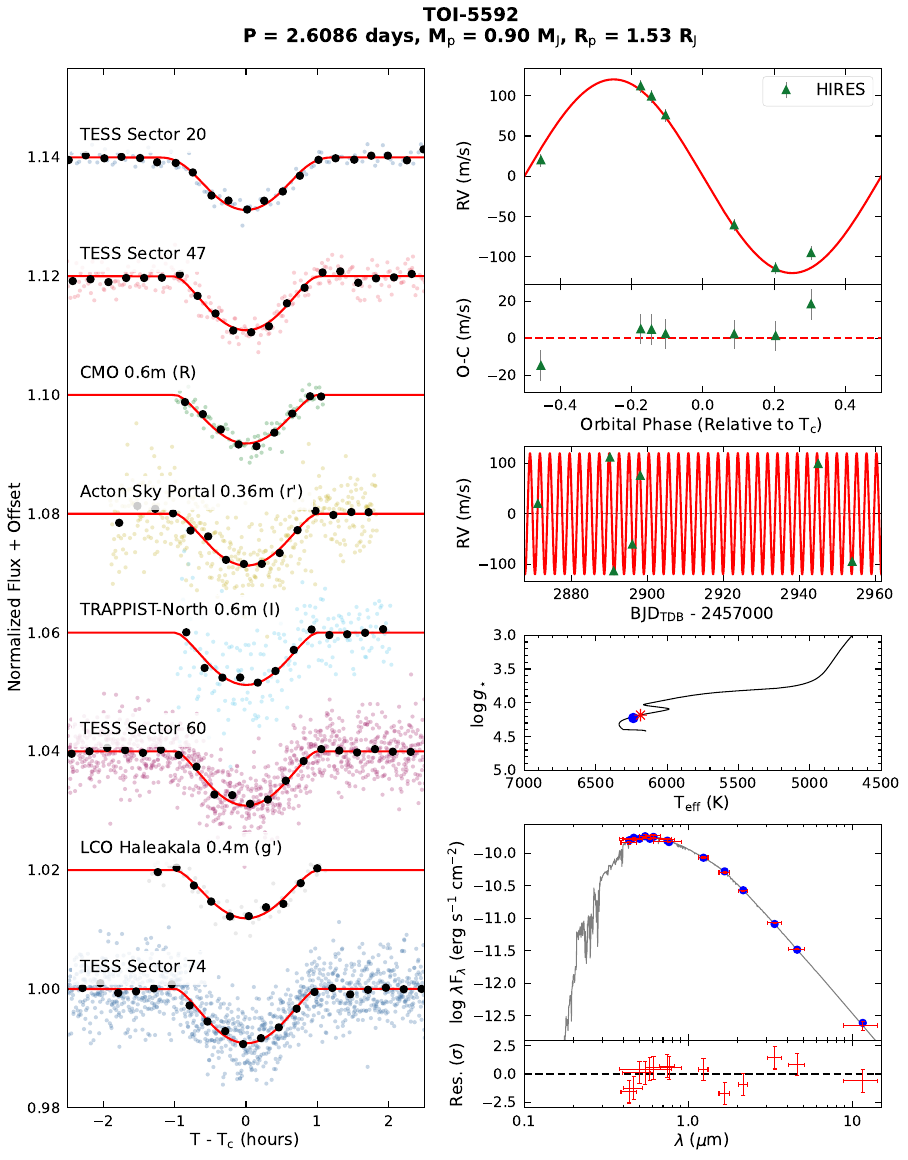}
\figsetgrpnote{Data and \Exofast fit results for TOI-5592\,b.
\textbf{Left:} \TESS and ground-based light-curves, phase-folded onto the best-fit period and time of conjunction.
Faint colored points represent the unbinned data, while large black circles show the time-series data binned to 30-min cadence.
The best-fit transit model in each band is shown as the red line.
\textbf{Top right:} RV observations, also phased onto the best-fit orbital period.
Error bars represent the fitted per-instrument jitter term $\sigma_\mathrm{jit}$ added in quadrature to the instrumental uncertainties.
The red line shows the best-fit RV model.
We plot the residuals after subtracting the model in the middle subpanel, and the unphased RV data and model time-series in the lower subpanel.
\textbf{Middle right:} The best-fit MIST stellar evolution track (black line), with a red asterisk marking the position along the track corresponding to the best-fit stellar age.
The blue point represents the best-fit stellar \Teff and \logg.
The discrepancy between the blue point and red asterisk are well within the fitted uncertainties in each parameter, indicating no tension between the different constraints on the stellar properties.
\textbf{Bottom right:} The observed stellar fluxes from the \Gaia, UCAC, 2MASS and WISE catalogs are plotted in red, with horizontal error bars corresponding to the width of the photometric bandpass.
The blue points show the best-fit model flux derived from the stellar properties and MIST bolometric correction grid.
We plot in gray an atmospheric model from \citet{Kurucz1993} corresponding to the best-fit stellar parameters for illustrative purposes only, as the fit is performed directly to the MIST grid.
The TESS and ground-based time-series photometry, as well as the RV measurements, are available as Data behind the Figure.}
\figsetgrpend

\figsetend

\figsetend
\fi

\makeatletter\onecolumngrid@push\makeatother
\clearpage
\pauselinenumbers
\begin{figure*}
\addtocounter{figure}{1}
\pdfbookmark[1]{Figure Set \thefigure: Data and Global Fit}{toi_multiplots}
\addtocounter{figure}{-1}
\captionsetup{labelfont=bf,font=small,labelformat=simple,
labelsep=period,width=\textwidth,listofformat=subsimple,subrefformat=parentfig}
\centering
\pdfbookmark[2]{TOI-2031\xspace b}{toi2031_multiplot}%
\subfloat[Data and \Exofast fit results for TOI-2031\,b.
\textbf{Left:} \TESS and ground-based light-curves, phase-folded onto the best-fit period and time of conjunction.
Faint colored points represent the unbinned data, while large black circles show the time-series data binned to 30-min cadence.
The best-fit transit model in each band is shown as the red line.
\textbf{Top right:} RV observations, also phased onto the best-fit orbital period.
Error bars represent the fitted per-instrument jitter term $\sigma_\mathrm{jit}$ added in quadrature to the instrumental uncertainties.
The red line shows the best-fit RV model.
We plot the residuals after subtracting the model in the middle subpanel, and the unphased RV data and model time-series in the lower subpanel.
\textbf{Middle right:} The best-fit MIST stellar evolution track (black line), with a red asterisk marking the position along the track corresponding to the best-fit stellar age.
The blue point represents the best-fit stellar \Teff and \logg.
The discrepancy between the blue point and red asterisk are well within the fitted uncertainties in each parameter, indicating no tension between the different constraints on the stellar properties.
\textbf{Bottom right:} The observed stellar fluxes from the \Gaia, UCAC, 2MASS and WISE catalogs are plotted in red, with horizontal error bars corresponding to the width of the photometric bandpass.
The blue points show the best-fit model flux derived from the stellar properties and MIST bolometric correction grid.
We plot in gray an atmospheric model from \citet{Kurucz1993} corresponding to the best-fit stellar parameters for illustrative purposes only, as the fit is performed directly to the MIST grid.
The TESS and ground-based time-series photometry, as well as the RV measurements, are available as Data behind the Figure.
The complete figure set for all TOIs (\ntois\xspace images) is available in the available in the online journal.%
\label{fig:toi2031_multiplot}]{\includegraphics[width=365pt]{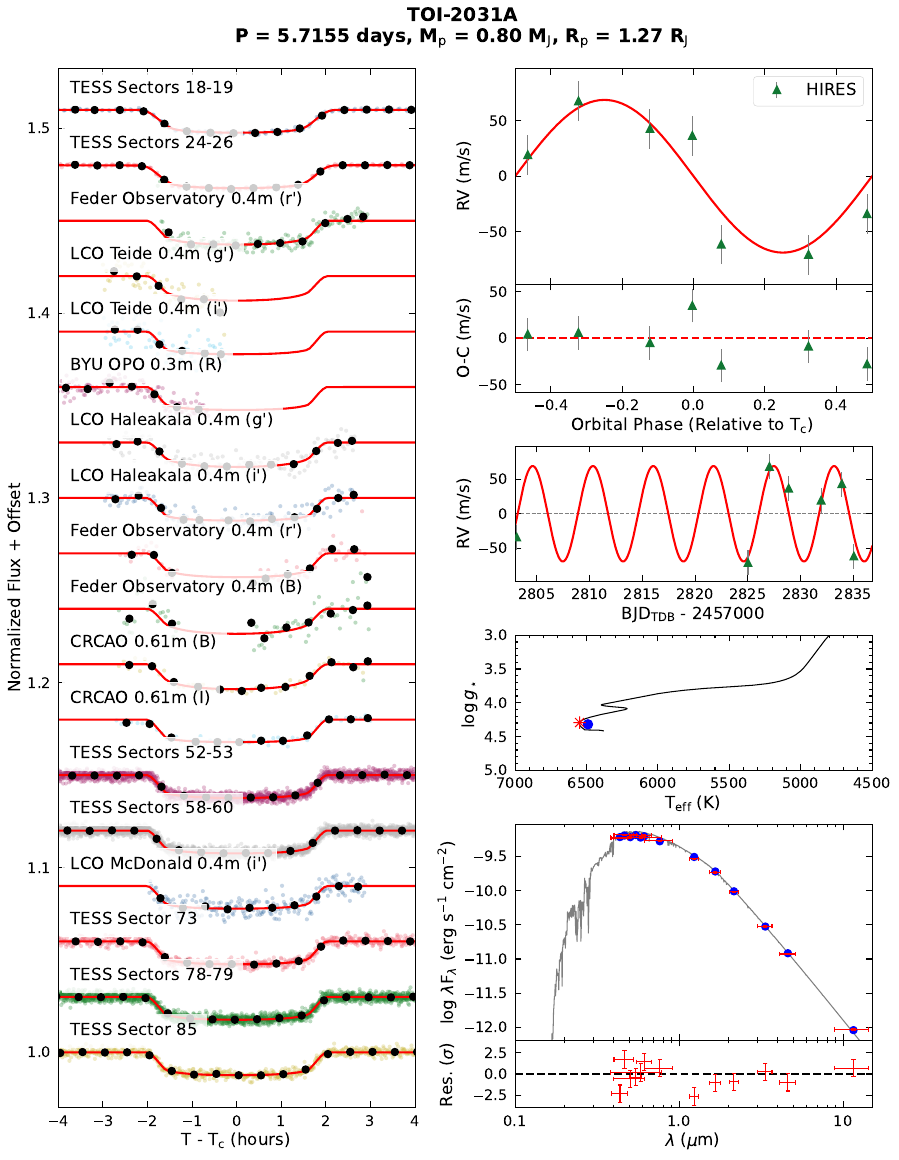}}
\addtocounter{figure}{1}
\end{figure*}
\resumelinenumbers

\ifjournal
\else
\pauselinenumbers
\def\ExToiMultiplot{2031}
\foreach \toi in 2031
2169
2346
2382
2876
2886
2986
2992
3135
3160
3464
3474
3486
3523
3593
3682
3856
3877
3980
4214
4487
4734
4794
4961
5181
5210
5322
5340
5386
5592%

{
\unless\ifnum\toi=\ExToiMultiplot{
\IfFileExists{toi\toi_multiplot.pdf}{
\begin{figure*}[p]
\ContinuedFloat
\pdfbookmark[2]{TOI-\toi\xspace b}{toi\toi_multiplot}%
\subfloat[Same as above, but for TOI-\toi\,b.
\label{fig:toi\toi_multiplot}]{\includegraphics[width=480pt]{toi\toi_multiplot.pdf}}
\addtocounter{figure}{1}
\end{figure*}
}{\textcolor{red}{Missing figure for TOI-\toi.}}
}
\resumelinenumbers
\fi
}
\fi

\ifSubfilesClassLoaded{%
\clearpage
\bibliography{software}
}{}
\end{document} \label{sf:multiplots}%

\foreach \n [count=\ni] in {0} {%
\begin{deluxetable*}{l>{\centering}cccccc}
\ifnum\ni=1%
\tablecaption{Median Values and 68\% Confidence Intervals for Fitted Parameters of Secondary Stellar Components \label{tab:comp_fitted_props}}
\addtocounter{table}{1}
\pdfbookmark[1]{Table \thetable: Fitted Parameters of Stellar Companions}{comp_fitted_props}%
\addtocounter{table}{-1}
\else
\tablecaption{\textit{(Continued)}}
\fi
\tabletypesize{\footnotesize}
\input{comp_fit_results_\n}

\ifnum\ni=1%
\tablecommentsmod{
Best-fit stellar properties for the secondary components of systems containing close stellar companions.\\
$^a$\,These parameters are required to be the same as for the primary star.}
\fi
\end{deluxetable*}
\addtocounter{table}{-1}
}
\addtocounter{table}{1}


 \label{sf:comp_fit_results}%
\makeatletter\onecolumngrid@pop\makeatother
\begin{acknowledgements} \label{sec:acknowledgments}

This paper includes data collected by the \TESS mission that are publicly available from the Mikulski Archive for Space Telescopes (MAST).
The raw \TESS data can be accessed at \dataset[10.17909/t9-nmc8-f686]{http://dx.doi.org/10.17909/t9-nmc8-f686} (SPOC 120-second light curves), \dataset[10.17909/t9-wpz1-8s54]{https://dx.doi.org/10.17909/t9-wpz1-8s54} (TESS-SPOC full-frame image light curves), and \dataset[10.17909/t9-r086-e880]{https://dx.doi.org/10.17909/t9-r086-e880} (QLP light curves).
The Data Validation reports are available at \dataset[10.17909/t9-2tc5-a751]{http://dx.doi.org/10.17909/t9-2tc5-a751} and \dataset[10.17909/t9-yjj5-4t42]{http://dx.doi.org/10.17909/t9-yjj5-4t42}.
The \TESS Input Catalog is available on MAST at \dataset[10.17909/fwdt-2x66]{https://doi.org/10.17909/fwdt-2x66}
Funding for the \TESS mission is provided by NASA's Science Mission Directorate. We acknowledge the use of public \TESS data from pipelines at the \TESS Science Office and at the \TESS Science Processing Operations Center.
Resources supporting this work were provided by the NASA High-End Computing (HEC) Program through the NASA Advanced Supercomputing (NAS) Division at Ames Research Center for the production of the SPOC data products.
We also acknowledge the use of data from the Exoplanet Follow-up Observation Program website, which is operated by the California Institute of Technology, under contract with the National Aeronautics and Space Administration under the Exoplanet Exploration Program \citep{ExoFoP,ExoFoPTESS}.

This research made use of Lightkurve, a Python package for \Kepler and \TESS data analysis \citep{Lightkurve18}.
We acknowledge that the work reported on in this paper was substantially performed using the Princeton Research Computing resources at Princeton University which is consortium of groups led by the Princeton Institute for Computational Science and Engineering (PICSciE) and Office of Information Technology's Research Computing.

Some of the data presented herein were obtained at the W. M. Keck Observatory, which is operated as a scientific partnership among the California Institute of Technology, the University of California and the National Aeronautics and Space Administration. The Observatory was made possible by the generous financial support of the W. M. Keck Foundation.
Part of the Keck telescope time was granted by NOIRLab (Prop. ID 2022A-543544, PI: Yee) through the Mid-Scale Innovations Program (MSIP). MSIP is funded by NSF.
This work was supported by a NASA Keck PI Data Award, administered by the NASA Exoplanet Science Institute.
The authors wish to recognize and acknowledge the very significant cultural role and reverence that the summit of Maunakea has always had within the indigenous Hawaiian community.  We are most fortunate to have the opportunity to conduct observations from this mountain.

This paper contains data taken with the NEID instrument, which was funded by the NASA-NSF Exoplanet Observational Research (NN-EXPLORE) partnership and built by Pennsylvania State University. NEID is installed on the WIYN telescope, which is operated by the National Optical Astronomy Observatory, and the NEID archive is operated by the NASA Exoplanet Science Institute at the California Institute of Technology. NN-EXPLORE is managed by the Jet Propulsion Laboratory, California Institute of Technology under contract with the National Aeronautics and Space Administration.
Some of the data presented herein were obtained at the WIYN Observatory from telescope time allocated to NN-EXPLORE through the scientific partnership of the National Aeronautics and Space Administration, the National Science Foundation, and NOIRLab.
This work was supported by a NASA WIYN PI Data Award, administered by the NASA Exoplanet Science Institute.
The authors are honored to be permitted to conduct astronomical research on Iolkam Du’ag (Kitt Peak), a mountain with particular significance to the Tohono O’odham.

This paper includes data gathered with the 6.5 meter Magellan Telescopes located at Las Campanas Observatory, Chile.

This research has used data from the CTIO/SMARTS 1.5m telescope, which is operated as part of the SMARTS Consortium by RECONS (\url{www.recons.org}) members Todd Henry, Hodari James, Wei-Chun Jao, and Leonardo Paredes. At the telescope, observations were carried out by Roberto Aviles and Rodrigo Hinojosa.
The CHIRON data were obtained from telescope time allocated under the NN-EXPLORE program with support from the National Aeronautics and Space Administration.

This work makes use of observations from the LCOGT network. Part of the LCOGT telescope time was granted by NOIRLab through the Mid-Scale Innovations Program (MSIP). MSIP is funded by NSF. This paper is based on observations made with the Las Cumbres Observatory’s education network telescopes that were upgraded through generous support from the Gordon and Betty Moore Foundation.

KAC acknowledges support from the TESS mission via subaward s3449 from MIT. JDH acknowledges support from NASA grant 80NSSC22K0315.

The research leading to these results has received funding from  the ARC grant for Concerted Research Actions, financed by the Wallonia-Brussels Federation. TRAPPIST is funded by the Belgian Fund for Scientific Research (Fond National de la Recherche Scientifique, FNRS) under the grant PDR T.0120.21. TRAPPIST-North is a project funded by the University of Liege (Belgium), in collaboration with Cadi Ayyad University of Marrakech (Morocco).
The postdoctoral fellowship of KB is funded by F.R.S.-FNRS grant T.0109.20 and by the Francqui Foundation.
MG and EJ are F.R.S.-FNRS Research Directors. 
This publication benefits from the support of the French Community of Belgium in the context of the FRIA Doctoral Grant awarded to M.T.

This work is based in part on data collected under the NGTS project at the ESO La Silla Paranal Observatory. The NGTS facility is operated by a consortium institutes with support from the UK Science and Technology Facilities Council (STFC) under projects ST/M001962/1, ST/S002642/1 and ST/W003163/1.
A.K. and S.M.O. are supported by UK Science and Technology Facilities Council (STFC) Studentships (ST/T506242/1 and ST/W507751/1 respectively).

This article is based on observations made with the MuSCAT2 instrument, developed by ABC, at Telescopio Carlos Sánchez operated on the island of Tenerife by the IAC in the Spanish Observatorio del Teide.
This work is partly supported by JSPS KAKENHI Grant Numbers JP24K00689, JP24H00017, and JP24K17083; JST SPRING, Grant Number JPMJSP2108; JSPS Grant-in-Aid for JSPS Fellows Grant Number JP24KJ0241; JSPS Bilateral Program Number JPJSBP120249910.
We acknowledge financial support from the Agencia Estatal de Investigaci\'on of the Ministerio de Ciencia e Innovaci\'on MCIN/AEI/10.13039/501100011033 and the ERDF “A way of making Europe” through project PID2021-125627OB-C32, and from the Centre of Excellence “Severo Ochoa” award to the Instituto de Astrofisica de Canarias.
The authors acknowledge support from the Swiss NCCR PlanetS and the Swiss National Science Foundation. This work has been carried out within the framework of the NCCR PlanetS supported by the Swiss National Science Foundation under grants 51NF40182901 and 51NF40205606. J.K. acknowledges support of the Swiss National Science Foundation under grant number TMSGI2\_211697.

DRC acknowledges partial support from NASA Grant 18-2XRP18\_2-0007. This research has made use of the Exoplanet Follow-up Observation Program (ExoFOP; DOI: 10.26134/ExoFOP5) website, which is operated by the California Institute of Technology, under contract with the National Aeronautics and Space Administration under the Exoplanet Exploration Program. Based on observations obtained at the Hale Telescope, Palomar Observatory, as part of a collaborative agreement between the Caltech Optical Observatories and the Jet Propulsion Laboratory operated by Caltech for NASA.

The work of B.S.S., N.A.M., and A.A.B. was conducted under the state assignment of Lomonosov Moscow State University.

In this study, the observational data obtained within the scope of the project numbered 22BT100-1958 carried out using the T100 telescope at the Antalya site of the T\"{u}rkiye National Observatories have been utilised, and we express our gratitude for the invaluable support provided by the T\"{u}rkiye National Observatories and all its staff.

\end{acknowledgements}

\facilities{
TESS, Keck:I (HIRES), Magellan:Clay (PFS), WIYN (NEID, NESSI), CTIO:1.5m (CHIRON), FLWO:1.5m (TRES), FLWO:1.2m (KeplerCam), LCOGT.
}

\software{
astropy \citep{Astropy13,Astropy18,Astropy2022}, lightkurve \citep{Lightkurve18}, EXOFASTv2 \citep{ExoFAST_Eastman2013,ExoFASTv2_Eastman19}, SpecMatch-Emp \citep{SpecMatchEmp_Yee2017}, SpecMatch-Synth \citep{SpecMatchSynth_Petigura2015}, AstroImageJ \citep{AstroImageJ_Collins17}, TAPIR \citep{TAPIR_Jensen2013}, numpy \citep{Numpy}, scipy \citep{Scipy}, pandas \citep{Pandas20,Pandas_McKinney10}, matplotlib \citep{Matplotlib}.
}

\appendix
\section{Additional Fit Parameters}
We present in Table \ref{tab:additional_fit_params} the median and 68\% confidence intervals for additional fit parameters not listed in Table \ref{tab:fitted_props} for the adopted fits.
These are: the RV slope $\dot{\gamma}$ (\msday) and reference time $T_\mathrm{ref}$ (BJD) if used in the fit; the linear and quadratic limb-darkening parameters $(u_1, u_2)$ in each band; additional flux dilution from neighboring stars in each band $(A_D)$; the relative RV offset for each instrument $\gamma_\mathrm{rel}$ (\ms); and the RV jitter for each instrument $\sigma_J$ (\ms).
\begin{rotatepage}
\begin{longrotatetable}
\begin{deluxetable*}{lccccccc} \label{tab:additional_fit_params}
\tablecaption{Additional Fit Parameters (Median and 68\% Confidence Intervals)}
\addtocounter{table}{1}
\pdfbookmark[2]{Table \thetable: Additional Fit Parameters}{additional_fit_params}%
\addtocounter{table}{-1}
\tabletypesize{\footnotesize}
\tablehead{\colhead{Parameter}}
\startdata
\\[-\normalbaselineskip]\multicolumn{2}{l}{\textbf{TOI-2031}}\\
 &B&I&R&g'&i'&r'&TESS\\
~~~~$u_{1}$ &$0.507\pm0.041$&$0.176\pm0.046$&$0.260^{+0.051}_{-0.050}$&$0.424\pm0.038$&$0.205\pm0.029$&$0.305\pm0.036$&$0.222\pm0.014$\\
~~~~$u_{2}$ &$0.294\pm0.038$&$0.290\pm0.048$&$0.319\pm0.049$&$0.294\pm0.036$&$0.299\pm0.028$&$0.345\pm0.035$&$0.302\pm0.018$\\
~~~~$A_D$ &--&--&--&--&--&--&$0.0002^{+0.0018}_{-0.0019}$\\
 &HIRES\\
~~~~$\gamma_{\rm rel}$ (\ms) &$-3\pm13$\\
~~~~$\sigma_J$ (\ms) &$32^{+23}_{-11}$\\
\hline\\[-\normalbaselineskip]\multicolumn{2}{l}{\textbf{TOI-2169}}\\
 &B&I&r'&z'&TESS\\
~~~~$u_{1}$ &$0.545\pm0.043$&$0.223\pm0.052$&$0.292\pm0.052$&$0.190\pm0.031$&$0.230^{+0.026}_{-0.025}$\\
~~~~$u_{2}$ &$0.229\pm0.040$&$0.313\pm0.050$&$0.299\pm0.050$&$0.310\pm0.029$&$0.301\pm0.023$\\
~~~~$A_D$ &--&--&--&--&$0.015\pm0.018$\\
 &NEID1&NEID2\\
~~~~$\gamma_{\rm rel}$ (\ms) &$12938.5^{+5.6}_{-5.4}$&$12938.3^{+5.4}_{-3.9}$\\
~~~~$\sigma_J$ (\ms) &$0.00^{+19}_{-0.00}$&$0.00^{+12}_{-0.00}$\\
\hline\\[-\normalbaselineskip]\multicolumn{2}{l}{\textbf{TOI-2346}}\\
~~~~$\dot{\gamma}$ (\msday) & $2.06\pm0.61$\\
~~~~$T_\mathrm{ref}$ & 2459864.007887\\
 &B&R&g'&TESS\\
~~~~$u_{1}$ &$0.492\pm0.058$&$0.255\pm0.051$&$0.452\pm0.051$&$0.240\pm0.031$\\
~~~~$u_{2}$ &$0.262\pm0.054$&$0.314\pm0.049$&$0.297\pm0.050$&$0.316\pm0.034$\\
~~~~$A_D$ &--&--&--&$0.085^{+0.037}_{-0.039}$\\
 &HIRES\\
~~~~$\gamma_{\rm rel}$ (\ms) &$5\pm18$\\
~~~~$\sigma_J$ (\ms) &$47^{+28}_{-15}$\\
\hline\\[-\normalbaselineskip]\multicolumn{2}{l}{\textbf{TOI-2382}}\\
 &Kepler&g'&i'&TESS\\
~~~~$u_{1}$ &$0.380^{+0.043}_{-0.044}$&$0.536\pm0.049$&$0.283\pm0.048$&$0.267\pm0.030$\\
~~~~$u_{2}$ &$0.294\pm0.046$&$0.228\pm0.049$&$0.305\pm0.048$&$0.290^{+0.028}_{-0.029}$\\
~~~~$A_D$ &--&--&--&$0.000000\pm0.000080$\\
 &CHIRON\\
~~~~$\gamma_{\rm rel}$ (\ms) &$-5455\pm20$\\
~~~~$\sigma_J$ (\ms) &$53^{+21}_{-18}$\\
\hline\\[-\normalbaselineskip]\multicolumn{2}{l}{\textbf{TOI-2876}}\\
 &R&g'&TESS\\
~~~~$u_{1}$ &$0.515^{+0.053}_{-0.054}$&$0.746^{+0.057}_{-0.058}$&$0.396^{+0.033}_{-0.034}$\\
~~~~$u_{2}$ &$0.226\pm0.051$&$0.076^{+0.056}_{-0.055}$&$0.221\pm0.031$\\
~~~~$A_D$ &--&--&$0.001\pm0.017$\\
 &PFS\\
~~~~$\gamma_{\rm rel}$ (\ms) &$-15.1^{+4.3}_{-4.2}$\\
~~~~$\sigma_J$ (\ms) &$7.1^{+7.7}_{-3.4}$\\
\hline\\[-\normalbaselineskip]\multicolumn{2}{l}{\textbf{TOI-2886}}\\
~~~~$\dot{\gamma}$ (\msday) & $-0.20\pm0.13$\\
~~~~$T_\mathrm{ref}$ & 2459759.895080\\
 &B&R&i'&z'&TESS\\
~~~~$u_{1}$ &$0.565\pm0.047$&$0.296\pm0.050$&$0.269\pm0.047$&$0.213\pm0.041$&$0.239^{+0.027}_{-0.026}$\\
~~~~$u_{2}$ &$0.219^{+0.053}_{-0.054}$&$0.312\pm0.050$&$0.319\pm0.048$&$0.319\pm0.046$&$0.294\pm0.029$\\
~~~~$A_D$ &--&--&--&--&$0.039\pm0.014$\\
 &TRES\\
~~~~$\gamma_{\rm rel}$ (\ms) &$120^{+30}_{-29}$\\
~~~~$\sigma_J$ (\ms) &$70^{+18}_{-19}$\\
\hline\\[-\normalbaselineskip]\multicolumn{2}{l}{\textbf{TOI-2986}}\\
 &i'&TESS\\
~~~~$u_{1}$ &$0.311\pm0.050$&$0.275\pm0.030$\\
~~~~$u_{2}$ &$0.308\pm0.050$&$0.282\pm0.029$\\
~~~~$A_D$ &--&$0.0028\pm0.0092$\\
 &NEID1&NEID2\\
~~~~$\gamma_{\rm rel}$ (\ms) &$5329.1^{+5.0}_{-5.8}$&$5314.8\pm9.6$\\
~~~~$\sigma_J$ (\ms) &$7.7^{+12}_{-7.7}$&$20.0^{+16}_{-9.0}$\\
\hline\\[-\normalbaselineskip]\multicolumn{2}{l}{\textbf{TOI-2992}}\\
 &R&TESS\\
~~~~$u_{1}$ &$0.313\pm0.050$&$0.272^{+0.029}_{-0.028}$\\
~~~~$u_{2}$ &$0.286^{+0.048}_{-0.049}$&$0.296\pm0.029$\\
~~~~$A_D$ &--&$-0.0001\pm0.0012$\\
 &CHIRON\\
~~~~$\gamma_{\rm rel}$ (\ms) &$140^{+35}_{-34}$\\
~~~~$\sigma_J$ (\ms) &$61^{+26}_{-33}$\\
\hline\\[-\normalbaselineskip]\multicolumn{2}{l}{\textbf{TOI-3135}}\\
 &B&R&TESS\\
~~~~$u_{1}$ &$0.618\pm0.060$&$0.331\pm0.041$&$0.304^{+0.032}_{-0.031}$\\
~~~~$u_{2}$ &$0.163^{+0.056}_{-0.057}$&$0.273\pm0.037$&$0.294^{+0.026}_{-0.027}$\\
~~~~$A_D$ &--&--&$0.059^{+0.020}_{-0.021}$\\
 &CHIRON\\
~~~~$\gamma_{\rm rel}$ (\ms) &$-4861^{+35}_{-31}$\\
~~~~$\sigma_J$ (\ms) &$45^{+77}_{-45}$\\
\hline\\[-\normalbaselineskip]\multicolumn{2}{l}{\textbf{TOI-3160}}\\
 &B&R&g'&i'&TESS\\
~~~~$u_{1}$ &$0.557^{+0.061}_{-0.060}$&$0.293^{+0.054}_{-0.053}$&$0.483^{+0.054}_{-0.053}$&$0.233\pm0.050$&$0.236\pm0.032$\\
~~~~$u_{2}$ &$0.242^{+0.055}_{-0.056}$&$0.309\pm0.050$&$0.257\pm0.050$&$0.280\pm0.048$&$0.303^{+0.029}_{-0.030}$\\
 &PFS\\
~~~~$\gamma_{\rm rel}$ (\ms) &$-75\pm15$\\
~~~~$\sigma_J$ (\ms) &$32^{+26}_{-14}$\\
\hline\\[-\normalbaselineskip]\multicolumn{2}{l}{\textbf{TOI-3464}}\\
~~~~$\dot{\gamma}$ (\msday) & $-0.38^{+0.20}_{-0.22}$\\
~~~~$T_\mathrm{ref}$ & 2459921.655350\\
 &r'&TESS\\
~~~~$u_{1}$ &$0.307\pm0.056$&$0.223^{+0.038}_{-0.036}$\\
~~~~$u_{2}$ &$0.324\pm0.051$&$0.307\pm0.030$\\
~~~~$A_D$ &--&$-0.007\pm0.026$\\
 &CHIRON\\
~~~~$\gamma_{\rm rel}$ (\ms) &$55157\pm23$\\
~~~~$\sigma_J$ (\ms) &$43^{+33}_{-44}$\\
\hline\\[-\normalbaselineskip]\multicolumn{2}{l}{\textbf{TOI-3474}}\\
~~~~$\dot{\gamma}$ (\msday) & $-0.39^{+0.13}_{-0.12}$\\
~~~~$T_\mathrm{ref}$ & 2459925.618385\\
 &R&g'&i'&TESS\\
~~~~$u_{1}$ &$0.373^{+0.054}_{-0.053}$&$0.503^{+0.060}_{-0.058}$&$0.281\pm0.054$&$0.253^{+0.035}_{-0.034}$\\
~~~~$u_{2}$ &$0.330\pm0.050$&$0.222^{+0.055}_{-0.056}$&$0.298\pm0.051$&$0.280\pm0.030$\\
~~~~$A_D$ &--&--&--&$0.113^{+0.027}_{-0.028}$\\
 &CHIRON\\
~~~~$\gamma_{\rm rel}$ (\ms) &$-39915\pm14$\\
~~~~$\sigma_J$ (\ms) &$26^{+21}_{-24}$\\
\hline\\[-\normalbaselineskip]\multicolumn{2}{l}{\textbf{TOI-3486}}\\
 &R&g'&TESS\\
~~~~$u_{1}$ &$0.535^{+0.052}_{-0.053}$&$0.821^{+0.057}_{-0.059}$&$0.463\pm0.033$\\
~~~~$u_{2}$ &$0.140^{+0.052}_{-0.051}$&$-0.026^{+0.058}_{-0.057}$&$0.200^{+0.032}_{-0.031}$\\
~~~~$A_D$ &--&--&$-0.023\pm0.025$\\
 &PFS\\
~~~~$\gamma_{\rm rel}$ (\ms) &$-26.4^{+7.5}_{-7.4}$\\
~~~~$\sigma_J$ (\ms) &$16.3^{+17}_{-6.6}$\\
\hline\\[-\normalbaselineskip]\multicolumn{2}{l}{\textbf{TOI-3523}}\\
 &B&R&g'&i'&r'&z'&TESS\\
~~~~$u_{1}$ &$0.491\pm0.052$&$0.280^{+0.050}_{-0.051}$&$0.405^{+0.028}_{-0.027}$&$0.165\pm0.029$&$0.227\pm0.036$&$0.167^{+0.030}_{-0.031}$&$0.206\pm0.018$\\
~~~~$u_{2}$ &$0.272\pm0.051$&$0.335\pm0.050$&$0.275\pm0.033$&$0.283\pm0.033$&$0.312\pm0.045$&$0.293\pm0.033$&$0.300\pm0.020$\\
 &HIRES\\
~~~~$\gamma_{\rm rel}$ (\ms) &$-29.3\pm7.7$\\
~~~~$\sigma_J$ (\ms) &$16.9^{+13}_{-7.0}$\\
\hline\\[-\normalbaselineskip]\multicolumn{2}{l}{\textbf{TOI-3593}}\\
~~~~$\dot{\gamma}$ (\msday) & $-0.08^{+0.39}_{-0.37}$\\
~~~~$T_\mathrm{ref}$ & 2459813.440116\\
 &R&g'&i'&r'&TESS\\
~~~~$u_{1}$ &$0.419^{+0.041}_{-0.040}$&$0.695\pm0.047$&$0.339\pm0.046$&$0.442\pm0.051$&$0.349^{+0.028}_{-0.027}$\\
~~~~$u_{2}$ &$0.229^{+0.037}_{-0.038}$&$0.131\pm0.053$&$0.244\pm0.048$&$0.223\pm0.051$&$0.252^{+0.026}_{-0.027}$\\
~~~~$A_D$ &--&--&--&--&$-0.003\pm0.017$\\
 &HIRES\\
~~~~$\gamma_{\rm rel}$ (\ms) &$37.7^{+6.5}_{-8.9}$\\
~~~~$\sigma_J$ (\ms) &$15^{+23}_{-11}$\\
\hline\\[-\normalbaselineskip]\multicolumn{2}{l}{\textbf{TOI-3682}}\\
 &g'&i'&TESS\\
~~~~$u_{1}$ &$0.645\pm0.055$&$0.373^{+0.047}_{-0.048}$&$0.318\pm0.023$\\
~~~~$u_{2}$ &$0.171^{+0.053}_{-0.054}$&$0.304^{+0.048}_{-0.049}$&$0.262\pm0.022$\\
~~~~$A_D$ &--&--&$0.0010\pm0.0040$\\
 &NEID\\
~~~~$\gamma_{\rm rel}$ (\ms) &$36089.8^{+7.3}_{-7.7}$\\
~~~~$\sigma_J$ (\ms) &$14.7^{+17}_{-8.5}$\\
\hline\\[-\normalbaselineskip]\multicolumn{2}{l}{\textbf{TOI-3856}}\\
~~~~$\dot{\gamma}$ (\msday) & $-0.106^{+0.067}_{-0.066}$\\
~~~~$T_\mathrm{ref}$ & 2459867.344371\\
 &I&R&g'&i'&r'&z'&TESS\\
~~~~$u_{1}$ &$0.350\pm0.049$&$0.415^{+0.046}_{-0.047}$&$0.618\pm0.034$&$0.315\pm0.044$&$0.463\pm0.033$&$0.266\pm0.033$&$0.405\pm0.032$\\
~~~~$u_{2}$ &$0.277\pm0.049$&$0.253\pm0.048$&$0.091\pm0.035$&$0.275\pm0.047$&$0.259\pm0.034$&$0.256\pm0.034$&$0.269\pm0.032$\\
~~~~$A_D$ &--&--&--&--&--&--&$0.0073^{+0.0045}_{-0.0046}$\\
 &HIRES\\
~~~~$\gamma_{\rm rel}$ (\ms) &$31^{+12}_{-11}$\\
~~~~$\sigma_J$ (\ms) &$28.8^{+13}_{-7.8}$\\
\hline\\[-\normalbaselineskip]\multicolumn{2}{l}{\textbf{TOI-3877}}\\
~~~~$\dot{\gamma}$ (\msday) & $-0.29\pm0.13$\\
~~~~$T_\mathrm{ref}$ & 2459684.446074\\
 &R&g'&i'&TESS\\
~~~~$u_{1}$ &$0.400\pm0.052$&$0.627^{+0.052}_{-0.053}$&$0.306\pm0.036$&$0.339\pm0.050$\\
~~~~$u_{2}$ &$0.276\pm0.050$&$0.194\pm0.052$&$0.261\pm0.036$&$0.291\pm0.049$\\
~~~~$A_D$ &--&--&--&$0.000000^{+0.000010}_{-0.000011}$\\
 &HIRES\\
~~~~$\gamma_{\rm rel}$ (\ms) &$-5.4\pm5.5$\\
~~~~$\sigma_J$ (\ms) &$14.3^{+8.8}_{-4.9}$\\
\hline\\[-\normalbaselineskip]\multicolumn{2}{l}{\textbf{TOI-3980}}\\
 &g'&i'&z'&TESS\\
~~~~$u_{1}$ &$0.586^{+0.054}_{-0.055}$&$0.287^{+0.051}_{-0.052}$&$0.302\pm0.053$&$0.287\pm0.030$\\
~~~~$u_{2}$ &$0.217\pm0.055$&$0.272\pm0.049$&$0.332\pm0.049$&$0.275^{+0.024}_{-0.025}$\\
~~~~$A_D$ &--&--&--&$-0.006\pm0.014$\\
 &NEID\\
~~~~$\gamma_{\rm rel}$ (\ms) &$-43043^{+14}_{-15}$\\
~~~~$\sigma_J$ (\ms) &$29^{+25}_{-14}$\\
\hline\\[-\normalbaselineskip]\multicolumn{2}{l}{\textbf{TOI-4214}}\\
~~~~$\dot{\gamma}$ (\msday) & $-0.23^{+0.10}_{-0.12}$\\
~~~~$T_\mathrm{ref}$ & 2459792.152157\\
 &B&R&TESS\\
~~~~$u_{1}$ &$0.529\pm0.066$&$0.295^{+0.045}_{-0.044}$&$0.216^{+0.035}_{-0.034}$\\
~~~~$u_{2}$ &$0.230^{+0.059}_{-0.060}$&$0.329^{+0.037}_{-0.038}$&$0.300\pm0.030$\\
~~~~$A_D$ &--&--&$0.0021\pm0.0074$\\
 &CHIRON&PFS\\
~~~~$\gamma_{\rm rel}$ (\ms) &$51584^{+24}_{-28}$&$26^{+32}_{-26}$\\
~~~~$\sigma_J$ (\ms) &$0.00$&$16.7^{+13}_{-6.6}$\\
\hline\\[-\normalbaselineskip]\multicolumn{2}{l}{\textbf{TOI-4487}}\\
 &R&TESS\\
~~~~$u_{1}$ &$0.305^{+0.051}_{-0.052}$&$0.245\pm0.021$\\
~~~~$u_{2}$ &$0.315\pm0.050$&$0.305\pm0.023$\\
~~~~$A_D$ &--&$0.045^{+0.096}_{-0.097}$\\
 &HIRES\\
~~~~$\gamma_{\rm rel}$ (\ms) &$9^{+11}_{-12}$\\
~~~~$\sigma_J$ (\ms) &$25^{+22}_{-11}$\\
\hline\\[-\normalbaselineskip]\multicolumn{2}{l}{\textbf{TOI-4734}}\\
~~~~$\dot{\gamma}$ (\msday) & $-0.148^{+0.018}_{-0.019}$\\
~~~~$T_\mathrm{ref}$ & 2460081.996933\\
 &g'&i'&TESS\\
~~~~$u_{1}$ &$0.514\pm0.043$&$0.267\pm0.042$&$0.262\pm0.030$\\
~~~~$u_{2}$ &$0.228^{+0.037}_{-0.038}$&$0.304\pm0.037$&$0.298\pm0.023$\\
~~~~$A_D$ &--&--&$-0.008\pm0.016$\\
 &HIRES&NEID\\
~~~~$\gamma_{\rm rel}$ (\ms) &$-35.3^{+3.8}_{-3.9}$&$-15454.9^{+3.6}_{-3.5}$\\
~~~~$\sigma_J$ (\ms) &$2.0^{+3.7}_{-2.0}$&$8.1^{+4.9}_{-4.3}$\\
\hline\\[-\normalbaselineskip]\multicolumn{2}{l}{\textbf{TOI-4794}}\\
 &R&TESS\\
~~~~$u_{1}$ &$0.283^{+0.052}_{-0.051}$&$0.209\pm0.036$\\
~~~~$u_{2}$ &$0.326^{+0.049}_{-0.050}$&$0.297^{+0.035}_{-0.036}$\\
~~~~$A_D$ &--&$0.010\pm0.036$\\
 &CHIRON\\
~~~~$\gamma_{\rm rel}$ (\ms) &$5953^{+21}_{-22}$\\
~~~~$\sigma_J$ (\ms) &$22^{+33}_{-22}$\\
\hline\\[-\normalbaselineskip]\multicolumn{2}{l}{\textbf{TOI-4961}}\\
 &R&z'&TESS\\
~~~~$u_{1}$ &$0.421^{+0.059}_{-0.060}$&$0.299\pm0.054$&$0.378^{+0.038}_{-0.039}$\\
~~~~$u_{2}$ &$0.189\pm0.055$&$0.232\pm0.051$&$0.246^{+0.029}_{-0.028}$\\
 &PFS\\
~~~~$\gamma_{\rm rel}$ (\ms) &$-9\pm15$\\
~~~~$\sigma_J$ (\ms) &$36^{+28}_{-15}$\\
\hline\\[-\normalbaselineskip]\multicolumn{2}{l}{\textbf{TOI-5181}}\\
 &g'&i'&r'&z'&TESS\\
~~~~$u_{1}$ &$0.549^{+0.056}_{-0.054}$&$0.280^{+0.039}_{-0.038}$&$0.359\pm0.051$&$0.235^{+0.051}_{-0.050}$&$0.243^{+0.029}_{-0.027}$\\
~~~~$u_{2}$ &$0.249^{+0.051}_{-0.052}$&$0.301\pm0.035$&$0.313\pm0.049$&$0.306^{+0.048}_{-0.049}$&$0.280^{+0.023}_{-0.024}$\\
 &HIRES\\
~~~~$\gamma_{\rm rel}$ (\ms) &$-10\pm12$\\
~~~~$\sigma_J$ (\ms) &$30^{+20}_{-10.000000}$\\
\hline\\[-\normalbaselineskip]\multicolumn{2}{l}{\textbf{TOI-5210}}\\
 &I&R&TESS\\
~~~~$u_{1}$ &$0.298\pm0.049$&$0.413\pm0.054$&$0.325^{+0.027}_{-0.026}$\\
~~~~$u_{2}$ &$0.263^{+0.049}_{-0.050}$&$0.277\pm0.051$&$0.270^{+0.026}_{-0.027}$\\
~~~~$A_D$ &--&--&$0.0005\pm0.0045$\\
 &HIRES\\
~~~~$\gamma_{\rm rel}$ (\ms) &$-5.7^{+3.2}_{-3.3}$\\
~~~~$\sigma_J$ (\ms) &$7.3^{+4.9}_{-3.5}$\\
\hline\\[-\normalbaselineskip]\multicolumn{2}{l}{\textbf{TOI-5322}}\\
 &B&TESS\\
~~~~$u_{1}$ &$0.651^{+0.056}_{-0.059}$&$0.287^{+0.023}_{-0.025}$\\
~~~~$u_{2}$ &$0.141^{+0.055}_{-0.054}$&$0.264\pm0.023$\\
~~~~$A_D$ &--&$0.00000\pm0.00038$\\
 &HIRES\\
~~~~$\gamma_{\rm rel}$ (\ms) &$1.2\pm4.3$\\
~~~~$\sigma_J$ (\ms) &$7.7^{+11}_{-6.2}$\\
\hline\\[-\normalbaselineskip]\multicolumn{2}{l}{\textbf{TOI-5340}}\\
 &i'&TESS\\
~~~~$u_{1}$ &$0.309\pm0.053$&$0.275^{+0.029}_{-0.028}$\\
~~~~$u_{2}$ &$0.323^{+0.051}_{-0.050}$&$0.291^{+0.027}_{-0.028}$\\
~~~~$A_D$ &--&$-0.0001\pm0.0021$\\
 &NEID&TRES\\
~~~~$\gamma_{\rm rel}$ (\ms) &$26284^{+24}_{-22}$&$-40\pm24$\\
~~~~$\sigma_J$ (\ms) &$41^{+30}_{-22}$&$56^{+21}_{-17}$\\
\hline\\[-\normalbaselineskip]\multicolumn{2}{l}{\textbf{TOI-5386}}\\
 &B&R&g'&i'&r'&z'&TESS\\
~~~~$u_{1}$ &$0.607\pm0.051$&$0.327\pm0.051$&$0.526\pm0.031$&$0.257\pm0.046$&$0.331\pm0.045$&$0.259^{+0.046}_{-0.047}$&$0.245\pm0.025$\\
~~~~$u_{2}$ &$0.209^{+0.050}_{-0.051}$&$0.308\pm0.050$&$0.252\pm0.030$&$0.282\pm0.047$&$0.296\pm0.047$&$0.327\pm0.047$&$0.286\pm0.024$\\
 &HIRES&NEID\\
~~~~$\gamma_{\rm rel}$ (\ms) &$1.5\pm5.5$&$-17437.2^{+5.4}_{-5.7}$\\
~~~~$\sigma_J$ (\ms) &$12.7^{+8.3}_{-4.7}$&$9.2^{+13}_{-8.2}$\\
\hline\\[-\normalbaselineskip]\multicolumn{2}{l}{\textbf{TOI-5592}}\\
 &I&R&g'&r'&TESS\\
~~~~$u_{1}$ &$0.228\pm0.050$&$0.313\pm0.046$&$0.464\pm0.047$&$0.303^{+0.050}_{-0.049}$&$0.238\pm0.026$\\
~~~~$u_{2}$ &$0.309\pm0.049$&$0.333\pm0.048$&$0.261\pm0.048$&$0.306\pm0.049$&$0.303^{+0.025}_{-0.024}$\\
~~~~$A_D$ &--&--&--&--&$0.0001^{+0.0062}_{-0.0063}$\\
 &HIRES\\
~~~~$\gamma_{\rm rel}$ (\ms) &$-0.8\pm5.7$\\
~~~~$\sigma_J$ (\ms) &$12.7^{+10.}_{-5.7}$\\
\hline
\enddata
\end{deluxetable*}
\end{longrotatetable}
\end{rotatepage}
\pdfpageattr{}

\section{High Angular Resolution Imaging with No Detected Companions} \label{sec:ao_no_comp}

Figure \ref{fig:high_res_imaging_nocomp} shows the high angular resolution imaging and contrast curves for those objects that did not have detected companions.

\ifjournal
\figsetstart
\figsetnum{5}
\figsettitle{High-resolution imaging of hot Jupiter hosts, without detected companions.}

\figsetgrpstart
\figsetgrpnum{5.1}
\figsetgrptitle{Gemini-N/'Alopeke observations of TOI-2031}
\figsetplot{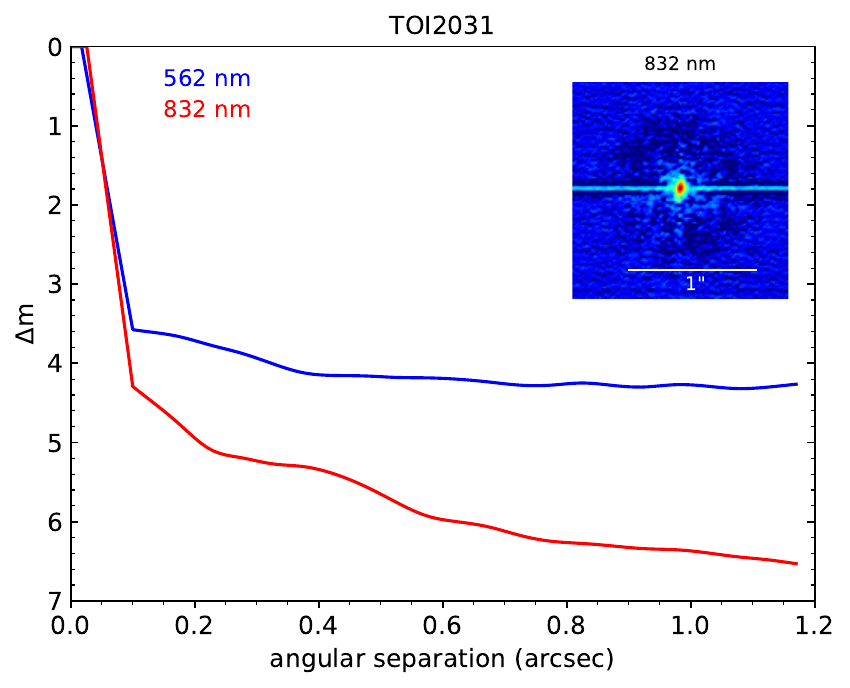}
\figsetgrpnote{Sensitivity limits and reconstructed image (inset) from Gemini-N/'Alopeke observations of TOI-2031.}
\figsetgrpend

\figsetgrpstart
\figsetgrpnum{5.2}
\figsetgrptitle{Palomar/PHARO Br$\gamma$ observation of TOI-2169}
\figsetplot{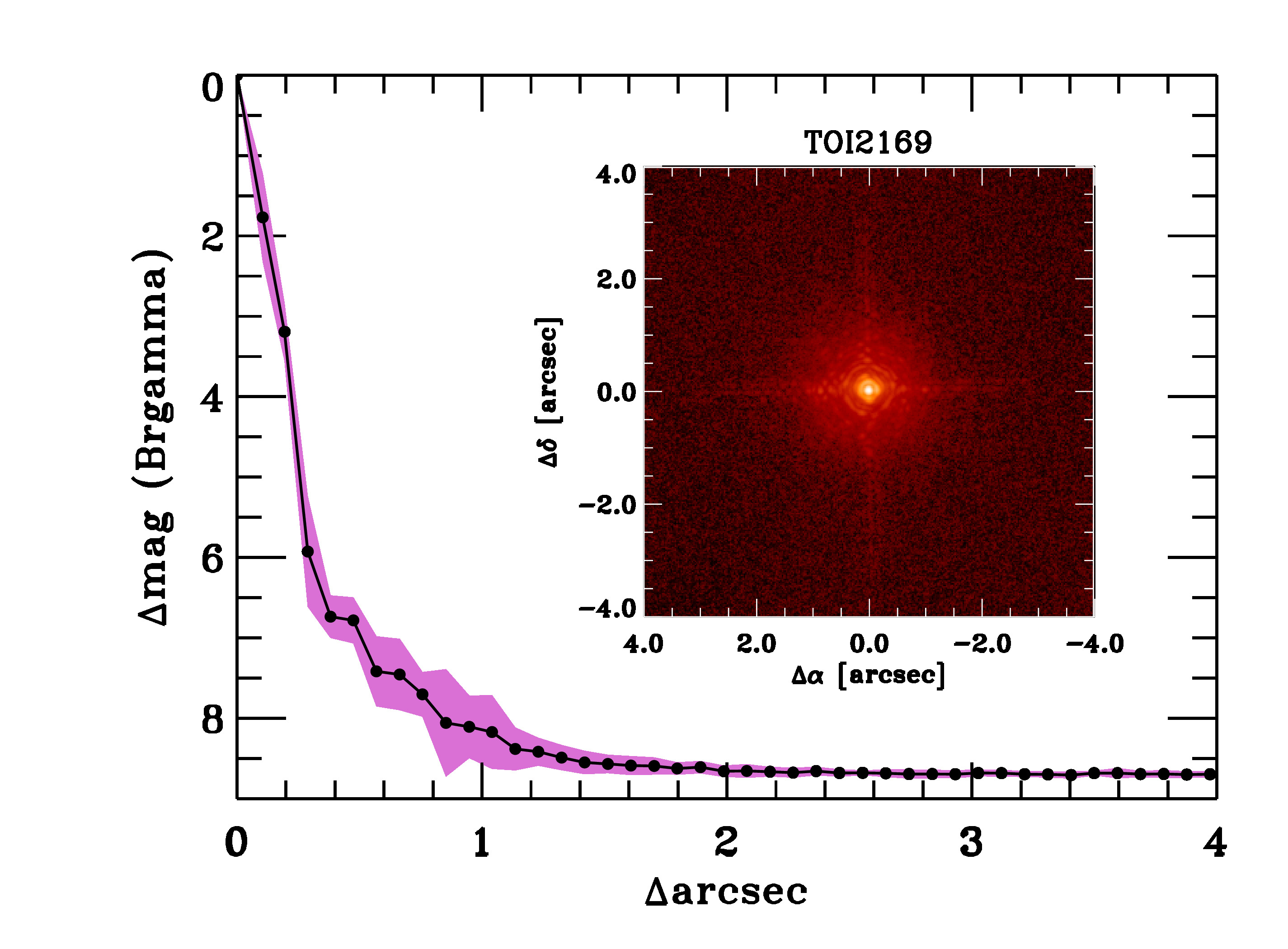}
\figsetgrpnote{Sensitivity limits and adaptive optics image (inset) from Palomar/PHARO observations of TOI-2169.}
\figsetgrpend

\figsetgrpstart
\figsetgrpnum{5.3}
\figsetgrptitle{Palomar/PHARO H\textit{{cont}} observation of TOI-2169}
\figsetplot{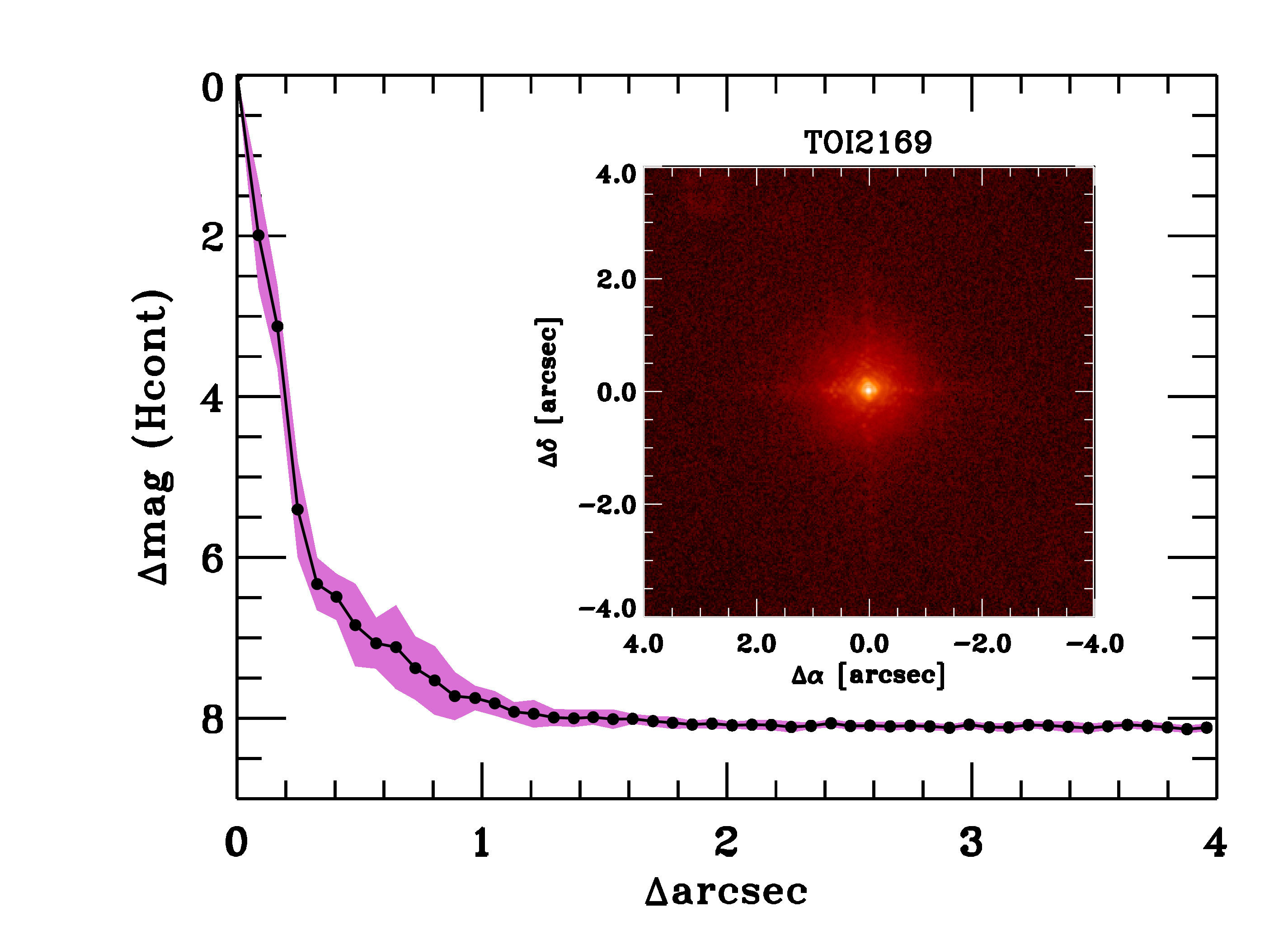}
\figsetgrpnote{Sensitivity limits and adaptive optics image (inset) from Palomar/PHARO observations of TOI-2169.}
\figsetgrpend

\figsetgrpstart
\figsetgrpnum{5.4}
\figsetgrptitle{SAI/Speckle Polarimeter observation of TOI-2169}
\figsetplot{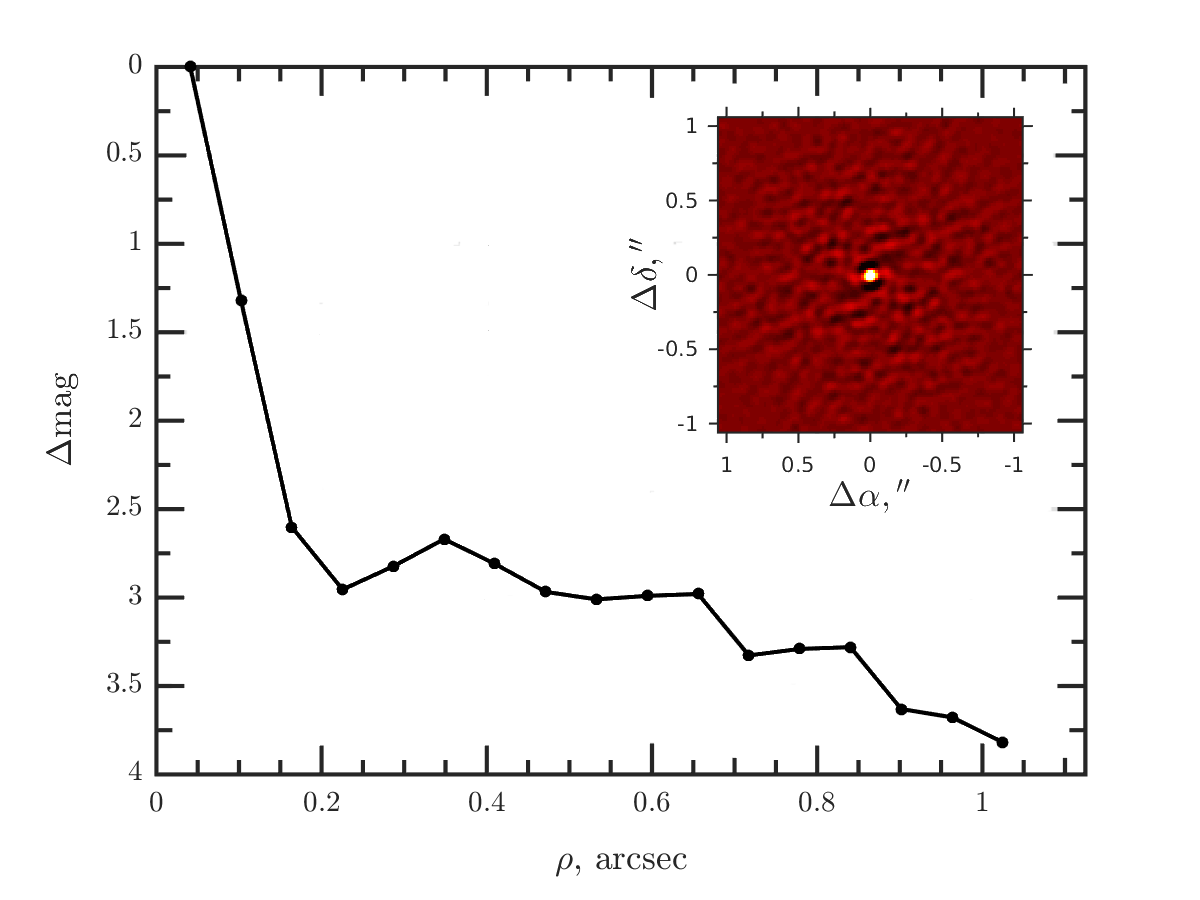}
\figsetgrpnote{Sensitivity limits and reconstructed image (inset) from SAI speckle observations of TOI-2169.}
\figsetgrpend

\figsetgrpstart
\figsetgrpnum{5.5}
\figsetgrptitle{SOAR/HRCam observation of TOI-2169}
\figsetplot{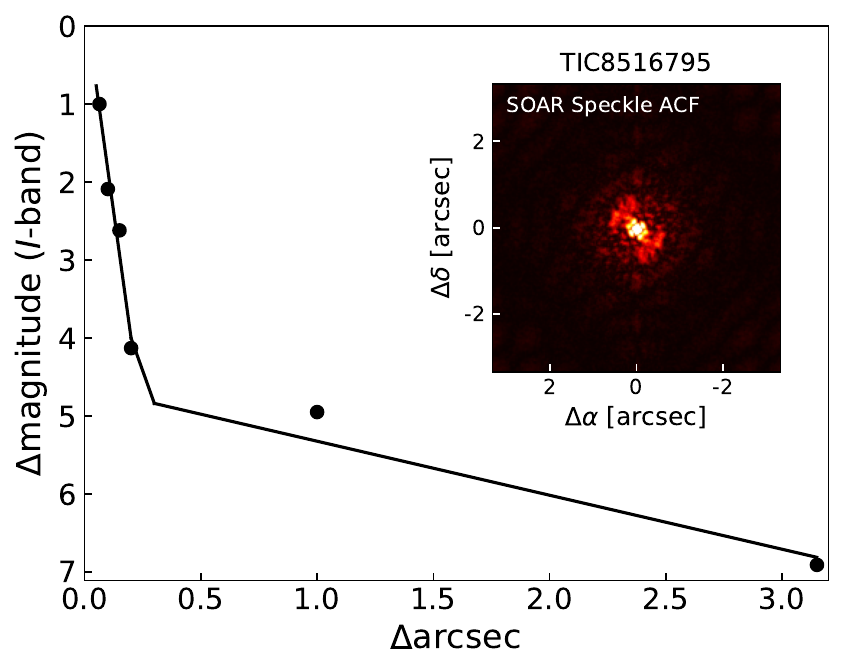}
\figsetgrpnote{Speckle sensitivity curve and auto-correlation function (inset) from SOAR/HRCam observations of TOI-2169.}
\figsetgrpend

\figsetgrpstart
\figsetgrpnum{5.6}
\figsetgrptitle{Shane/ShARCS J observation of TOI-2169}
\figsetplot{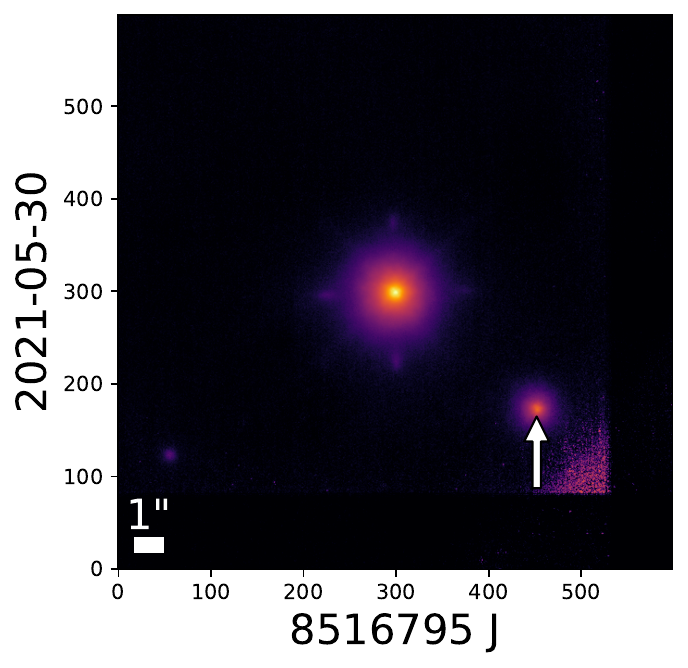}
\figsetgrpnote{Adaptive optics image from Shane/ShARCS observations of TOI-2169.}
\figsetgrpend

\figsetgrpstart
\figsetgrpnum{5.7}
\figsetgrptitle{Shane/ShARCS K$_s$ observation of TOI-2169}
\figsetplot{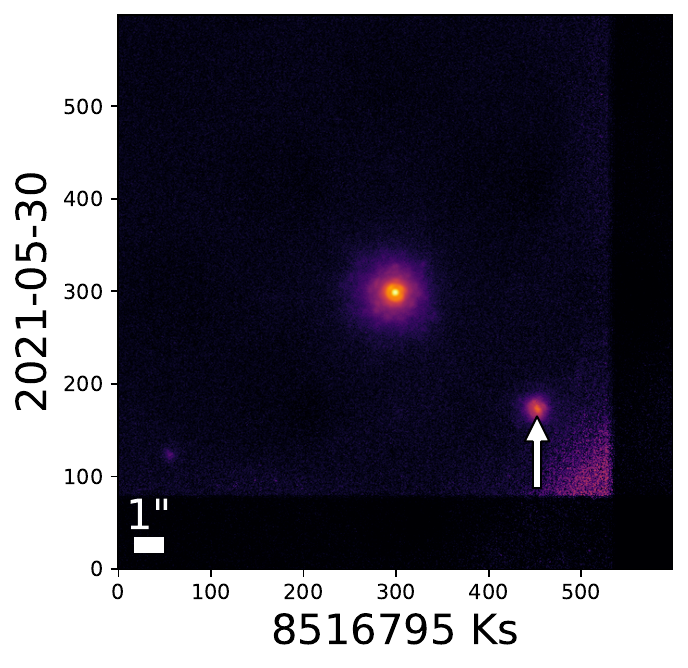}
\figsetgrpnote{Adaptive optics image from Shane/ShARCS observations of TOI-2169.}
\figsetgrpend

\figsetgrpstart
\figsetgrpnum{5.8}
\figsetgrptitle{SOAR/HRCam observation of TOI-2346}
\figsetplot{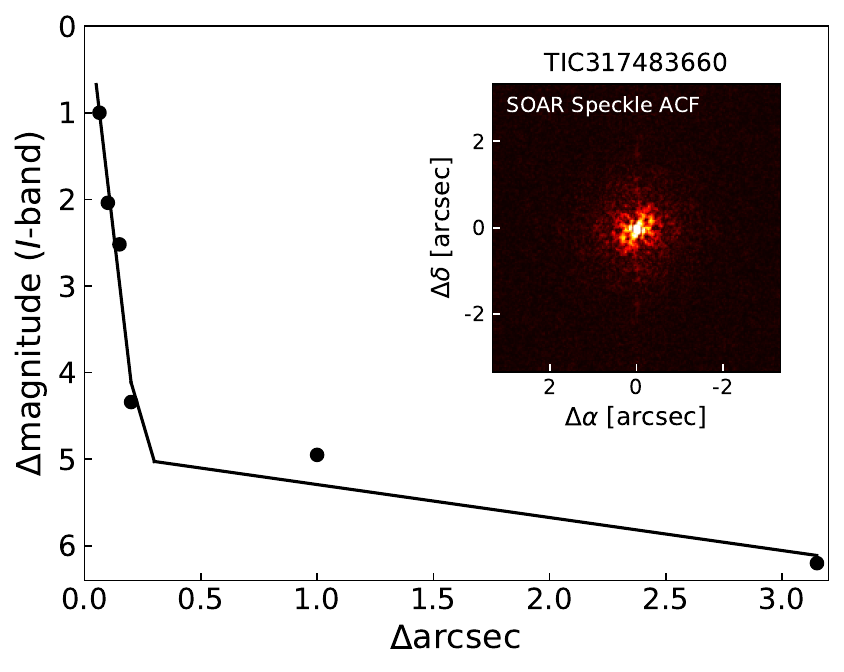}
\figsetgrpnote{Speckle sensitivity curve and auto-correlation function (inset) from SOAR/HRCam observations of TOI-2346.}
\figsetgrpend

\figsetgrpstart
\figsetgrpnum{5.9}
\figsetgrptitle{SOAR/HRCam observation of TOI-2382}
\figsetplot{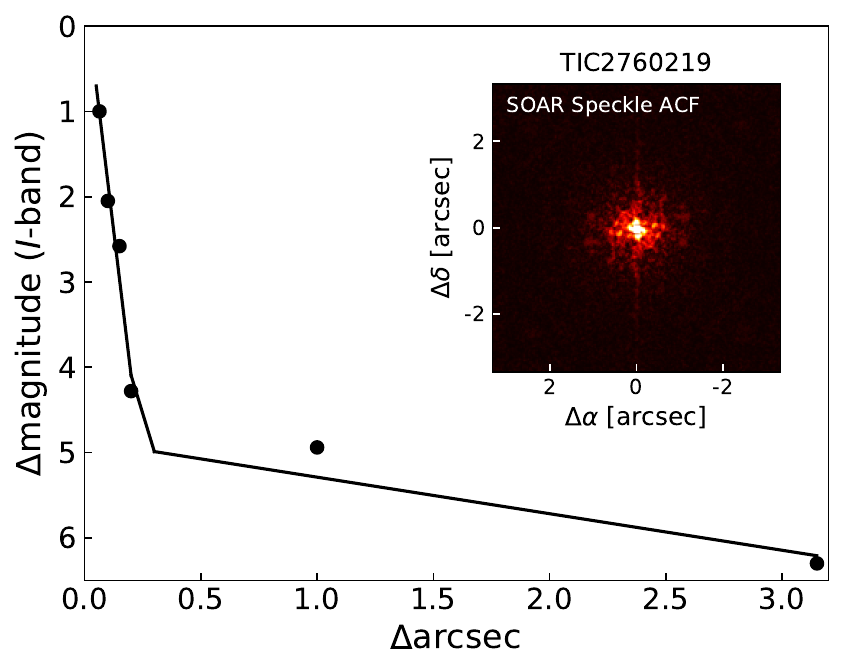}
\figsetgrpnote{Speckle sensitivity curve and auto-correlation function (inset) from SOAR/HRCam observations of TOI-2382.}
\figsetgrpend

\figsetgrpstart
\figsetgrpnum{5.10}
\figsetgrptitle{SOAR/HRCam observation of TOI-2876}
\figsetplot{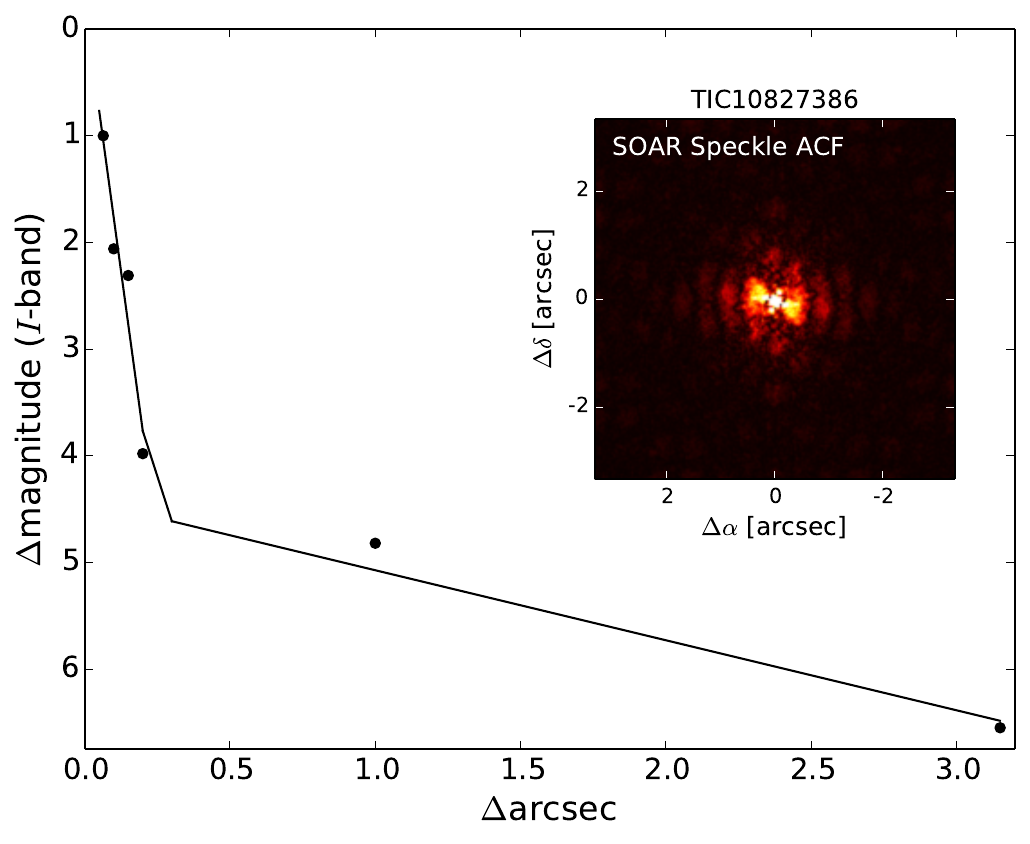}
\figsetgrpnote{Speckle sensitivity curve and auto-correlation function (inset) from SOAR/HRCam observations of TOI-2876.}
\figsetgrpend

\figsetgrpstart
\figsetgrpnum{5.11}
\figsetgrptitle{SOAR/HRCam observation of TOI-2886}
\figsetplot{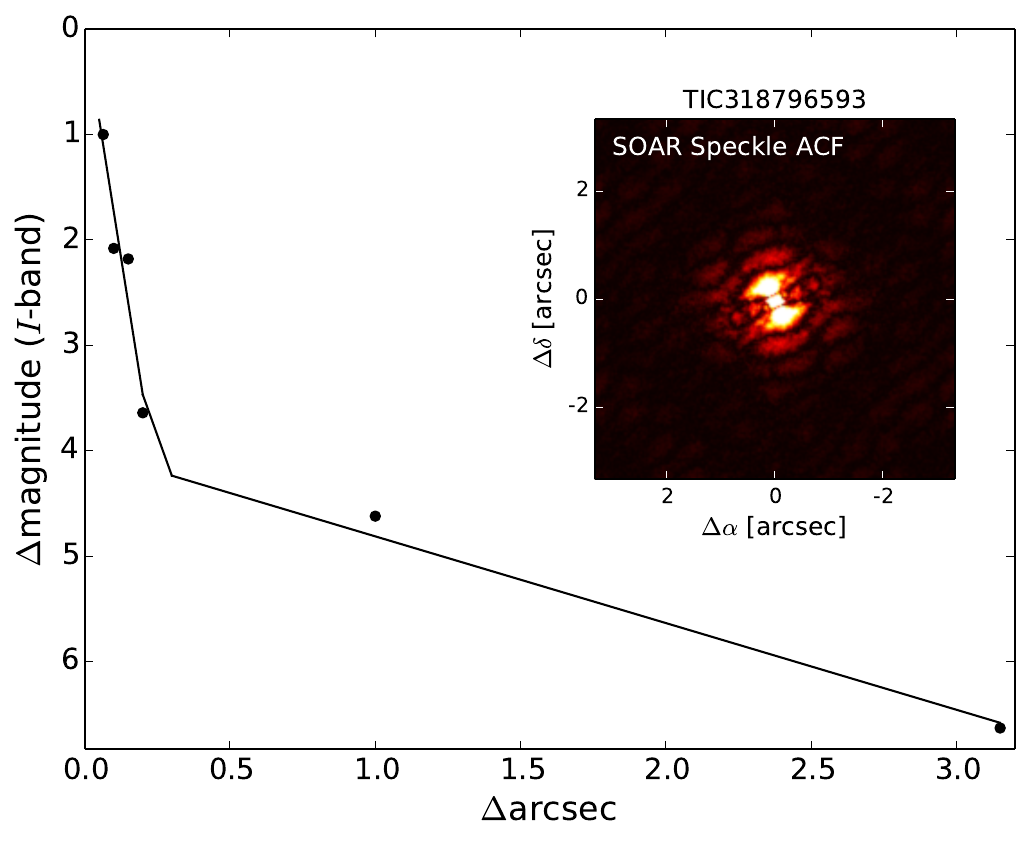}
\figsetgrpnote{Speckle sensitivity curve and auto-correlation function (inset) from SOAR/HRCam observations of TOI-2886.}
\figsetgrpend

\figsetgrpstart
\figsetgrpnum{5.12}
\figsetgrptitle{SOAR/HRCam observation of TOI-2986}
\figsetplot{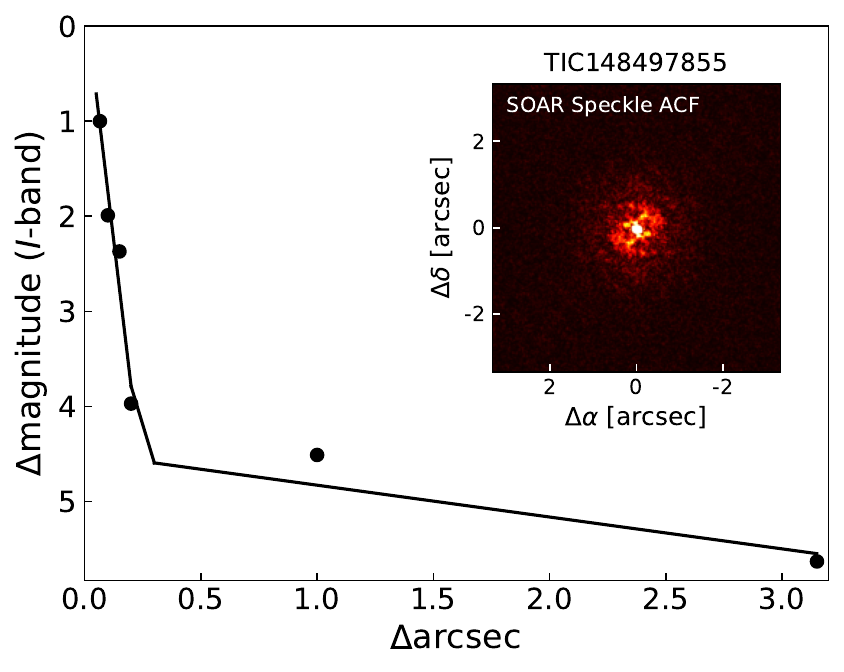}
\figsetgrpnote{Speckle sensitivity curve and auto-correlation function (inset) from SOAR/HRCam observations of TOI-2986.}
\figsetgrpend

\figsetgrpstart
\figsetgrpnum{5.13}
\figsetgrptitle{Shane/ShARCS J observation of TOI-2986}
\figsetplot{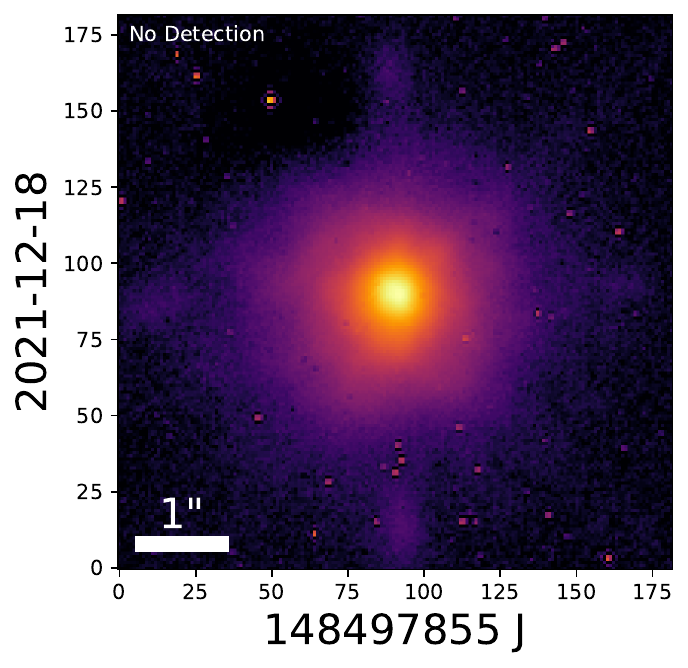}
\figsetgrpnote{Adaptive optics image from Shane/ShARCS observations of TOI-2986.}
\figsetgrpend

\figsetgrpstart
\figsetgrpnum{5.14}
\figsetgrptitle{Shane/ShARCS K$_s$ observation of TOI-2986}
\figsetplot{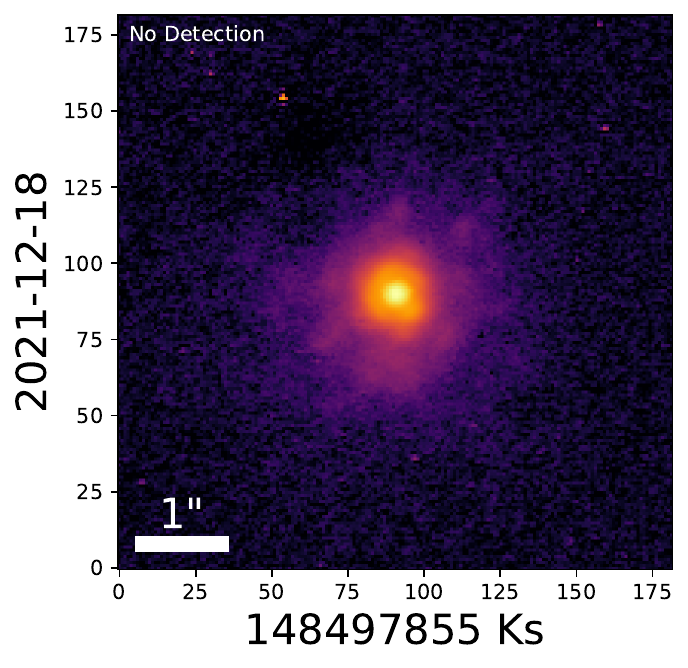}
\figsetgrpnote{Adaptive optics image from Shane/ShARCS observations of TOI-2986.}
\figsetgrpend

\figsetgrpstart
\figsetgrpnum{5.15}
\figsetgrptitle{SOAR/HRCam observation of TOI-2992}
\figsetplot{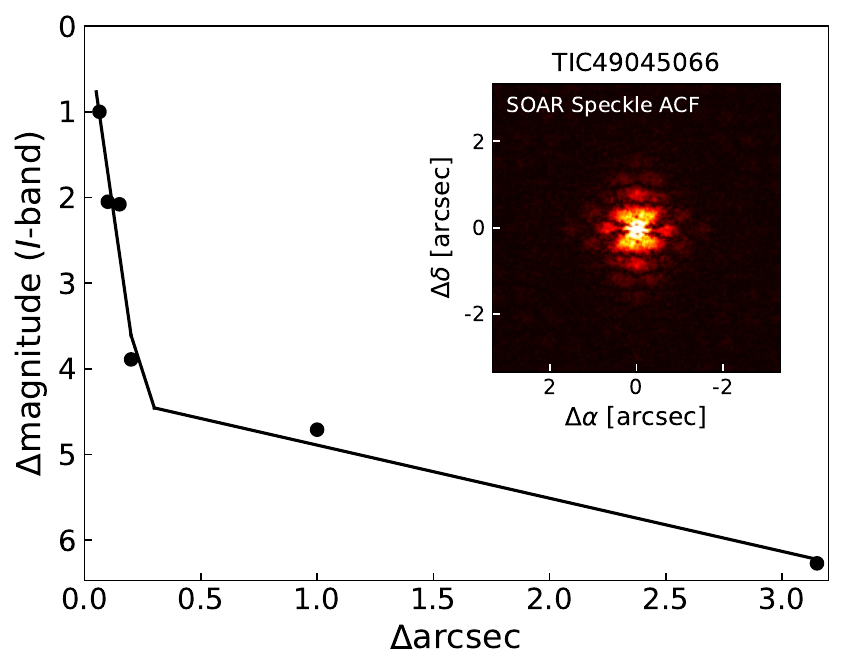}
\figsetgrpnote{Speckle sensitivity curve and auto-correlation function (inset) from SOAR/HRCam observations of TOI-2992.}
\figsetgrpend

\figsetgrpstart
\figsetgrpnum{5.16}
\figsetgrptitle{SOAR/HRCam observation of TOI-3135}
\figsetplot{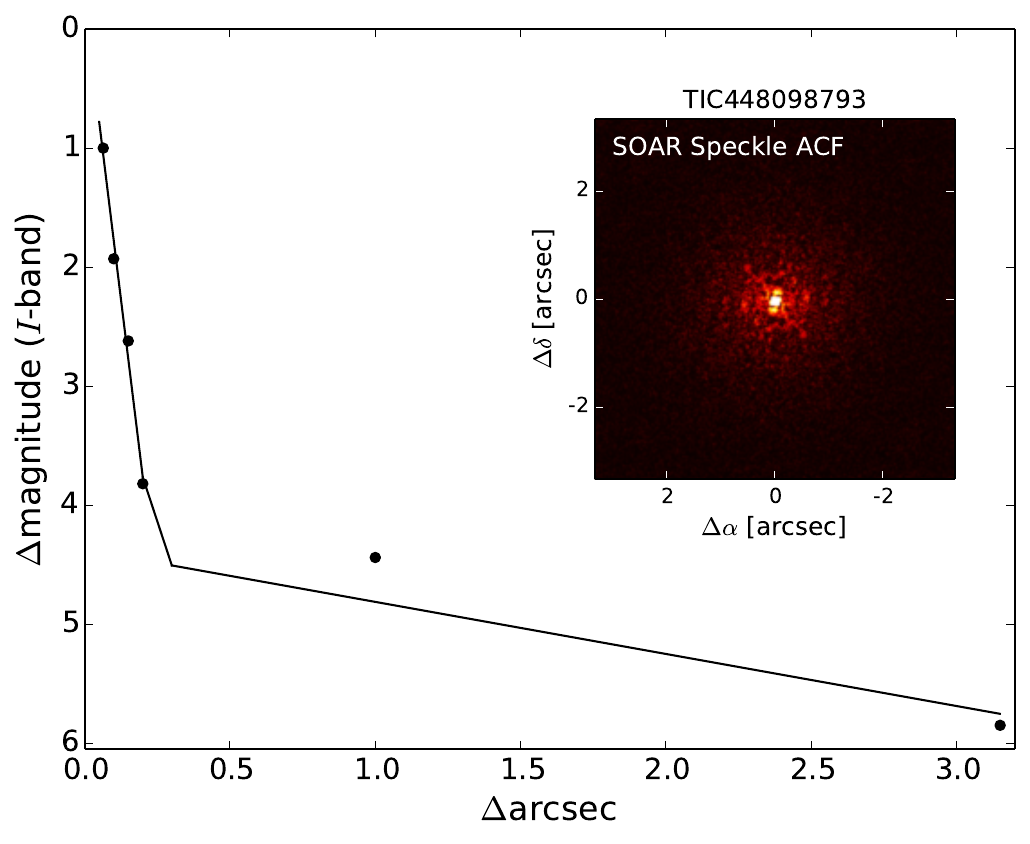}
\figsetgrpnote{Speckle sensitivity curve and auto-correlation function (inset) from SOAR/HRCam observations of TOI-3135.}
\figsetgrpend

\figsetgrpstart
\figsetgrpnum{5.17}
\figsetgrptitle{SOAR/HRCam observation of TOI-3474}
\figsetplot{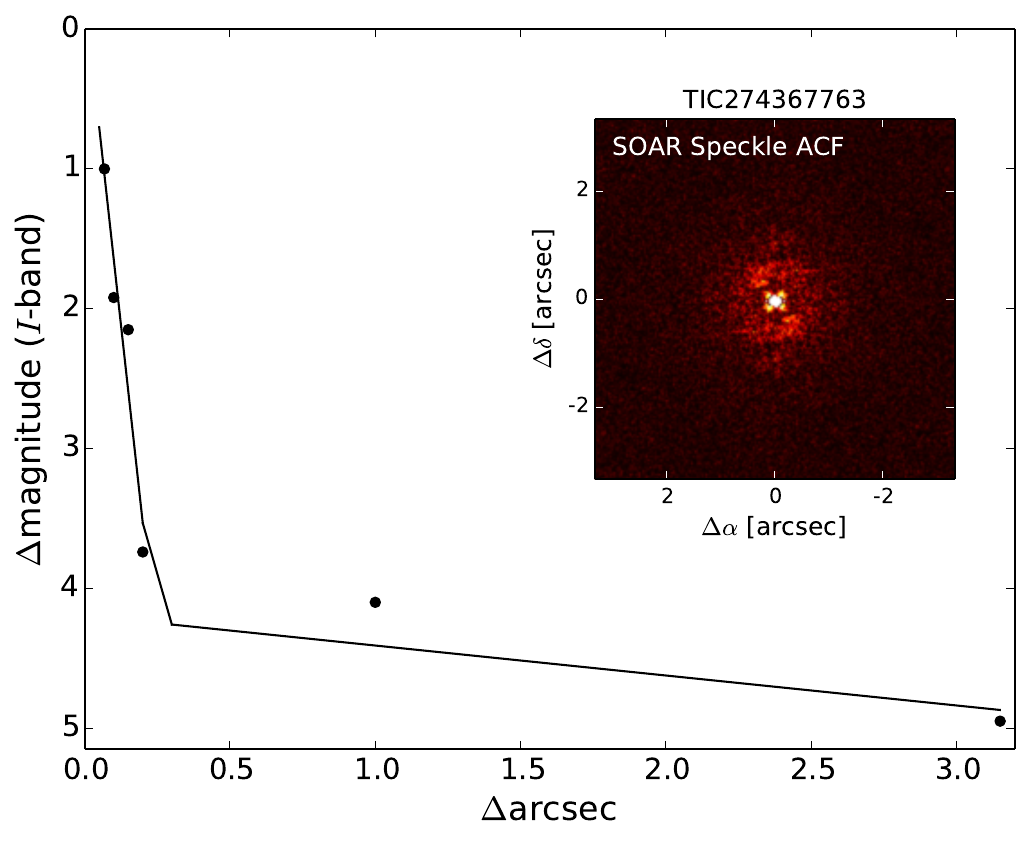}
\figsetgrpnote{Speckle sensitivity curve and auto-correlation function (inset) from SOAR/HRCam observations of TOI-3474.}
\figsetgrpend

\figsetgrpstart
\figsetgrpnum{5.18}
\figsetgrptitle{SOAR/HRCam observation of TOI-3486}
\figsetplot{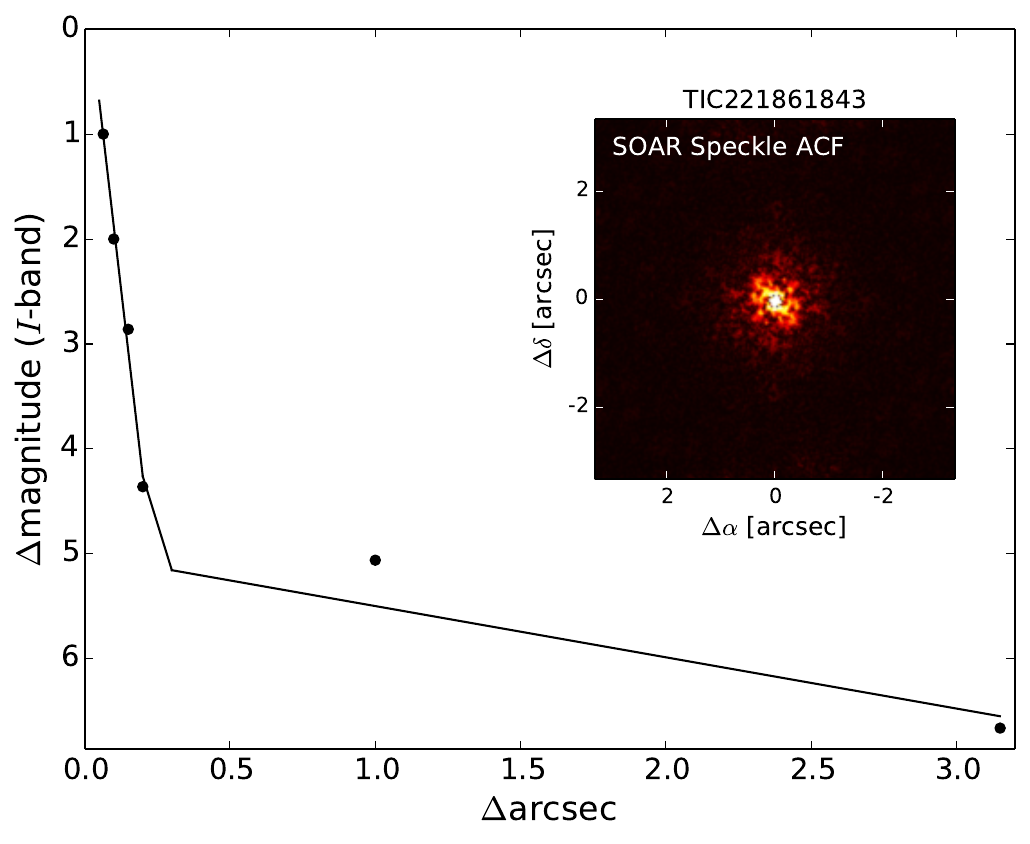}
\figsetgrpnote{Speckle sensitivity curve and auto-correlation function (inset) from SOAR/HRCam observations of TOI-3486.}
\figsetgrpend

\figsetgrpstart
\figsetgrpnum{5.19}
\figsetgrptitle{Palomar/PHARO observation of TOI-3593}
\figsetplot{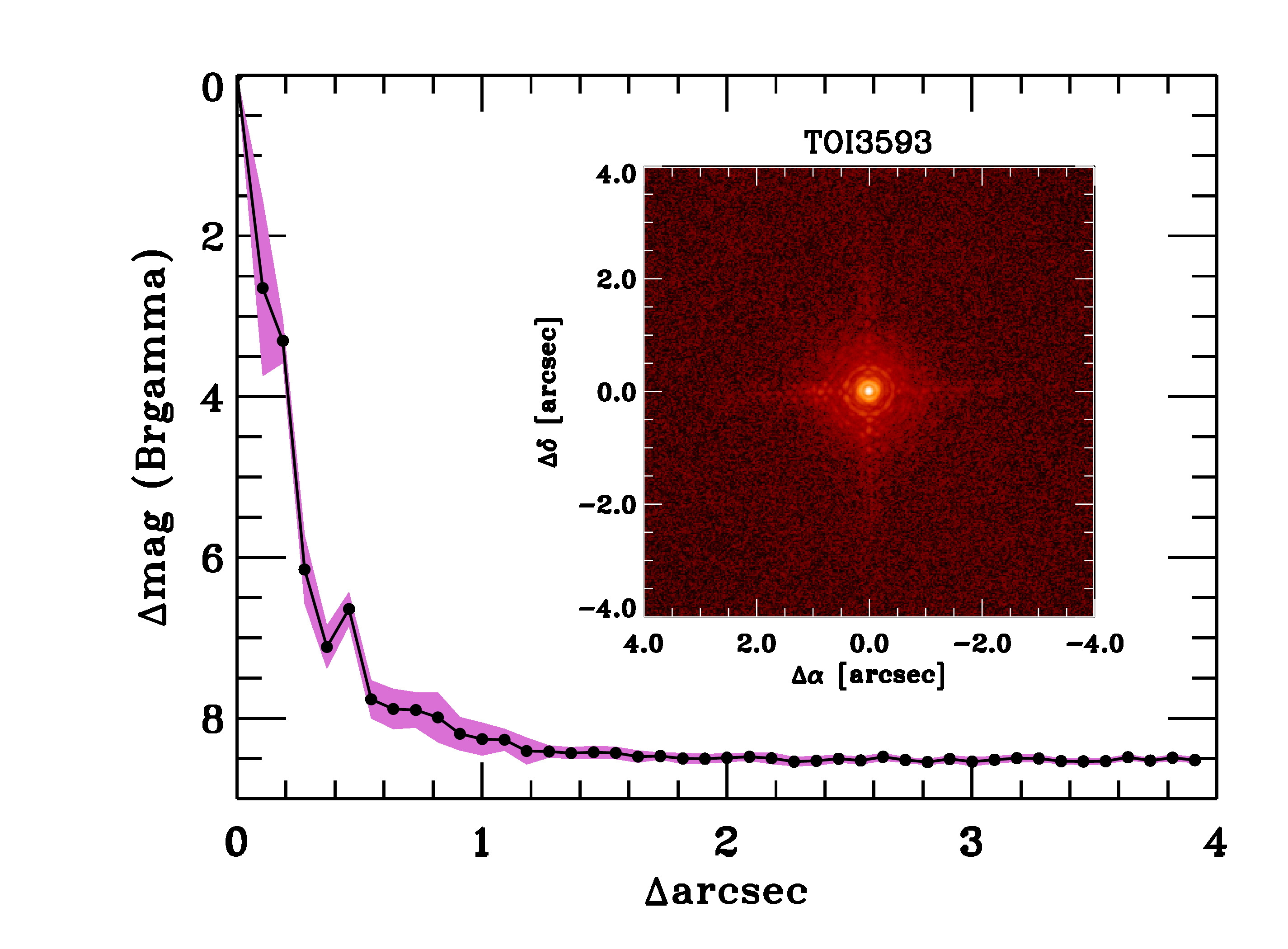}
\figsetgrpnote{Sensitivity limits and adaptive optics image (inset) from Palomar/PHARO observations of TOI-3593.}
\figsetgrpend

\figsetgrpstart
\figsetgrpnum{5.20}
\figsetgrptitle{Palomar/PHARO observation of TOI-3682}
\figsetplot{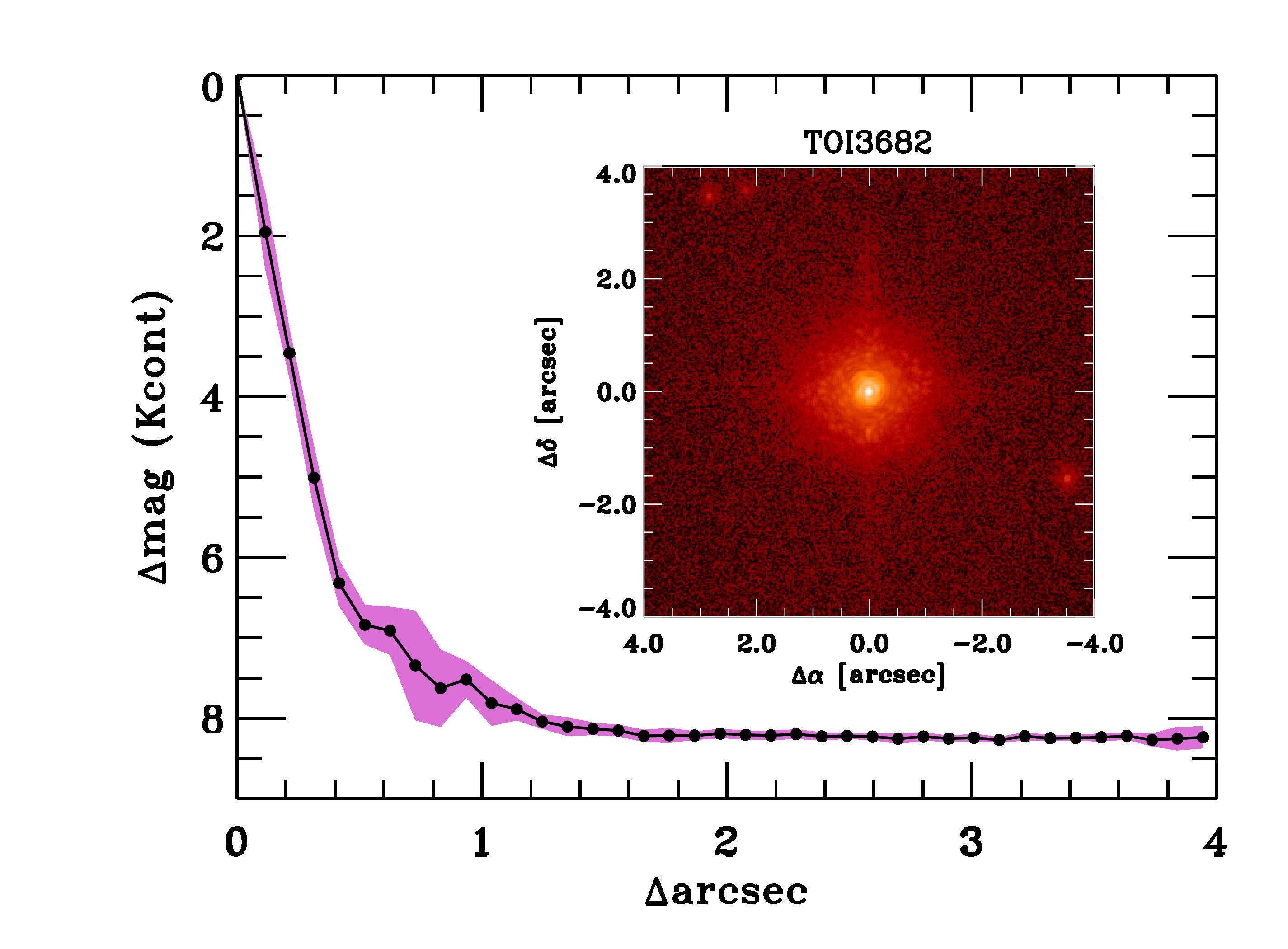}
\figsetgrpnote{Sensitivity limits and adaptive optics image (inset) from Palomar/PHARO observations of TOI-3682.}
\figsetgrpend

\figsetgrpstart
\figsetgrpnum{5.21}
\figsetgrptitle{SAI/Speckle Polarimeter observation of TOI-3682}
\figsetplot{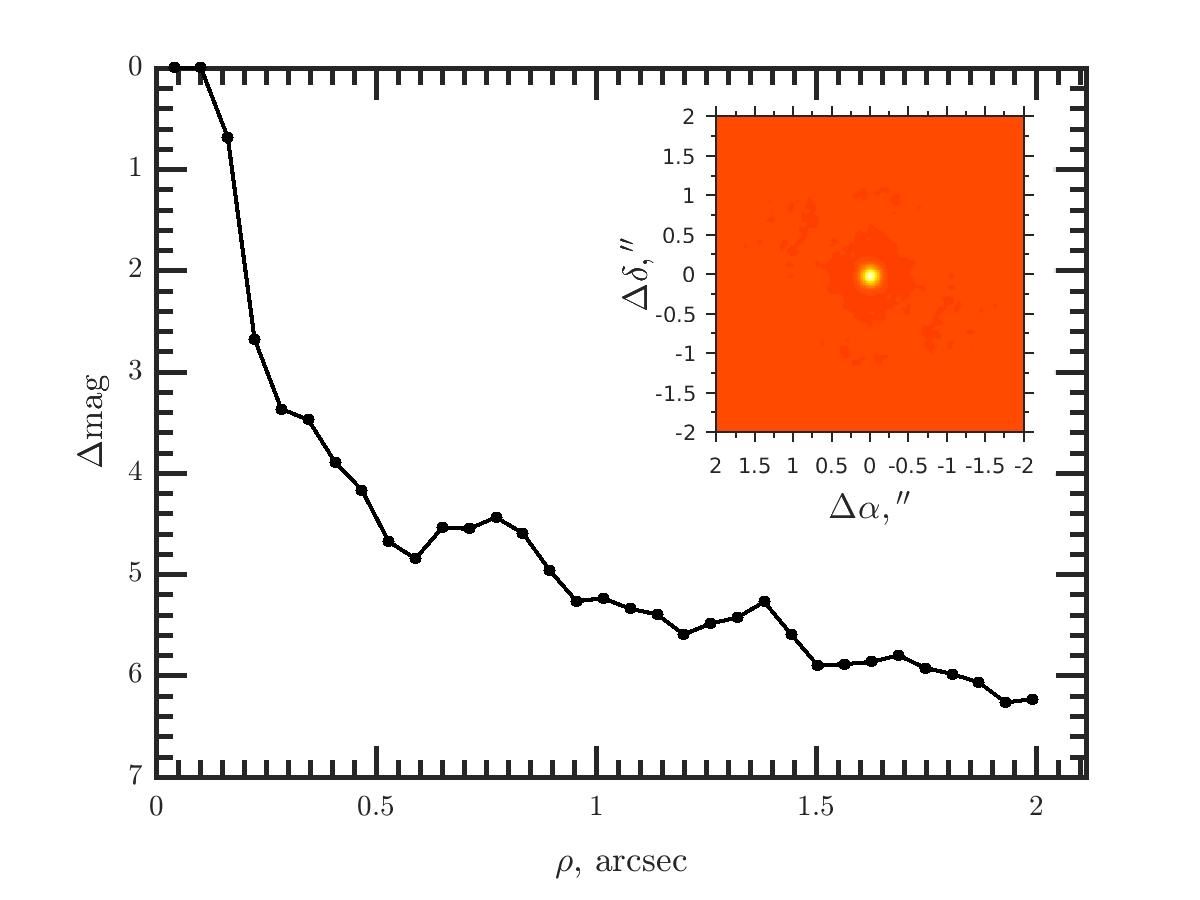}
\figsetgrpnote{Sensitivity limits and reconstructed image (inset) from SAI speckle observations of TOI-3682.}
\figsetgrpend

\figsetgrpstart
\figsetgrpnum{5.22}
\figsetgrptitle{SOAR/HRCam observation of TOI-3682}
\figsetplot{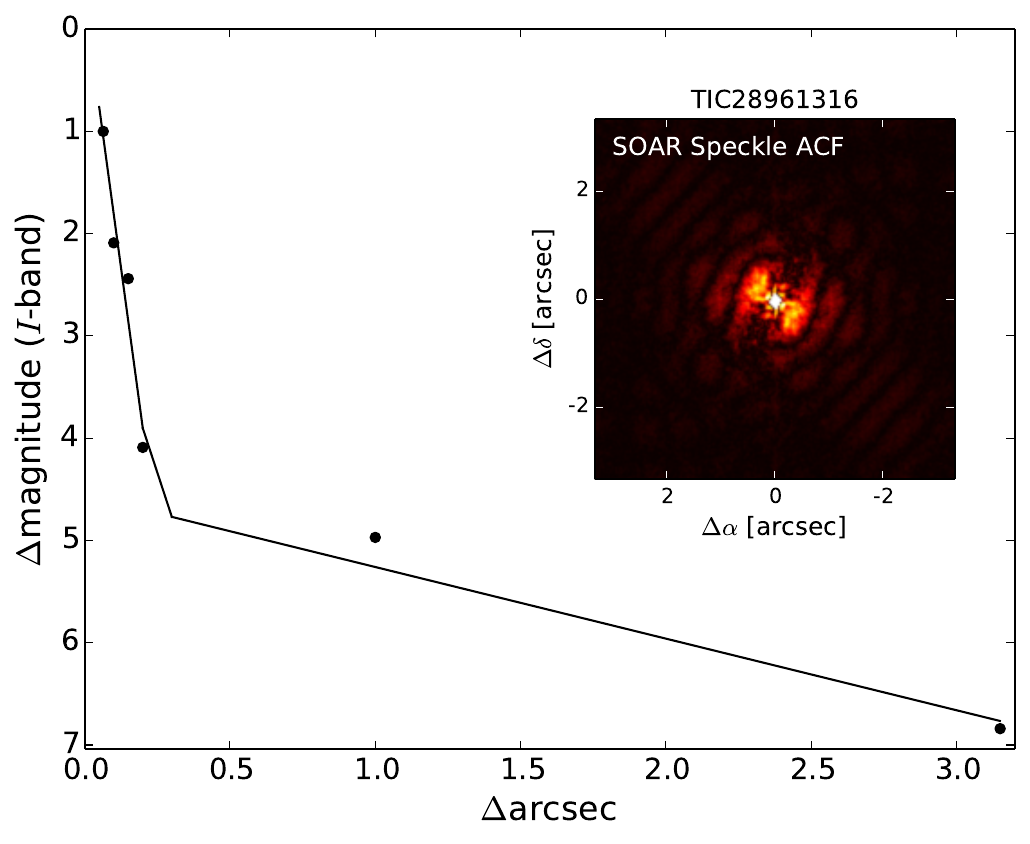}
\figsetgrpnote{Speckle sensitivity curve and auto-correlation function (inset) from SOAR/HRCam observations of TOI-3682.}
\figsetgrpend

\figsetgrpstart
\figsetgrpnum{5.23}
\figsetgrptitle{WIYN/NESSI observations of TOI-3682}
\figsetplot{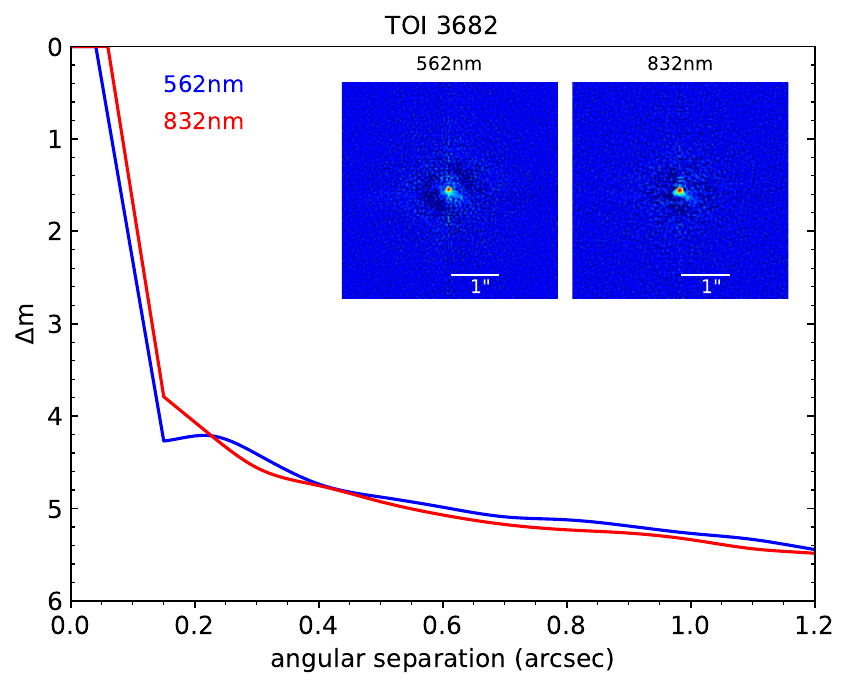}
\figsetgrpnote{Sensitivity limits and reconstructed image (inset) from WIYN/NESSI observations of TOI-3682.}
\figsetgrpend

\figsetgrpstart
\figsetgrpnum{5.24}
\figsetgrptitle{Palomar/PHARO observation of TOI-3856}
\figsetplot{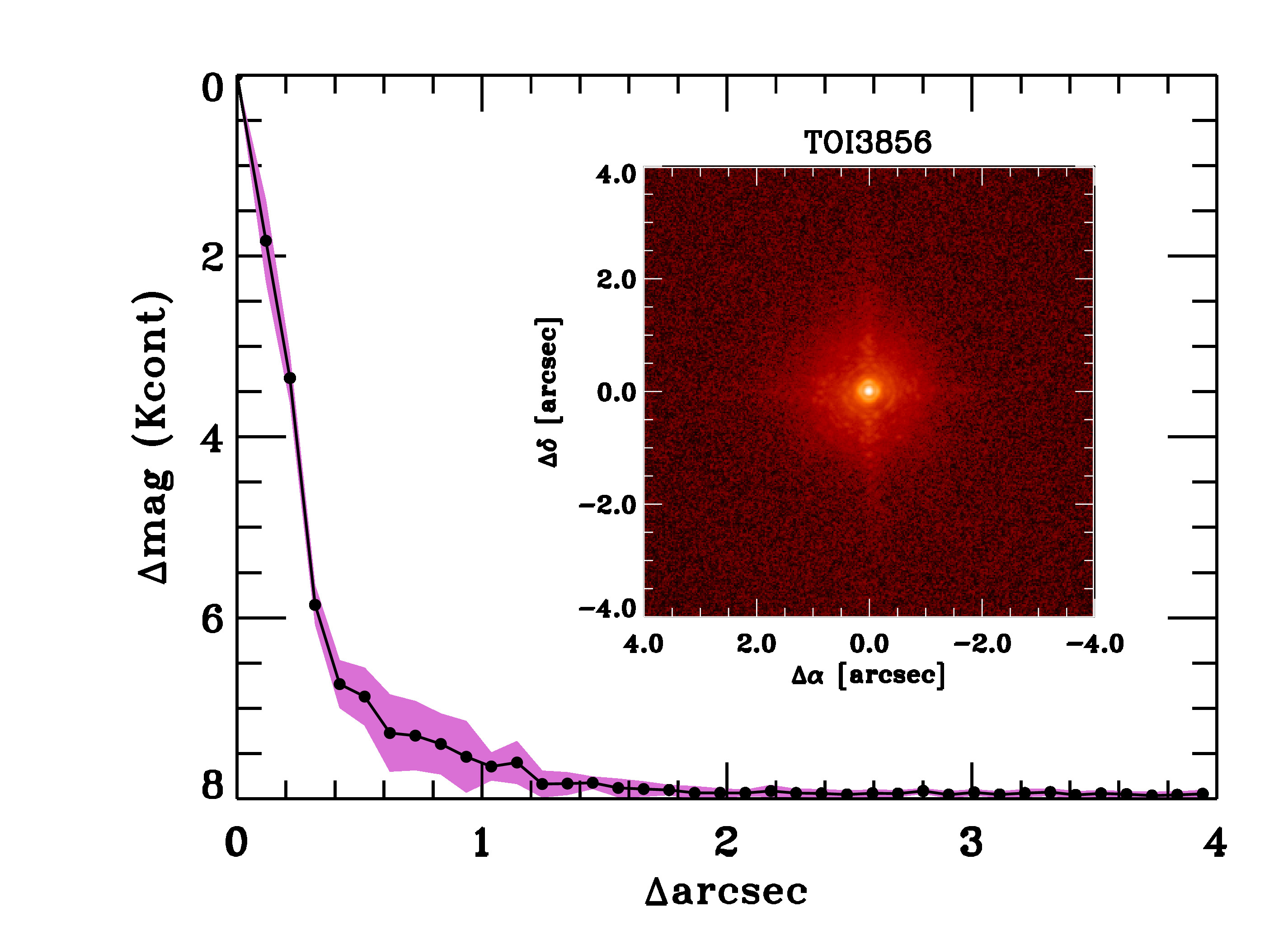}
\figsetgrpnote{Sensitivity limits and adaptive optics image (inset) from Palomar/PHARO observations of TOI-3856.}
\figsetgrpend

\figsetgrpstart
\figsetgrpnum{5.25}
\figsetgrptitle{SAI/Speckle Polarimeter observation of TOI-3856}
\figsetplot{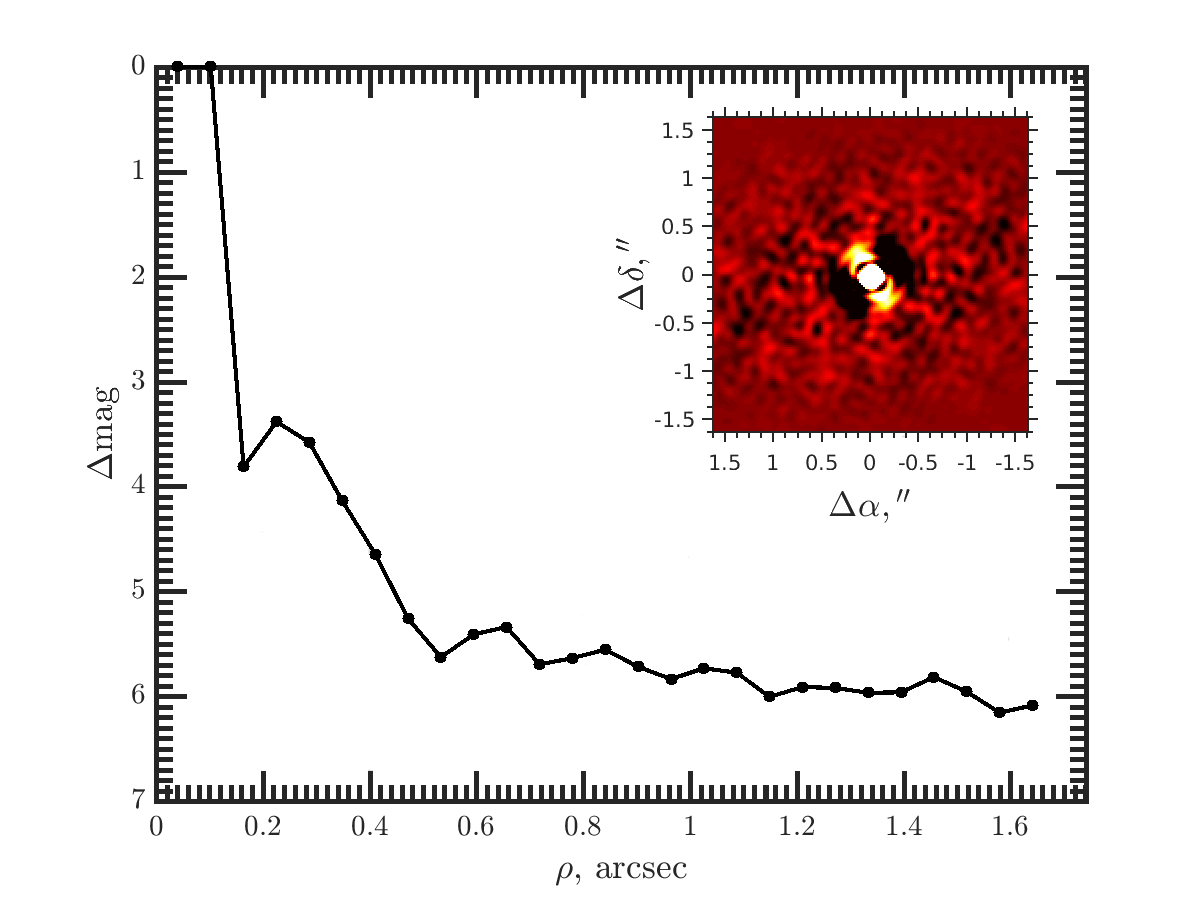}
\figsetgrpnote{Sensitivity limits and reconstructed image (inset) from SAI speckle observations of TOI-3856.}
\figsetgrpend

\figsetgrpstart
\figsetgrpnum{5.26}
\figsetgrptitle{WIYN/NESSI observations of TOI-3856}
\figsetplot{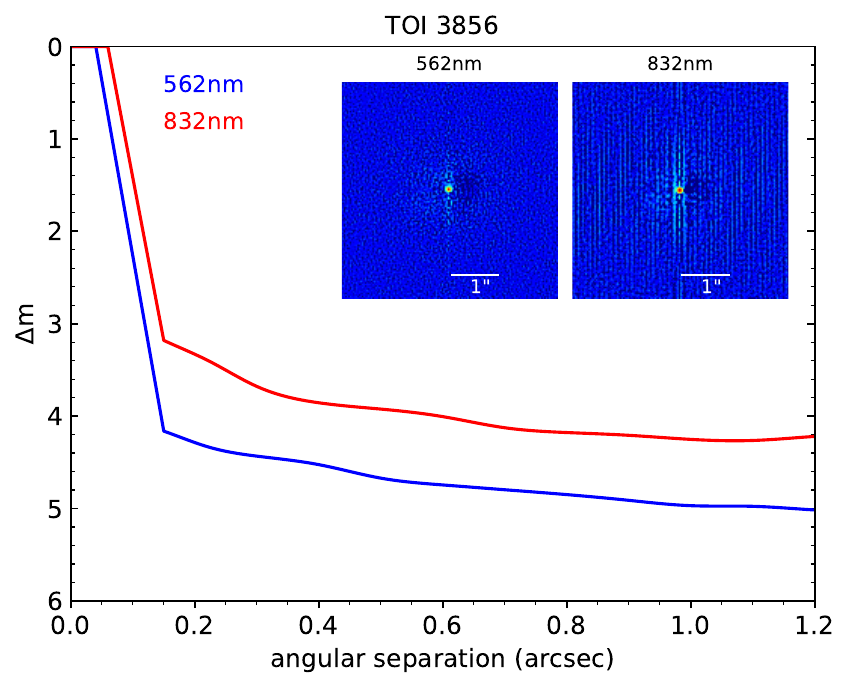}
\figsetgrpnote{Sensitivity limits and reconstructed image (inset) from WIYN/NESSI observations of TOI-3856.}
\figsetgrpend

\figsetgrpstart
\figsetgrpnum{5.27}
\figsetgrptitle{SAI/Speckle Polarimeter observation of TOI-3877}
\figsetplot{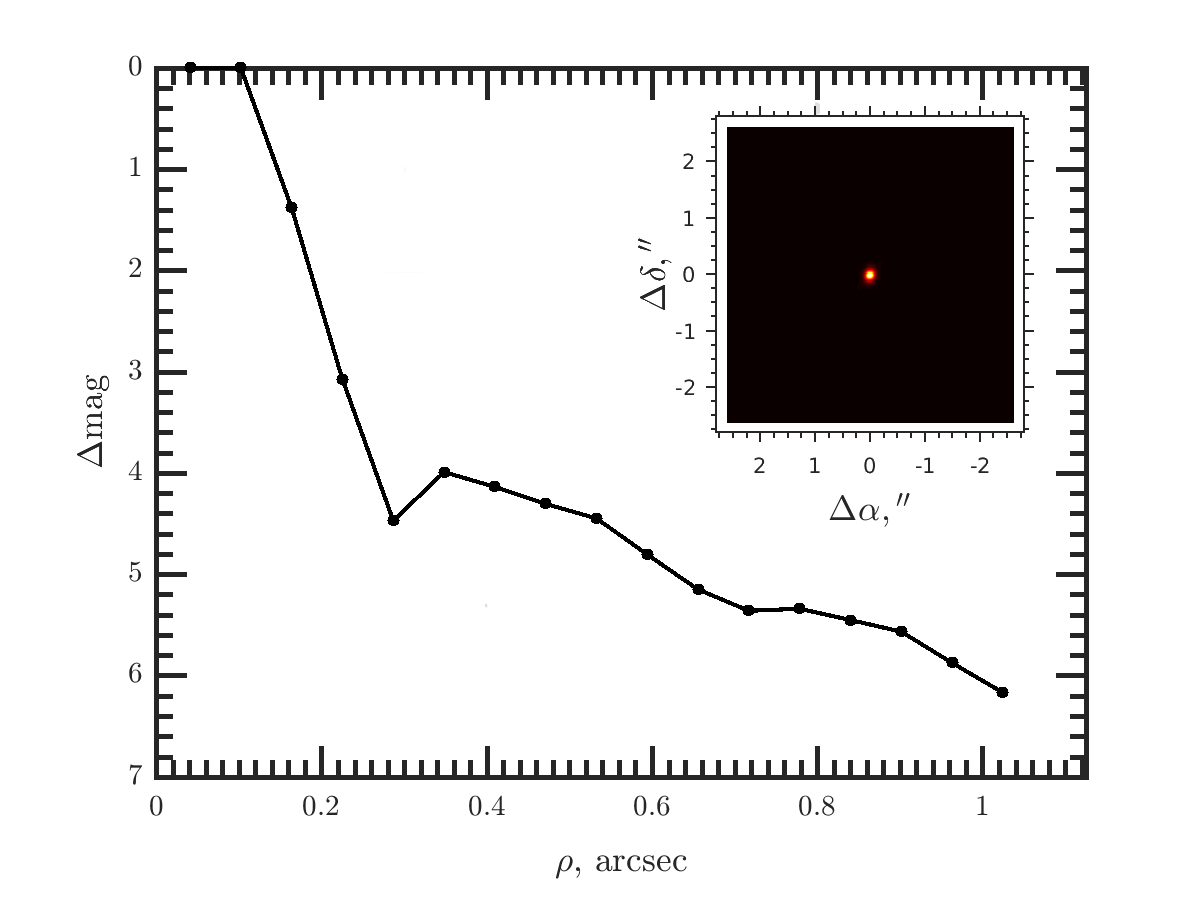}
\figsetgrpnote{Sensitivity limits and reconstructed image (inset) from SAI speckle observations of TOI-3877.}
\figsetgrpend

\figsetgrpstart
\figsetgrpnum{5.28}
\figsetgrptitle{Shane/ShARCS J observation of TOI-3877}
\figsetplot{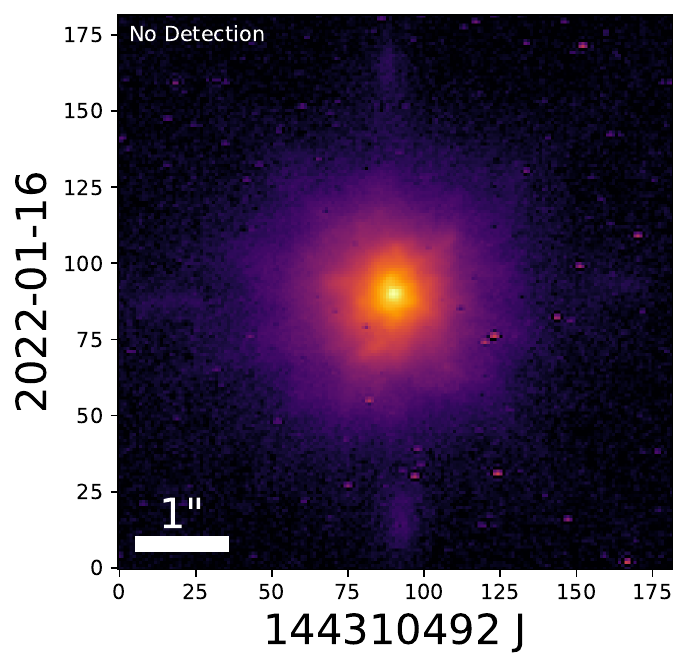}
\figsetgrpnote{Adaptive optics image from Shane/ShARCS observations of TOI-3877.}
\figsetgrpend

\figsetgrpstart
\figsetgrpnum{5.29}
\figsetgrptitle{Shane/ShARCS K$_s$ observation of TOI-3877}
\figsetplot{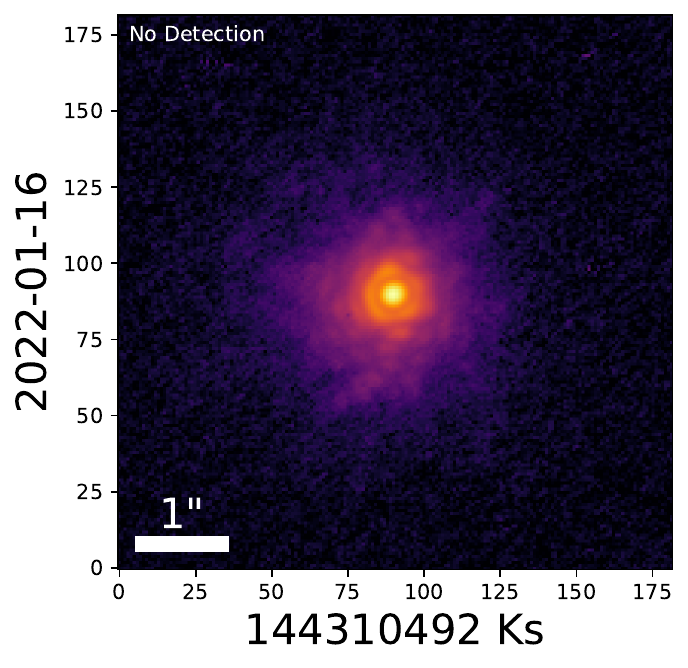}
\figsetgrpnote{Adaptive optics image from Shane/ShARCS observations of TOI-3877.}
\figsetgrpend

\figsetgrpstart
\figsetgrpnum{5.30}
\figsetgrptitle{WIYN/NESSI observation of TOI-3877}
\figsetplot{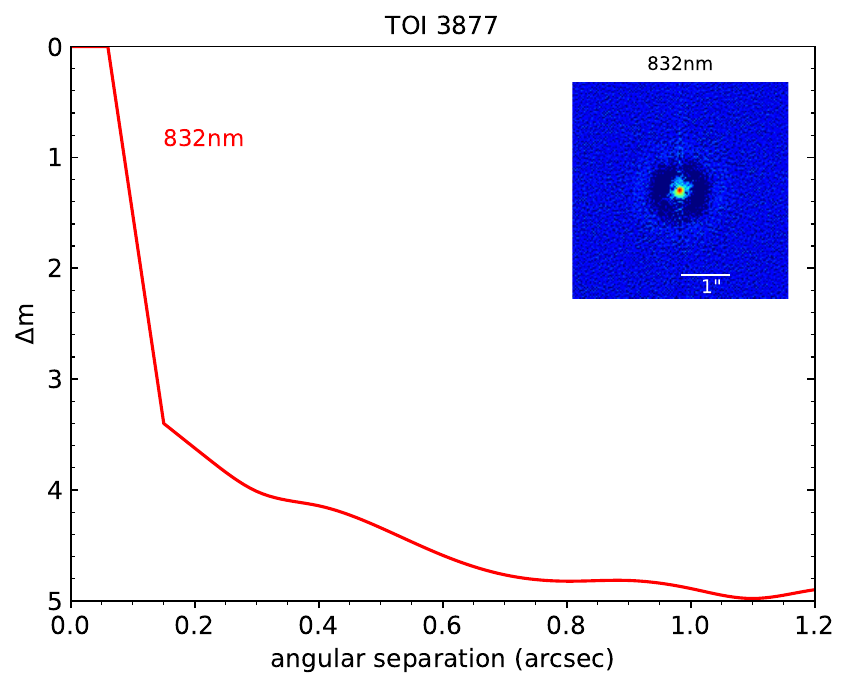}
\figsetgrpnote{Sensitivity limits and reconstructed image (inset) from WIYN/NESSI observations of TOI-3877.}
\figsetgrpend

\figsetgrpstart
\figsetgrpnum{5.31}
\figsetgrptitle{Gemini-N/'Alopeke observations of TOI-3980}
\figsetplot{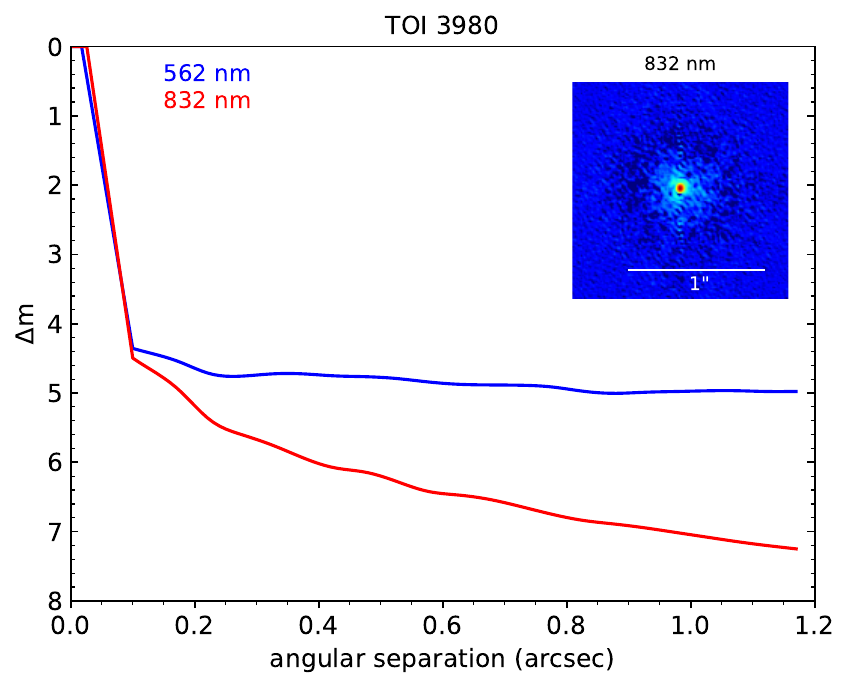}
\figsetgrpnote{Sensitivity limits and reconstructed image (inset) from Gemini-N/'Alopeke observations of TOI-3980.}
\figsetgrpend

\figsetgrpstart
\figsetgrpnum{5.32}
\figsetgrptitle{SAI/Speckle Polarimeter observation of TOI-3980}
\figsetplot{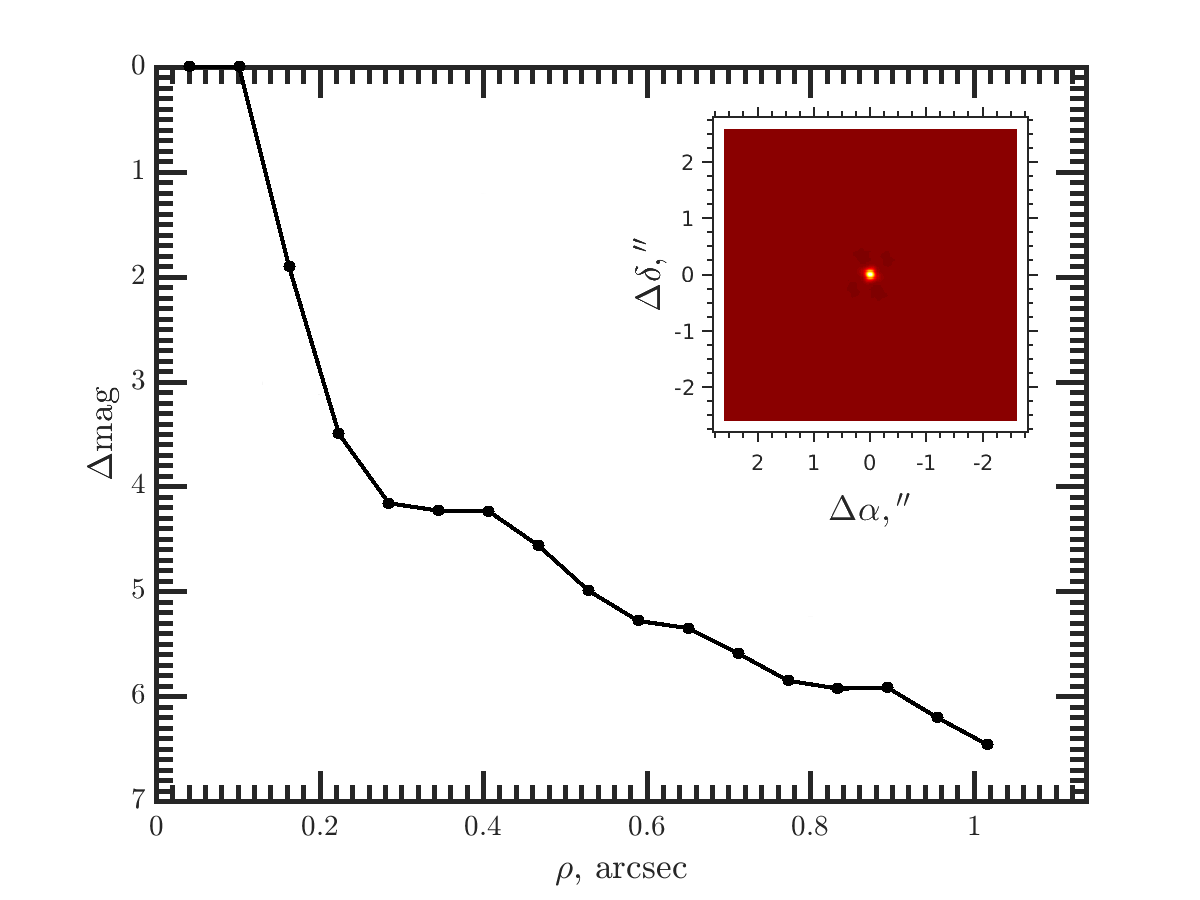}
\figsetgrpnote{Sensitivity limits and reconstructed image (inset) from SAI speckle observations of TOI-3980.}
\figsetgrpend

\figsetgrpstart
\figsetgrpnum{5.33}
\figsetgrptitle{SOAR/HRCam observation of TOI-4214}
\figsetplot{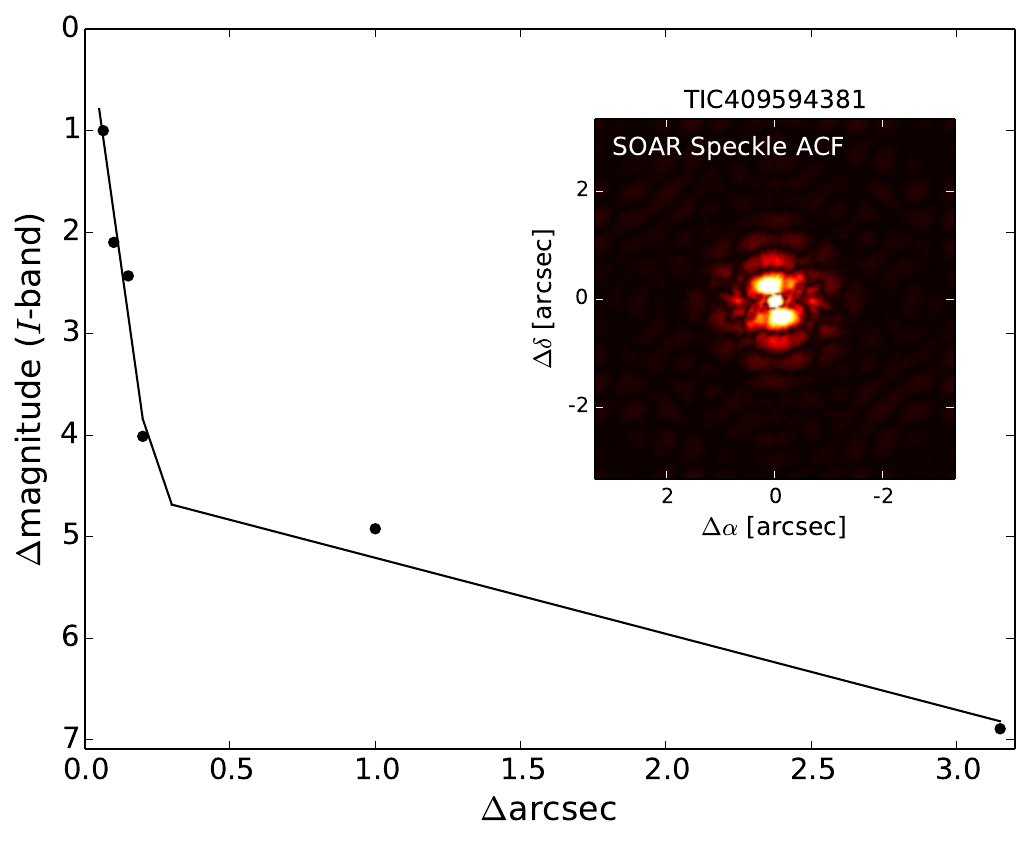}
\figsetgrpnote{Speckle sensitivity curve and auto-correlation function (inset) from SOAR/HRCam observations of TOI-4214.}
\figsetgrpend

\figsetgrpstart
\figsetgrpnum{5.34}
\figsetgrptitle{Shane/ShARCS J observation of TOI-4214}
\figsetplot{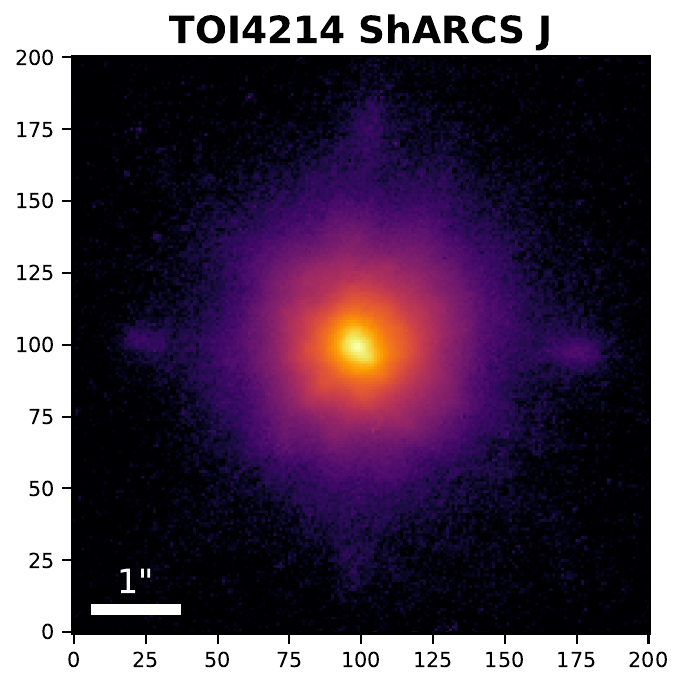}
\figsetgrpnote{Adaptive optics image from Shane/ShARCS observations of TOI-4214.}
\figsetgrpend

\figsetgrpstart
\figsetgrpnum{5.35}
\figsetgrptitle{Shane/ShARCS K$_s$ observation of TOI-4214}
\figsetplot{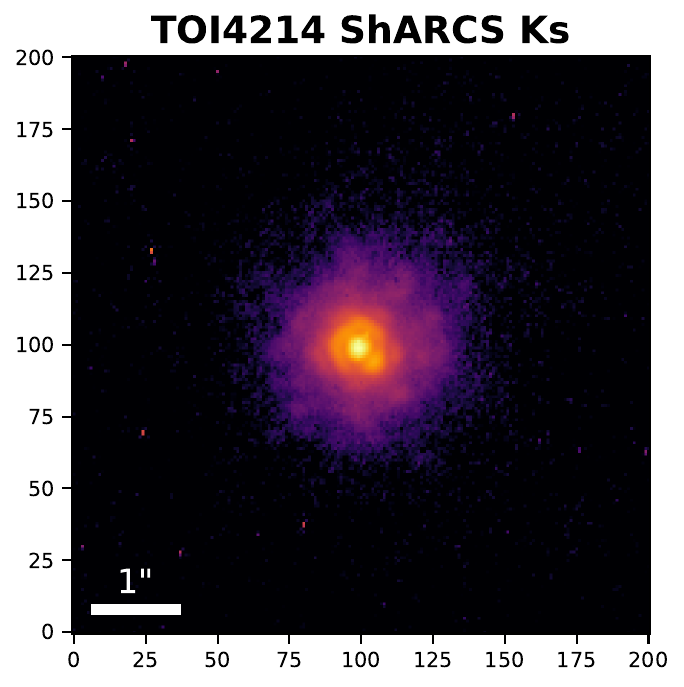}
\figsetgrpnote{Adaptive optics image from Shane/ShARCS observations of TOI-4214.}
\figsetgrpend

\figsetgrpstart
\figsetgrpnum{5.36}
\figsetgrptitle{SAI/Speckle Polarimeter observation of TOI-4487}
\figsetplot{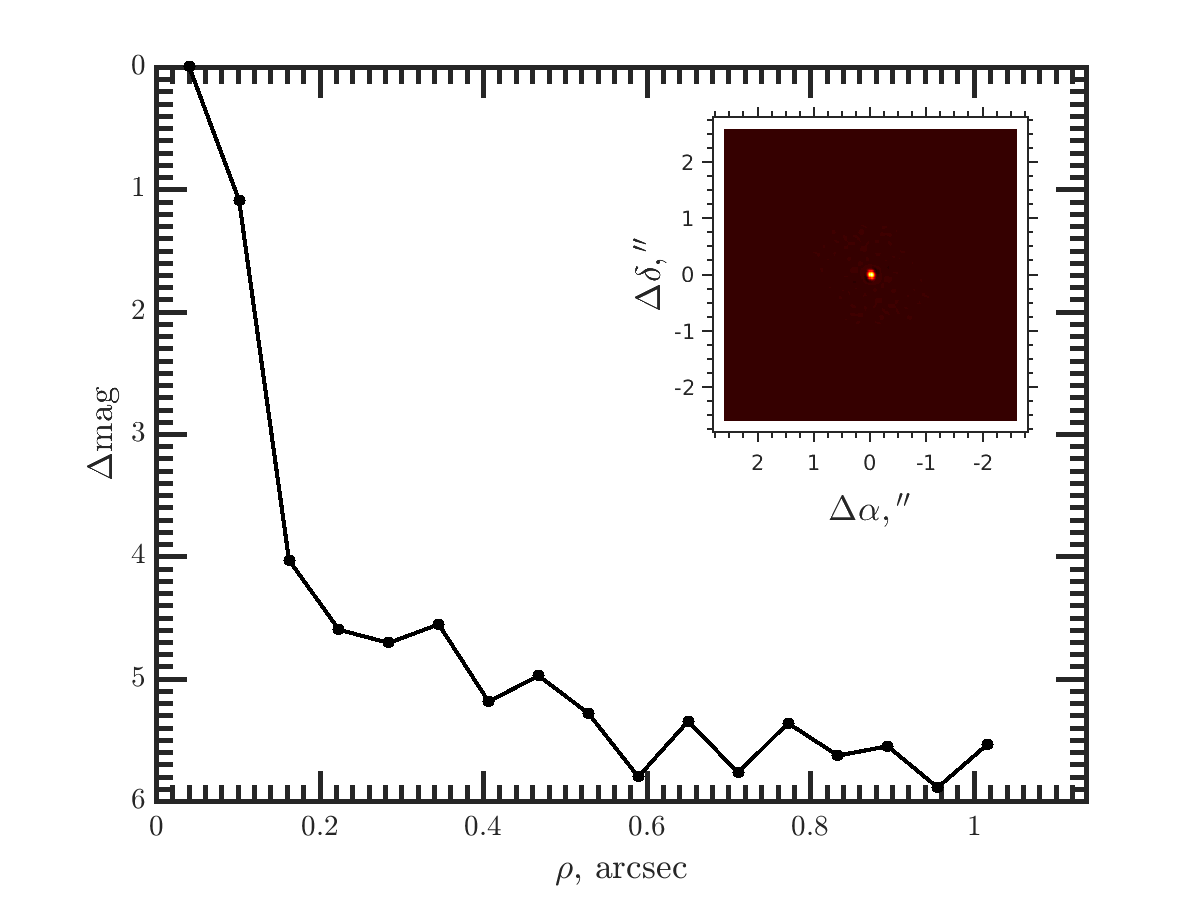}
\figsetgrpnote{Sensitivity limits and reconstructed image (inset) from SAI speckle observations of TOI-4487.}
\figsetgrpend

\figsetgrpstart
\figsetgrpnum{5.37}
\figsetgrptitle{WIYN/NESSI observation of TOI-4487}
\figsetplot{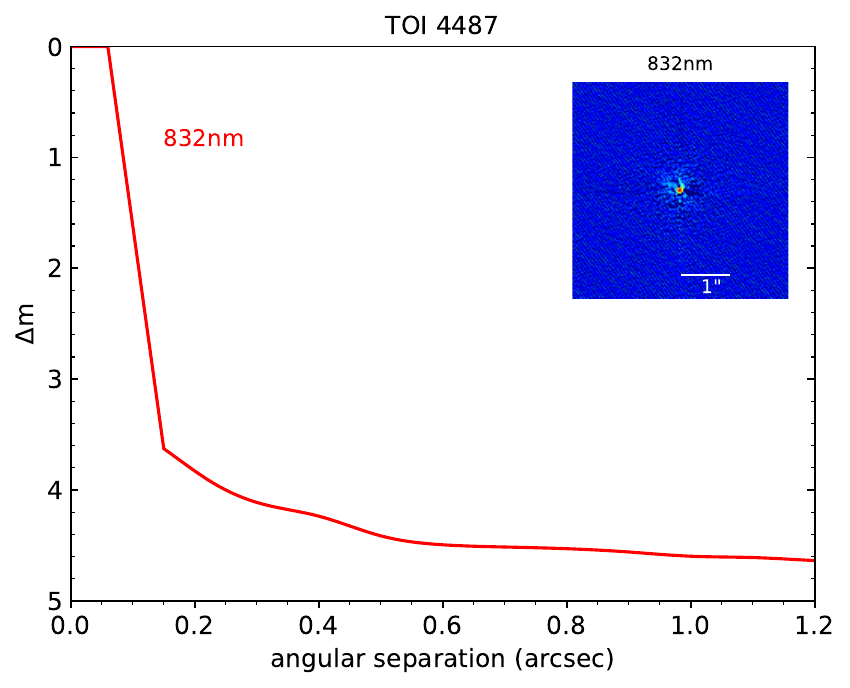}
\figsetgrpnote{Sensitivity limits and reconstructed image (inset) from WIYN/NESSI observations of TOI-4487.}
\figsetgrpend

\figsetgrpstart
\figsetgrpnum{5.38}
\figsetgrptitle{Gemini-S/Zorro observations of TOI-4734}
\figsetplot{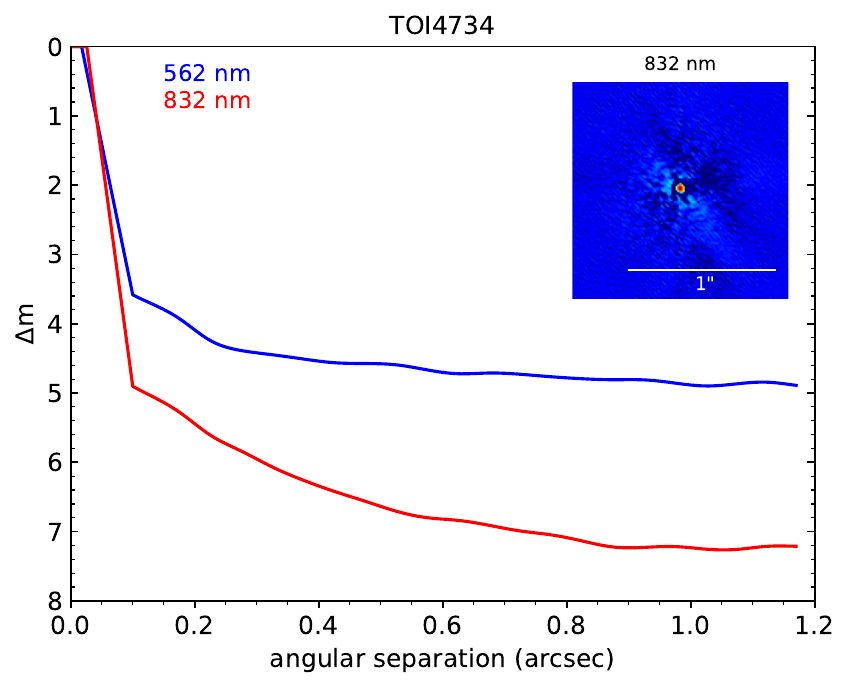}
\figsetgrpnote{Sensitivity limits and reconstructed image (inset) from Gemini-S/Zorro observations of TOI-4734.}
\figsetgrpend

\figsetgrpstart
\figsetgrpnum{5.39}
\figsetgrptitle{SAI/Speckle Polarimeter observation of TOI-4734}
\figsetplot{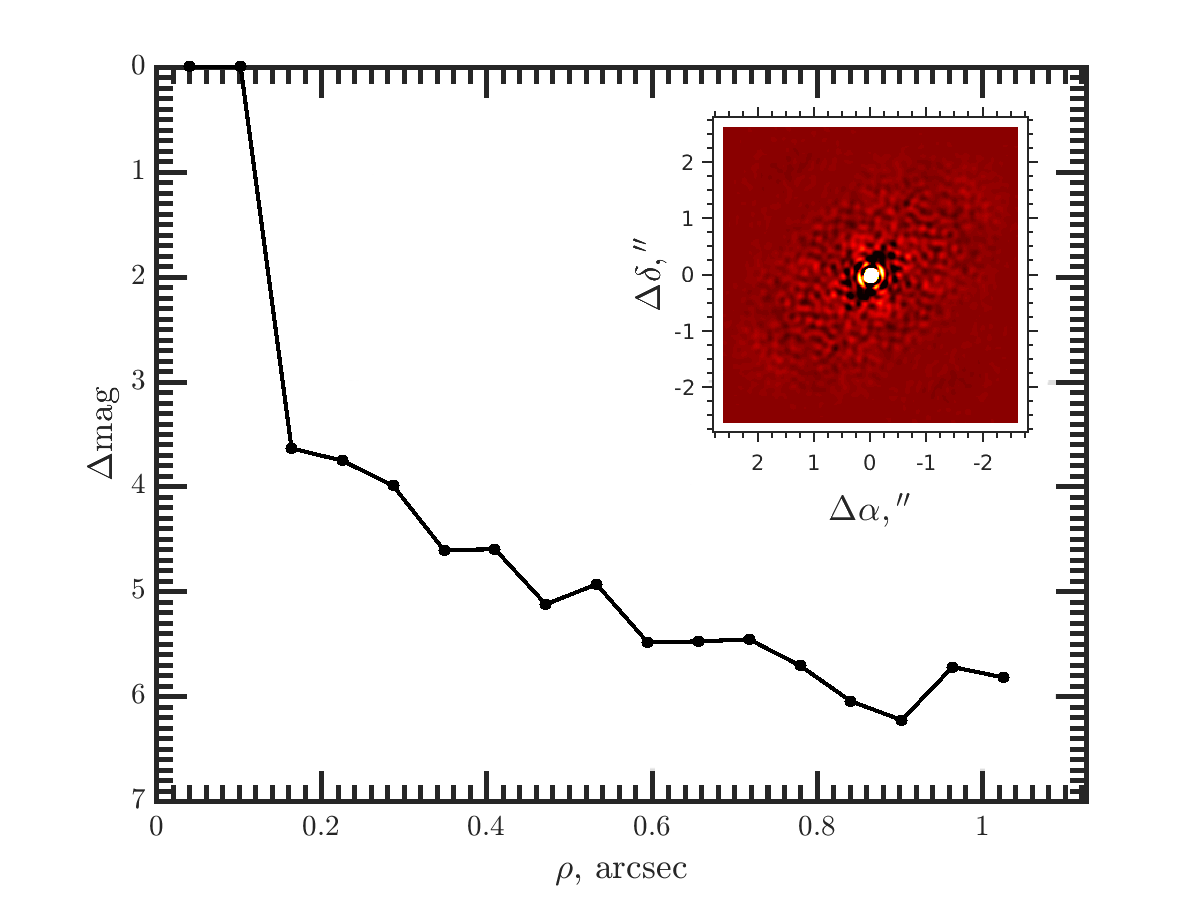}
\figsetgrpnote{Sensitivity limits and reconstructed image (inset) from SAI speckle observations of TOI-4734.}
\figsetgrpend

\figsetgrpstart
\figsetgrpnum{5.40}
\figsetgrptitle{WIYN/NESSI observations of TOI-4734}
\figsetplot{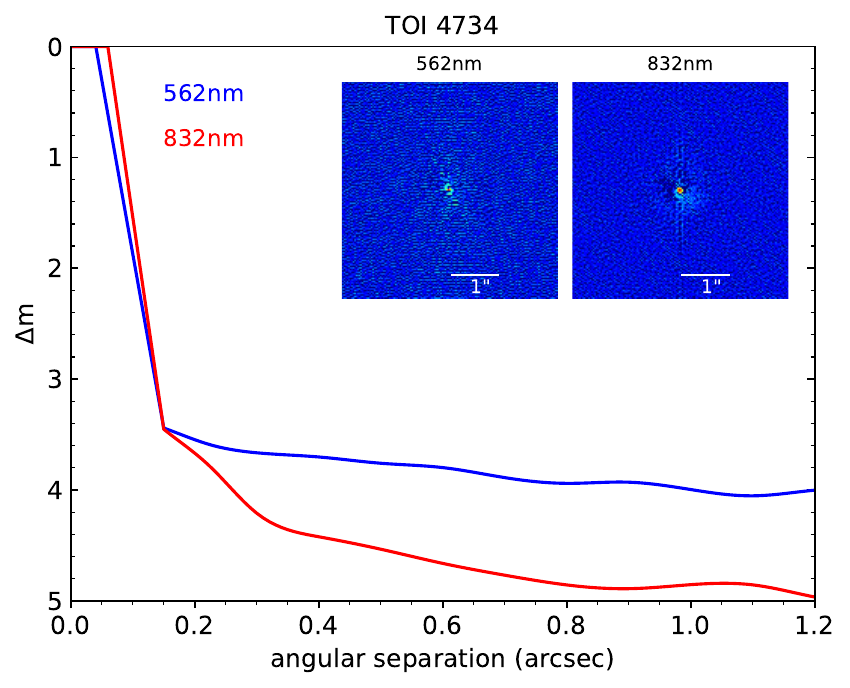}
\figsetgrpnote{Sensitivity limits and reconstructed image (inset) from WIYN/NESSI observations of TOI-4734.}
\figsetgrpend

\figsetgrpstart
\figsetgrpnum{5.41}
\figsetgrptitle{WIYN/NESSI observations of TOI-4734}
\figsetplot{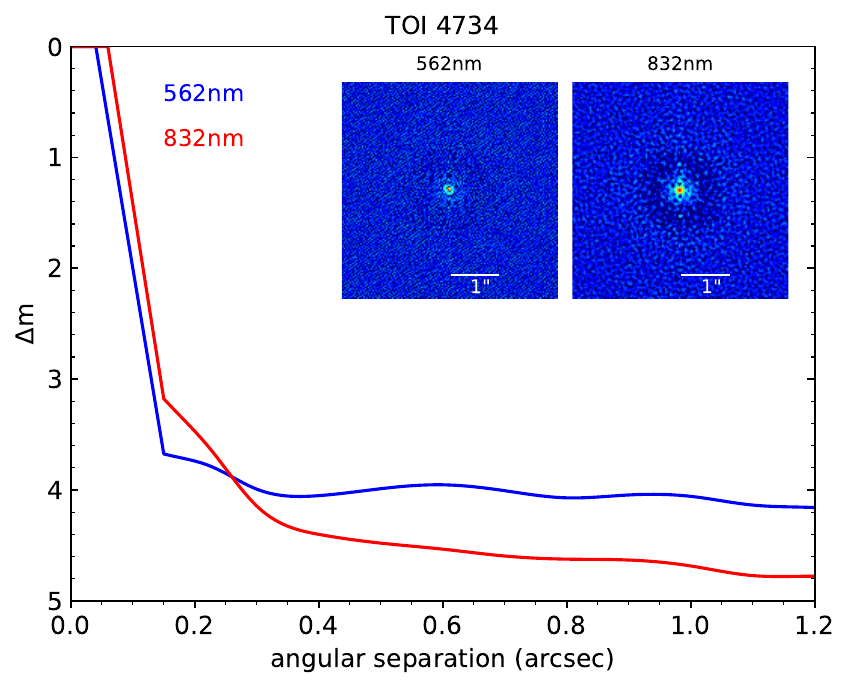}
\figsetgrpnote{Sensitivity limits and reconstructed image (inset) from WIYN/NESSI observations of TOI-4734.}
\figsetgrpend

\figsetgrpstart
\figsetgrpnum{5.42}
\figsetgrptitle{SOAR/HRCam observation of TOI-4794}
\figsetplot{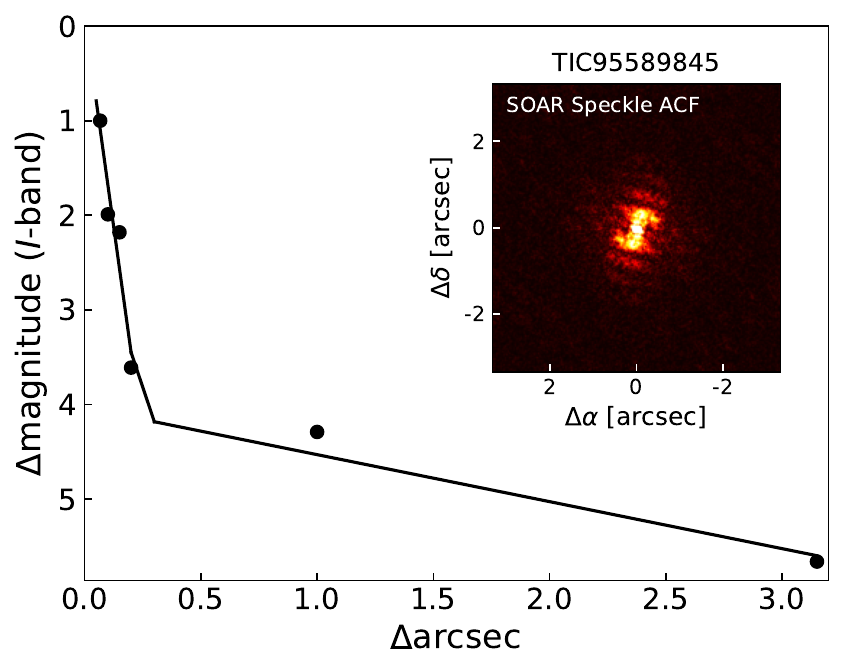}
\figsetgrpnote{Speckle sensitivity curve and auto-correlation function (inset) from SOAR/HRCam observations of TOI-4794.}
\figsetgrpend

\figsetgrpstart
\figsetgrpnum{5.43}
\figsetgrptitle{SOAR/HRCam observation of TOI-4961}
\figsetplot{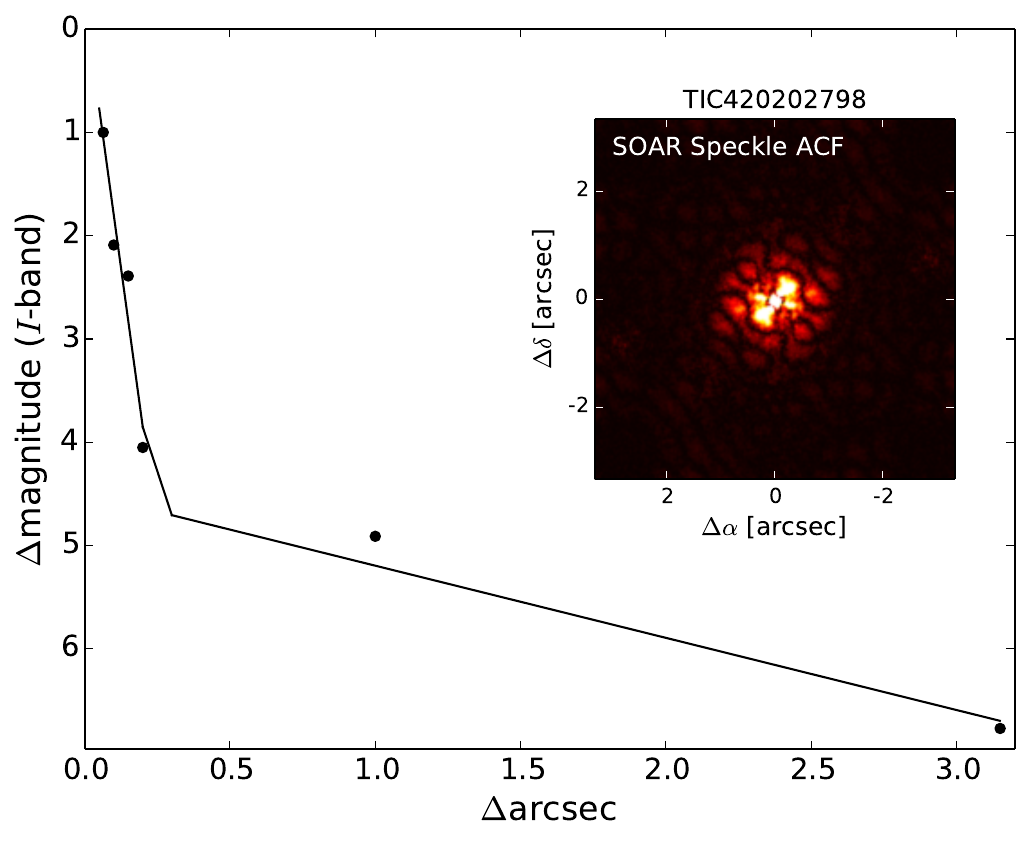}
\figsetgrpnote{Speckle sensitivity curve and auto-correlation function (inset) from SOAR/HRCam observations of TOI-4961.}
\figsetgrpend

\figsetgrpstart
\figsetgrpnum{5.44}
\figsetgrptitle{SAI/Speckle Polarimeter observation of TOI-5210}
\figsetplot{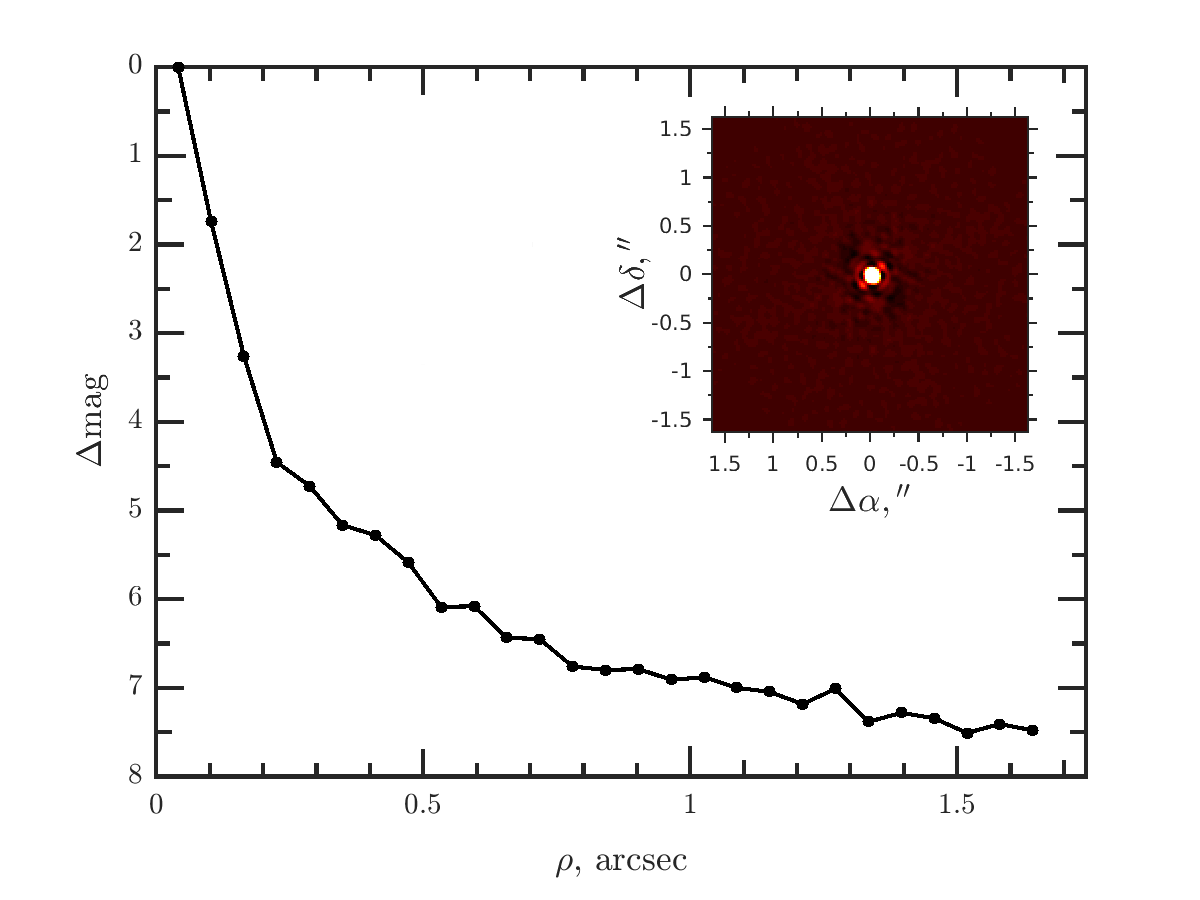}
\figsetgrpnote{Sensitivity limits and reconstructed image (inset) from SAI speckle observations of TOI-5210.}
\figsetgrpend

\figsetgrpstart
\figsetgrpnum{5.45}
\figsetgrptitle{SOAR/HRCam observation of TOI-5210}
\figsetplot{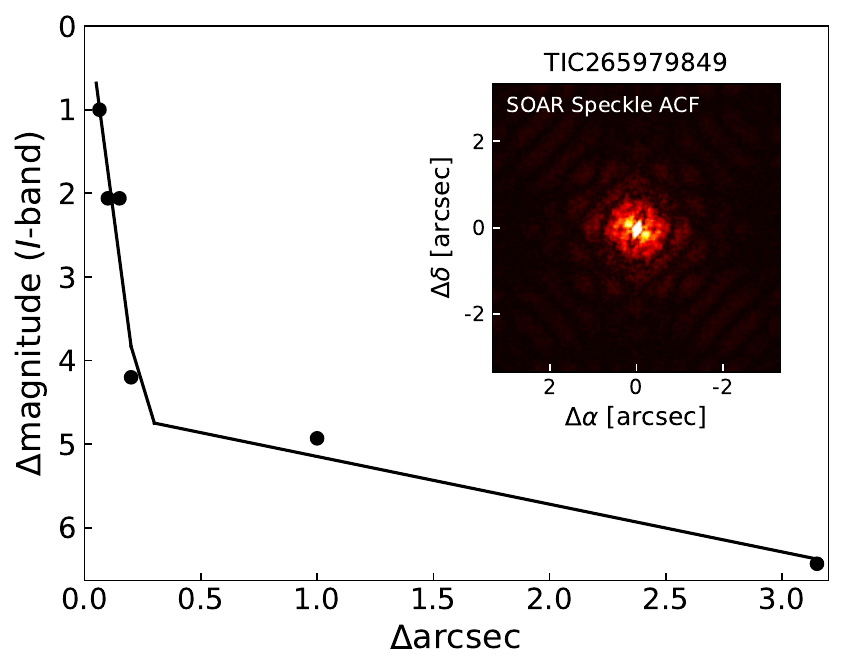}
\figsetgrpnote{Speckle sensitivity curve and auto-correlation function (inset) from SOAR/HRCam observations of TOI-5210.}
\figsetgrpend

\figsetgrpstart
\figsetgrpnum{5.46}
\figsetgrptitle{WIYN/NESSI observation of TOI-5210}
\figsetplot{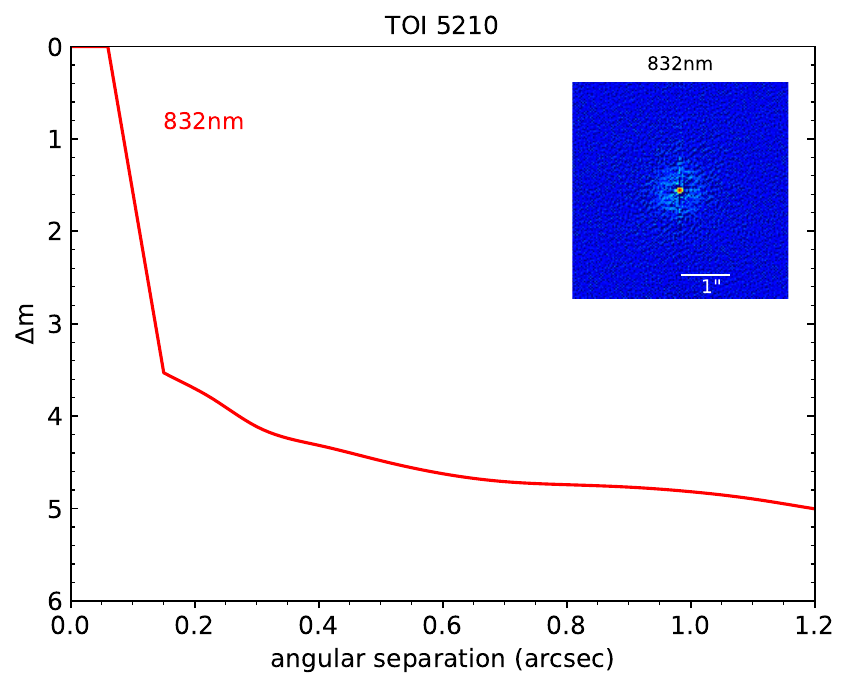}
\figsetgrpnote{Sensitivity limits and reconstructed image (inset) from WIYN/NESSI observations of TOI-5210.}
\figsetgrpend

\figsetgrpstart
\figsetgrpnum{5.47}
\figsetgrptitle{SAI/Speckle Polarimeter observation of TOI-5322}
\figsetplot{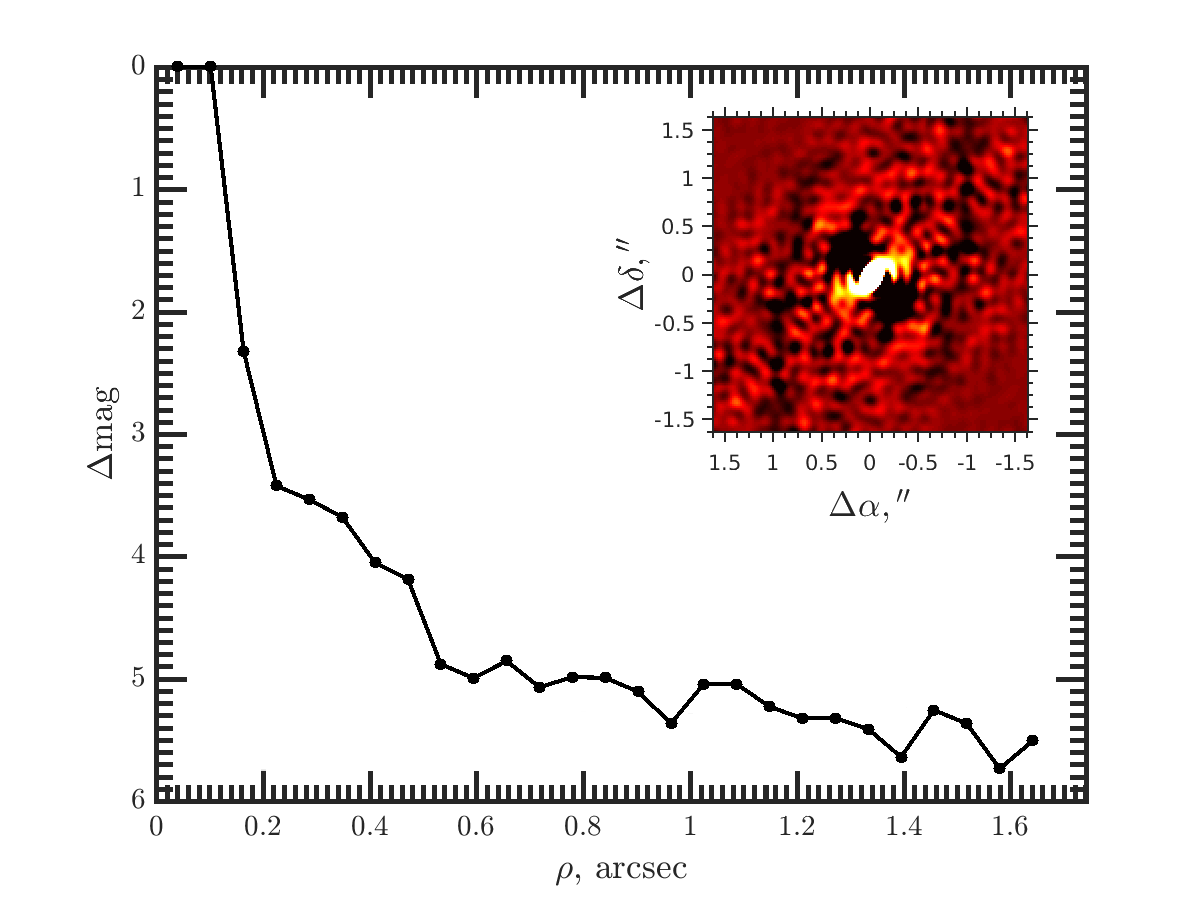}
\figsetgrpnote{Sensitivity limits and reconstructed image (inset) from SAI speckle observations of TOI-5322.}
\figsetgrpend

\figsetgrpstart
\figsetgrpnum{5.48}
\figsetgrptitle{WIYN/NESSI observations of TOI-5322}
\figsetplot{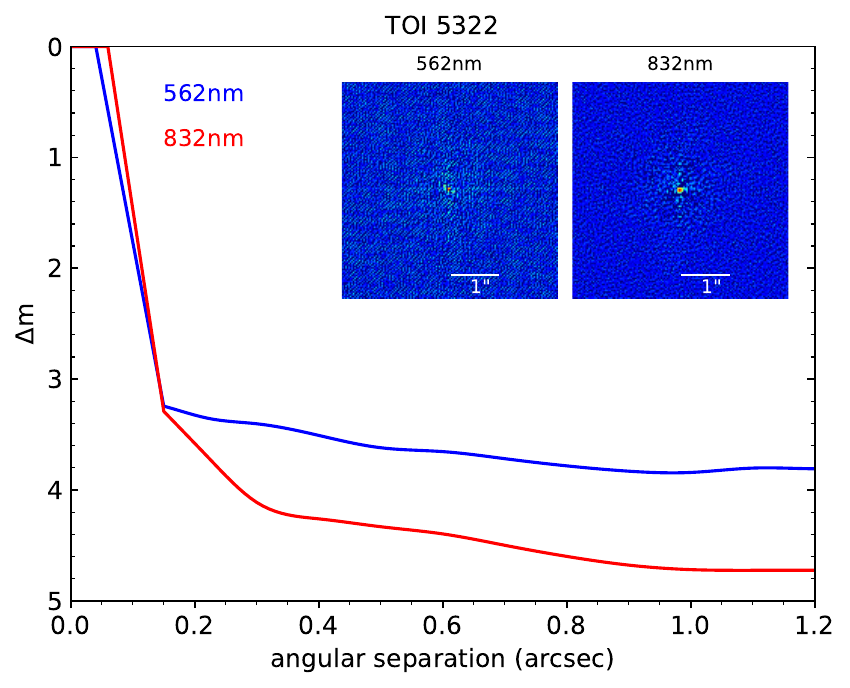}
\figsetgrpnote{Sensitivity limits and reconstructed image (inset) from WIYN/NESSI observations of TOI-5322.}
\figsetgrpend

\figsetgrpstart
\figsetgrpnum{5.49}
\figsetgrptitle{SAI/Speckle Polarimeter observation of TOI-5340}
\figsetplot{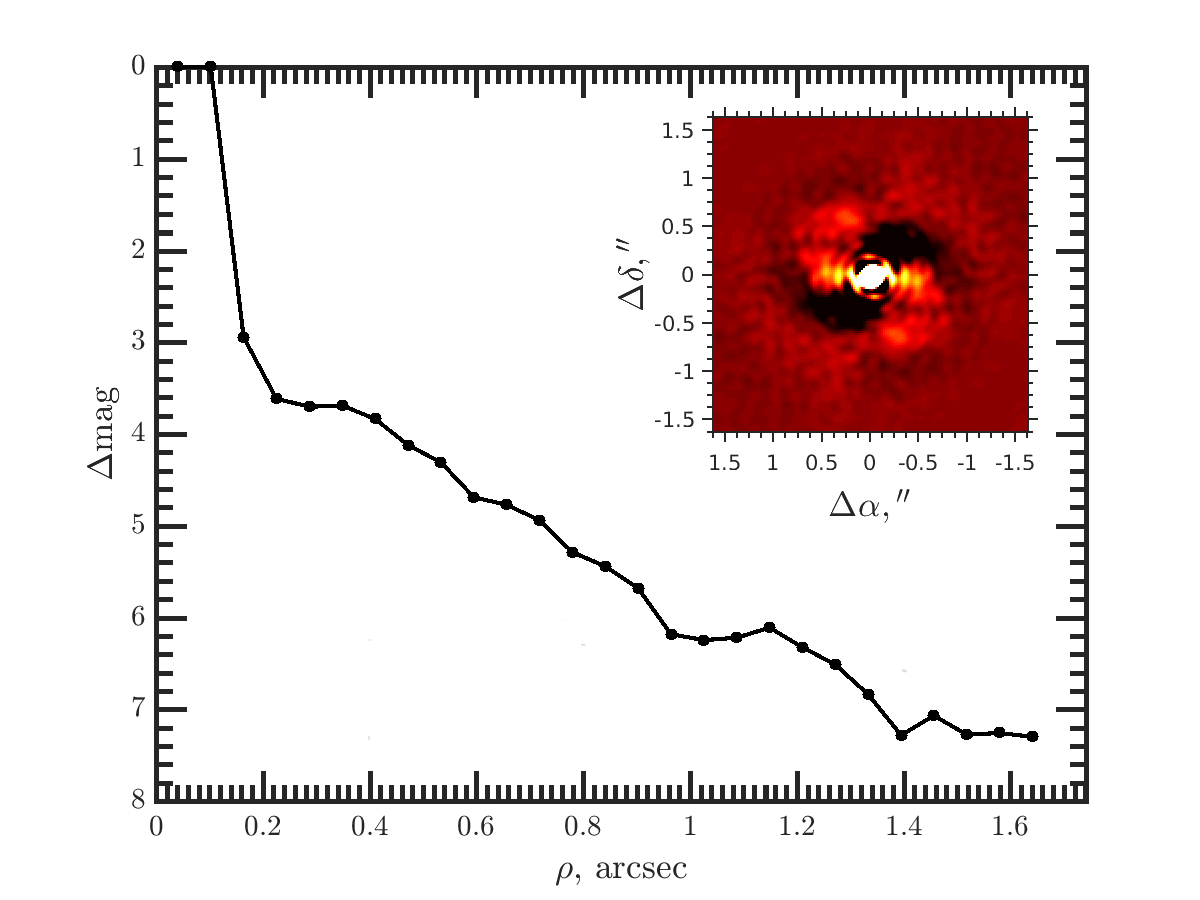}
\figsetgrpnote{Sensitivity limits and reconstructed image (inset) from SAI speckle observations of TOI-5340.}
\figsetgrpend

\figsetgrpstart
\figsetgrpnum{5.50}
\figsetgrptitle{WIYN/NESSI observations of TOI-5340}
\figsetplot{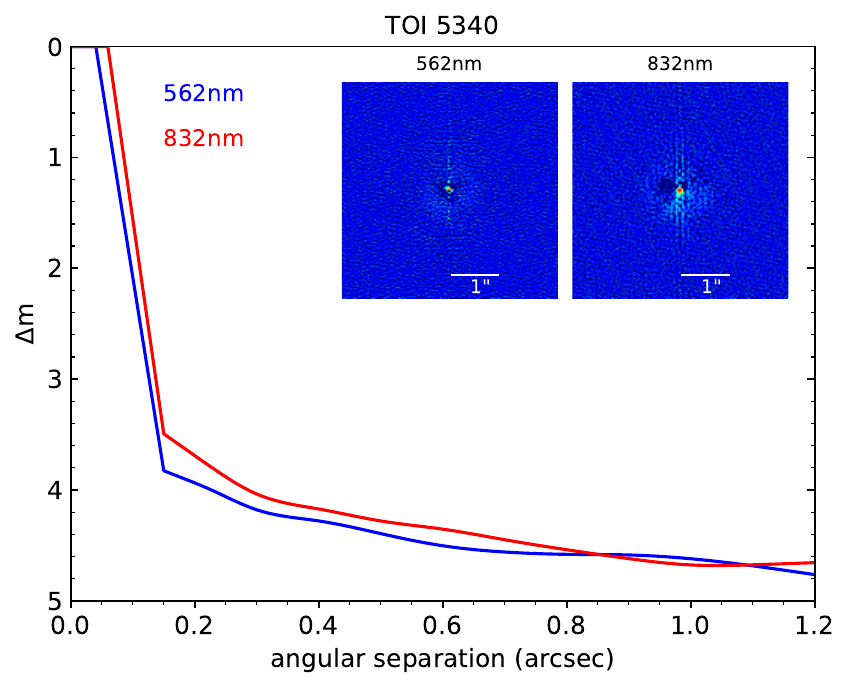}
\figsetgrpnote{Sensitivity limits and reconstructed image (inset) from WIYN/NESSI observations of TOI-5340.}
\figsetgrpend

\figsetgrpstart
\figsetgrpnum{5.51}
\figsetgrptitle{SAI/Speckle Polarimeter observation of TOI-5592}
\figsetplot{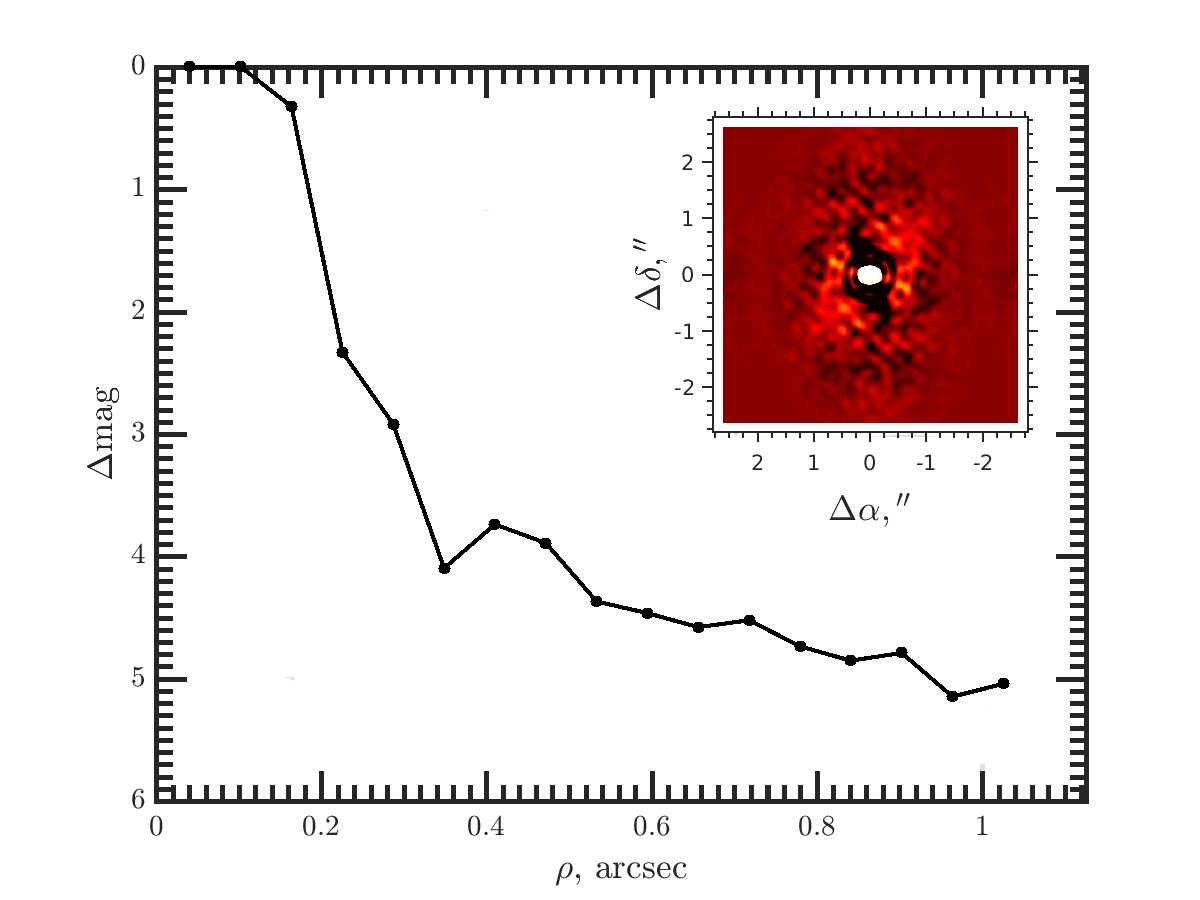}
\figsetgrpnote{Sensitivity limits and reconstructed image (inset) from SAI speckle observations of TOI-5592.}
\figsetgrpend

\figsetend

\begin{figure}
\centering
\includegraphics[width=0.5\linewidth]{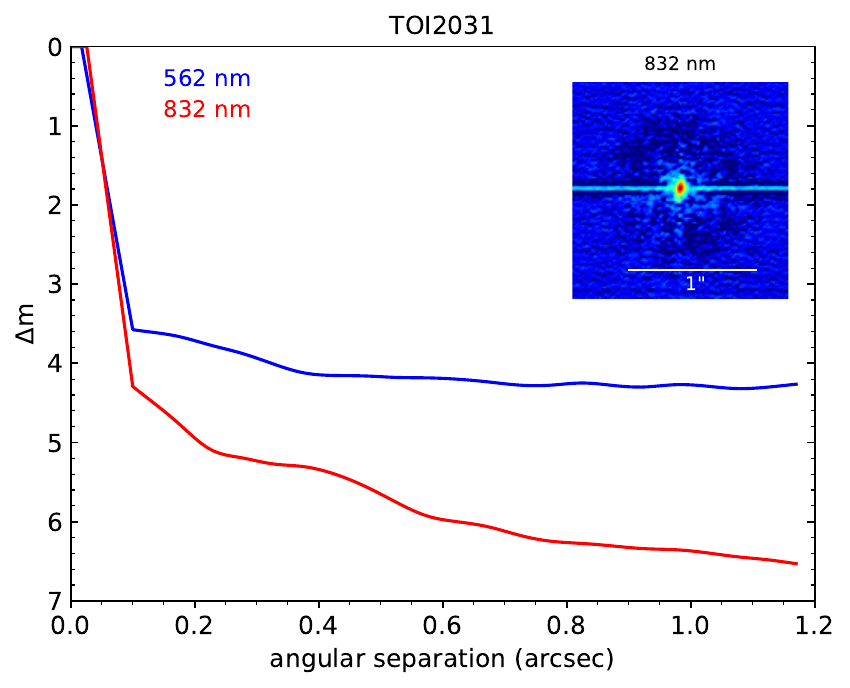}
\caption{High-Resolution imaging of hot Jupiter hosts described in this paper, with no detected companions.
This image shows Gemini-N/'Alopeke observations of TOI-2031.
The complete figure set for all TOIs (51 images) is available in the available in the online journal.
}
\label{fig:high_res_imaging_nocomp}
\end{figure}
\else

\begin{subfigures}
\label{fig:high_res_imaging_nocomp}

\makeatletter\onecolumngrid@push\makeatother
\begin{figure*}
\centering

\includegraphics[width=0.32\linewidth]{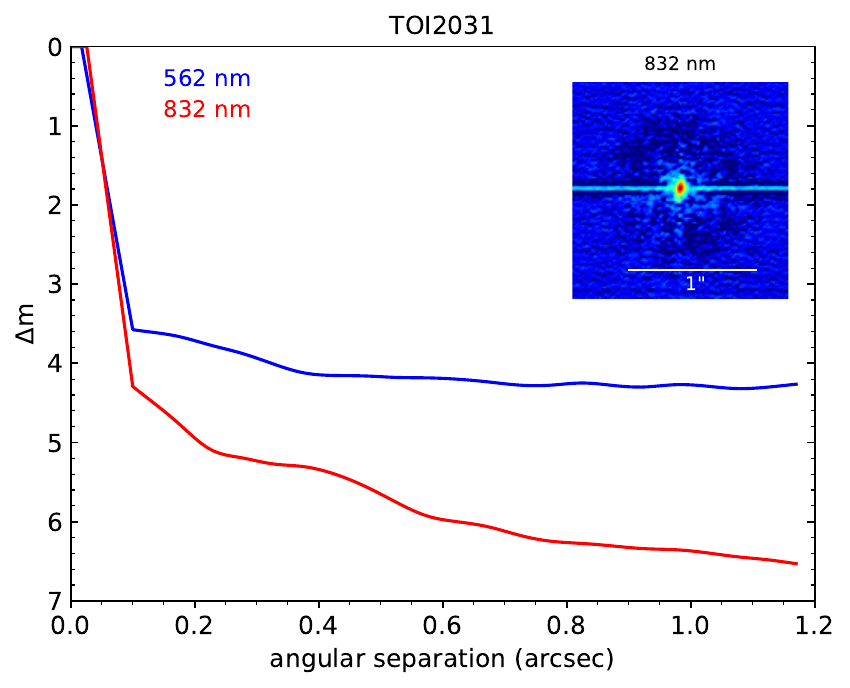}
\includegraphics[width=0.32\linewidth]{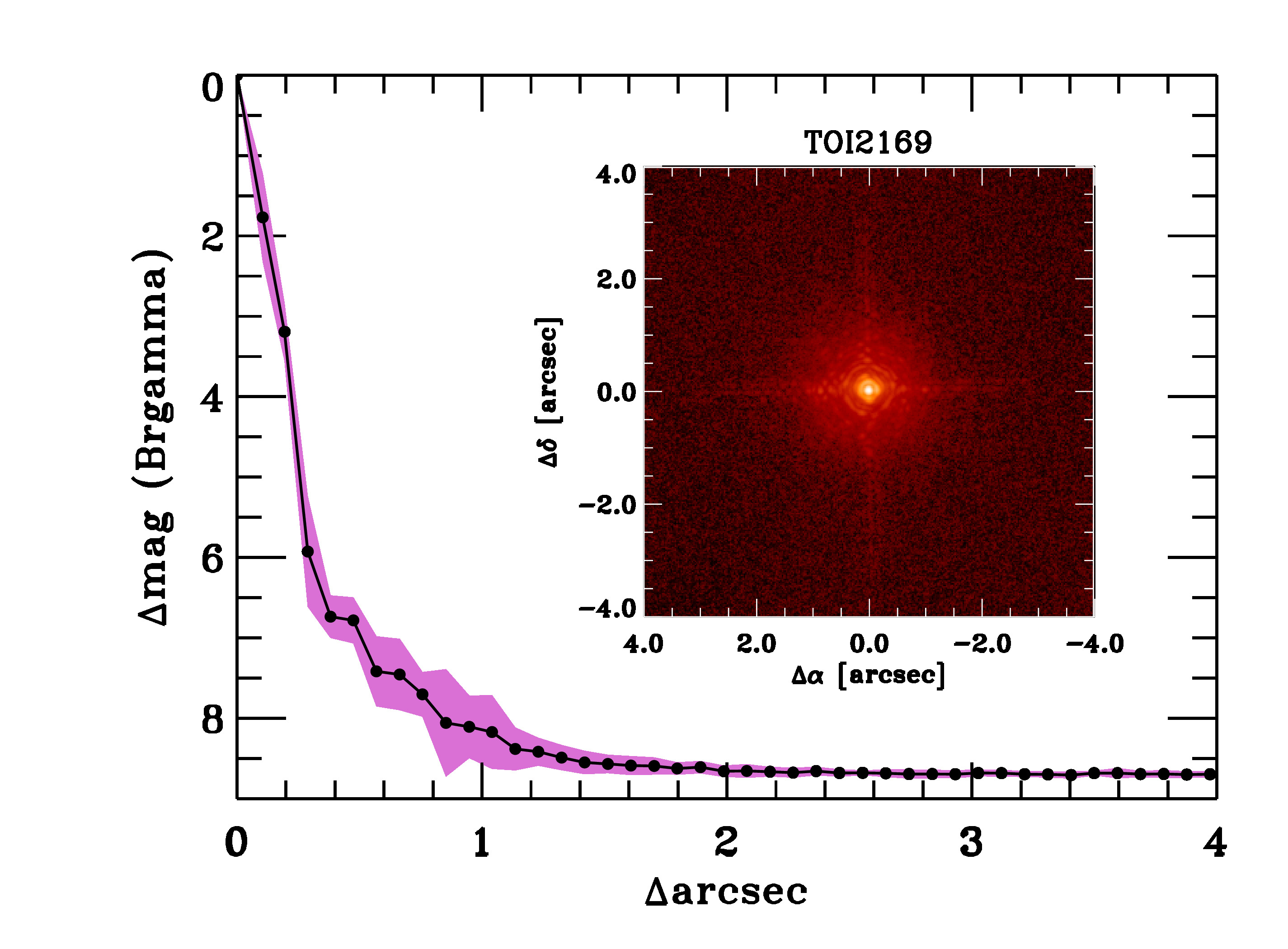}
\includegraphics[width=0.32\linewidth]{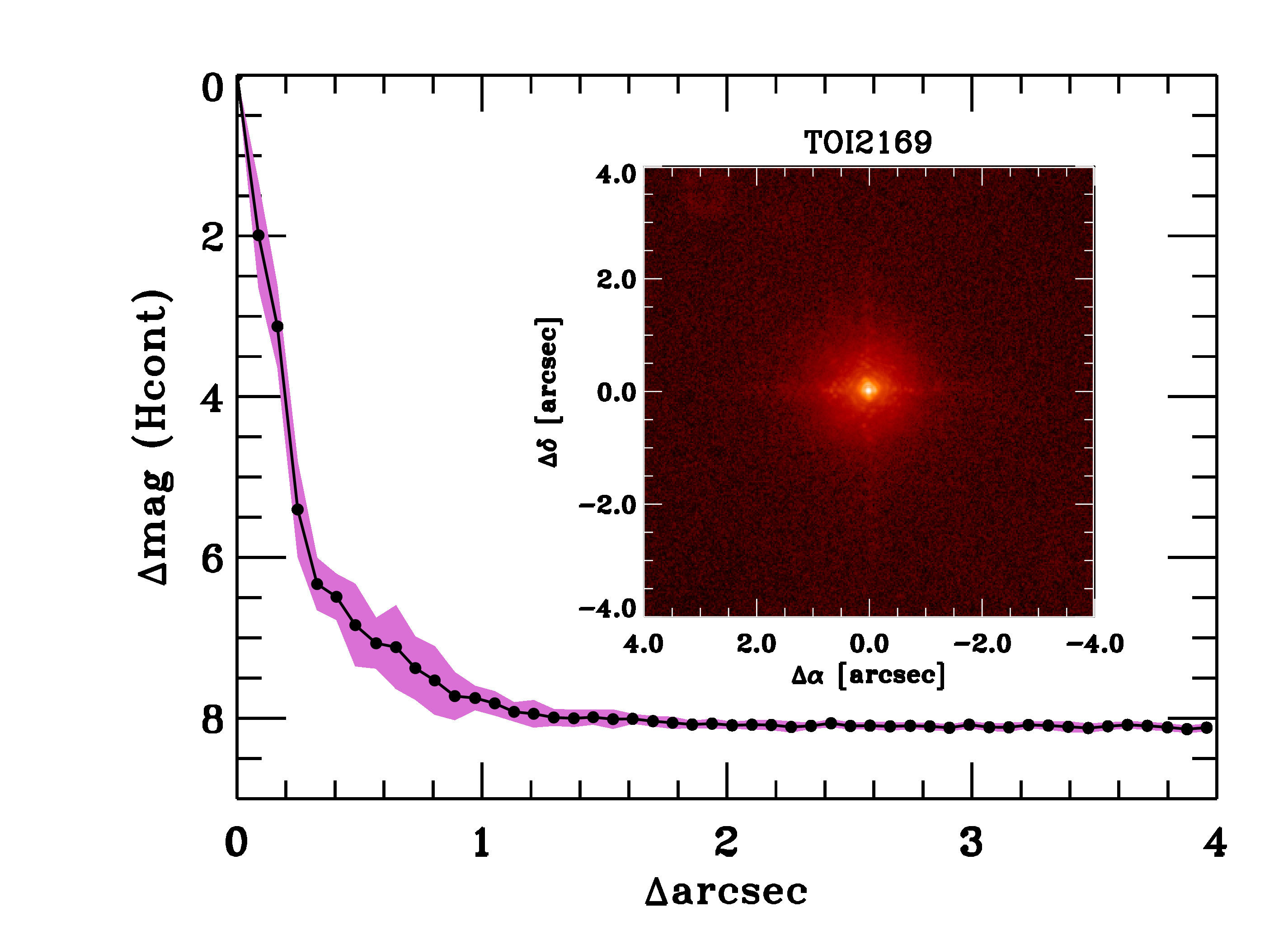}\\
\includegraphics[width=0.32\linewidth]{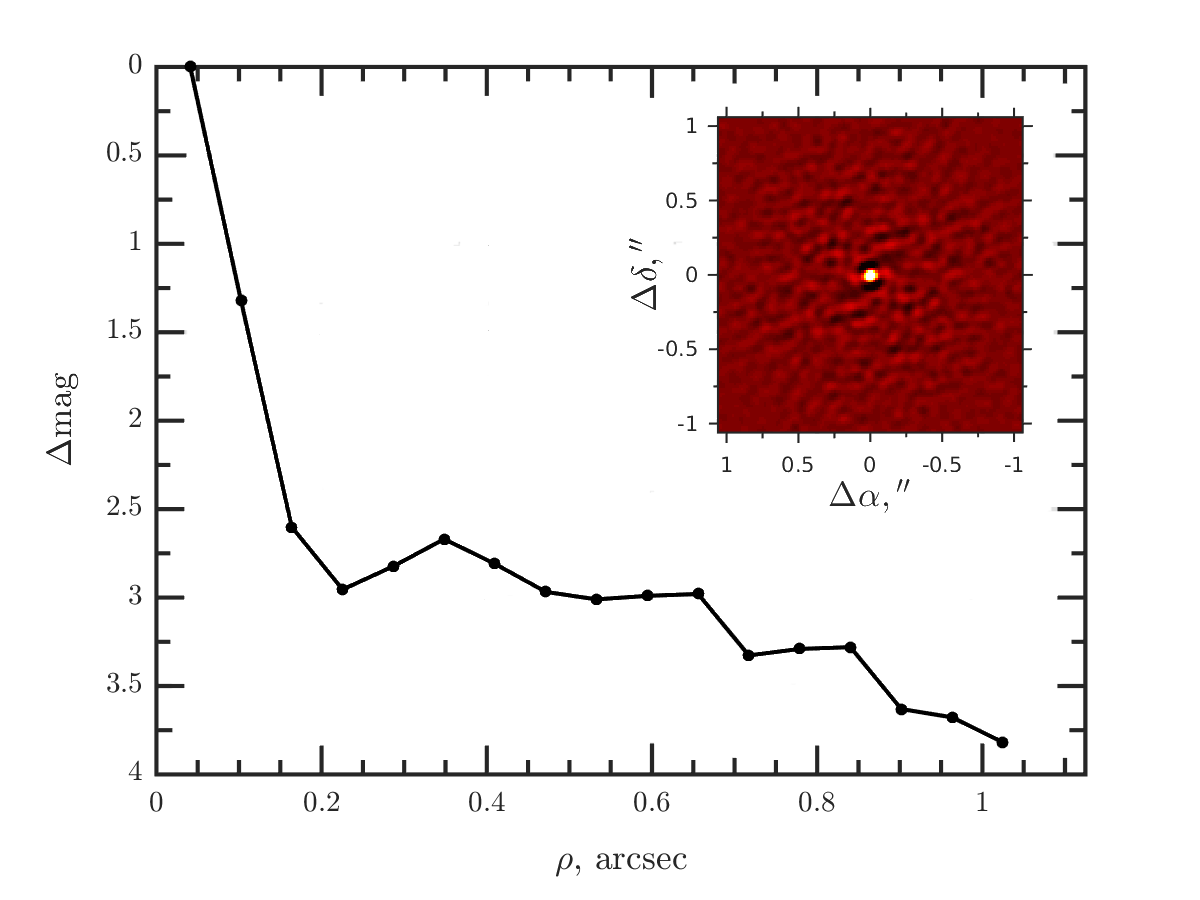}
\includegraphics[width=0.32\linewidth]{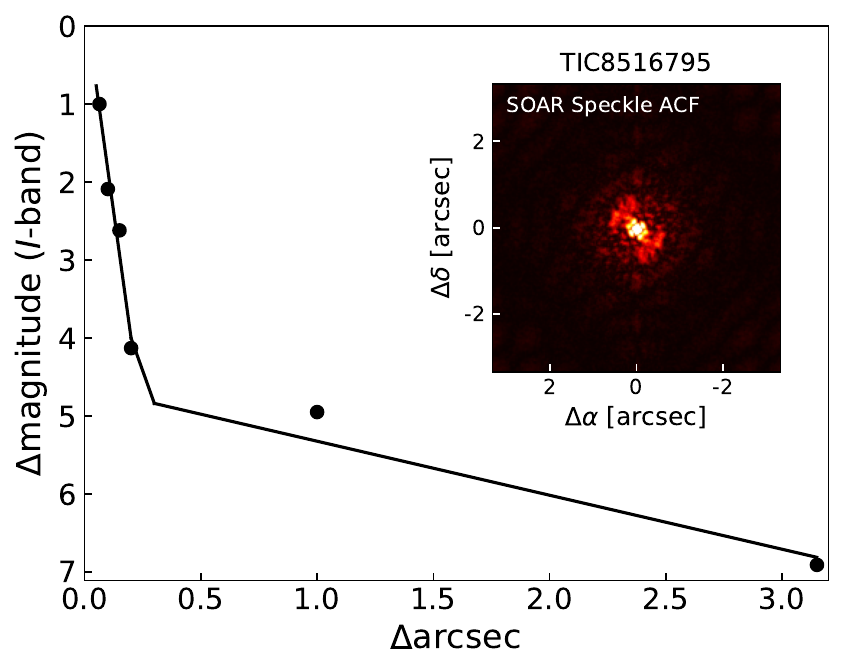}
\includegraphics[width=0.32\linewidth]{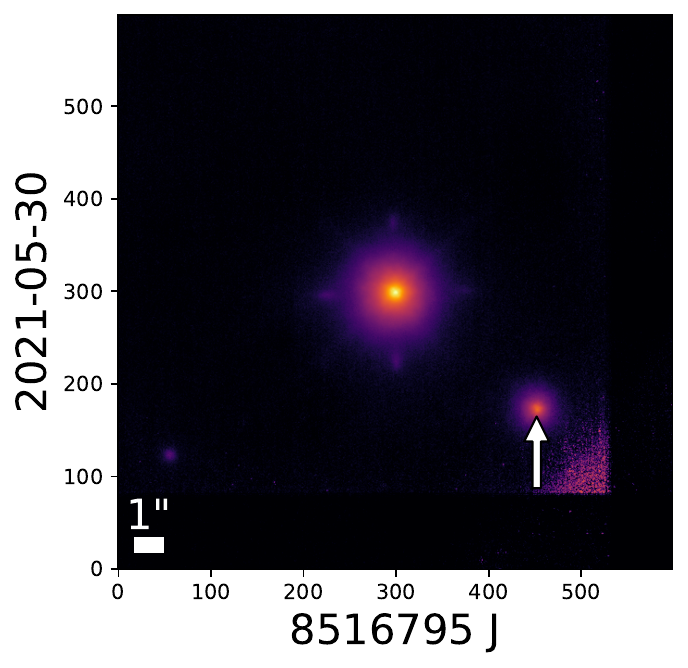}\\
\includegraphics[width=0.32\linewidth]{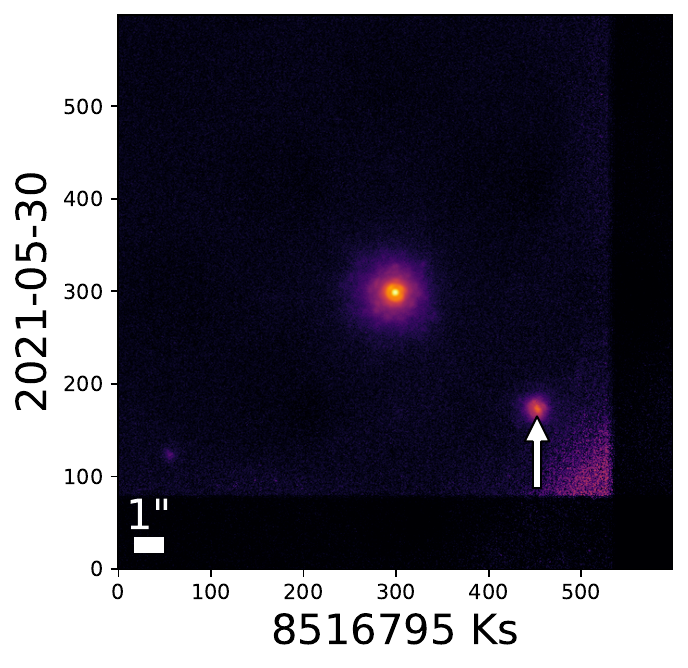}
\includegraphics[width=0.32\linewidth]{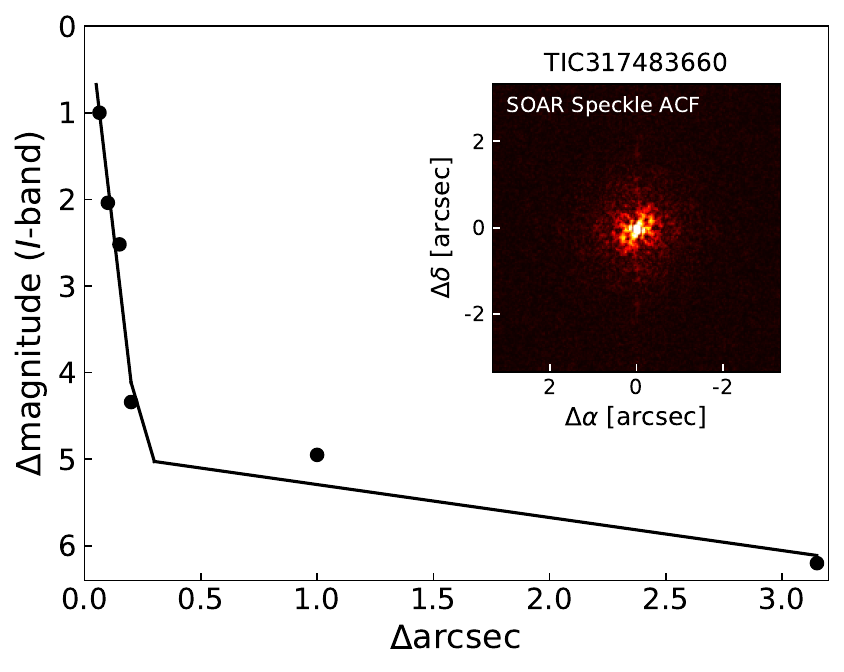}
\includegraphics[width=0.32\linewidth]{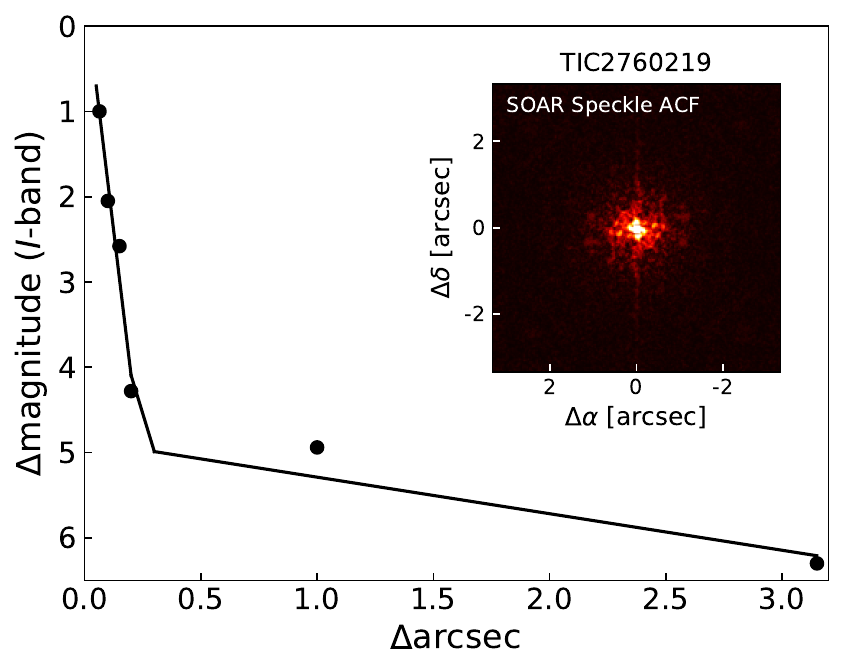}\\
\includegraphics[width=0.32\linewidth]{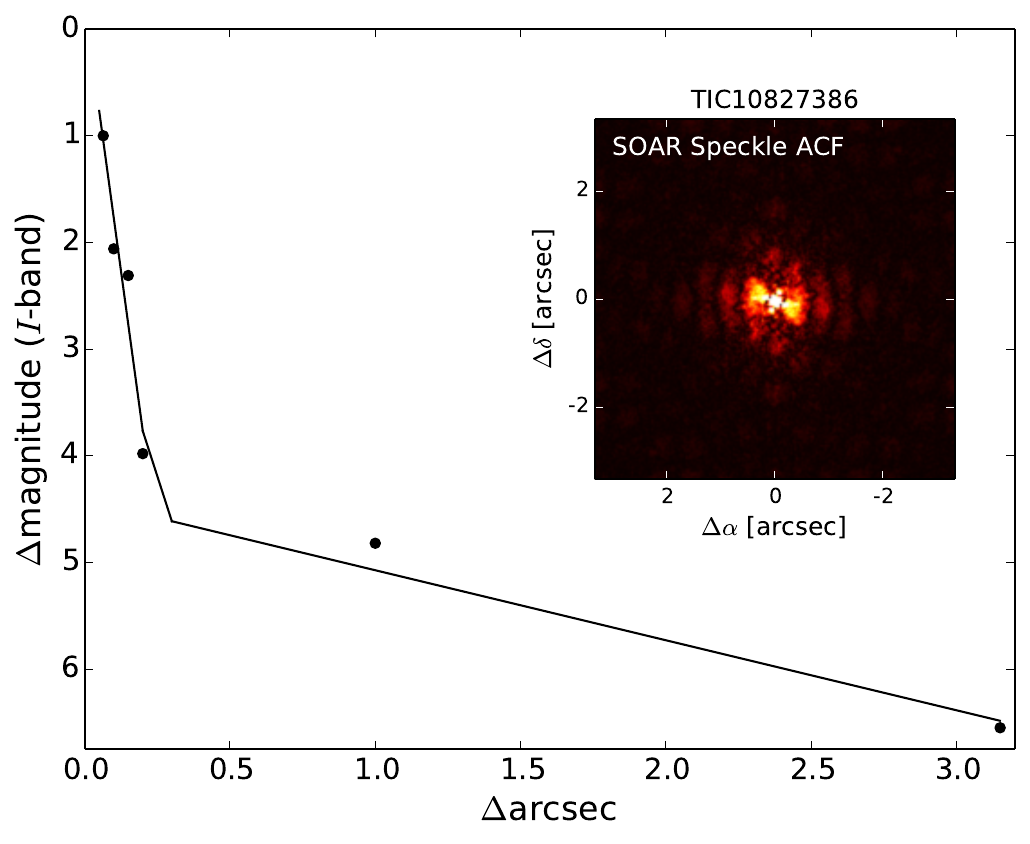}
\includegraphics[width=0.32\linewidth]{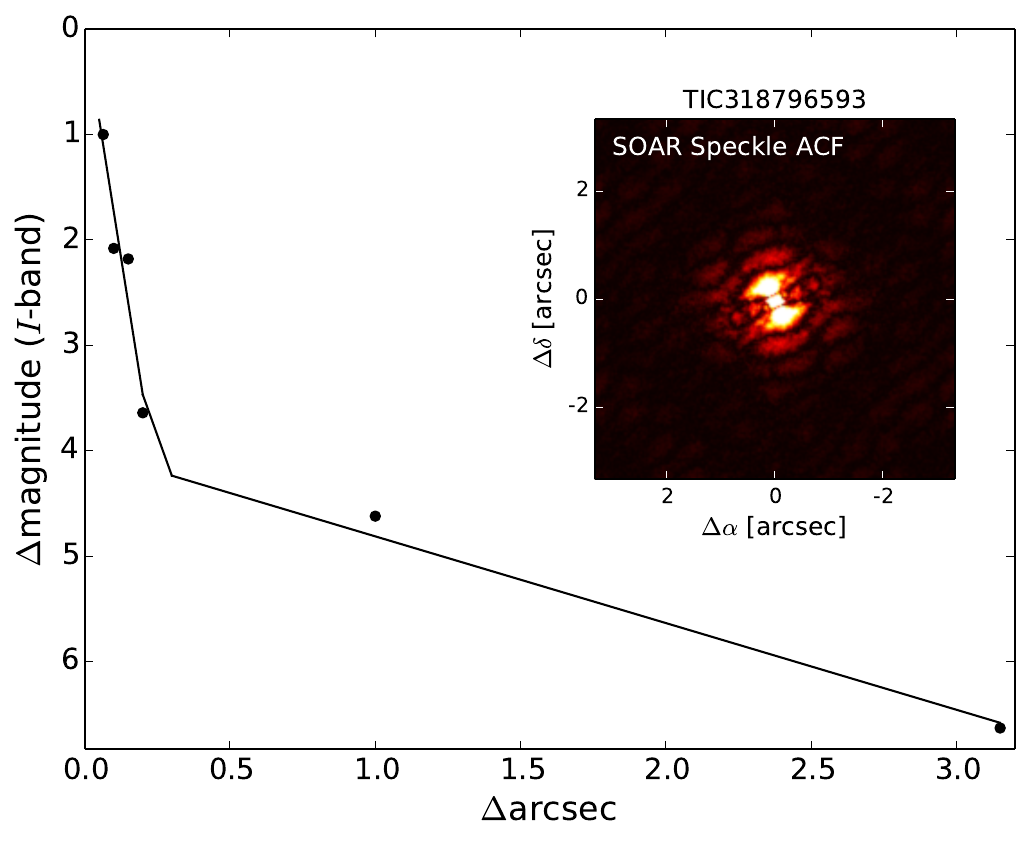}
\includegraphics[width=0.32\linewidth]{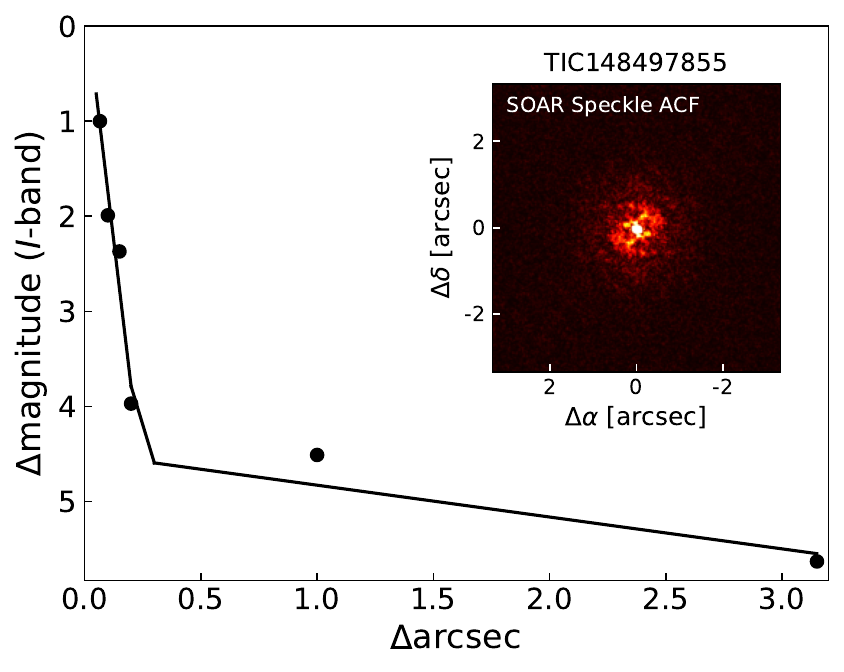}\\
\caption{High-Resolution imaging of hot Jupiter hosts described in this paper.
From top to bottom, left to right:
\textbf{Row 1:} Gemini-N/'Alopeke observations of TOI-2031; 
Palomar/PHARO Br$\gamma$ observation of TOI-2169; 
Palomar/PHARO H\textit{{cont}} observation of TOI-2169; 
\textbf{Row 2:} SAI/Speckle Polarimeter observation of TOI-2169; 
SOAR/HRCam observation of TOI-2169; 
Shane/ShARCS J observation of TOI-2169; 
\textbf{Row 3:} Shane/ShARCS K$_s$ observation of TOI-2169; 
SOAR/HRCam observation of TOI-2346; 
SOAR/HRCam observation of TOI-2382; 
\textbf{Row 4:} SOAR/HRCam observation of TOI-2876; 
SOAR/HRCam observation of TOI-2886; 
SOAR/HRCam observation of TOI-2986.
}
\label{fig:high_res_imaging_0}
\end{figure*}
\makeatletter\onecolumngrid@pop\makeatother
\clearpage

\makeatletter\onecolumngrid@push\makeatother
\begin{figure*}
\centering

\includegraphics[width=0.32\linewidth]{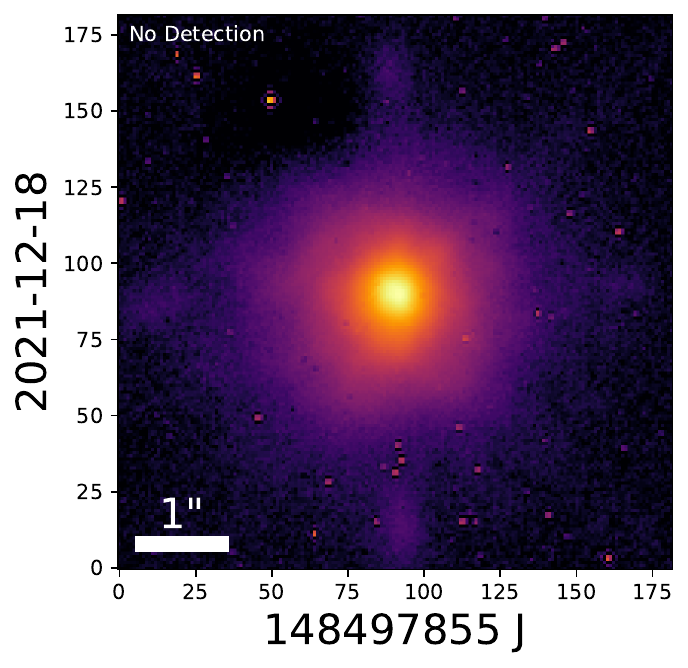}
\includegraphics[width=0.32\linewidth]{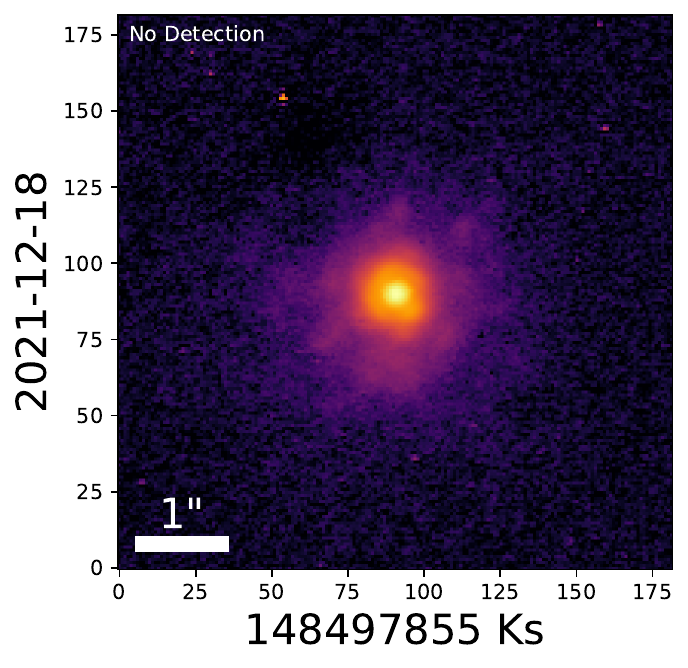}
\includegraphics[width=0.32\linewidth]{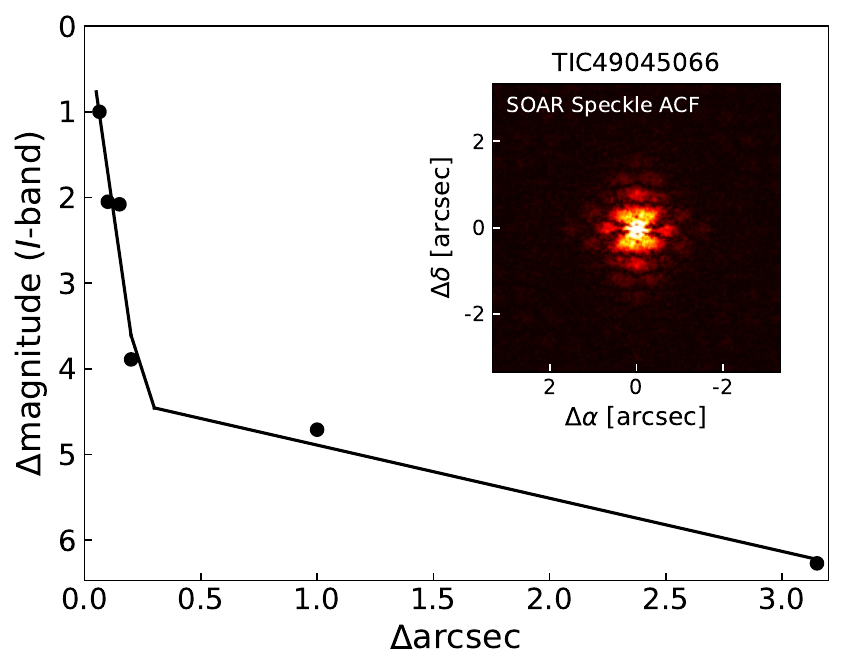}\\
\includegraphics[width=0.32\linewidth]{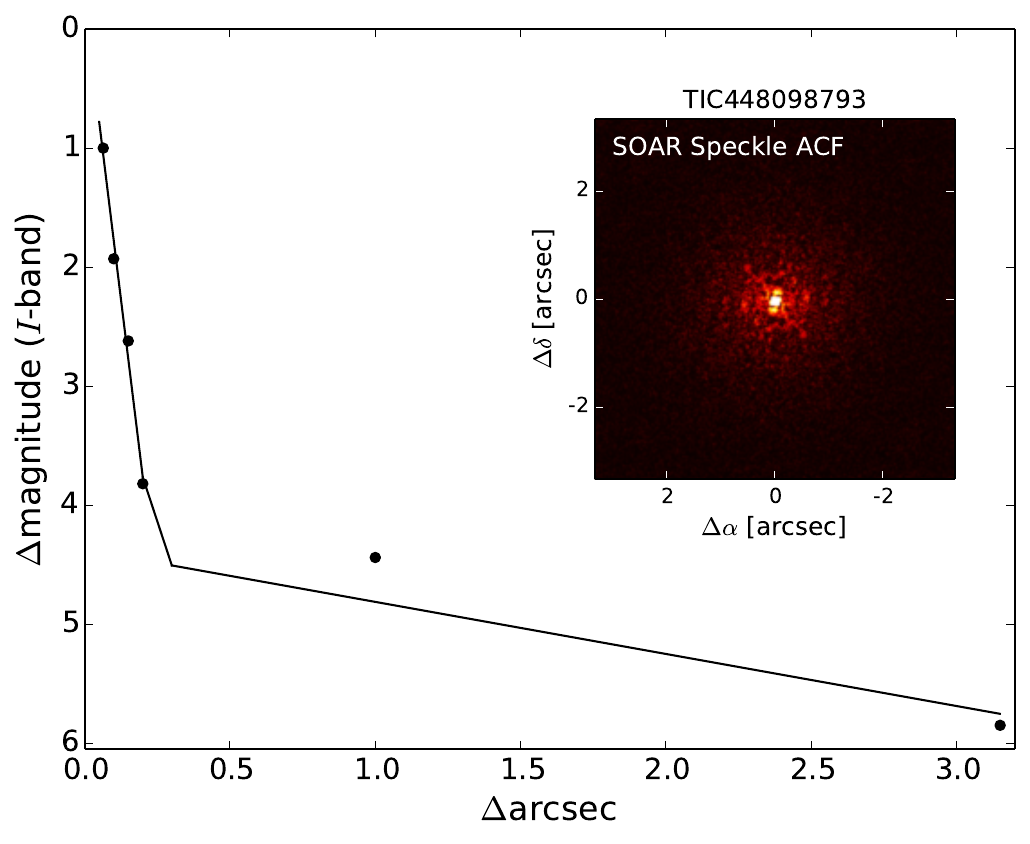}
\includegraphics[width=0.32\linewidth]{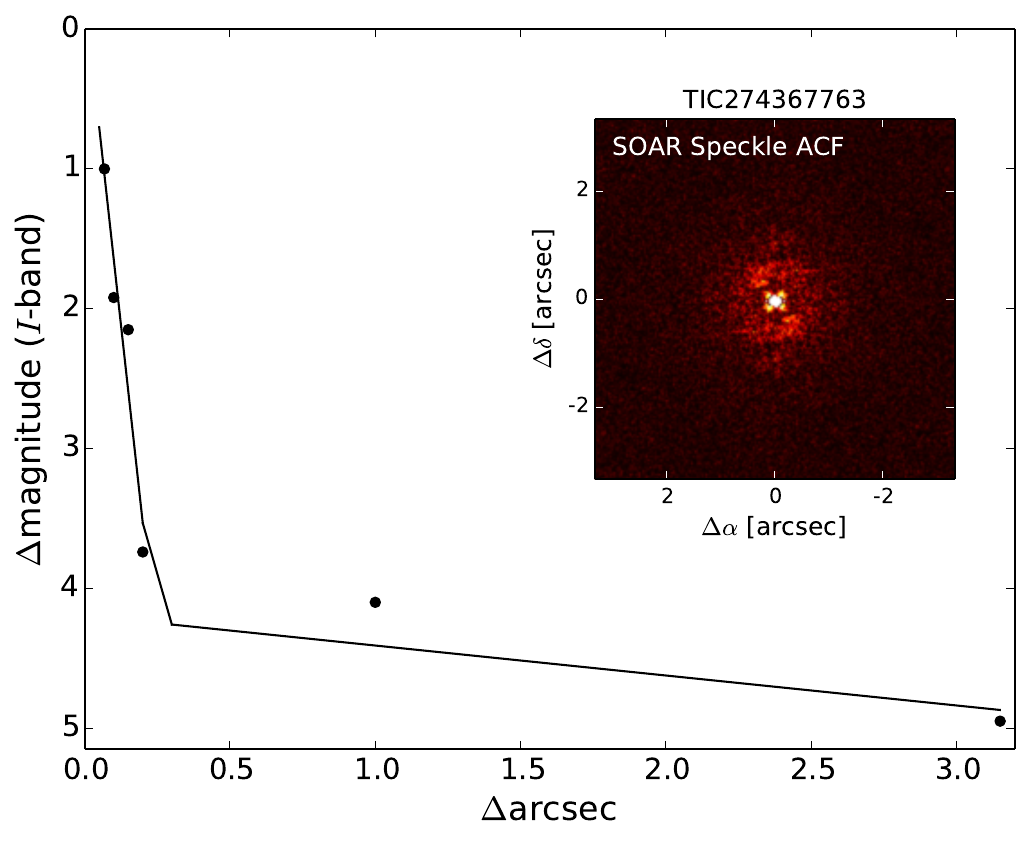}
\includegraphics[width=0.32\linewidth]{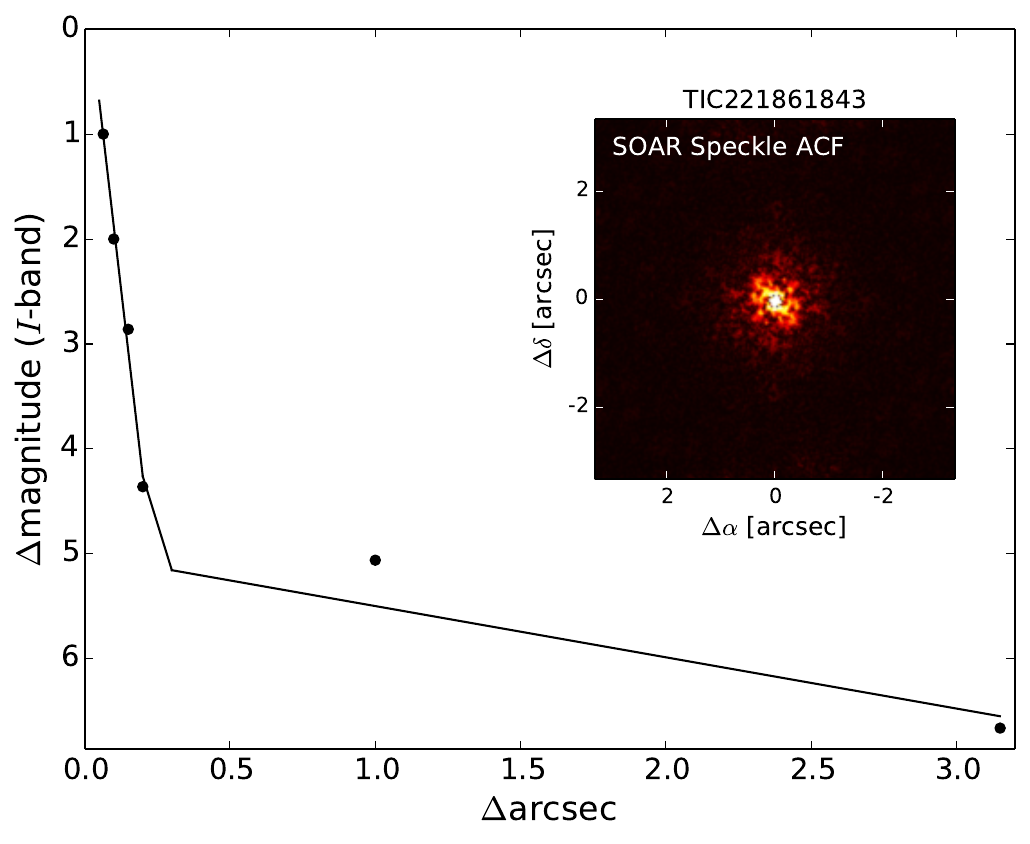}\\
\includegraphics[width=0.32\linewidth]{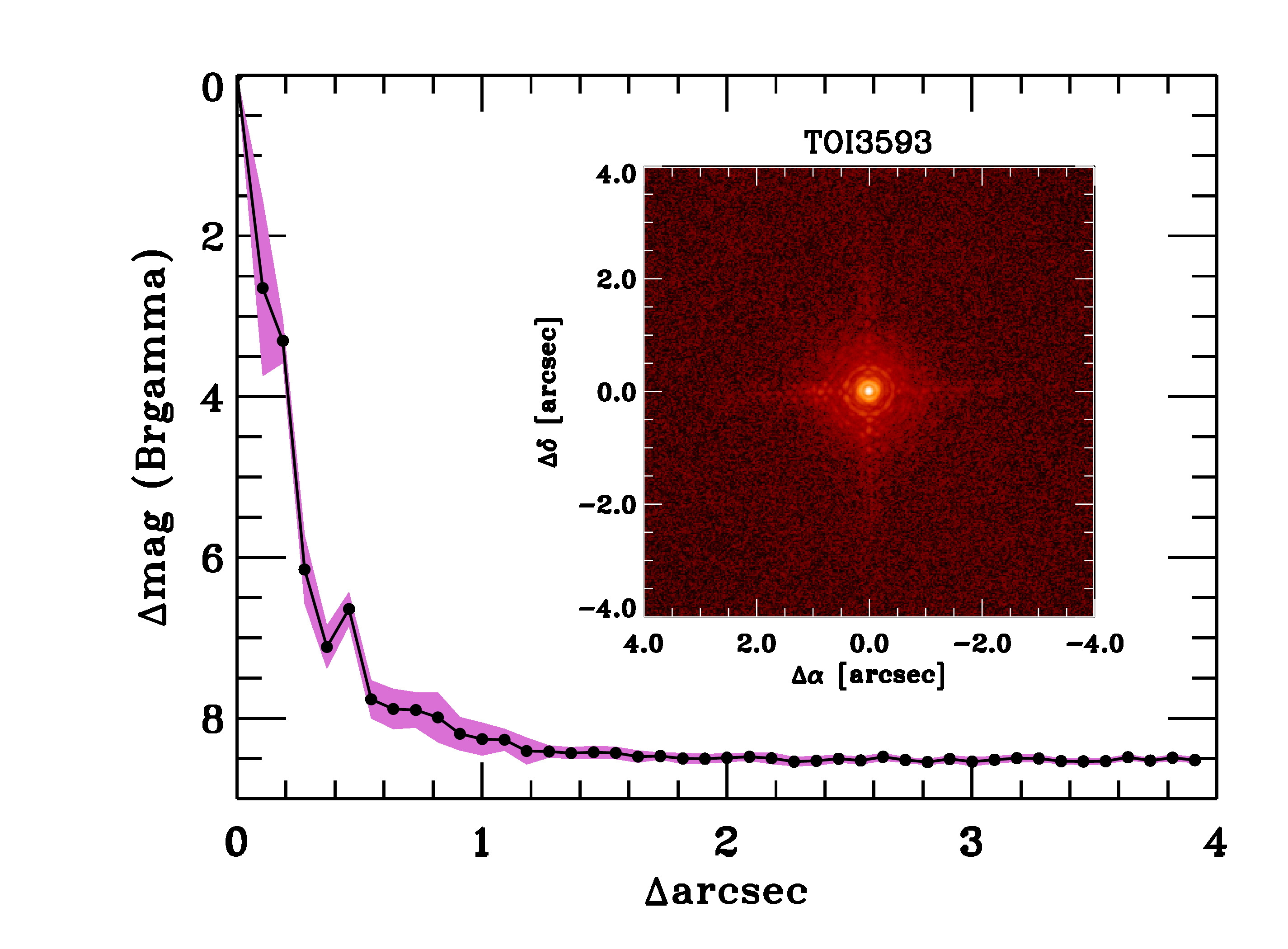}
\includegraphics[width=0.32\linewidth]{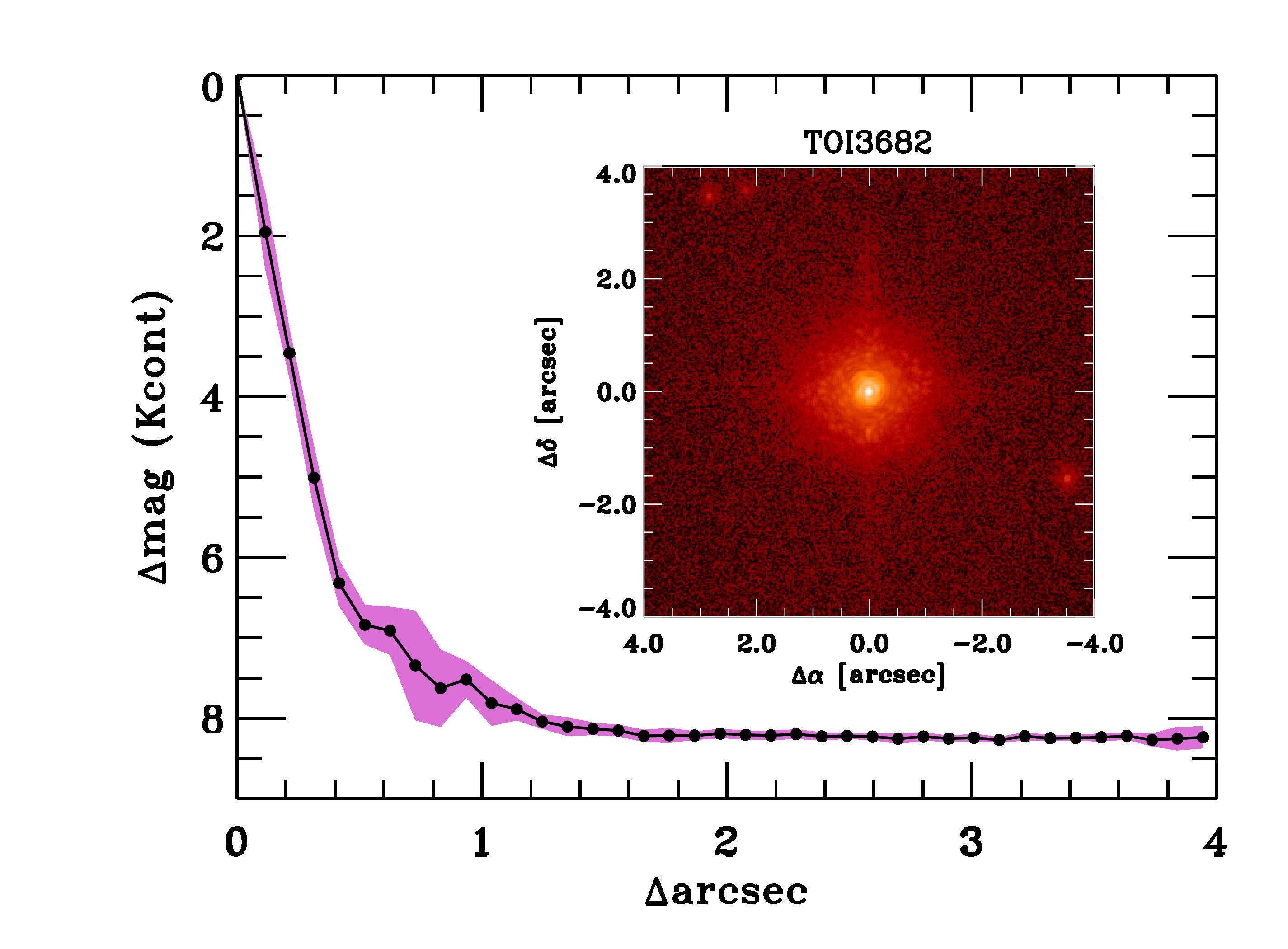}
\includegraphics[width=0.32\linewidth]{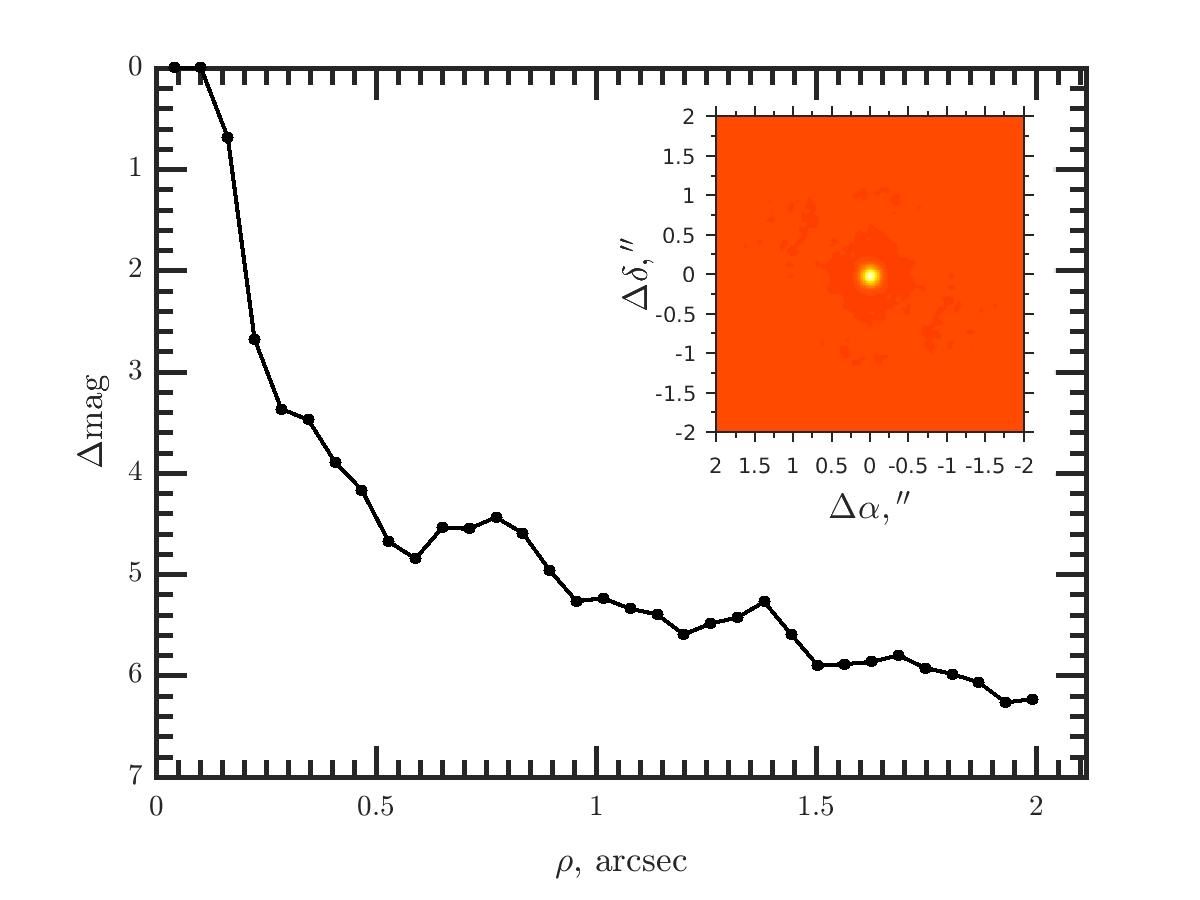}\\
\includegraphics[width=0.32\linewidth]{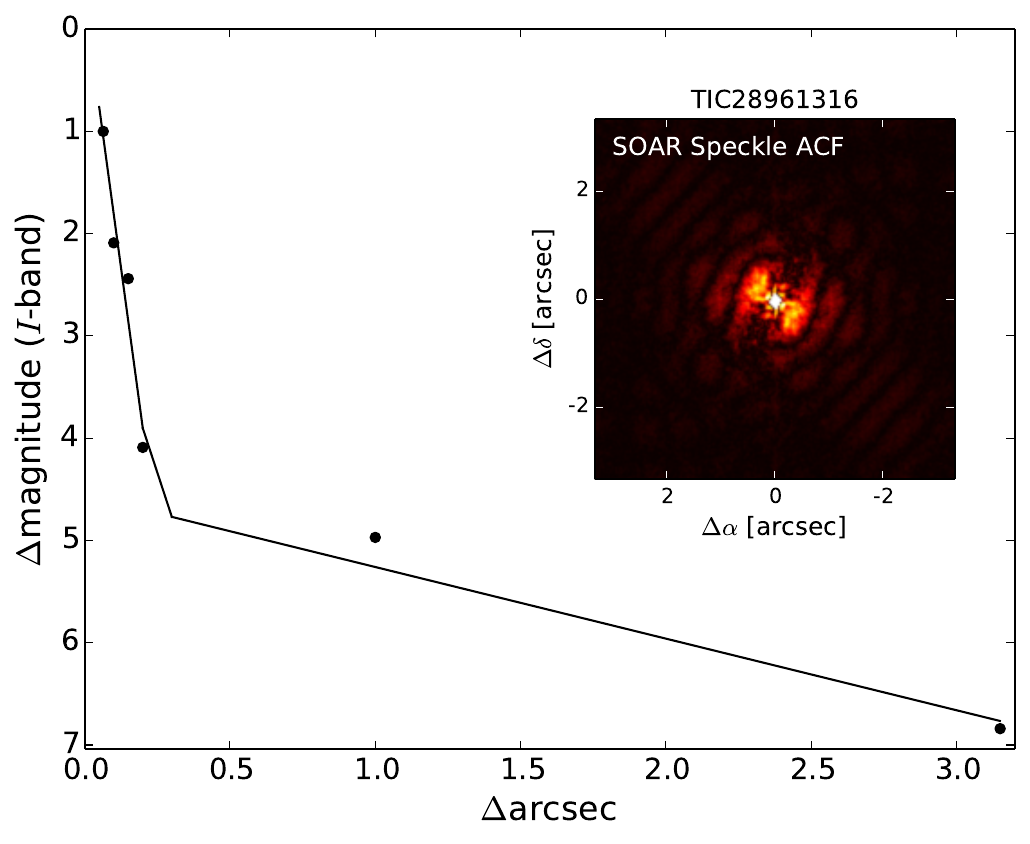}
\includegraphics[width=0.32\linewidth]{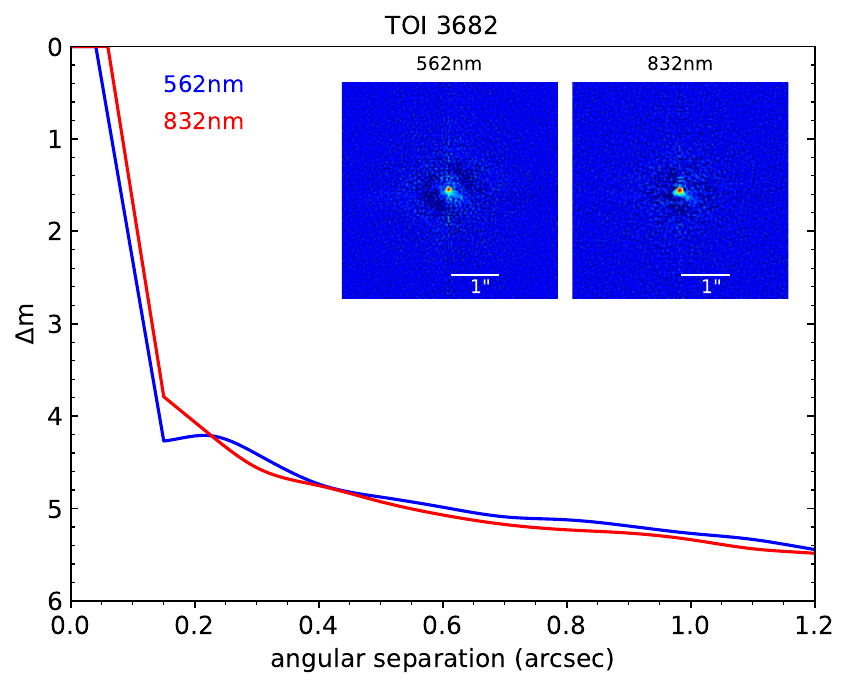}
\includegraphics[width=0.32\linewidth]{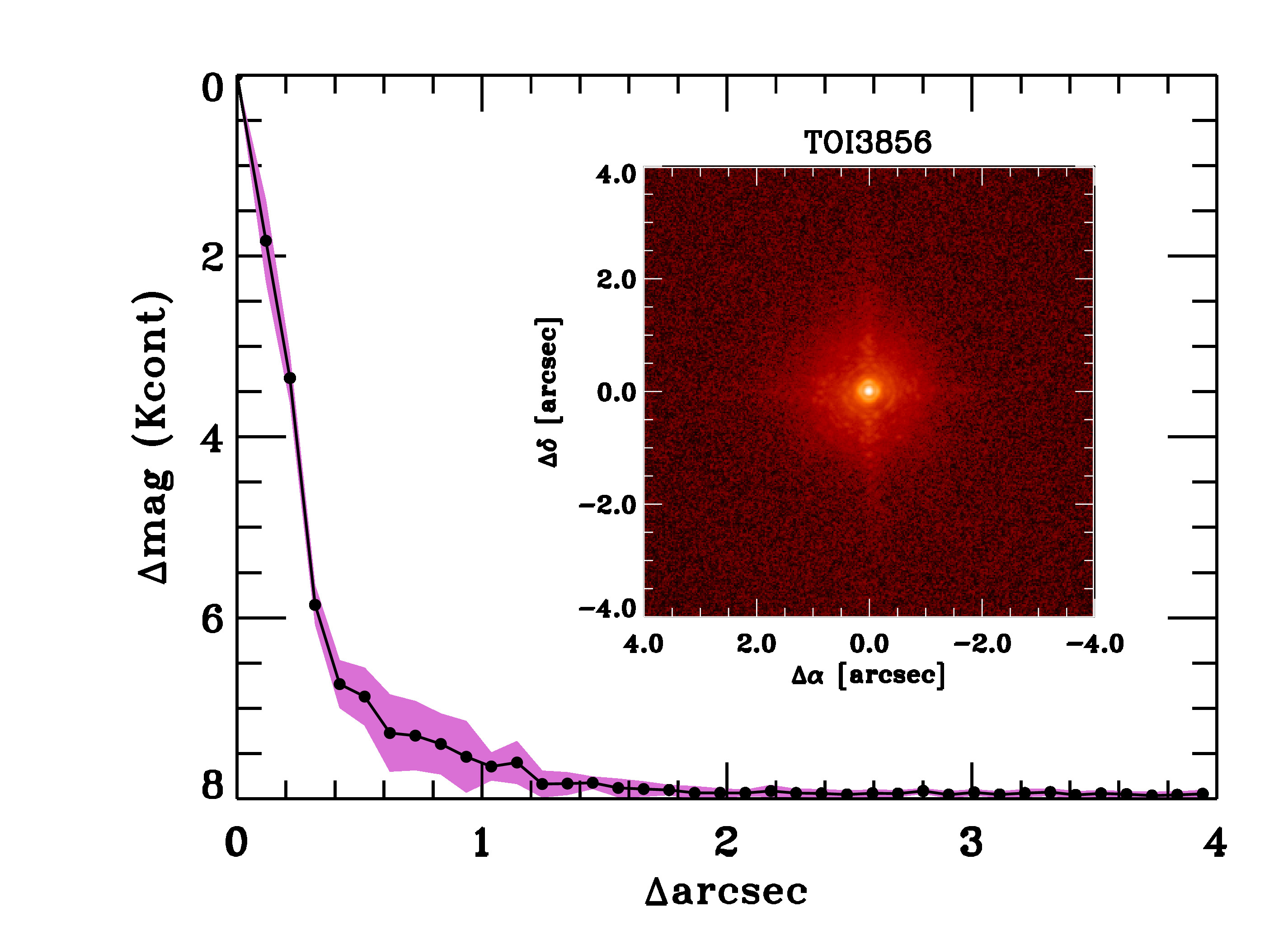}\\
\caption{High-Resolution imaging of hot Jupiter hosts described in this paper (continued).
From top to bottom, left to right:
\textbf{Row 1:} Shane/ShARCS J observation of TOI-2986; 
Shane/ShARCS K$_s$ observation of TOI-2986; 
SOAR/HRCam observation of TOI-2992; 
\textbf{Row 2:} SOAR/HRCam observation of TOI-3135; 
SOAR/HRCam observation of TOI-3474; 
SOAR/HRCam observation of TOI-3486; 
\textbf{Row 3:} Palomar/PHARO observation of TOI-3593; 
Palomar/PHARO observation of TOI-3682; 
SAI/Speckle Polarimeter observation of TOI-3682; 
\textbf{Row 4:} SOAR/HRCam observation of TOI-3682; 
WIYN/NESSI observations of TOI-3682; 
Palomar/PHARO observation of TOI-3856.
}
\label{fig:high_res_imaging_1}
\end{figure*}
\makeatletter\onecolumngrid@pop\makeatother
\clearpage

\makeatletter\onecolumngrid@push\makeatother
\begin{figure*}
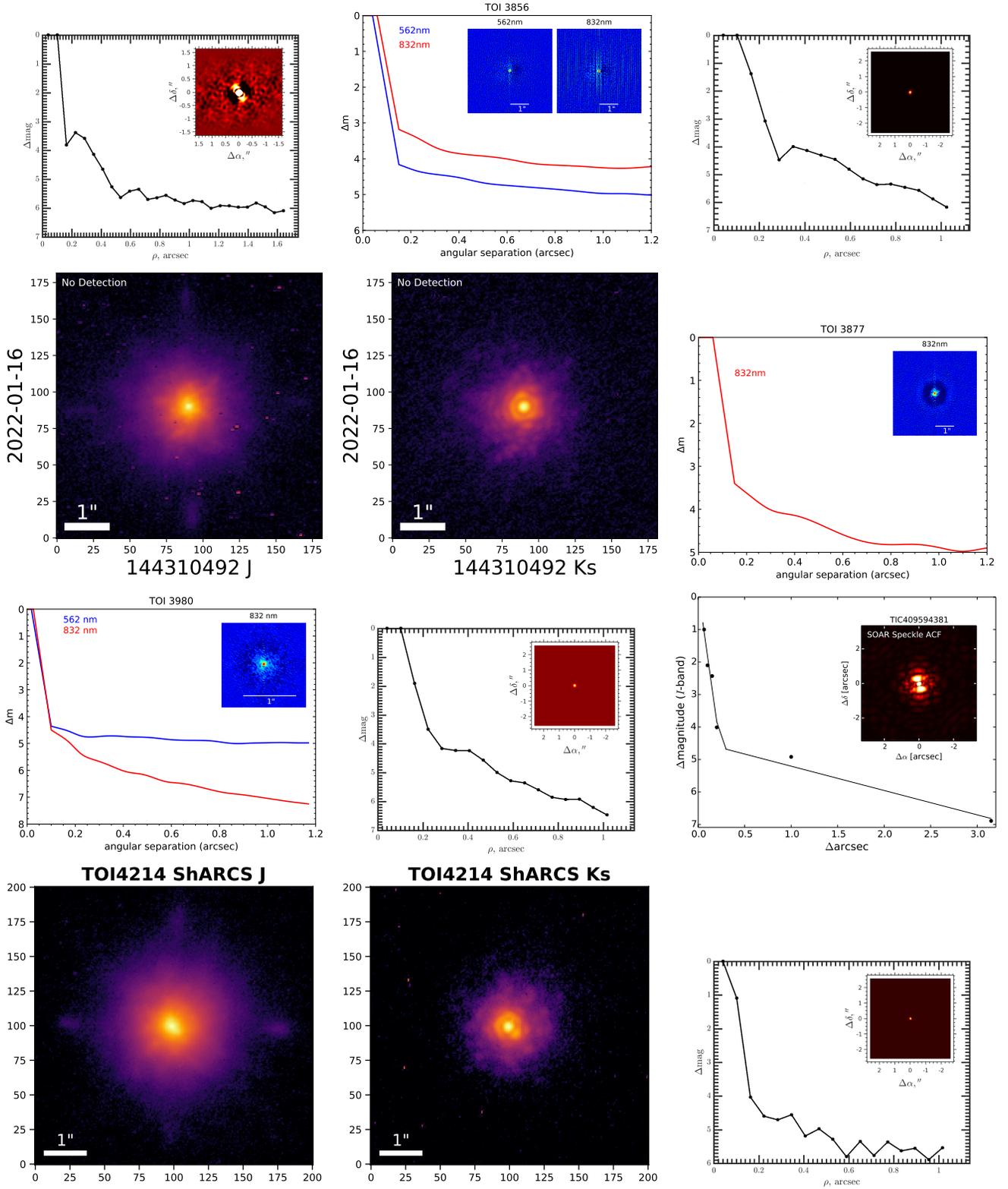

\centering

\includegraphics[width=0.32\linewidth]{toi3856_imaging_SAI_I_20221202.png}
\includegraphics[width=0.32\linewidth]{toi3856_imaging_WIYN_562nm_832nm_20230204.pdf}
\includegraphics[width=0.32\linewidth]{toi3877_imaging_SAI_I_20220321.png}\\
\includegraphics[width=0.32\linewidth]{toi3877_imaging_Shane_J_20220116.pdf}
\includegraphics[width=0.32\linewidth]{toi3877_imaging_Shane_K_20220116.pdf}
\includegraphics[width=0.32\linewidth]{toi3877_imaging_WIYN_832nm_20220421.pdf}\\
\includegraphics[width=0.32\linewidth]{toi3980_imaging_Gemini_562nm_832nm_20220914.pdf}
\includegraphics[width=0.32\linewidth]{toi3980_imaging_SAI_I_20211030.png}
\includegraphics[width=0.32\linewidth]{toi4214_imaging_SOAR_I_20211120.pdf}\\
\includegraphics[width=0.32\linewidth]{toi4214_imaging_Shane_J_20211218.pdf}
\includegraphics[width=0.32\linewidth]{toi4214_imaging_Shane_K_20211218.pdf}
\includegraphics[width=0.32\linewidth]{toi4487_imaging_SAI_I_20211030.png}\\
\caption{High-Resolution imaging of hot Jupiter hosts described in this paper (continued).
From top to bottom, left to right:
\textbf{Row 1:} SAI/Speckle Polarimeter observation of TOI-3856; 
WIYN/NESSI observations of TOI-3856; 
SAI/Speckle Polarimeter observation of TOI-3877; 
\textbf{Row 2:} Shane/ShARCS J observation of TOI-3877; 
Shane/ShARCS K$_s$ observation of TOI-3877; 
WIYN/NESSI observation of TOI-3877; 
\textbf{Row 3:} Gemini-N/'Alopeke observations of TOI-3980; 
SAI/Speckle Polarimeter observation of TOI-3980; 
SOAR/HRCam observation of TOI-4214; 
\textbf{Row 4:} Shane/ShARCS J observation of TOI-4214; 
Shane/ShARCS K$_s$ observation of TOI-4214; 
SAI/Speckle Polarimeter observation of TOI-4487.
}
\label{fig:high_res_imaging_2}
\end{figure*}
\makeatletter\onecolumngrid@pop\makeatother
\clearpage

\makeatletter\onecolumngrid@push\makeatother
\begin{figure*}
\centering

\includegraphics[width=0.32\linewidth]{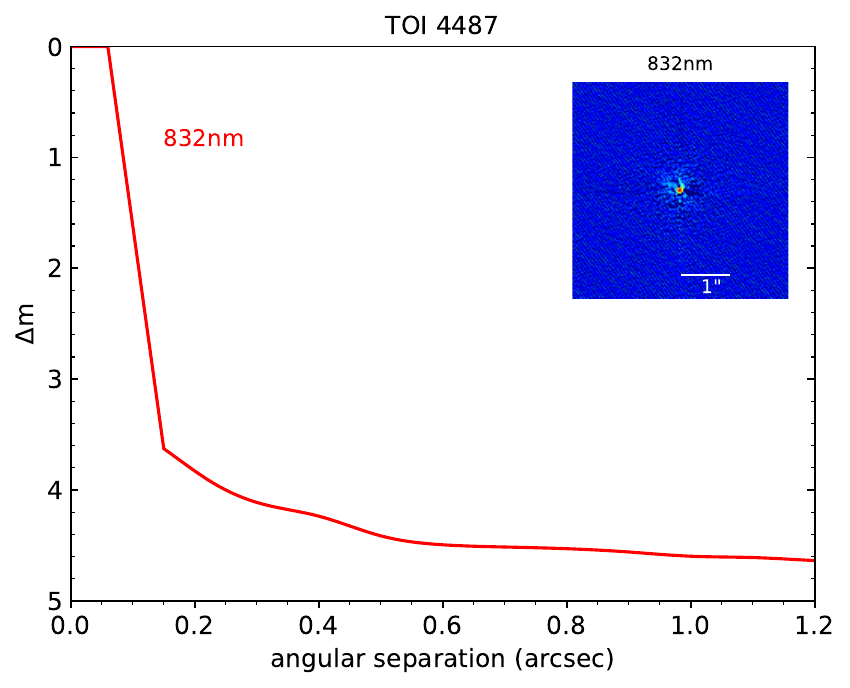}
\includegraphics[width=0.32\linewidth]{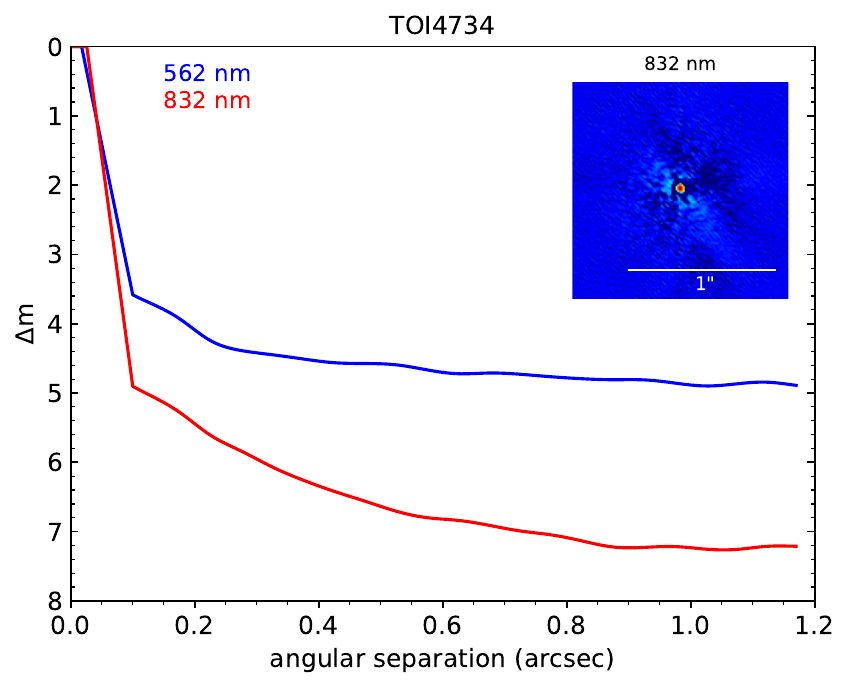}
\includegraphics[width=0.32\linewidth]{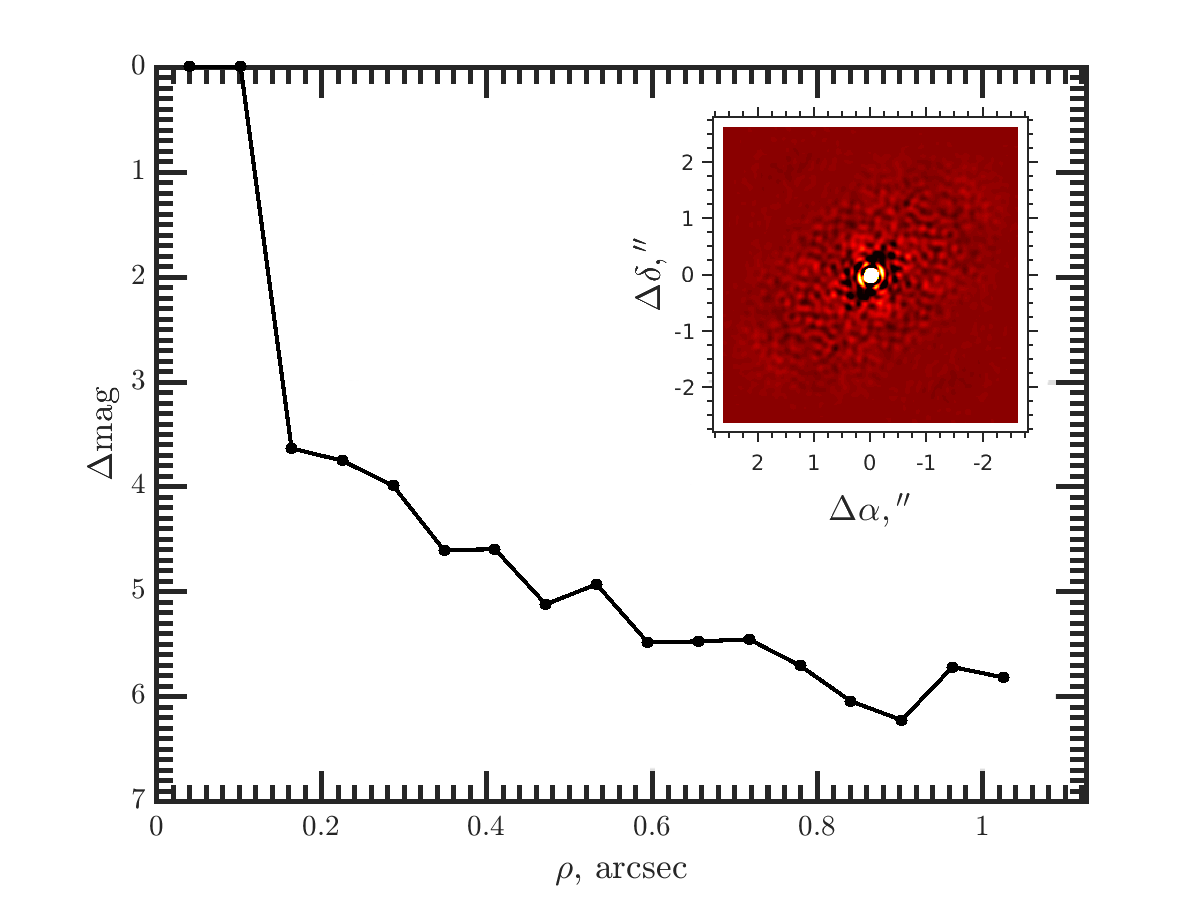}\\
\includegraphics[width=0.32\linewidth]{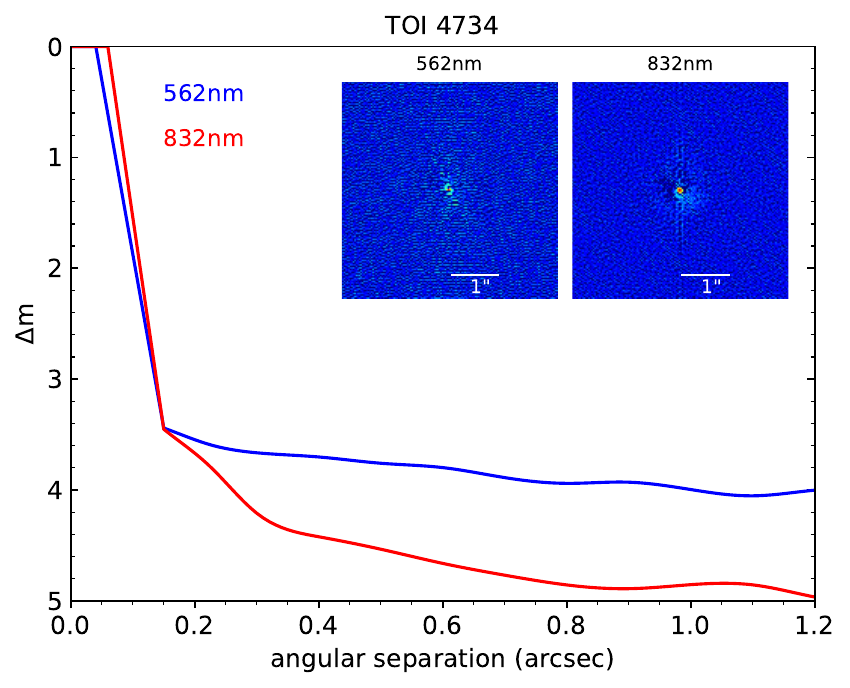}
\includegraphics[width=0.32\linewidth]{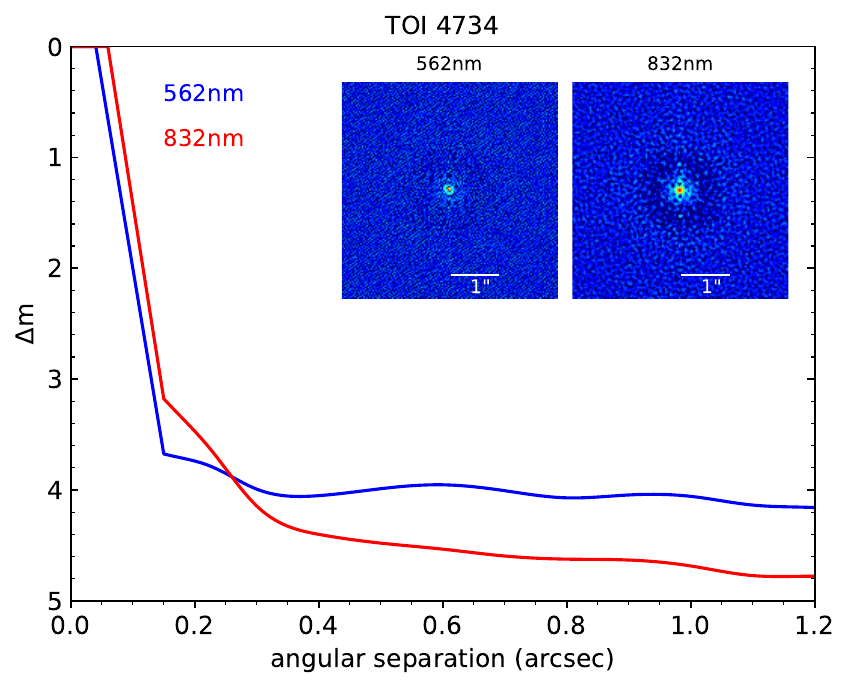}
\includegraphics[width=0.32\linewidth]{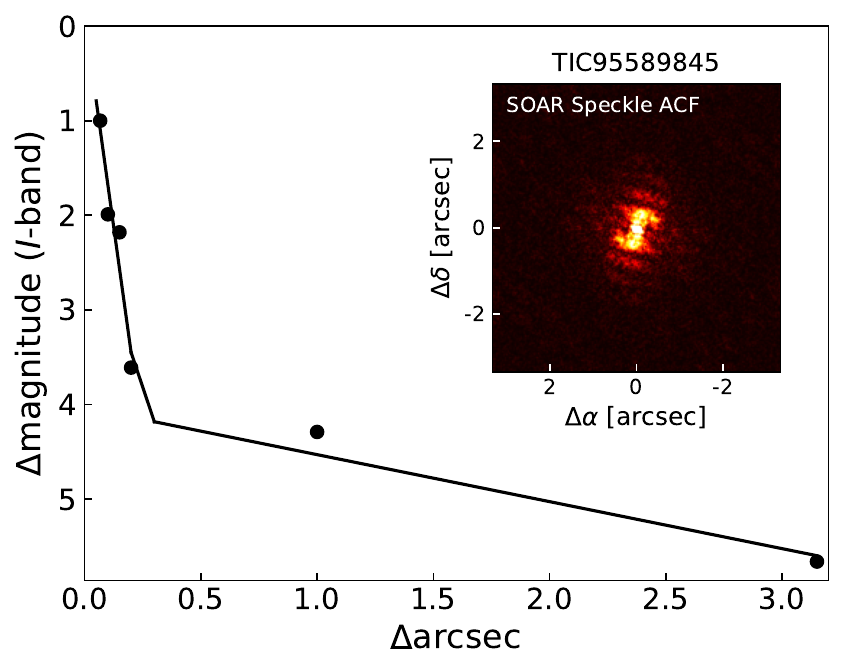}\\
\includegraphics[width=0.32\linewidth]{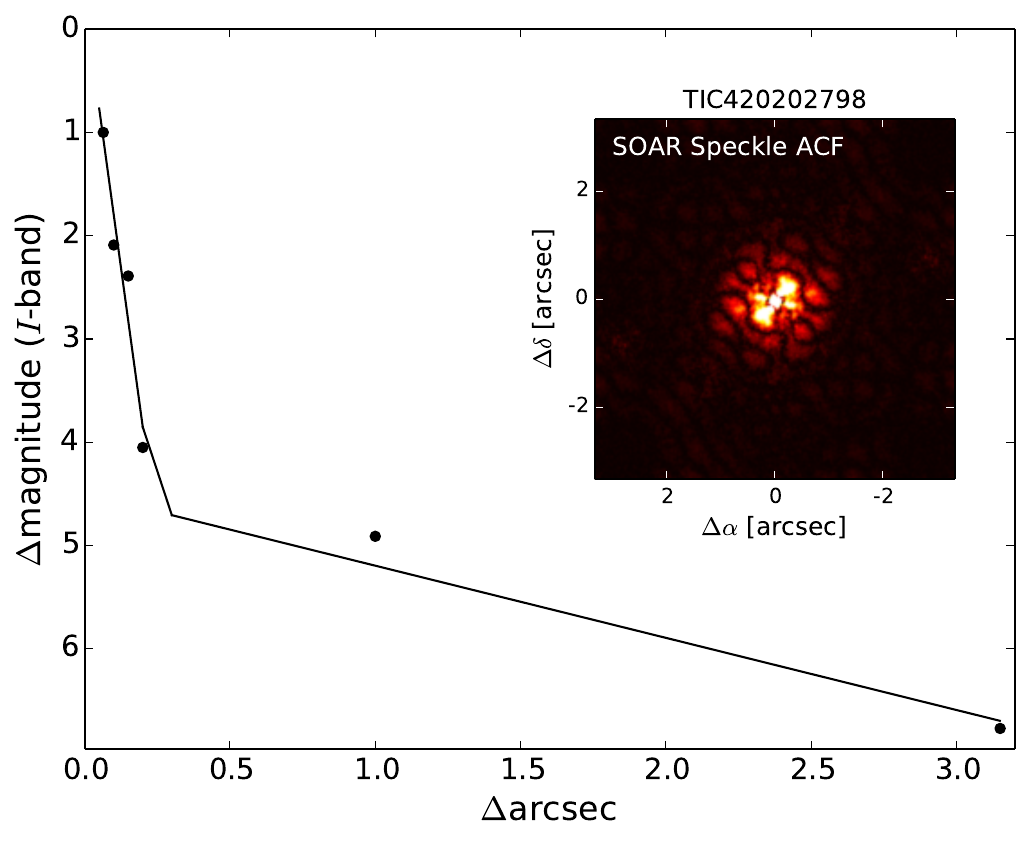}
\includegraphics[width=0.32\linewidth]{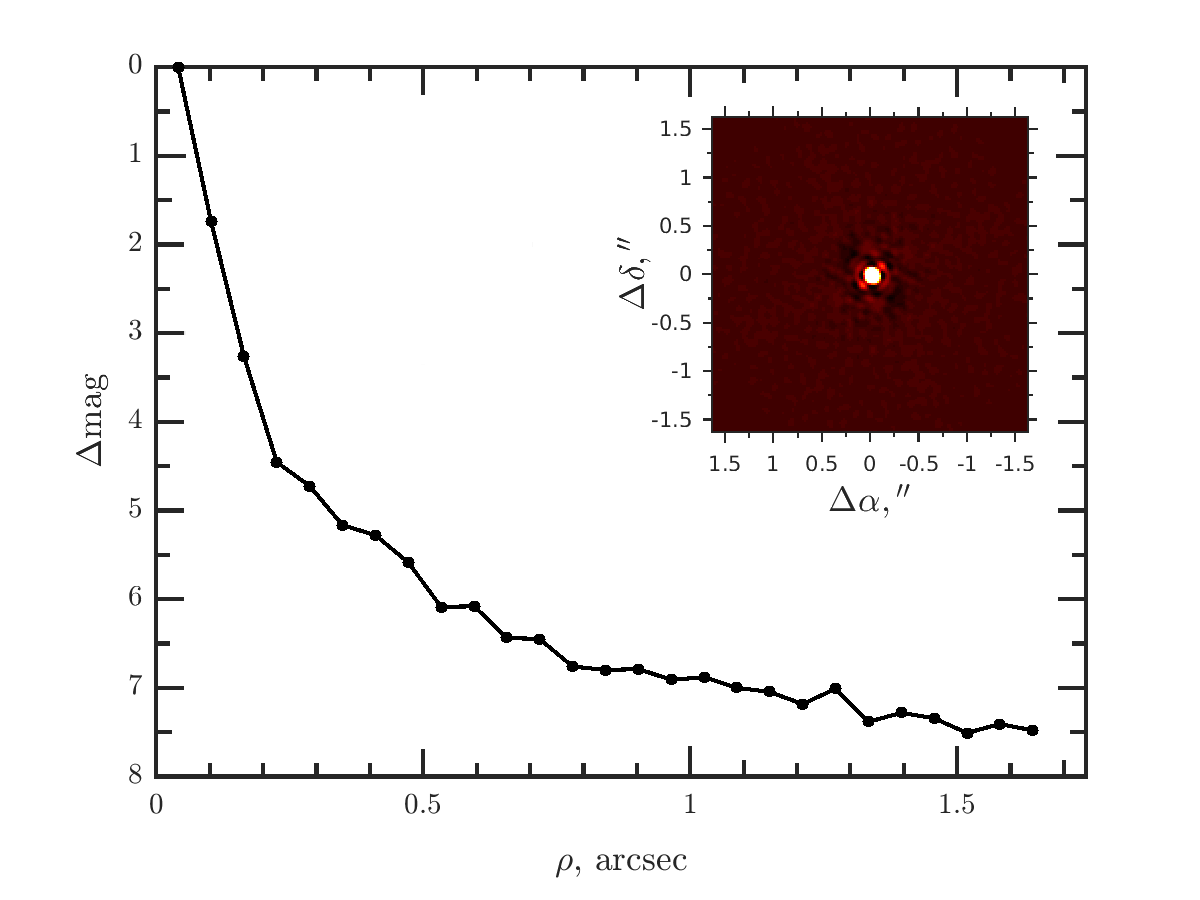}
\includegraphics[width=0.32\linewidth]{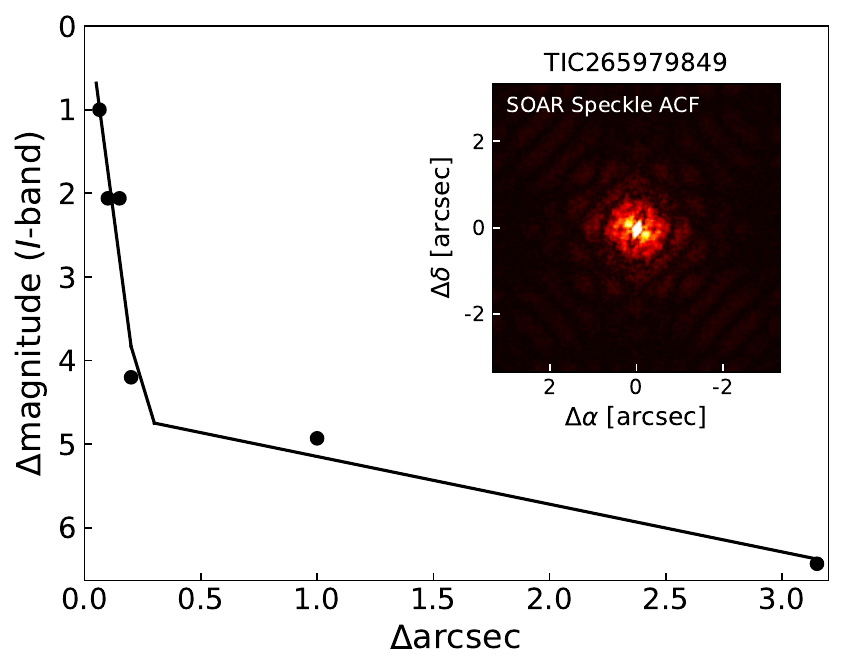}\\
\includegraphics[width=0.32\linewidth]{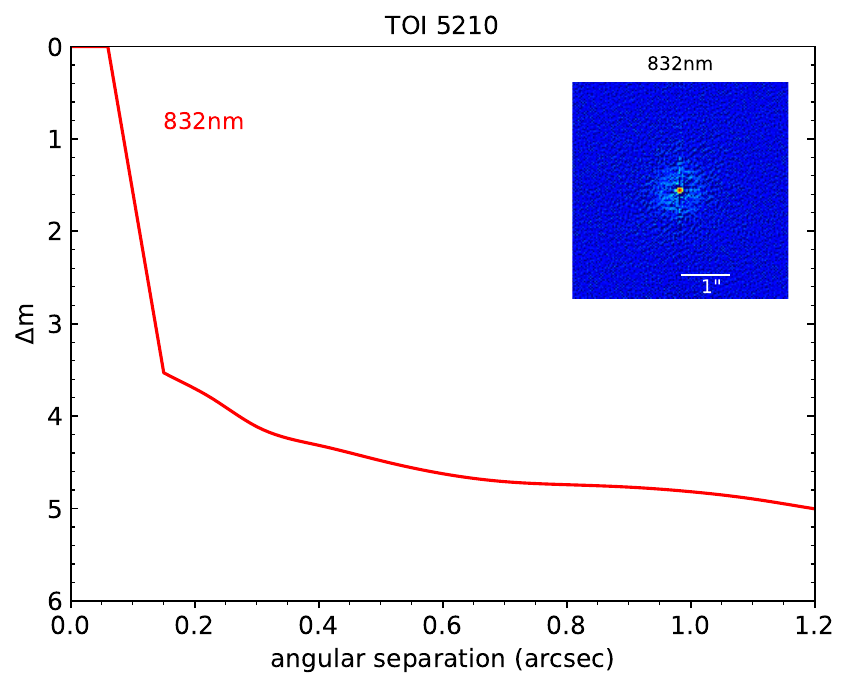}
\includegraphics[width=0.32\linewidth]{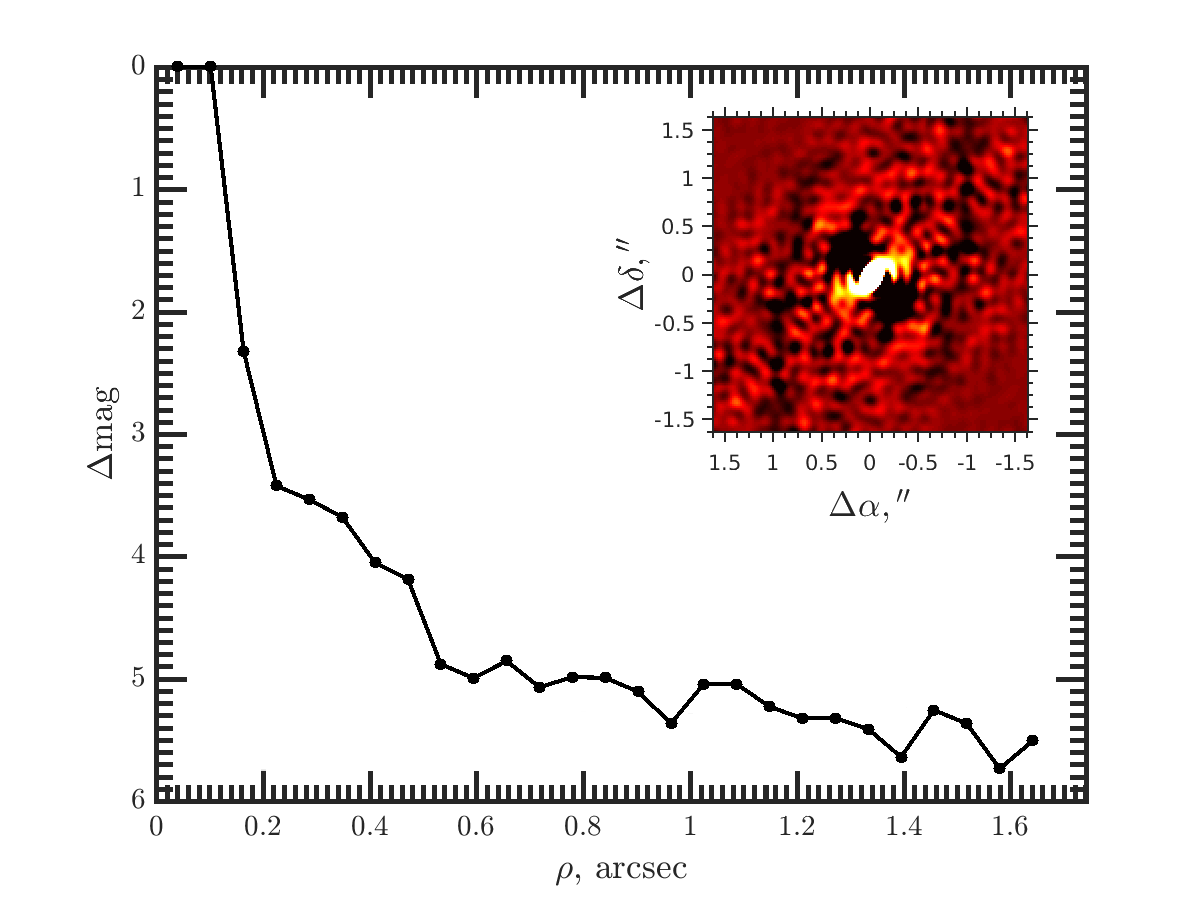}
\includegraphics[width=0.32\linewidth]{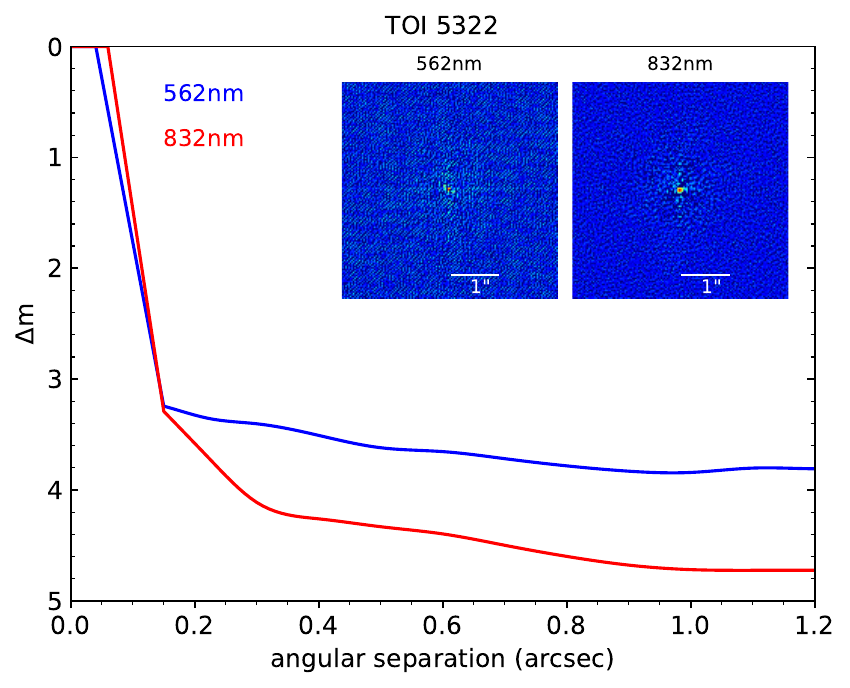}\\
\caption{High-Resolution imaging of hot Jupiter hosts described in this paper (continued).
From top to bottom, left to right:
\textbf{Row 1:} WIYN/NESSI observation of TOI-4487; 
Gemini-S/Zorro observations of TOI-4734; 
SAI/Speckle Polarimeter observation of TOI-4734; 
\textbf{Row 2:} WIYN/NESSI observations of TOI-4734; 
WIYN/NESSI observations of TOI-4734; 
SOAR/HRCam observation of TOI-4794; 
\textbf{Row 3:} SOAR/HRCam observation of TOI-4961; 
SAI/Speckle Polarimeter observation of TOI-5210; 
SOAR/HRCam observation of TOI-5210; 
\textbf{Row 4:} WIYN/NESSI observation of TOI-5210; 
SAI/Speckle Polarimeter observation of TOI-5322; 
WIYN/NESSI observations of TOI-5322.
}
\label{fig:high_res_imaging_3}
\end{figure*}
\makeatletter\onecolumngrid@pop\makeatother
\clearpage

\makeatletter\onecolumngrid@push\makeatother
\begin{figure*}
\centering

\includegraphics[width=0.32\linewidth]{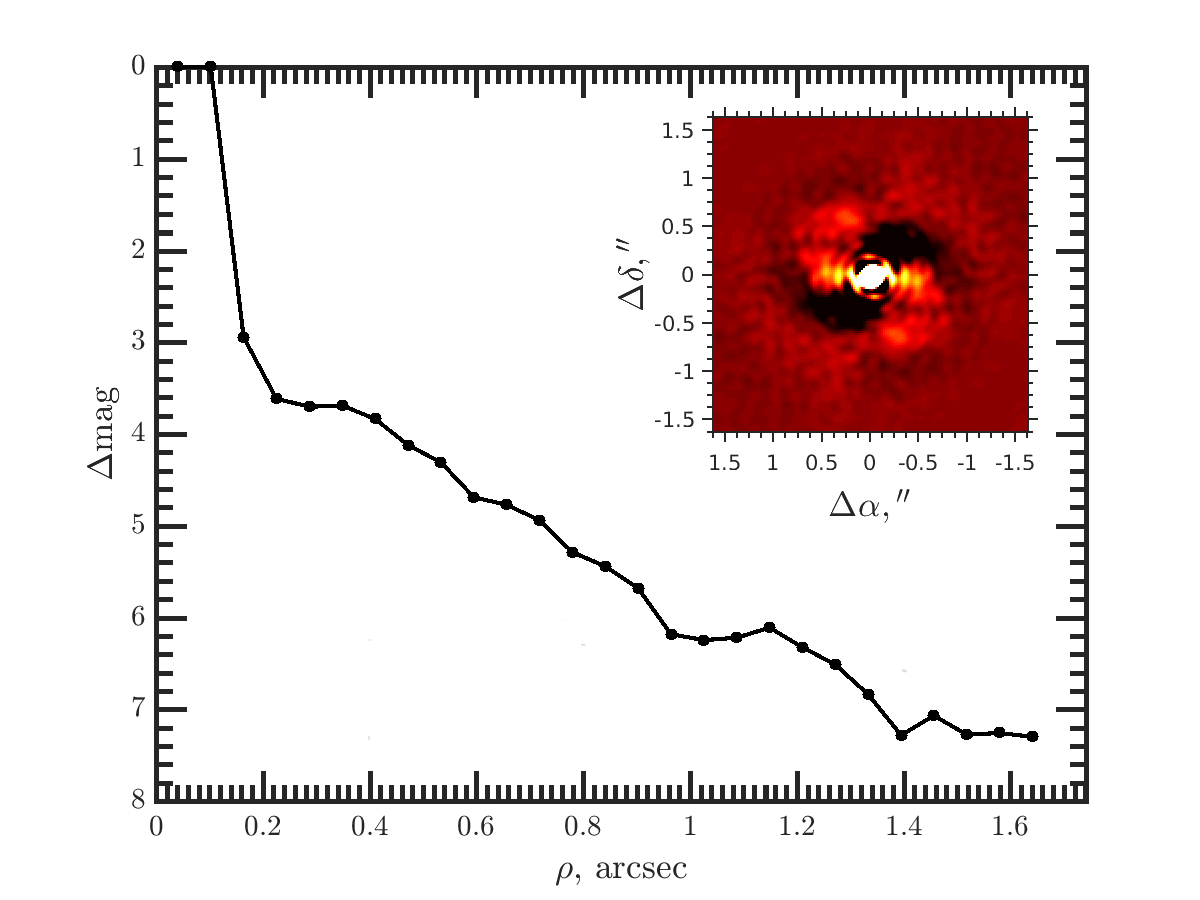}
\includegraphics[width=0.32\linewidth]{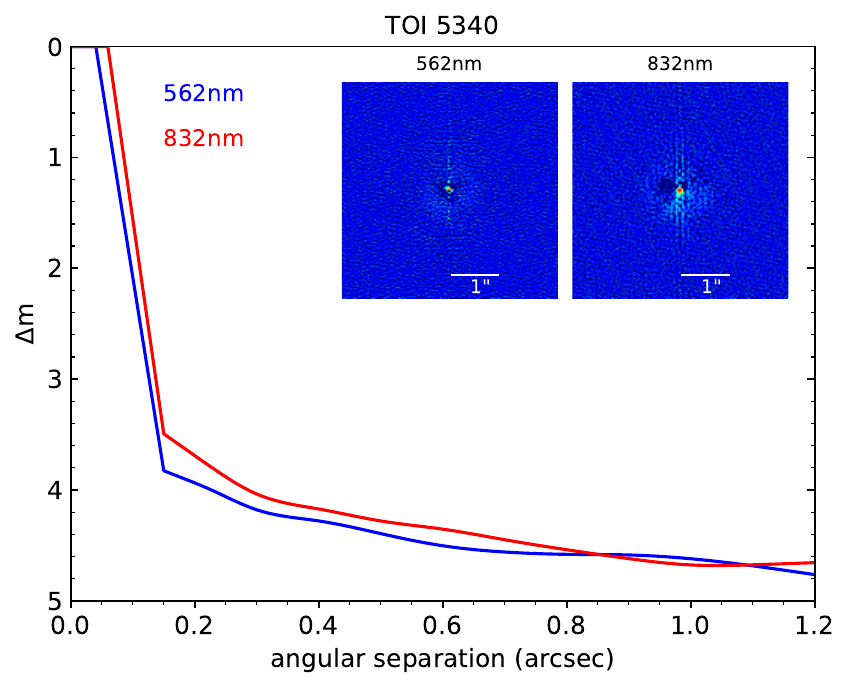}
\includegraphics[width=0.32\linewidth]{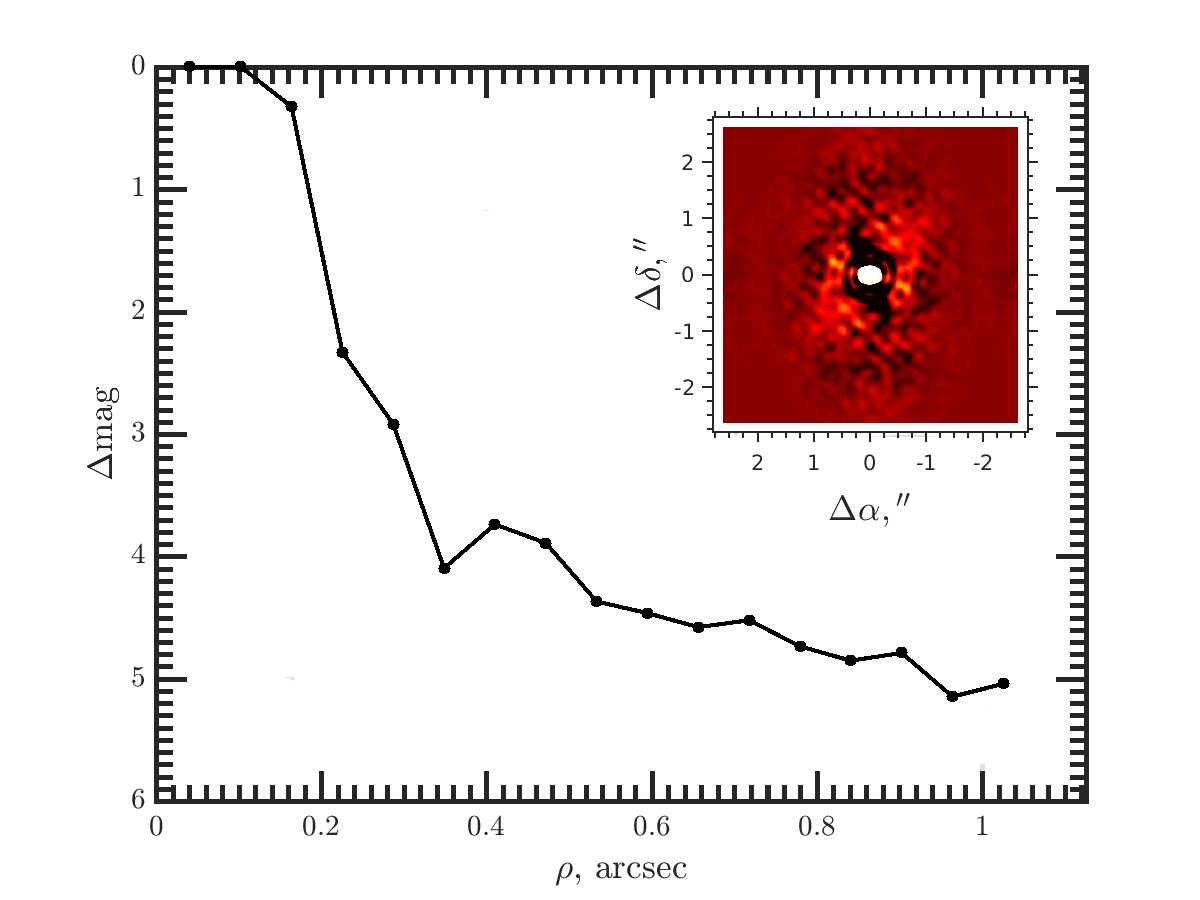}\\
\caption{High-Resolution imaging of hot Jupiter hosts described in this paper (continued).
From top to bottom, left to right:
\textbf{Row 1:} SAI/Speckle Polarimeter observation of TOI-5340; 
WIYN/NESSI observations of TOI-5340; 
SAI/Speckle Polarimeter observation of TOI-5592.
}
\label{fig:high_res_imaging_4}
\end{figure*}
\makeatletter\onecolumngrid@pop\makeatother
\clearpage

\end{subfigures}

\fi

\bibliography{manuscript,instruments,software,catalogs}
\bibliographystyle{aasjournal}

\end{document}